\documentclass[numberedappendix]{emulateapj}
\usepackage{apjfonts}
\usepackage{amsmath}
\usepackage{multirow}
\usepackage{array}
\usepackage[usenames]{color}
\allowdisplaybreaks[1]
\begin{document}
\newcommand{\comment}[1]{}
\newcommand{\risa}[1]{\textcolor{red}{(\bf #1)}}
\definecolor{purple}{RGB}{160,32,240}
\newcommand{\peter}[1]{}
\newcommand{\macc}{M_\mathrm{acc}}
\newcommand{\mpeak}{M_\mathrm{peak}}
\newcommand{\mnow}{M_\mathrm{now}}
\newcommand{\vacc}{v^\mathrm{acc}_\mathrm{max}}
\newcommand{\vpeak}{v^\mathrm{peak}_\mathrm{max}}
\newcommand{\vnow}{v^\mathrm{now}_\mathrm{max}}

\newcommand{\na}{NA}
\newcommand{\hinv}{h^{-1}}
\newcommand{\mpc}{\rm{Mpc}}
\newcommand{\hmpc}{$\hinv\mpc$}

\newcommand{\Msun}{M_{\odot}}

\shortauthors{BEHROOZI ET AL}
\shorttitle{Average Star Formation Histories From $Z$=0-8}

\title{The average star formation histories of galaxies in dark matter halos from  $z=0-8$}

\author{Peter S. Behroozi, Risa H. Wechsler}
\affil{Kavli Institute for Particle Astrophysics and Cosmology; Physics Department, Stanford University; 
Department of Particle Physics and Astrophysics, SLAC National  Accelerator Laboratory; 
Stanford, CA 94305}
\author{Charlie Conroy}
\affil{Department of Astronomy \& Astrophysics, University of California at Santa Cruz, Santa Cruz, CA 95064}

\begin{abstract}
We present a robust method to constrain average galaxy star formation rates, star formation histories, and the intracluster light as a function of halo mass.  Our results are consistent with observed galaxy stellar mass functions, specific star formation rates, and cosmic star formation rates from $z=0$ to $z=8$.  We consider the effects of a wide range of uncertainties on our results, including those affecting stellar masses, star formation rates, and the halo mass function at the heart of our analysis.  As they are relevant to our method, we also present new calibrations of the dark matter halo mass function, halo mass accretion histories, and halo-subhalo merger rates out to $z=8$.  We also provide new compilations of cosmic and specific star formation rates; more recent measurements are now consistent with the buildup of the cosmic stellar mass density at all redshifts.  Implications of our work include: halos near $10^{12}\Msun$ are the most efficient at forming stars at all redshifts, the baryon conversion efficiency of massive halos drops markedly after $z\sim 2.5$ (consistent with theories of cold-mode accretion), the ICL for massive galaxies is expected to be significant out to at least $z\sim 1-1.5$, and dwarf galaxies at low redshifts have higher stellar mass to halo mass ratios than previous expectations and form later than in most theoretical models.   Finally, we provide new fitting formulae for star formation histories that are more accurate than the standard declining tau model.  Our approach places a wide variety of observations relating to the star formation history of galaxies into a self-consistent framework based on the modern understanding of structure formation in $\Lambda$CDM.  Constraints on the stellar mass---halo mass relationship and star formation rates are available for download online.
\end{abstract}

\newcommand{\Mnfw}{M_\mathrm{NFW}}
\newcommand{\mvir}{M_\mathrm{vir}}
\newcommand{\rvir}{R_\mathrm{vir}}
\newcommand{\vmax}{v_\mathrm{max}}
\newcommand{\vmac}{v_\mathrm{max}^\mathrm{acc}}
\newcommand{\mvac}{M_\mathrm{vir}^\mathrm{acc}}
\newcommand{\sfr}{\mathrm{SFR}}
\newcommand{\plotgrace}[1]{\includegraphics[angle=-90,width=\columnwidth,type=eps,ext=.eps,read=.eps]{#1}}
\newcommand{\plotgraceflip}[1]{\includegraphics[angle=-90,width=\columnwidth,type=eps,ext=.eps,read=.eps]{#1}}
\newcommand{\plotlargegrace}[1]{\includegraphics[angle=-90,width=2\columnwidth,type=eps,ext=.eps,read=.eps]{#1}}
\newcommand{\plotlargegraceflip}[1]{\includegraphics[angle=-90,width=2\columnwidth,type=eps,ext=.eps,read=.eps]{#1}}
\newcommand{\plotminigrace}[1]{\includegraphics[angle=-90,width=0.5\columnwidth,type=eps,ext=.eps,read=.eps]{#1}}
\newcommand{\plotmicrograce}[1]{\includegraphics[angle=-90,width=0.25\columnwidth,type=eps,ext=.eps,read=.eps]{#1}}
\newcommand{\plotsmallgrace}[1]{\includegraphics[angle=-90,width=0.66\columnwidth,type=eps,ext=.eps,read=.eps]{#1}}

\section{Introduction}

Constraining the buildup of stellar mass in galaxies provides fundamental constraints on galaxy formation models.  An ever growing number of galaxy observations have been made, covering a time period from 500 Myr after the Big Bang \citep[e.g.][]{Zheng12} to the present day.  A model of galaxy formation that can match observations over this entire stretch of time would represent a significant aid to our understanding.  So far, matching the evolution of the stellar masses and star formation rates of galaxies over this entire epoch with has proved difficult \citep[see, e.g.,][]{Lu12,Borgani11, Weinmann12}.  Despite challenges of reproducing these observations with detailed models of galaxy formation, significant progress has been made in recent years with empirical models that connect the evolution of galaxy properties to the evolution of dark matter halos.

Over the past decade, a range of studies have associated galaxies with dark matter halos at a given epoch, using a variety of techniques, including Halo Occupation Distribution modeling \citep[e.g.,][]{Berlind02, Bullock02} the Conditional Luminosity Function modeling \citep[e.g.,][]{Yang03}, and variants of the abundance matching technique \citep[e.g.,][]{Colin99, kravtsov_klypin:99, Neyrinck04, Kravtsov04, Vale04, Vale06, Tasitsiomi04, conroy:06, Shankar06, Berrier06, Marin08, Guo-09, moster-09, Behroozi10}.  The simplest abundance matching models assign the most massive observed galaxies in rank order to the largest halos in an equal simulation volume.  With only slight modifications (e.g., using the peak mass for satellite halos), this technique accurately reproduces the redshift-- and scale--dependent clustering of galaxies \citep[e.g.,][]{conroy:06,Reddick12}.

Because the cosmological model provides a prediction for the buildup of dark matter halos, one can combine knowledge of the galaxy-halo connection at different epochs with knowledge of the mass accretion and merger histories of halos to constrain the buildup of stellar mass in galaxies over cosmic time.  This was first done in a comprehensive way by \cite{cw-08}, who provided an empirical constraint on the star formation histories and stellar mass growth in galaxies from $z=2$ to the present.  This approach has been explored in several studies using variants of the techniques \citep{zheng-07,White07,Firmani10,Leitner11,Bethermin12,Wang12, Moster12}, which together provide important constraints on the basic picture for the buildup of stars in galaxies and their connection to dark matter halos.

These studies represented important advances, albeit with shortcomings.  For example, most of these studies only modeled galaxy evolution from $z\sim 2$ to the present due to perceived conflicts between the integrated cosmic star formation and the cosmic stellar mass density at $z>1$ \citep{Wilkins08,Hopkins06b}.  In addition, because passive stellar mass loss (e.g., supernovae of massive stars) depends on galaxies' star formation histories, these studies had to make assumptions about the ages of stars already formed at $z>2$.  Finally, these studies paid limited attention to the vast array of observational uncertainties as well as modeling uncertainties affecting their derived constraints; indeed, many results are presented without error bars \citep{cw-08,Firmani10,Bethermin12,Wang12}.

Here we present a new technique and a new compilation of observations, aimed squarely at resolving these issues.  On the observational side, new observations of galaxies at very high redshifts \citep[e.g.][]{Bouwens11,BORG12,Bouwens11b,McLure11} have made it possible to place constraints on galaxy formation all the way to $z\sim8$.  At lower redshifts, we present the first constraints from stellar mass functions based on the PRIMUS survey, as well as from a new analysis at $z=0$ based on SDSS and GALEX data \citep{Moustakas12}. These are obtained from a consistent methodology from $z=0$ to $z=1$, and thus alleviate many concerns about matching inconsistently-identified galaxies at different epochs.  Using these combined data sets, we find that observations of the evolution of galaxy star formation rates and stellar masses can now be reconciled (see also \citealt{Bernardi10} and \citealt{Moster12}).

Our new method constrains the galaxy--halo relation using observed galaxy stellar mass functions \textit{as well as} specific star formation rates and cosmic star formation rates.  As with previous studies, we match observed galaxies to halos, but the additional information on star formation rates allows us to break degeneracies and directly constrain the buildup of the intracluster light, as well as the amount of stars that can transfer from satellites to the central galaxy during mergers.  We account for a number of statistical and systematic effects, including uncertainties from stellar population synthesis models, dust models, star formation history models, the faint-end slope of the stellar mass function, observational completeness, scatter between stellar mass and halo mass.

Given a parametrization for the intrinsic stellar mass---halo mass relationship as well as for these observational effects, we use a Markov Chain Monte Carlo (MCMC) method to find the allowed posterior distribution.  This is important not only for addressing concerns that our results may be biased by limited observational constraints at high redshifts, but it also gives us the power to determine which observations would best improve the resulting constraints on the relationship between stellar mass, star formation, and halo mass. The direct results of this analysis are constraints on the distribution of galaxy stellar masses as a function of halo mass and redshift.  From these, many other constraints relevant to galaxy formation are derived, including the average star formation rate as a function of halo mass and redshift, the average star formation history in galaxies at a given epoch as a function of galaxy stellar mass or host halo mass, the instantaneous baryon conversion efficiency of galaxies as a function of mass and redshift, the buildup of the intracluster light, and the evolution of the stellar mass to halo mass ratio for progenitors of today's galaxies, all including full uncertainties and covering a redshift range from $z=0$ to $z=8$.

We provide a broad overview of our methodology in \S \ref{s:methodology}, including our parametrization of the stellar mass -- halo mass relation and the relevant uncertainties, with additional details in Appendices \ref{a:av_afh}-\ref{a:mergers}.  We discuss the observational data sets relevant to our method in \S \ref{s:data}, with special attention to new measurements of cosmic star formation rates.  We discuss the simulations that we use in \S \ref{s:sim}, along with recalibrations of the halo mass function (Appendix \ref{a:tinker}), halo mass accretion rates (Appendix \ref{a:mah}), and subhalo merger/disruption rates (Appendix \ref{a:disruption}). We present our main results in \S \ref{s:results}, with discussion in \S \ref{s:discussion} and a summary of our conclusions in \S \ref{s:conclusions}.

Throughout this work, we assume a \cite{chabrier-2003-115} initial mass function, the stellar population synthesis model of \cite{bc-03}, and the dust model employed in \cite{blanton-roweis-07}.  We convert data sets from other papers to these models as necessary.  We additionally assume a flat, $\Lambda$CDM cosmology with parameters $\Omega_M = 0.27$, $\Omega_\Lambda = 0.73$, $h=0.7$, $n_s = 0.95$, and $\sigma_8 = 0.82$.

\section{Methodology}

\label{s:methodology}

\subsection{Overview}

Much previous work has gone into determining the relation between halo mass and stellar mass as a function of redshift (e.g., \citealt{Moster12,Yang11,Leauthaud11,Behroozi10,moster-09,Guo-09,cw-08,Wang09,zheng-07, mandelbaum-06,Gavazzi07,yang-08,Hansen09,LinMohr04}).  The redshift evolution of this relation is due to three contributing physical effects for stars in galaxies (new star formation, merging satellite galaxies, and stellar mass loss) and one effect for halos (continuing mass accretion).  Merging galaxies and mass accretion are well-constrained by dark matter simulations, and stellar mass loss is well-constrained with the assumption of an initial mass function (IMF) for stars.  The most uncertain of these effects, in relation to dark matter halos, are new star formation and stars ejected during galaxy mergers.  With observational constraints on the redshift evolution of the stellar mass and the star formation rate as well as computed constraints on satellite mass loss, we may effectively constrain both of these factors.

Schematically, the star formation rate for a given galaxy over a given timestep is constrained by:
\begin{eqnarray}
SFR \cdot \Delta t & = & \textrm{Expected Stellar Mass Now}\nonumber\\
& - & \textrm{Remaining Stellar Mass from Previous Timestep}\nonumber\\
& - & (\textrm{Stellar Mass from Mergers})\nonumber\\
&&\times(1 - \textrm{Fraction Ejected})\label{e:sfr_schem}.
\end{eqnarray}

In this way, we can combine a specific assignment of stellar mass to halos with the evolution of the halo mass function 
over time, to derive star formation rates for halo trajectories.  Our approach is to flexibly parametrize the possibilities for the stellar mass --- halo mass relation, $M_\ast(M_h,z)$, as well as the uncertainties affecting the remaining terms in Eq.\ \ref{e:sfr_schem}.  Then, using a Markov Chain Monte Carlo method, we can determine the allowed parameter space for $M_\ast(M_h,z)$ by comparing the implied SFRs and stellar mass abundances to observations across a wide range of redshifts.

Parametrizing $M_\ast(M_h,z)$ separately for individual halos is beyond the scope of this work (although it will be explored in future papers).  Here, we parametrize the median value of $M_\ast$ for halos at a given mass and redshift (i.e., a $M_\ast(M_h,z)$) as well as the scatter in stellar mass at fixed halo mass as a function of redshift.  By taking a summation of Eq.\ \ref{e:sfr_schem} over halos in a specific mass and redshift bin, we can derive the total amount of stars formed in that bin.  Thus, dividing by the number of halos in that bin gives the \textit{average} star formation rate, which is similar to approaches taken in previous works \citep{cw-08,Leitner11,Wang12,Moster12}.  This approach is accurate except in terms of small second-order effects that come from different stellar mass histories having a range of stellar mass loss rates.  This aspect is discussed in detail in Appendix \ref{a:av_afh}.  Further discussion of how individual galaxy star formation histories can differ from the average is presented in \S \ref{s:new_fits}.

In this section, we present our method for parametrizing the stellar mass -- halo mass relation (\S \ref{s:smhm}), our parametrization of the uncertainties affecting observational data (\S \ref{s:obs_syst}), our model for stellar mass loss (\S \ref{s:smhist}), and our model for ejection of stars into the ICL (\S \ref{s:icl}). To streamline the presentation of this section, we present the methodology for incorporating the effects of stellar mass accreted in mergers (which involves calibrating merger rates from simulations to $z=8$) and for calculating observables (such as the stellar mass function and specific star formation rates) to Appendices \ref{a:mergers} and \ref{a:observables}.  We present a summary of the methodology and details of the Markov Chain Monte Carlo method in \S \ref{s:methodology_summary}.

\subsection{Determining the Stellar Mass -- Halo Mass Relation to $z=8$}

\subsubsection{Intrinsic Relation}

\label{s:smhm}

As in \cite{Behroozi10}, we parametrize the stellar mass -- halo mass (SMHM) relation at a given epoch. The most commonly-used function is a double power law \citep{Yang11,Moster12}, which has a characteristic halo mass ($M_1$) and stellar mass ($M_{\ast,0}$), a low-mass slope ($\alpha$), and a high-mass slope ($\beta$).  Because we include data down to low stellar masses ($M_\ast \sim 10^{7.25} \Msun$) and have tight observational constraints from the SDSS, we find that a double power-law cannot accurately fit the unique shape of the stellar mass function (SMF).  As shown in Appendix \ref{a:dp_law}, a double power-law results in a stellar mass function off by as much as 0.1 dex at $z=0$.  For this reason, we choose a different, five-parameter form as the fitting function for the SMHM relation.  The chosen form retains the low-mass power-law slope ($\alpha$) as well as a characteristic stellar ($M_{\ast,0}$) and halo mass ($M_1$).  The high-mass behavior is trickier to fit.  We find the best match with a subpower (superlogarithm) function with index $\gamma$:
\begin{equation}
\log_{10}(M_\ast(M_h \to \infty)) \propto \left( \log_{10}\left(\frac{M_h}{M_1}\right)\right)^\gamma.
\end{equation}
If $\gamma$ is 1, this equation becomes an ordinary power law.  As $\gamma$ approaches $0$, this equation approaches a logarithm.  In between, this equation grows more slowly than any power law, but faster than any logarithm.  While this equation can become multiply-valued for $M_h < M_1$, it is straightforward to mitigate this behavior as well as allowing a smooth connection to a power-law for $M_h < M_1$.  The specific parametrization we adopt is:\footnote{This relation gives the median stellar mass $M_\ast$ for halos of mass $M_h$.  Because of scatter in the SMHM relation, the inverse of this function does \textit{not} give the average halo mass for a given stellar mass.}
\begin{eqnarray}
\log_{10}(M_\ast(M_h)) & = & \log_{10}(\epsilon M_1) + f\left(\log_{10}\left(\frac{M_h}{M_1}\right)\right) - f(0) \label{e:cosmic_sfh} \nonumber\\
f(x) & = & -\log_{10}(10^{\alpha x} + 1) + \delta \frac{(\log_{10}(1+\exp(x)))^\gamma}{1+\exp(10^{-x})}.
\end{eqnarray}
This is a power law with slope $-\alpha$ for $M_h \ll M_1$ and a subpower law with index $\gamma$ for $M_h \gg M_1$. The characteristic stellar mass to halo mass ratio is $\epsilon$ at the characteristic halo mass $M_1$.  The maximum errors of this fit to the SMHM relation are about 0.025 dex, or roughly four times better than a double power-law fit, as discussed in Appendix \ref{a:dp_law}.

We use the virial mass (as defined in \citealt{mvir_conv}) to define the halo mass of central galaxies, and for satellites, we use the peak progenitor virial mass ($\mpeak$).  As this paper was being prepared, \cite{Reddick12} found that the peak $\vmax$ for halos is an even better proxy for stellar mass, at least at $z=0.05$, in agreement with physical expectations that the depth of the halo potential well before it is impacted by stripping should be most correlated with the galaxy mass.  The use of $\mpeak$ is at present easier both conceptually (to consider mass ratios between galaxies and halos) and operationally (because it allows the use of several previously calibrated relationships in the literature), so we stick with this choice in the present work for simplicity.  The use of $\mpeak$ may result in slight underestimates of the clustering and the satellite fraction (see discussion in \citealt{Reddick12} for a comparison of these quantities in the local universe).  However, \S \ref{s:comparison} demonstrates that the differences between our results and the best-fitting stellar mass--halo mass relation of \cite{Reddick12} are small, and we expect that they are within the current systematic errors.

The redshift scaling of the relation poses a unique challenge.  At $z=0$, the parameter fits must be flexible enough to match the tight constraints from SDSS observations.   At higher redshifts, the fit must be flexible enough to allow exploration of possible star formation histories allowed by higher uncertainties, but not so flexible that it allows significant over-fitting of the systematic and statistical errors in the data.  Moreover, the scaling must be physical: obviously, negative star formation rates cannot be allowed, nor can stellar mass to halo mass ratios that exceed the halo's baryon fraction.  These physical constraints help ensure basic sanity for the functional form where no observational data exists (for redshifts $z>8$, as well as for faint galaxies).

By necessity, the error bars we obtain will be sensitive to the choice of the redshift fit.  This is not only so for the variation allowed in the stellar mass histories, but it is also true for the variation allowed in the nuisance parameters for systematic errors.  Because the latter are often poorly constrained by available data, the flexibility of the redshift fit also influences how much of the available nuisance parameter space can be explored.  For this reason, we have tried a large number of different redshift scalings, including cubic spline interpolation with many control points (which results in over-fitting) and many different choices of scaling parameters with $a$ or $z$---many of which have unphysical behavior at very high redshifts.  We have also tried fitting directly to abundance matching results (Appendix \ref{a:ab_matching}), but no robust trends with redshift emerged.

The choice that we settled on has several conventions.  For many variables in Eq.\ \ref{e:cosmic_sfh}, we have three parameters: one each for the low-redshift value, one for the scaling at intermediate redshifts ($0.5\lesssim z \lesssim 2$), and one for scaling at high redshifts ($z \gtrsim 2$).  We also apply an exponential shutoff to the redshift scaling towards lower redshift to isolate the highly constrained low-redshift parameters from higher-redshift ones.  Our final parametrization is the following:

\begin{eqnarray}
\label{e:redshift_scaling}
\nu(a) & = & \exp(-4 a^2)\nonumber\\
\log_{10}(M_1) & = & M_{1,0} + (M_{1,a}(a-1) + M_{1,z}z)\nu \nonumber\\
\log_{10}(\epsilon) & = & \epsilon_0 + (\epsilon_a (a-1) + \epsilon_z z)\nu + \epsilon_{a,2} (a-1)\nonumber\\
\alpha & = & \alpha_0 + (\alpha_a (a-1))\nu\nonumber\\
\delta & = &  \delta_0 + (\delta_a (a-1) + \delta_{z} z)\nu\nonumber\\
\gamma & = &  \gamma_0 + (\gamma_a (a-1) + \gamma_{z} z)\nu.
\end{eqnarray}

Additionally, there will be scatter in stellar mass at fixed halo mass ($\xi$).  While current studies have indicated that this is approximately 0.16--0.2 dex at $z=0$ \citep{more-09,yang-09, Reddick12} with no evidence for a mass trend at least down to $10^{12} \Msun$ in halo mass \citep{Reddick12}, little is known about the redshift evolution of this scatter.  For this work, we parametrize the possible evolution of scatter with redshift via a two-parameter scaling:
\begin{equation}
\label{e:scatter}
\xi = \xi_0 + \xi_a (a-1),
\end{equation}
and we take the prior on $\xi_0$ to be $0.20$ dex $\pm 0.03$ dex, as determined by \cite{Reddick12}
using a combination of constraints from the correlation function and conditional stellar mass function.

\subsubsection{Observational Systematics}

\label{s:obs_syst}

A full discussion of the systematic uncertainties affecting stellar mass functions may be found in \cite{Behroozi10}.
These include uncertainties from the Initial Mass Function, the stellar population synthesis model, the dust model, the star formation history model, sample variance, Eddington bias, redshift errors, and magnification bias.  
Most of these effects result in a constant systematic offset ($\mu$) in stellar masses:
\begin{equation}
\label{e:syst}
\log_{10}\left(\frac{M_{\ast,\mathrm{meas}}}{M_{\ast,\mathrm{true}}}\right) = \mu.
\end{equation}
However, some effects can have different behavior as a function of stellar mass, largely on account of assumptions for passive galaxies (e.g., declining star formation rates and low dust fractions) not being true for active galaxies.  We account for these effects by introducing an additional parameter ($\kappa$) to capture offsets in the stellar masses of active galaxies:
\begin{equation}
\label{e:syst2}
\log_{10}\left(\frac{M_{\ast,\mathrm{meas,active}}}{M_{\ast,\mathrm{true,active}}}\right) = \mu + \kappa.
\end{equation}
For the fraction of galaxies that are quiescent, we use a fitting formula for recent measurements from \cite{Brammer11}, corrected to the stellar mass estimates we use in this paper:
\begin{equation}
\label{e:passive}
f_\mathrm{passive}(M_{\ast,\mathrm{meas}},z) = \left[\left(\frac{M_{\ast,\mathrm{meas}}}{10^{10.2+0.5z}\Msun}\right)^{-1.3}+1\right]^{-1}.
\end{equation}
At low redshifts, most galaxies below $10^{10.2}\Msun$ in stellar mass are active; beyond $z=3$, nearly all galaxies are considered active.

As in \cite{Behroozi10}, we do not model uncertainties in the IMF in this work.  The most significant remaining uncertainties come from the SPS model ($\sim0.1$dex), the dust model ($\sim0.1$dex), and the star formation history ($\sim0.2$dex).  By definition, passive galaxies have had no recent star formation, so that the star formation history uncertainties are substantially smaller.  For that reason, we take the priors on $\mu$ to be a log-normal distribution centered at zero with width 0.14 dex (combined SPS and dust model errors), but the priors on $\kappa$ to be of width 0.24 dex (all three sources combined).  The functional form of how $\mu$ and $\kappa$ evolve with redshift is unknown.  We therefore adopt the following fiducial formulae:
\begin{eqnarray}
\mu & = & \mu_0 + (a-1)\mu_a \label{e:syst_mu}\\
\kappa & = & \kappa_0 + (a-1)\kappa_a \label{e:syst_kappa}.
\end{eqnarray}
At higher redshifts ($z\sim2.3$), \cite{Muzzin09} suggest that uncertainties from SPS modeling become more important, on the order of 0.2 dex instead of 0.1 dex.  We thus set the widths of the priors on $\mu_a$ and $\kappa_a$ to be 0.22 dex and 0.30 dex, respectively, which take these increased uncertainties into account.

Because stellar populations cannot be fully constrained with limited photometric information, stellar mass estimates for individual galaxies have intrinsic scatter relative to the true galaxy stellar mass.  This causes an Eddington bias \citep{Eddington13,Eddington40} in the stellar mass function: on the high-mass end of the SMF, many low-mass galaxies are upscattered, but there are a limited number of higher-mass galaxies that can be downscattered.  This results in a net increase in the observed numbers of high stellar-mass galaxies.  This effect is best estimated for each redshift range and modeling technique individually; while few authors have corrected for this error, many at least provide an estimate of its effect.  As in \cite{Behroozi10}, we find that these estimates depend most significantly on the redshift.  We thus model the distribution in the observed stellar mass estimates (compared to the true stellar masses) as a log-normal Gaussian with mean 0 and a redshift-dependent standard deviation given by:
\begin{equation}
\label{e:psf}
\sigma(z) = \sigma_0 + \sigma_z z.
\end{equation}
Following \cite{Conroy09,Behroozi10}, we take $\sigma_0 = 0.07$ and we take the prior on $\sigma_z$ to be $\sigma_z = 0.04\pm0.015$, consistent with estimates from the literature \citep{Conroy09,Kajisawa09,perezgonzalez-2008,marchesini-2008,Stark09,Caputi11,Lee11,Marchesini10}.

Finally, there are some concerns that, due to the Lyman break techniques used to detect high-redshift galaxies, not all high-redshift galaxies may be detected.  If galaxies have bursty star formation, or if some fraction of star forming galaxies have extreme quantities of dust, then not all galaxies will be detected in Lyman break surveys \citep[see also][]{Stark09,Lee09}.  We parametrize the stellar mass completeness of high-redshift surveys with a two-parameter fit, governing the amplitude ($A$) and redshift onset ($z_c$) of incompleteness:
\begin{equation}
\label{e:completeness_bare}
c_i(z) = 1 - \frac{A}{\exp(z_c-z)+1}.
\end{equation}
Because there is little reason to believe that surveys are missing a significant fraction of galaxies at $z<1$, we require $z_c>0.8$, and we set the completeness fraction to 1 for $z<1$.  This amounts to adopting the following formula for the completeness:
\begin{equation}
\label{e:completeness}
c(z) = \left\{ \begin{matrix} 1 & \textrm{if $z<1$}\\ c_i(z) + (1-c_i(1)) & \textrm{if $z>1$} \end{matrix} \right.
\end{equation}

We parametrize the effect of this incompleteness on the observed SFRs by a single parameter $b$, which sets the fraction of ``bursty'' vs. ``dusty'' star formation.  Fully bursty star formation ($b=1$) would have no effect on the total observed cosmic SFR, but the observed SSFRs would be boosted by the incompleteness factor compared to the SSFR averaged over longer periods of time.  Fully dusty star formation ($b=0$) would lower the observed cosmic SFR by the same factor as the cosmic SM density.  Presumably, a moderate amount of unobscured star formation would be enough to render a high-redshift galaxy observable---hence, dustiness will impact galaxy completeness only if star formation is largely obscured.  Given that models for SFRs from observed galaxies already attempt to correct for dust obscuration, the main effect of dustiness on SSFRs is that the SSFRs for very dusty galaxies will not be observed.  Therefore, in our current model, we assume that this has no effect on the average SSFR.

\subsection{Star Formation Histories}

\label{s:smhist}

Under our assumption of a \cite{chabrier-2003-115} IMF and \cite{bc-03} stellar evolution tracks, we use the FSPS package \citep{conroy-09,Conroy10} to calculate the rate of stellar mass loss.  We find the fraction of mass lost from a single stellar population as a function of the time since its formation to be well-fit by the following formula:\footnote{This represents a corrected calibration from the fit in \cite{cw-08}.}
\begin{equation}
f_\mathrm{loss}(t) = 0.05 \ln\left(1 + \frac{t}{1.4\; \mathrm{Myr}}\right) \label{e:sm_loss}
\end{equation}

\subsection{Extragalactic Light and Merging Galaxies}

\label{s:icl}

From Eq.\ \ref{e:sfr_schem}, galaxies can build up stars either through mergers or internal star formation.  Schematically, this implies that the unknown amount of stars deposited in mergers can be derived given separate constraints on the stellar mass growth of galaxies (i.e., from stellar mass functions) and their star formation rates:
\begin{equation}
\Delta SM_\mathrm{deposited} = \Delta SM_\mathrm{central} - SFR_\mathrm{central} \Delta t,
\label{e:mergers_deposited}
\end{equation}
where $\Delta SM_\mathrm{deposited}$ is the amount of stellar mass deposited from merging satellites, $\Delta SM_\mathrm{central}$ is the stellar mass growth of the central galaxy, and $SFR_\mathrm{central}$ is the star formation rate of the central galaxy over the chosen period $\Delta t$.  (Note that the real equation is slightly more complicated due to stellar mass loss from passive evolution; see Appendix \ref{a:mergers}).

When a galaxy merger occurs, the stars associated with the satellite galaxy may either be deposited onto the central galaxy or be ejected into the intrahalo light (IHL), also called the intracluster light (ICL) for galaxy clusters.  Some care is necessary when using these terms, because there exist two separate definitions of the ICL/IHL from simulations (i.e., stars not bound to the main galaxy) and from observations (stars not counted in the light profile of the main galaxy).   In this work, we use ``intracluster light'' (and ``ICL'') to mean any remnant stars from past galaxy mergers which are not counted as part of the light profile from the main galaxy, regardless of their boundedness.  We do not make a distinction between IHL and ICL; we use one term (``ICL'') regardless of the size of the host halo, as this usage is by far more common in the literature.

Because we can constrain the total amount of stars in merging satellites through the halo merger rate (Appendix \ref{a:disruption}) and the stellar mass -- halo mass relation, we can also derive the amount of stellar mass deposited into the ICL ($\Delta ICL$):
\begin{equation}
\Delta ICL = \Delta SM_\mathrm{incoming} - \Delta SM_\mathrm{deposited},
\label{e:mergers_icl}
\end{equation}
where $\Delta SM_\mathrm{incoming}$ is the total amount of stars in satellites that merge within $\Delta t$.  Note that this quantity includes all stars ever ejected from merging galaxies.  Some fraction of these ejected stars may have speeds higher than the escape velocity of the surrounding dark matter halo and will be scattered to very large distances; however, this fraction is expected to be on the order of a few percent or less \citep{BehrooziUnbound}.

This simple picture is made more complicated by inconsistencies in how galaxies and the ICL are separated.  The total luminosity recovered and the fraction of it attributed to the galaxy are both dependent on surface-brightness limits, sky subtraction methods, and galaxy fitting methods, which are different for the different data sets we use and for different redshift
ranges in the data sets themselves.  Over most mass ranges and redshifts, this is not a problem: the rate of star formation in many galaxies results in an overall change in stellar mass between redshifts that is much larger than that attributable to a change in the luminosity modeling method.  However, for massive galaxies ($M_\ast > 10^{11}\Msun$) at low redshifts ($z<1$), this is a more significant issue.  For example, including 5\% more light between $z=0.3$ and $z=0.1$ in a massive galaxy (0.05 magnitudes) would result in an equal buildup in stellar mass as would a specific star formation rate of about 2.5$\times 10^{-11}$ yr$^{-1}$ (the expected SSFR for such galaxies; \citealt{Salim07}).  As the true SSFR drops to $10^{-12}$ yr$^{-1}$, the galaxy models must be self-consistent to about 0.2\% (0.002 magnitudes) to avoid introducing a comparable error, which is well beyond current calibration methods \citep{Bernardi10}.

This said, changes in the mass of the galaxy because of definitional issues are completely degenerate with incoming mass from mergers in Eqs.\ \ref{e:mergers_deposited} and \ref{e:mergers_icl}: $\Delta SM_\mathrm{deposited}$ is replaced by the sum of the definitional and deposited stellar mass changes ($\Delta SM_\mathrm{definitional} + \Delta SM_\mathrm{deposited}$) everywhere it appears.  Notably, Eq.\ \ref{e:mergers_icl} still gives us a way to robustly determine the buildup of stars in the ICL.

If we used Eqs.\ \ref{e:mergers_icl} and \ref{e:mergers_deposited} directly, the star formation rate would have to be parametrized separately from the stellar mass growth.  Instead, we rewrite Eq.\ \ref{e:mergers_deposited} in terms of the fraction of stellar mass growth that comes from the star formation rate:
\begin{eqnarray}
\Delta SM_\mathrm{deposited} + \Delta SM_\mathrm{definitional} & = & (1-f_{SFR}) \Delta SM_\mathrm{central}\\
f_{SFR} & = & \frac{SFR_\mathrm{central} \Delta t}{\Delta SM_\mathrm{central}}.
\label{e:f_sfr}
\end{eqnarray}
This is useful because there are strong physical priors on what $f_{SFR}$ can be.  Low-mass galaxies do not have enough incoming stellar mass in satellites to account for a significant part of their stellar mass growth, so $f_{SFR}$ asymptotes to 100\% for such galaxies.  On the other hand, the highest-mass galaxies experience almost no internal star formation \citep{Salim07}, meaning that $f_{SFR}$ must approach 0 for such galaxies.  So, we can approximate the halo mass dependence of $f_{SFR}$ using a double power law as
\begin{equation}
f_{SFR}(M_h) = \left[\left(\frac{M_h}{M_{h,ICL}}\right)^\beta + 1\right]^{-1}.
\end{equation}
We allow $M_{h,ICL}$ to be redshift-dependent using a two-parameter fit:
\begin{equation}
M_{h,ICL} = M_{h,ICL,0} + (a-1) M_{h,ICL,a}.
\end{equation}
Rather than add extra parameters for $\beta$, we make the assumption that $f_{SFR}(10^{16}\Msun)$ is 1\%.  This is equivalent to the constraint
\begin{equation}
\beta = \frac{\log_{10}(99)}{16 - \log_{10}(M_{h,ICL})}.
\end{equation}

\begin{figure*}[t]
\begin{center}
\includegraphics[width=1.4\columnwidth]{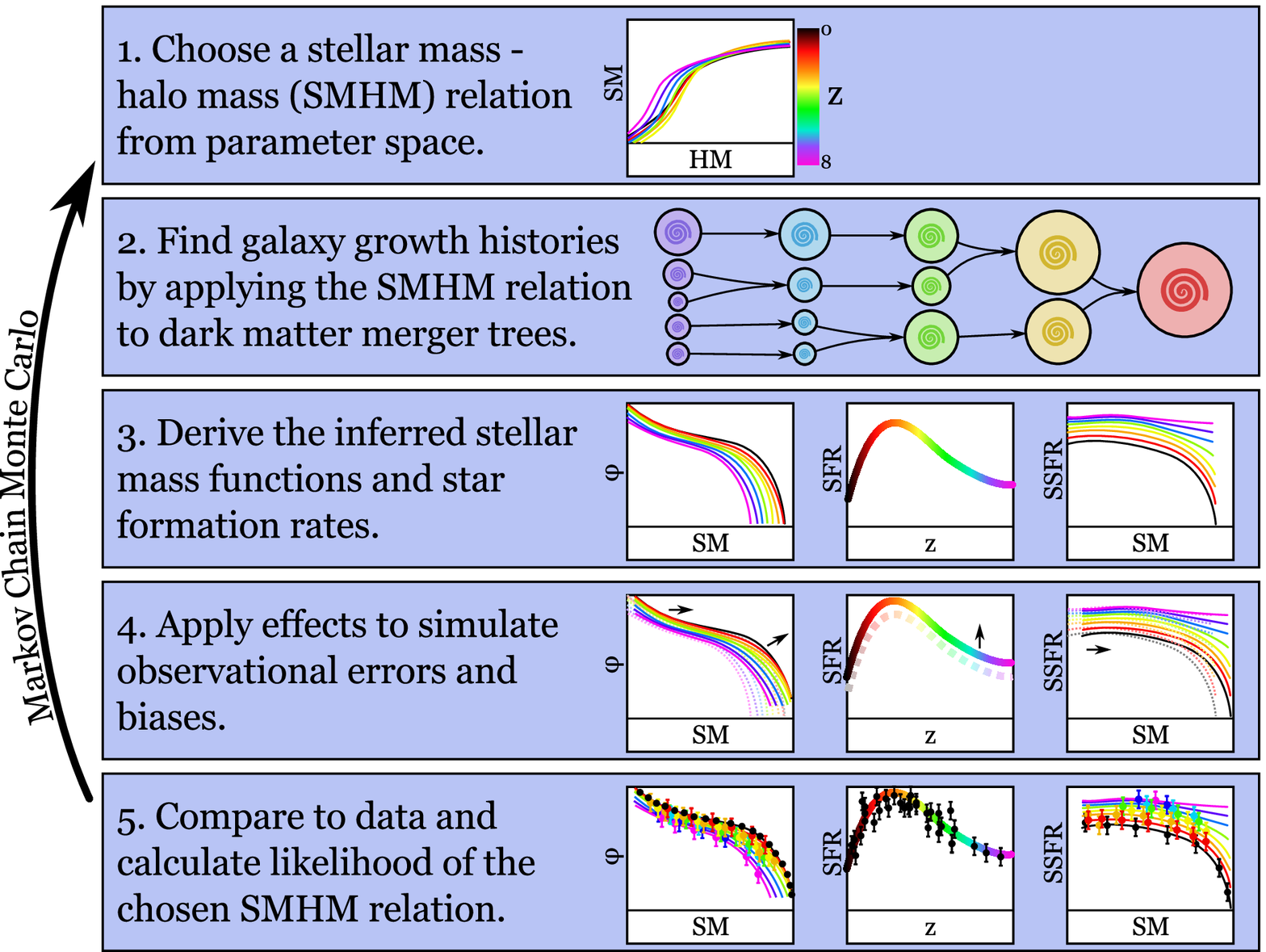}
\end{center}
\vspace{-2ex}
\caption{Visual summary of the methodology used to constrain the stellar mass -- halo mass relation.}
\label{f:methodology_summary}
\end{figure*}

\begin{table*}
\vspace{-4ex}
\begin{center}
\caption{Table of Parameters}
\label{t:parameters}
\begin{tabular}{rccc}
\hline
\hline
Symbol & Description & Equation & Section\\
\hline
$M_1$ & Characteristic halo mass & \ref{e:cosmic_sfh} & \ref{s:smhm}\\
$\epsilon$ & Characteristic stellar mass to halo mass ratio & \ref{e:cosmic_sfh} & \ref{s:smhm}\\
$\alpha$ & Faint-end slope of SMHM relation & \ref{e:cosmic_sfh} & \ref{s:smhm}\\
$\delta$ & Strength of subpower law at massive end of SMHM relation & \ref{e:cosmic_sfh}& \ref{s:smhm}\\
$\gamma$ & Index of subpower law at massive end of SMHM relation & \ref{e:cosmic_sfh}& \ref{s:smhm}\\
$\nu$ & Exponential cutoff of evolution of $M_\ast(M_h)$ with scale factor & \ref{e:redshift_scaling} & \ref{s:smhm}\\
$\xi$ & Scatter in dex of true stellar mass at fixed halo mass & \ref{e:scatter} & \ref{s:smhm}\\
$M_{h,ICL}$ & Characteristic halo mass at which half of stellar mass growth is due to mergers & \ref{e:f_sfr} & \ref{s:icl}\\
\hline
$\mu$ & Systematic offset in stellar masses and SFRs & \ref{e:syst} & \ref{s:obs_syst}\\
$\kappa$ & Systematic offset in stellar masses for active galaxies & \ref{e:syst2} & \ref{s:obs_syst}\\
$\sigma$ & Scatter in measured stellar mass at fixed true stellar mass & \ref{e:psf} & \ref{s:obs_syst}\\
$c$ & Galaxy detection completeness & \ref{e:completeness} & \ref{s:obs_syst}\\
$b$ & Fraction of incompleteness due to burstiness (as opposed to dustiness) & \ref{e:burstiness_correction} & \ref{s:obs_syst}\\
$\rho$ & Correlation of SFR to stellar mass at fixed halo mass & \ref{e:rho_def} & \ref{s:obs_ssfr_csfr}\\
\hline
\end{tabular}
\end{center}
\end{table*}

\begin{table*}
\begin{center}
\caption{Table of Priors}
\label{t:priors}
\begin{tabular}{rccc}
\hline
\hline
Symbol & Description & Equation & Prior\\
\hline
$\xi_0$ & Value of $\xi$ at $z=0$, in dex & \ref{e:scatter} & $G(0.16,0.04)$\\
$\xi_a$ & Redshift scaling of $\xi$, in dex & \ref{e:scatter} & $G(0,0.16)$\\
$\mu_0$ & Value of $\mu$ at $z=0$, in dex & \ref{e:syst} & $G(0,0.14)$\\
$\mu_a$ & Redshift scaling of $\mu$, in dex & \ref{e:syst_mu} & $G(0,0.22)$\\
$\kappa_0$ & Value of $\kappa$ at $z=0$, in dex & \ref{e:syst2} & $G(0,0.24)$\\
$\kappa_a$ & Redshift scaling of $\kappa$, in dex & \ref{e:syst_kappa} & $G(0,0.3)$\\
$\sigma_z$ & Redshift scaling of $\sigma$, in dex & \ref{e:psf} & $G(0.05,0.015)$\\
$A$ & Amplitude of galaxy detection completeness & \ref{e:completeness_bare} & $U(0,1)$\\
$z_c$ & Onset redshift of galaxy detection completeness & \ref{e:completeness_bare} & $z_c>0.8$\\
$b$ & Fraction of incompleteness due to burstiness (as opposed to dustiness) & - & $U(0,1)$\\
$\rho_{0.5}$ & Correlation of SFR to stellar mass at fixed halo mass at $a=0.5$ & \ref{e:rho_def} & $U(0.23,1.0)$\\
\hline
\end{tabular}
\end{center}
\tablecomments{$G(x,y)$ denotes a Gaussian distribution with center $x$ and width $y$.  $U(x_1,x_2)$ denotes a uniform distribution from $x_1$ to $x_2$.  The remaining parameters have no explicit priors; $M_1$ and $\epsilon$ are explored in logarithmic space, whereas $\alpha,\delta$, and $\gamma$ are explored in linear space.}
\end{table*}

\subsection{Methodology Summary}

\label{s:methodology_summary}

Although the equations involved are somewhat complicated, the logical steps involved in our method are straightforward and are shown visually in Fig.\ \ref{f:methodology_summary}.  These steps include:
\begin{enumerate}
\item Choose parameters for the stellar mass -- halo mass relation as well as observational and methodology uncertainties (see Table \ref{t:parameters} for full list; see Table \ref{t:priors} for adopted priors) using a Markov Chain Monte Carlo method.
\item Use the chosen stellar mass -- halo mass relation to populate halos with galaxies, and use merger rates and mass accretion histories calculated from dark matter simulations to infer galaxy growth rates.
\item From the previous step, calculate the average star formation rates as a function of halo mass and redshift, and use those to calculate the average specific star formation rate and the cosmic star formation rate.  In addition, using the abundances of halos, calculate the stellar mass function.
\item Apply corrections for observational errors and biases to the derived stellar mass function and star formation rates.
\item Compare with observational measures of the stellar mass function and star formation rates (i.e., sum $\chi^2$ errors for each data point) to calculate the likelihood that the chosen stellar mass -- halo mass relation matches observed results (i.e., $\exp(-0.5\chi^2)$).
\item Return to step \#1 until the MCMC algorithm has converged.
\end{enumerate}
To ensure convergence on a space with such a large number of parameters, we use the Adaptive Metropolis exploration method \citep{Haario01}.  Although the algorithm of \cite{dunkley-2005} indicates that we converge after only $5\times10^{5}$ points, we continue to run for $4\times10^6$ total points to minimize the chance of unexplored regions and to ensure that the adaptive updates of the step covariance matrix converge and do not bias our final results.

\section{Observational Constraints and Uncertainties}

\label{s:data}

We use three types of observational constraints on star formation in galaxies: the stellar mass function (SMF), the specific star formation rate of galaxies (SSFR, star formation rate per unit stellar mass), and the cosmic star formation rate (CSFR).  Stellar mass functions are discussed in \S \ref{s:sm_func}, cosmic star formation rates are discussed in \S \ref{s:ssfr}, and specific star formation rates are discussed in \S \ref{s:csfr}.

To ensure a fair comparison between the different data sets, we convert each measurement to the same set of systematic assumptions.  For the initial mass function (IMF), we convert all results to that of \cite{chabrier-2003-115}.  We do not consider evolving IMFs, nor do we consider the uncertainties for a non-universal IMF, as both uncertainties are beyond the scope of this paper.  For the stellar population synthesis and dust models, we convert all results to the models of \cite{bc-03} and \cite{blanton-roweis-07}.  In all cases, we convert authors' supplied units to physical units under the assumption that $h = 0.7$.

\subsection{Stellar Mass Functions}

\begin{table}
\begin{center}
\caption{Observational Constraints on the Stellar Mass Function}
\label{t:smf_data}
\begin{tabular}{rcccc}
\hline
\hline
Publication & Redshifts & Colors & Area & Notes\\
\hline
\cite{Baldry08} & 0.003-0.05 & \textit{ugriz} & 4783 deg$^2$ & VBD\\
\cite{Moustakas12} &  0.05-1 & UV-MIR & 9 deg$^2$ & D\\
\cite{perezgonzalez-2008} &  0.2-1.6 & UV-MIR & 0.184 deg$^2$ & ISD\\
\cite{Mortlock11}  & 1-3 & \textit{BVizH}$_{160}$ & 0.0125 deg$^2$ & VID\\ 
\cite{marchesini-2008} & 1.3-4 & \textit{B}-MIR & 0.142 deg$^2$ & ID\\
\cite{Marchesini10} & 3-4 & UV-MIR & 0.43 deg$^2$ & ID\\
\cite{Lee11b} & 4-5 & \textit{B}-MIR & 0.089 deg$^2$ & SD\\
\cite{Stark09} & 6 & \textit{B}-MIR & 0.089 deg$^2$ & SD\\
\cite{Bouwens11} & 7-8 & \textit{B-H}$_{160}$ & 0.0148 deg$^2$ & UD\\
\cite{BORG12} & 8 & \textit{MIR} & 0.076 deg$^2$ & UD\\
\hline
\end{tabular}
\end{center}
\tablecomments{Letters correspond to the following corrections made to the published results: I (Initial Mass Function), S (Stellar Population Synthesis model), D (Dust model), V (Sample Variance), B (Surface Brightness incompleteness), C (Cosmology, specifically $h$), U (UV to stellar mass conversion according to \citealt{Gonzalez10}).  The local results ($z<0.2$) in the \cite{Moustakas12} mass functions are taken from the Sloan Digital Sky Survey and cover an area of 2505 deg$^2$.}
\end{table}

\label{s:sm_func}

We combine many overlapping data sets \citep{Baldry08,Moustakas12,perezgonzalez-2008,Mortlock11,marchesini-2008,Stark09,Bouwens11} to constrain the evolution of the stellar mass function from $z=0$ to $z=8$.  These are shown in Table \ref{t:smf_data}.

Different observational challenges present themselves in each redshift range.  At low redshifts, the statistics are sufficient that systematic errors in calculating stellar masses are the largest source of uncertainty \citep{Behroozi10}.  At higher redshifts, this is no longer the case; statistical and sample variance errors can be of equal magnitude to estimated systematic errors.  Nonetheless, under the assumption of a smoothly varying stellar mass -- halo mass (SMHM) relation, our approach (which simultaneously constrains the SMHM relation against many stellar mass functions at different redshifts) can constrain the statistical errors better than the individual error bars on stellar mass functions might suggest.  We refer the interested reader to \cite{Behroozi10} for a discussion of the most common uncertainties affecting observation constraints on the stellar mass function and to \S \ref{s:smhm} for our parametrization of their effects.  In this section, therefore, we limit ourselves to discussing special concerns relevant to using stellar mass functions over this broad range of redshifts.

Questions about the accuracy of stellar mass estimates over this redshift range come from several previously published results that suggest that the evolution in the cosmic SFR density is inconsistent with the estimated cosmic stellar mass density at $z>1$ \citep{Nagamine06, Hopkins06b, perezgonzalez-2008, Wilkins08}.  One possible explanation is an evolving or a non-universal IMF; a number of different lines of evidence exist in support of this theory \citep[e.g.,][]{Dokkum10,Lucatello05, Tumlinson07a, Tumlinson07b, vanDokkum08}.  Another possible reason is that star formation at high redshift could be bursty or dust-obscured, yielding an incomplete census of galaxies in Lyman-break surveys \citep{Lee09,Stark09}.  On the other hand, \citet{Reddy09} offer a simpler explanation.  They appeal to luminosity--dependent reddening corrections in the ultraviolet luminosity functions at high redshift as well as steeper faint-end slopes for stellar mass functions, and demonstrate that the purported discrepancy then largely vanishes.

In our analysis of recent cosmic SFR literature (\S \ref{s:csfr}), we find that newer estimates of the cosmic SFR are systematically lower at $z>3$  than the data in \cite{Hopkins06b} (see Fig.\ \ref{f:csfr_comp}).  A similar conclusion is reached in \cite{Bernardi10} and \cite{Moster12}.  In addition, deep probes of the stellar mass function at lower redshifts ($z<1.5$) have resulted in findings of steep faint-end slopes \citep{Baldry08,Drory09,Mortlock11}.  We find this to be strong evidence in support of the explanation in \cite{Reddy09}.  Nonetheless, we also consider simple models for the effects of bursty or dusty modes of star formation; an analysis of models for an evolving or non-universal IMF is beyond the scope of this paper.

The highest redshifts present unique challenges.  Herein, for redshifts $7< z < 8.5$, we convert the UV luminosity function given in \cite{Bouwens11} to a stellar mass function according to the recipe in \cite{Gonzalez10}.  It has been suggested that this conversion is reasonable given the blue UV continuum slopes of such high-redshift galaxies, and follow-up observations from IRAC that imply that these early galaxies are nearly dust-free \citep{Labbe10}.  At the same time, concerns arise because the UV luminosity function presented in \cite{Bouwens11} is steep ($\alpha \approx -1.6$ to $-2.0$, instead of the shallower $\alpha \approx -1.5$ to $-1.7$ at lower redshifts).  One interpretation of this result is that assumptions for galaxy morphology (and, therefore, the ability to measure estimate the full UV flux or half-light radii) can bias the calculated UV luminosity function to have a steeper faint-end slope \citep{Grazian10}.  Indeed, any systematic effect that interferes more with brighter galaxies than with dimmer ones (e.g., dust) will tend to cause an overestimation of the slope of the UV luminosity function.  \cite{Gonzalez10} assume that luminosity-dependent corrections for the UV-to-stellar-mass conversion do not evolve substantially from $z=4$ to $z=7$.  This may in fact overestimate the dust corrections, an error that would \textit{lower} the faint-end slope, partially correcting for a failure to account for galaxy morphology.  In addition, recent work has shown that many estimates of stellar masses at $z>5$ may have been overestimated due to nebular line emission \citep{Stark12}; however, the conversions in \cite{Gonzalez10} do not share this problem (D.\ Stark, priv.\ comm.).  As the use of UV-only stellar masses therefore remains controversial, we present results both with and without the converted \cite{Bouwens11} stellar mass functions for $z=7-8$ in Appendix \ref{a:models}.

\subsection{Cosmic Star Formation Rates}

To assemble constraints on cosmic star formation rates, we conducted a comprehensive literature search on the astro-ph arXiv for papers posted within the past 6 years (2006-2012); details of the search are in Appendix \ref{a:csfr_ssfr_uncert}.  
The selected papers are summarized in Table \ref{t:csfr_data}.  Many different types of surveys were conducted, including estimates of the cosmic SFR from narrowband (e.g., H$\alpha$), broadband (UV-IR), and radio (1.4 GHz) surveys.

\begin{figure}
\vspace{-6ex}
\plotgrace{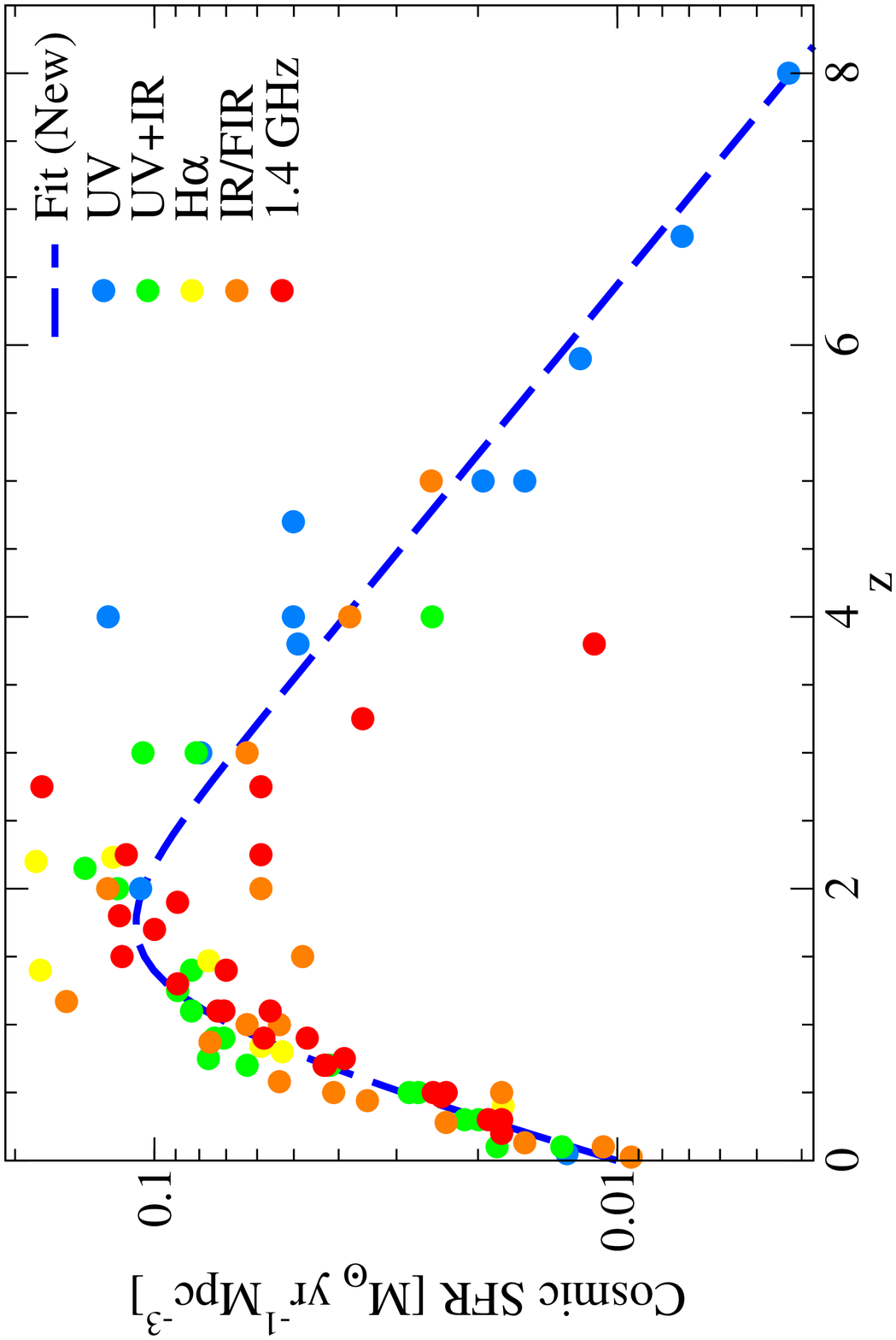}\\[-6ex]
\plotgrace{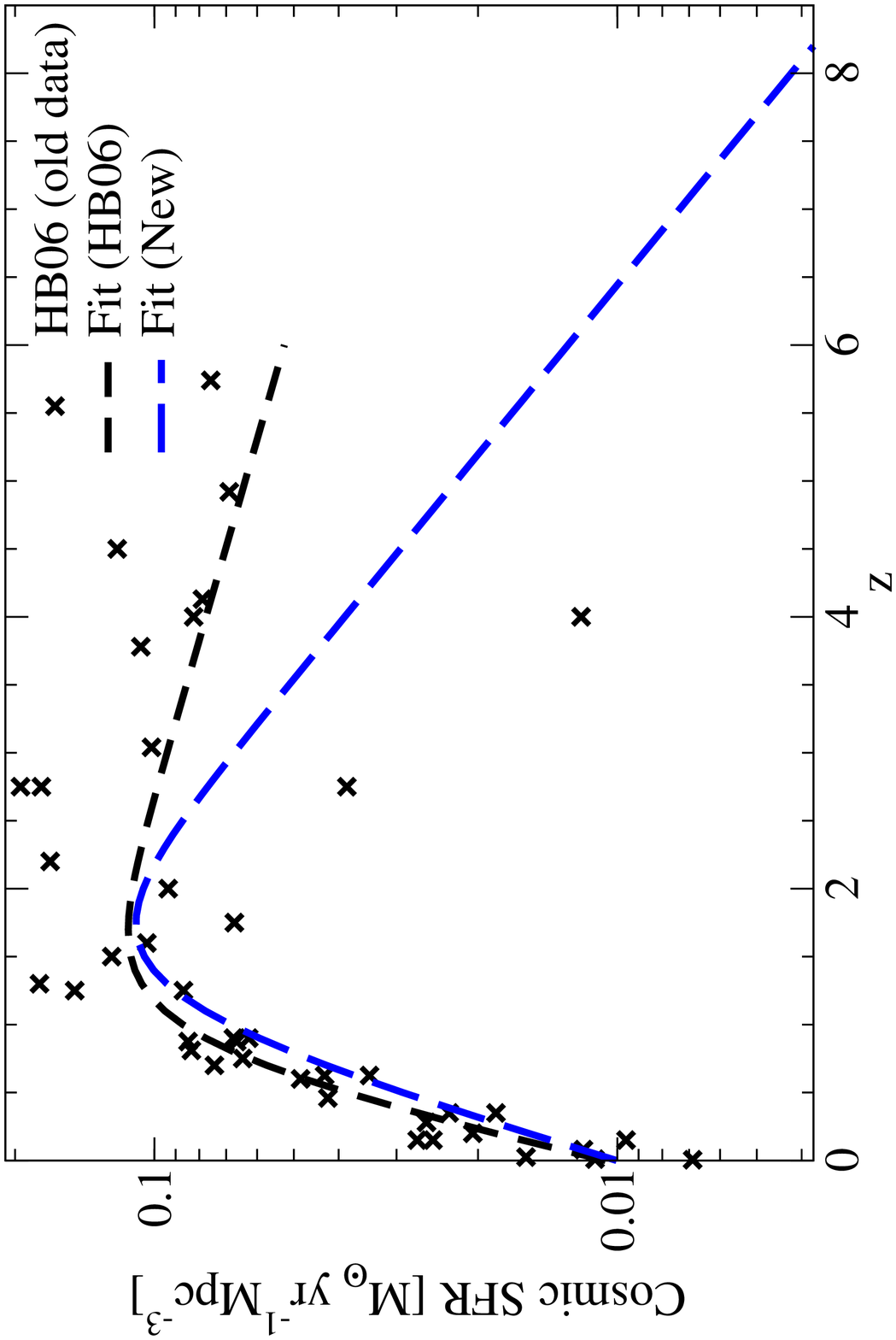}\\[-3ex]
\caption{Cosmic star formation rates reported in the past six years have been dramatically lower at high redshifts ($z>2$) than previous results.  \textbf{Top} panel: our new compilation of cosmic star formation rates, which excludes those reported before 2006 (see Table \ref{t:csfr_data}), along with a best-fitting double power law.  \textbf{Bottom} panel: the compilation of cosmic star formation rates reported in \citep{Hopkins06b} along with a best-fitting double power law (black dashed line) compared to the best-fitting double power law from our new compilation (blue dashed line).   The largest differences at high redshifts result from changing assumptions about the amount of dust present at $z>3$.  Fits (Eq.\ \ref{e:csfr_fit}) to the new and older data sets are given in Table \ref{t:csfr_fit}.  Error bars are not shown since publications have often dramatically underestimated the true magnitude of the systematic errors (see \S \ref{s:csfr_errs}).}
\label{f:csfr_comp}
\end{figure}

As shown in Fig.\ \ref{f:csfr_comp}, newer estimates of the cosmic SFR are systematically and substantially lower than previous compilations (such as \citealt{Hopkins06b}) at $z>3$.  At high redshifts, the predominant method of probing star formation is through UV emission; however, conversion to a star formation rate requires quantifying dust obscuration systematics.  Previously, \cite{Hopkins06b} assumed fixed dust obscuration for $z> 2.5$.  However, new measurements have shown that the amount of dust is likely to decrease substantially with redshift beyond $z\sim 3$ \citep{Bouwens11b,Reddy09}.  The inferred star formation rates thus were likely to have been overcorrected in \cite{Hopkins06b}, although substantial uncertainty about the cosmic star formation rate at high redshifts still remains. 

\label{s:csfr}

\begin{table}
\begin{center}
\caption{Observational Constraints on the Cosmic Star Formation Rate}
\label{t:csfr_data}
\begin{tabular}{rcccc}
\hline
\hline
Publication & Redshifts & Technique & Area & Notes\\
\hline
\cite{Robotham11} & $0.013-0.1$ & UV & 833 deg$^2$ & I\\
\cite{Salim07} & $0.005-0.2$ & UV & 741 deg$^2$ & A\\
\cite{Ly10} & 0.8 & H$\alpha$ & 0.8 deg$^2$ & I\\
\cite{Zheng07} & $0.2-1$ & UV/IR & 0.458 deg$^2$ & I\\
\cite{Rujopakarn10} & $0-1.2$ & FIR & 0.389 - 9 deg$^2$ & I\\
\cite{Smolcic09} & $0.2-1.3$ & 1.4 GHz & 2 deg$^2$ & I\\
\cite{Shim09}  & $0.7-1.9$ & H$\alpha$ & 0.029 deg$^2$ & I\\
\cite{Tadaki11} & 2.2 & H$\alpha$ & 0.0156 deg$^2$ & I\\
\cite{Sobral12} & 0.4-2.3 & H$\alpha$ & 0.016-1.7 deg$^2$ & I\\
\cite{Magnelli11} & 1.3-2.3 & IR & 0.0786 deg$^2$ & I\\
\cite{Karim11} & $0.2-3$ & 1.4 GHz & 2 deg$^2$ &\\
\cite{Ly11} & $1-3$ & UV & 0.241 deg$^2$ & AI\\
\cite{Kajisawa10} & $0.5-3.5$ & UV/IR & 0.029 deg$^2$ & AI\\
\cite{Dunne09} & $0-4$ & 1.4 GHz & 0.8 deg$^2$ & I\\
\cite{Cucciati11} & $0-5$ & UV & 0.611 deg$^2$ & I\\
\cite{LeBorgne09} & $0-5$ & IR-mm & varies & I\\
\cite{vdBurg10} & $3-5$ & UV & 4 deg$^2$ & I\\
\cite{Yoshida06} & $4-5$ & UV & 0.243 deg$^2$ & I\\
\cite{Bouwens11b} & $4-8$ & UV & 0.040 deg$^2$ & I\\
\hline
\end{tabular}
\end{center}
\tablecomments{Letters correspond to the following corrections made to the published results: I (Initial Mass Function), A (Intrinsic Scatter Correction; see Appendix \ref{a:csfr_ssfr_uncert}).  The technique of \cite{LeBorgne09} (parametric derivation of the cosmic SFH from counts of IR-sub mm sources) uses multiple surveys with different areas.}
\end{table}

\label{s:csfr_errs}

Because most survey authors do not fully consider systematic errors, most reported error estimates (1-10\%) are too small to explain the observed variance in published estimates of the cosmic star formation rate (see Fig.\ \ref{f:csfr_comp}).  Therefore, instead of using authors' estimates for systematic uncertainties, we estimate the true systematic errors by computing the variance of our collection of published SFR estimates, as detailed in Appendix \ref{a:csfr_ssfr_uncert}.  Our results are given in Table \ref{t:csfr_errs}.  We find the average systematic error to range from 0.13 dex at $z=0$ to 0.27 dex for $z>3$.

\subsection{Specific Star Formation Rates}

\label{s:ssfr}

\begin{table}
\begin{center}
\caption{Observational Constraints on the Specific Star Formation Rate}
\label{t:ssfr_data}
\begin{tabular}{rcccc}
\hline
\hline
Publication & Redshifts & Technique & Area & Notes\\
\hline
\cite{Salim07} & $0.005-0.2$ & UV & 741 deg$^2$ & \\
\cite{Zheng07} & $0.2-1$ & UV/IR & 0.458 deg$^2$ & I\\
\cite{Twite12} & $1$ & H$\alpha$ & 1.4 deg$^2$ &\\
\cite{Noeske07} & $0.2-1.1$ & UV/IR & 0.5 deg$^2$ & I\\
\cite{Tadaki11} & 2.2 & H$\alpha$ & 0.0156 deg$^2$ & I\\
\cite{Whitaker12} & $0-2.5$ & UV/IR & 0.4 deg$^2$ &\\
\cite{Daddi07} & $1.4-2.5$ & UV-1.4GHz & <0.025 deg$^2$ & I\\
\cite{Salmi12} & $0.9-1.3$ & U-IR & <0.025 deg$^2$ & \\
\cite{Karim11} & $0.2-3$ & 1.4 GHz & 2 deg$^2$ &\\
\cite{Kajisawa10} & $0.5-3.5$ & UV/IR & 0.029 deg$^2$ & I\\
\cite{Reddy12} & $1.4-3.7$ & UV/IR & 0.44 deg$^2$ & I\\
\cite{Lee11} & $3.3-4.3$ & UV/IR & 5.3 deg$^2$ &\\
\cite{Feulner08} & $0.4-5$ & UV/IR & 0.025 deg$^2$ & I\\
\cite{Gonzalez12} & $4-6$ & UV/IR & 0.040 deg$^2$ & I\\
\hspace{-3ex}\cite{Schaerer10} & $6-8$ & UV & 2 deg$^2$ & I\\
\cite{Labbe12} & $8$ & UV/IR & 0.040 deg$^2$ & I\\
\cite{McLure11} & $6-8.7$ & UV & 0.0125 deg$^2$ & IN\\
\hline
\end{tabular}
\end{center}
\tablecomments{The letter ``I'' corresponds to an Initial Mass Function correction; ``N'' corresponds to a correction for nebular emission lines \citep{Stark12}.  \cite{Daddi07} report using a fraction of the GOODS-N and GOODS-S fields, but did not report an exact area.  \cite{Schaerer10} use a combination of fields, the largest of which is COSMOS (with 2 deg$^2$ area).  However, this has a comparatively bright limiting magnitude.}
\end{table}

We conducted a literature survey in an almost identical manner as for the cosmic star formation rate to assemble a collection of specific star formation rate (SSFR) constraints as a function of stellar mass, again in October 2012 (see Appendix \ref{a:csfr_ssfr_uncert}). The selected papers are summarized in Table \ref{t:ssfr_data}.  Most authors do not consider the full range of systematics that could affect their stellar masses or SFR estimates.  To account for this we take a similar approach as in \S \ref{s:csfr_errs} and estimate systematic errors by calculating inter-publication variances.  Details of our approach are given in Appendix \ref{a:csfr_ssfr_uncert}.  We did not find evidence for mass or redshift trends in the variance, and have adopted a uniform systematic uncertainty of 0.28 dex for SSFRs.

Besides these observational constraints, we apply one modeling constraint to SFRs in high-mass halos at low redshifts.  The slope of the stellar mass --- halo mass relationship flattens towards high halo masses \citep{Behroozi10,moster-09}, meaning that central galaxy stellar mass becomes a poor indicator of the host halo mass.  Consequently, observed SSFRs (reported as functions of stellar mass) offer poor constraints on the star formation rate of high-mass halos.  Nonetheless, observations of galaxy clusters have indicated that the central galaxies generally have very low star formation rates \citep[e.g.][]{Donahue10,Rawle12}.  To account for this we apply a weak prior based on the \cite{Wetzel11} group catalog. The recent star formation history of $M_h > 10^{14}\Msun$ halos is averaged from $z=0$ to $z=0.2$ (corresponding to the range of the SDSS survey used in \citealt{Wetzel11}); if it is above $1 \Msun$ yr$^{-1}$, the the total $\chi^2$ error of the fit is increased by $(\langle SFH\rangle - 1\Msun$ yr$^{-1})^2 (\Msun$ yr$^{-1})^{-2}$.

\section{Simulation Data}

\label{s:sim}

We rely mainly on the high-resolution \textit{Bolshoi} simulation, described in \cite{Bolshoi}.  Bolshoi follows 2048$^3$ ($\approx 8.6$ billion) particles in a comoving, periodic box with side length 250 $h^{-1}$ Mpc from $z=80$ to the present day.  Its mass resolution ($1.9 \times 10^8$ $\Msun$) and force resolution (1 $h^{-1}$ kpc) make it ideal for studying the evolution of halos from $10^{10}$ $\Msun$ (e.g., satellites of the Milky Way) to the largest clusters in the universe ($10^{15}$ $\Msun$).  Bolshoi was run as a collisionless dark matter simulation with the Adaptive Refinement Tree Code \citep[\textsc{art}; ][]{kravtsov_etal:97,kravtsov_klypin:99} assuming a flat, $\Lambda$CDM cosmology ($\Omega_M = 0.27$, $\Omega_\Lambda = 0.73$, $h = 0.7$, $\sigma_8 = 0.82$, and $n_s = 0.95$).  These cosmological parameters
are consistent with results from both WMAP5 \citep{wmap5} and the latest WMAP7+BAO+H$_0$ results \citep{wmap7}.

Halos in Bolshoi were identified using \textsc{Rockstar}, a seven-dimensional halo finder that uses phase space plus temporal information \citep{Rockstar}.  Merger trees were generated using a new algorithm presented in \cite{BehrooziTree},  which enforces physically consistent evolution of halo properties across timesteps to provide increased robustness against anomalies in the halo finder.  Halo masses are calculated using spherical overdensities, according to the virial overdensity criterion of \cite{mvir_conv}.

To supplement results from Bolshoi where there exist concerns about statistics (i.e., for high-mass halos) or for initial conditions (as Bolshoi uses Zel'dovich/1LPT initial conditions), we also make use of two larger simulations.  \textit{MultiDark} follows the same number of particles in a much larger volume (1000 $h^{-1}$ Mpc) with 7 times worse force resolution and 64 times worse mass resolution, but using the identical simulation code and cosmology \citep{Riebe11}.  \textit{Consuelo} follows $1400^3$ particles in a somewhat larger volume (420 $h^{-1}$ Mpc) than Bolshoi; it has eight times worse force resolution and fourteen times worse mass resolution (McBride et al, in preparation; see also \citealt{BehrooziTree}).\footnote{{\tt http://lss.phy.vanderbilt.edu/lasdamas/}}   It has a slightly different cosmology ($\Omega_m = 0.25$, $\Omega_\Lambda = 0.75$, $h = 0.7$, $n_s = 1.0$, and $\sigma = 0.8$) and was run using the GADGET-2 code \citep{Springel05}.  Notably, however, it uses the more accurate 2LPT initial conditions instead of Zel'dovich initial conditions, so it serves as a way to correct for inaccuracies in the high-redshift mass function for Bolshoi.

Results from these three simulations are used to calibrate the halo mass function (Appendix \ref{a:tinker}), halo mass accretion histories (Appendix \ref{a:mah}), and subhalo disruption rates (Appendix \ref{a:disruption}) used in this work.
\begin{figure}
\plotgrace{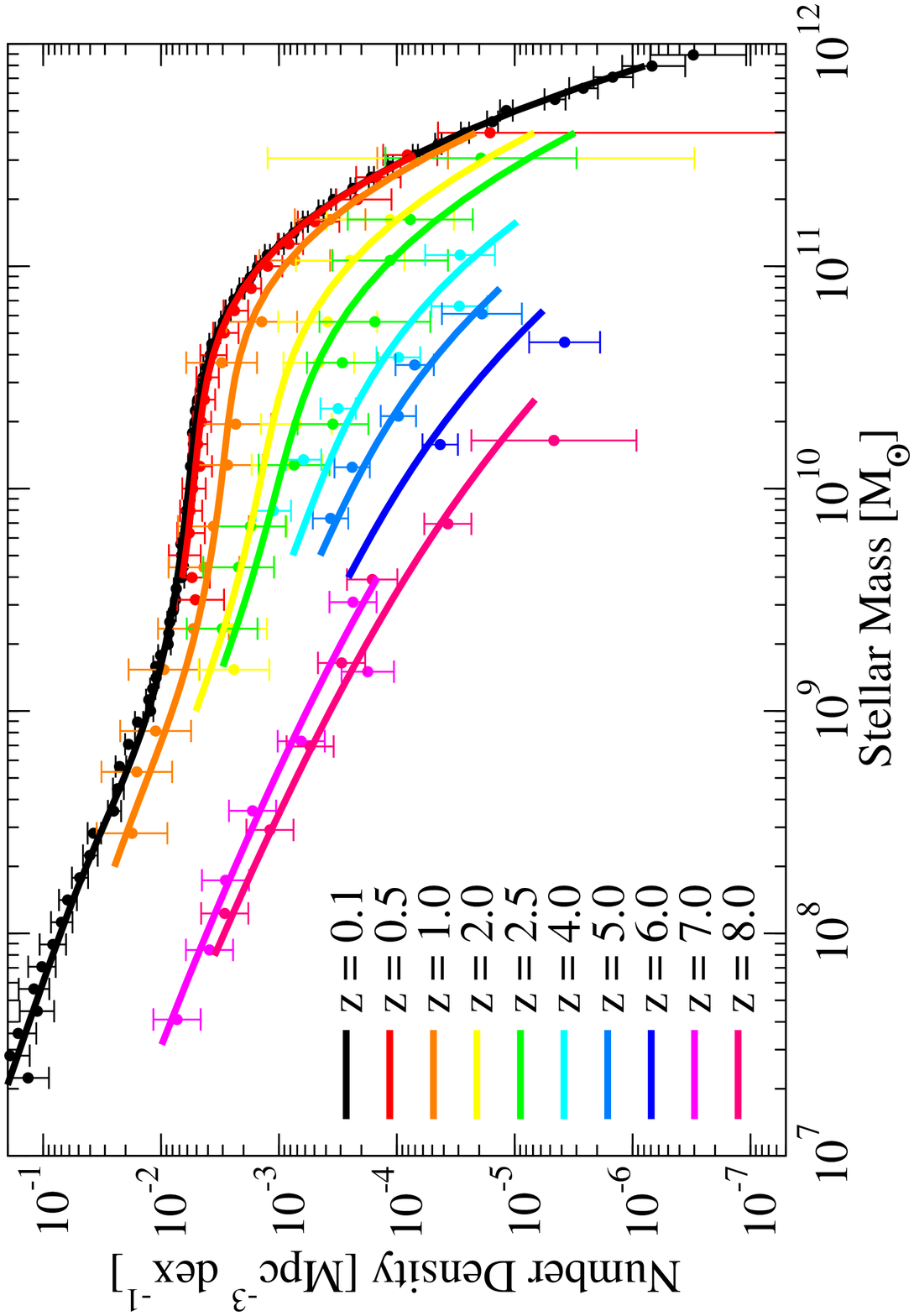}\\[-5ex]
\plotgrace{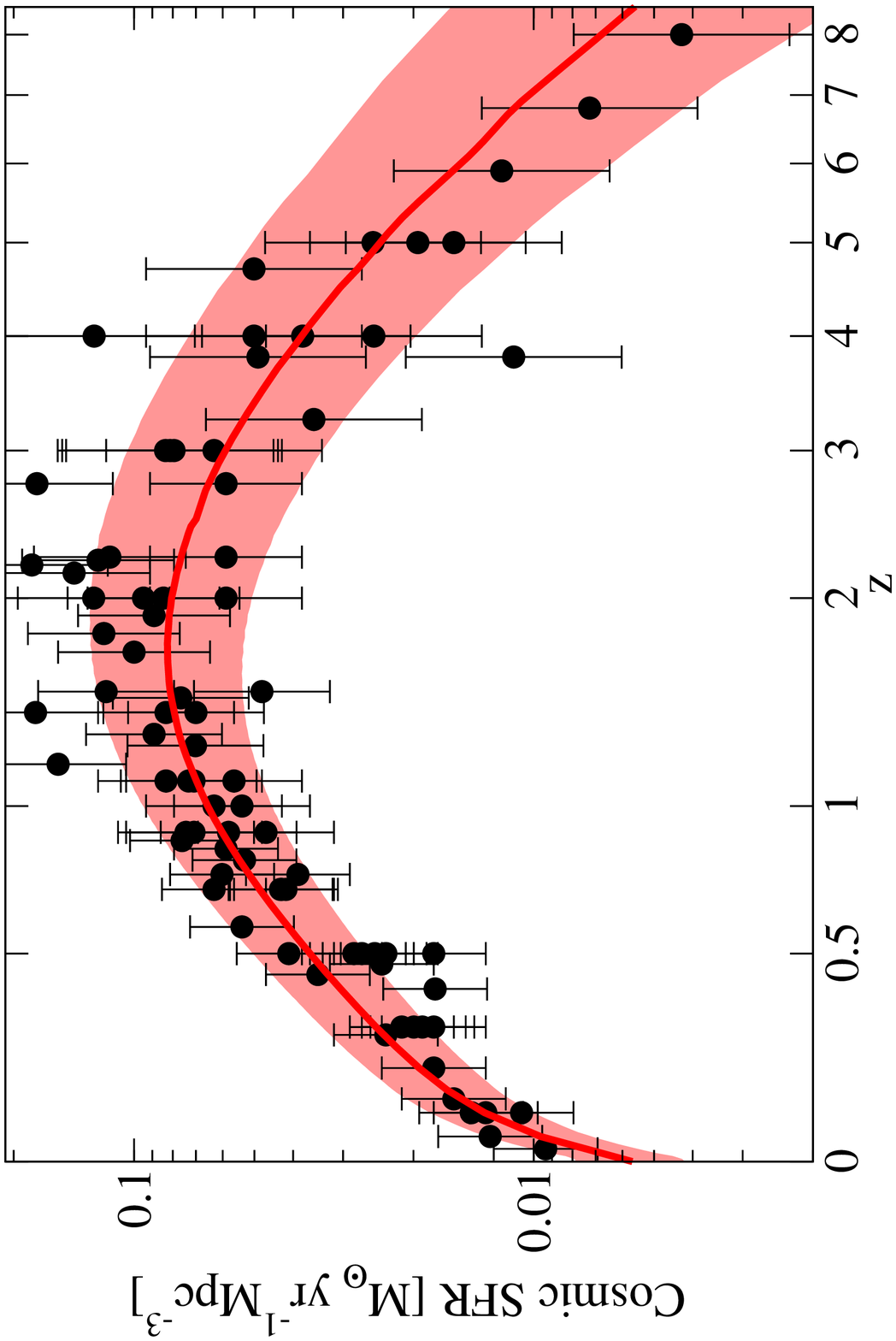}\\[-5ex]
\caption{\textbf{Top} panel: Evolution of the stellar mass function from $z=0$ to $z=8$ in the best fitting model (colored lines), compared to observations (points with error bars; for clarity not all data is shown).  \textbf{Bottom} panel: Observational constraints on the cosmic star formation rate (black points),  compared to the best-fit model (red solid line) and the posterior one-sigma distribution (red shaded region).}\label{f:constraints}
\end{figure}

\begin{figure}
\vspace{-5ex}
\plotgrace{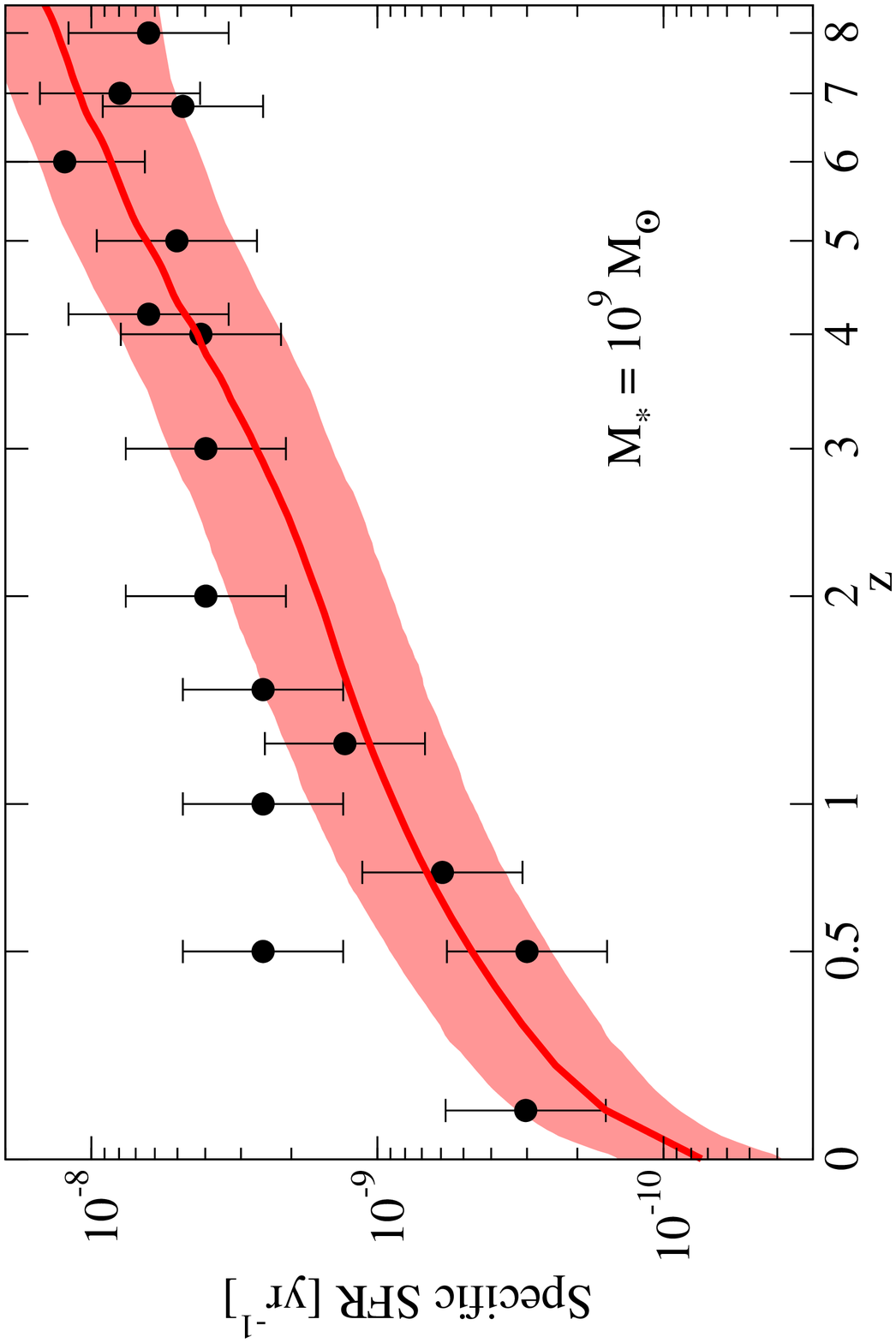}\\[-10ex]
\plotgrace{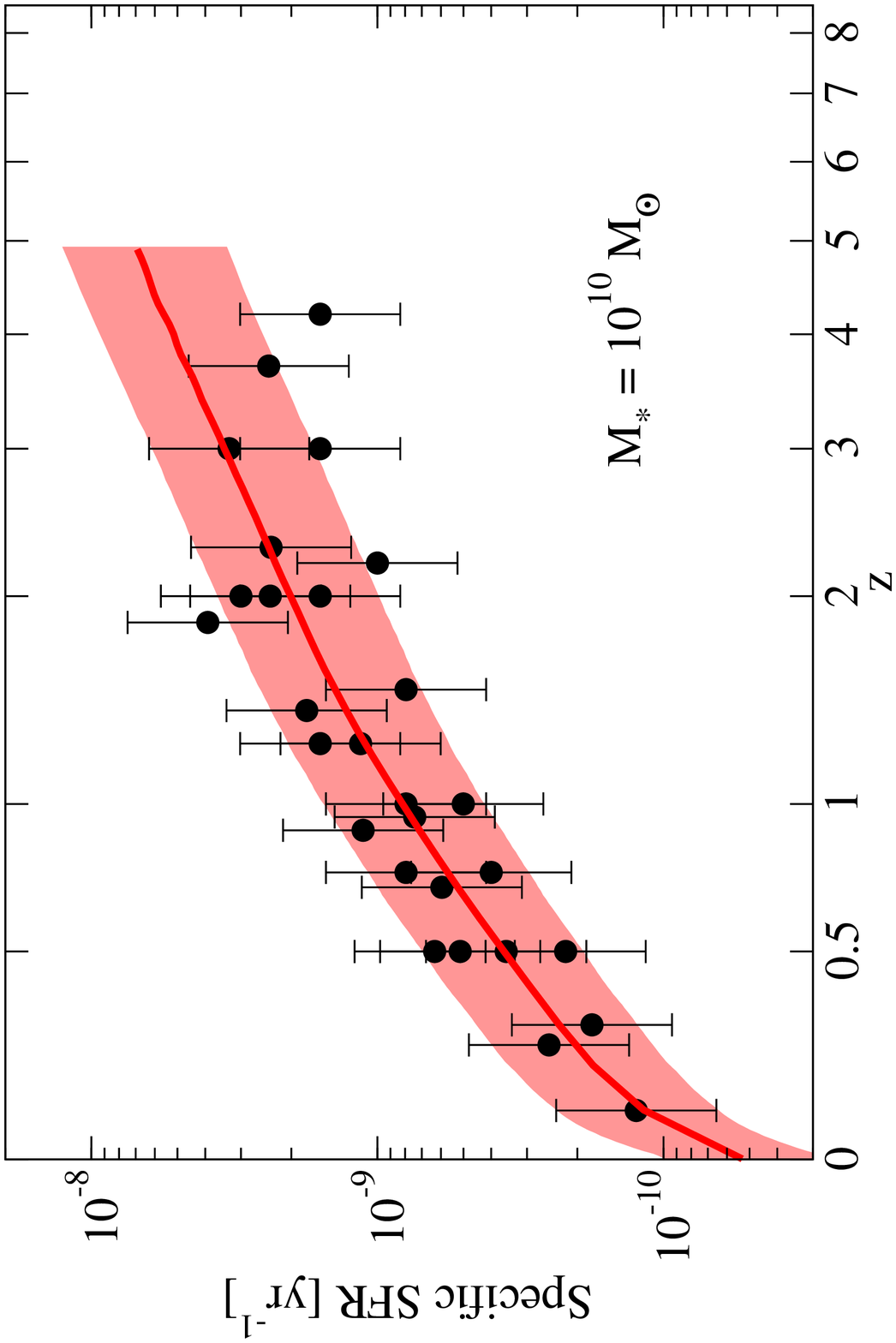}\\[-10ex]
\plotgrace{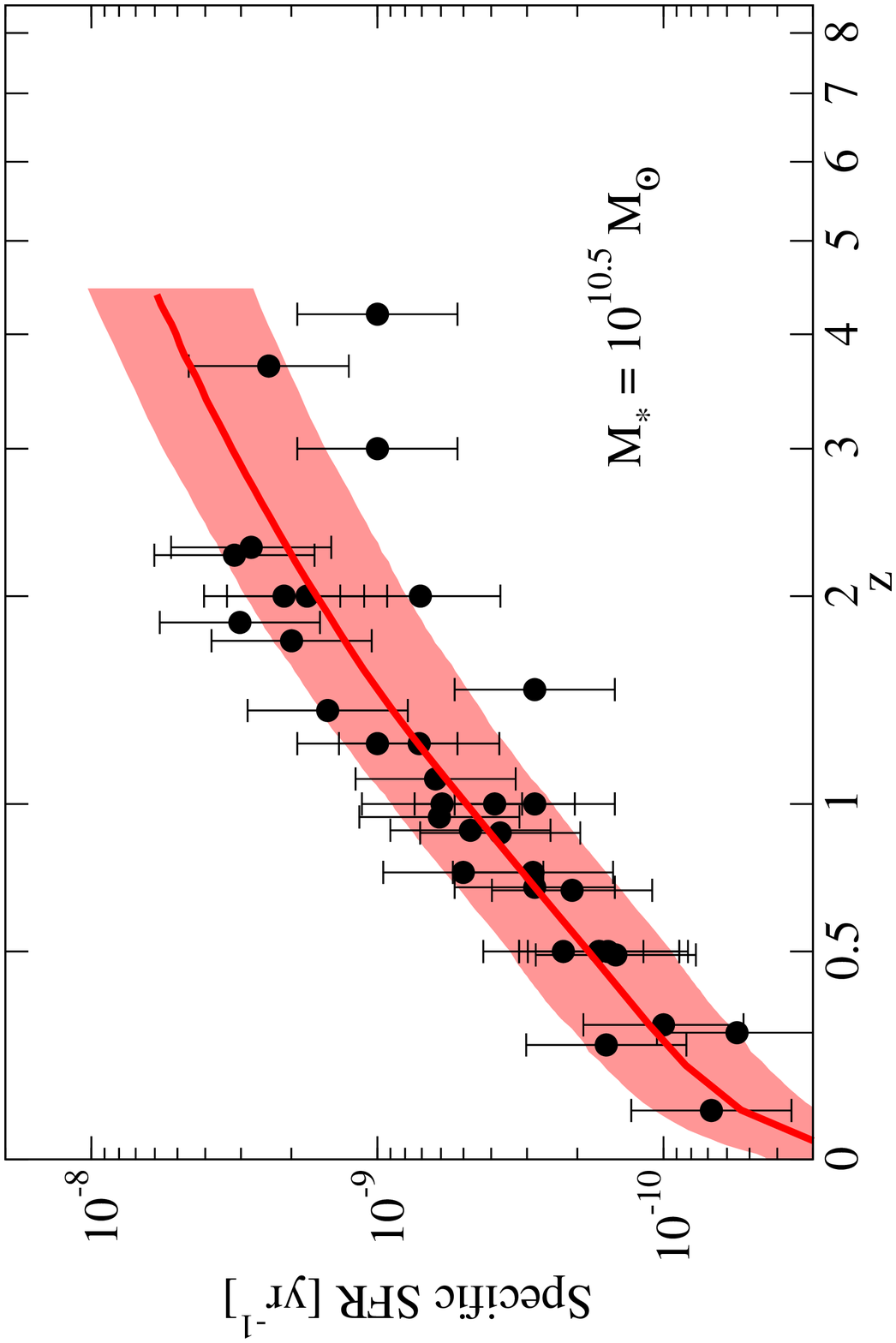}\\[-5ex]
\caption{The best fitting model (red line) and posterior one-sigma distribution (red shaded region) for the evolution of the specific star formation rate from $z=0$ to $z=8$, compared to observational estimates (black points).}
\label{f:constraints2}
\end{figure}

\begin{figure*}[h]
\plotgrace{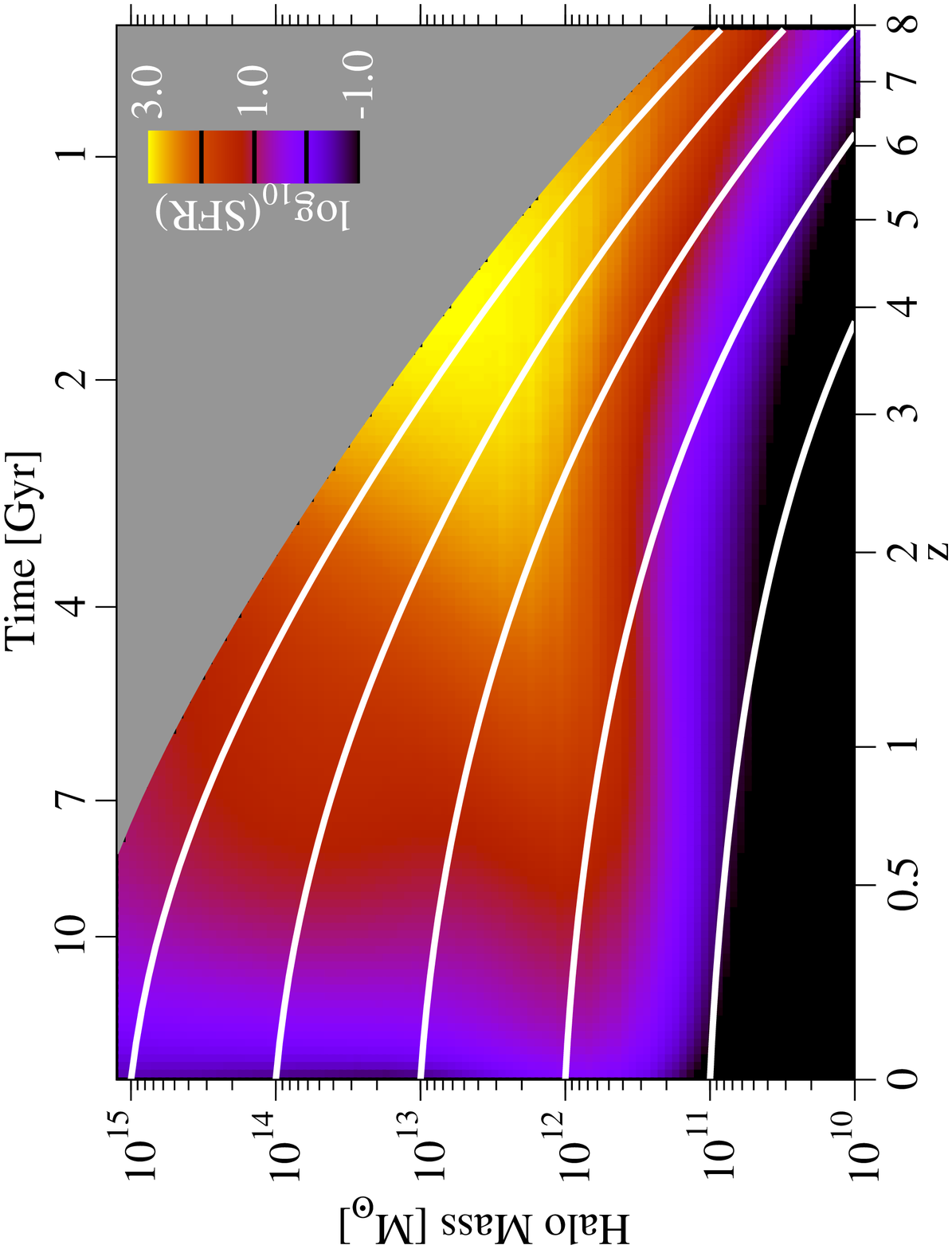}\plotgrace{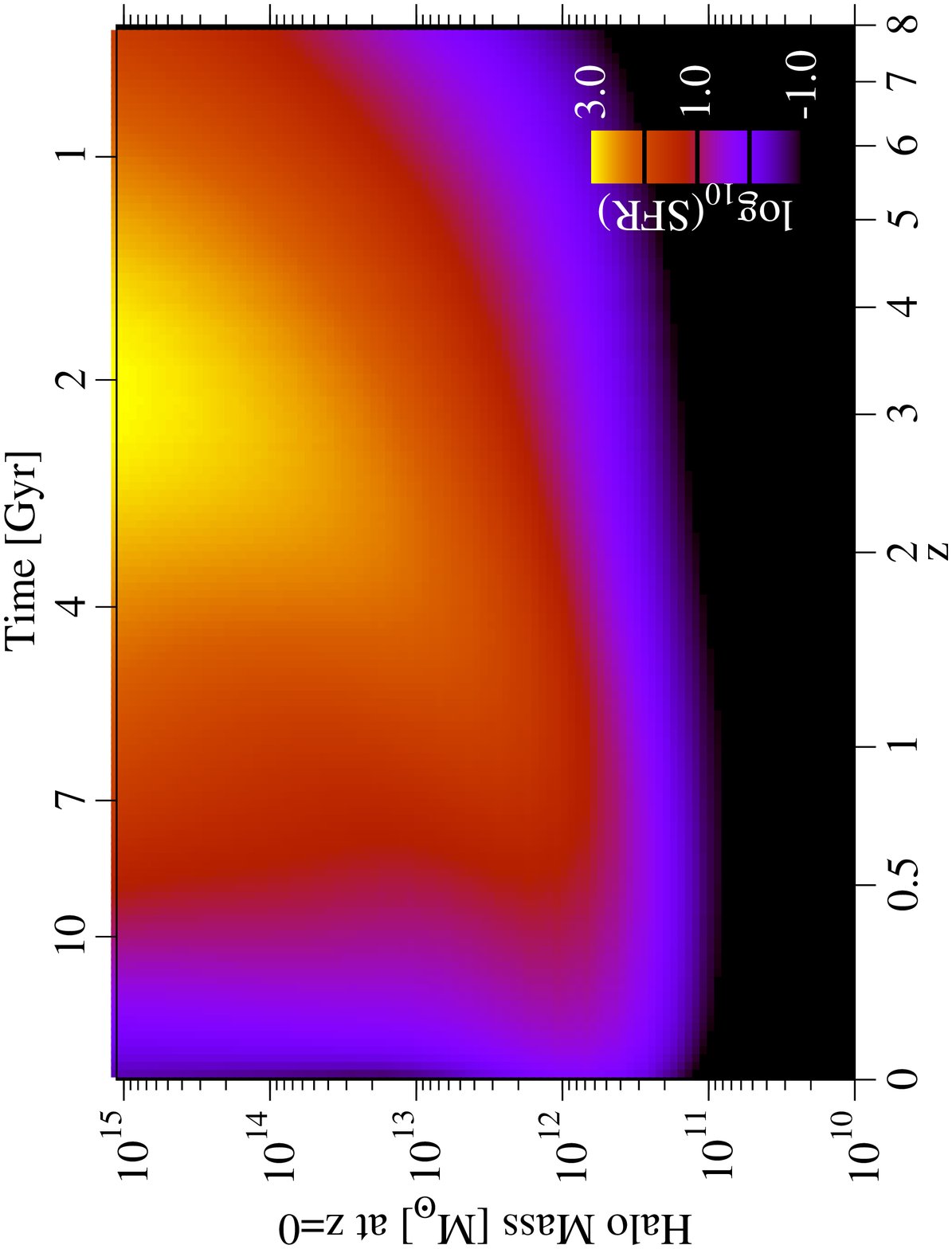}\\
\caption{\textbf{Left} panel: Average star formation rates as a function of halo mass and redshift.  The overlaid white lines show average mass accretion histories for halos as a function of redshift for comparison.  The grey area shows halos that would have a mass of $>10^{15.5}\Msun$ at $z=0$ and therefore are not expected to exist.
\textbf{Right} panel: Star formation histories (SFH) as a function of present-day halo mass and redshift, for galaxies at $z=0$.  This figure shows the historical star formation rate for stars in the galaxy at the present day. Since the contribution of stars from merging galaxies is so low, this is equivalent to the star formation rate traced along the white mass accretion trajectories in the left panel.}
\label{f:sfr_sfh}
\plotgrace{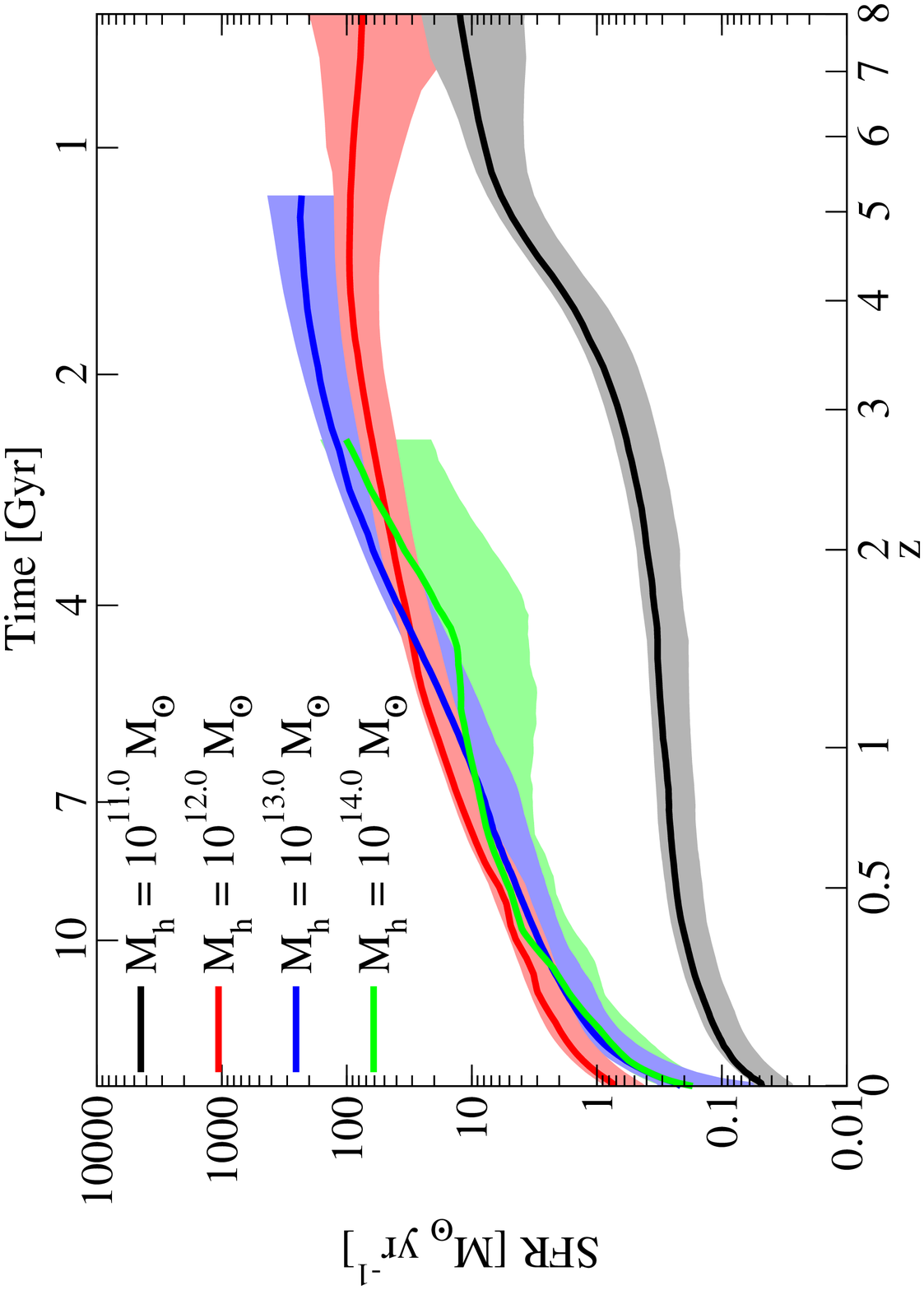}\plotgrace{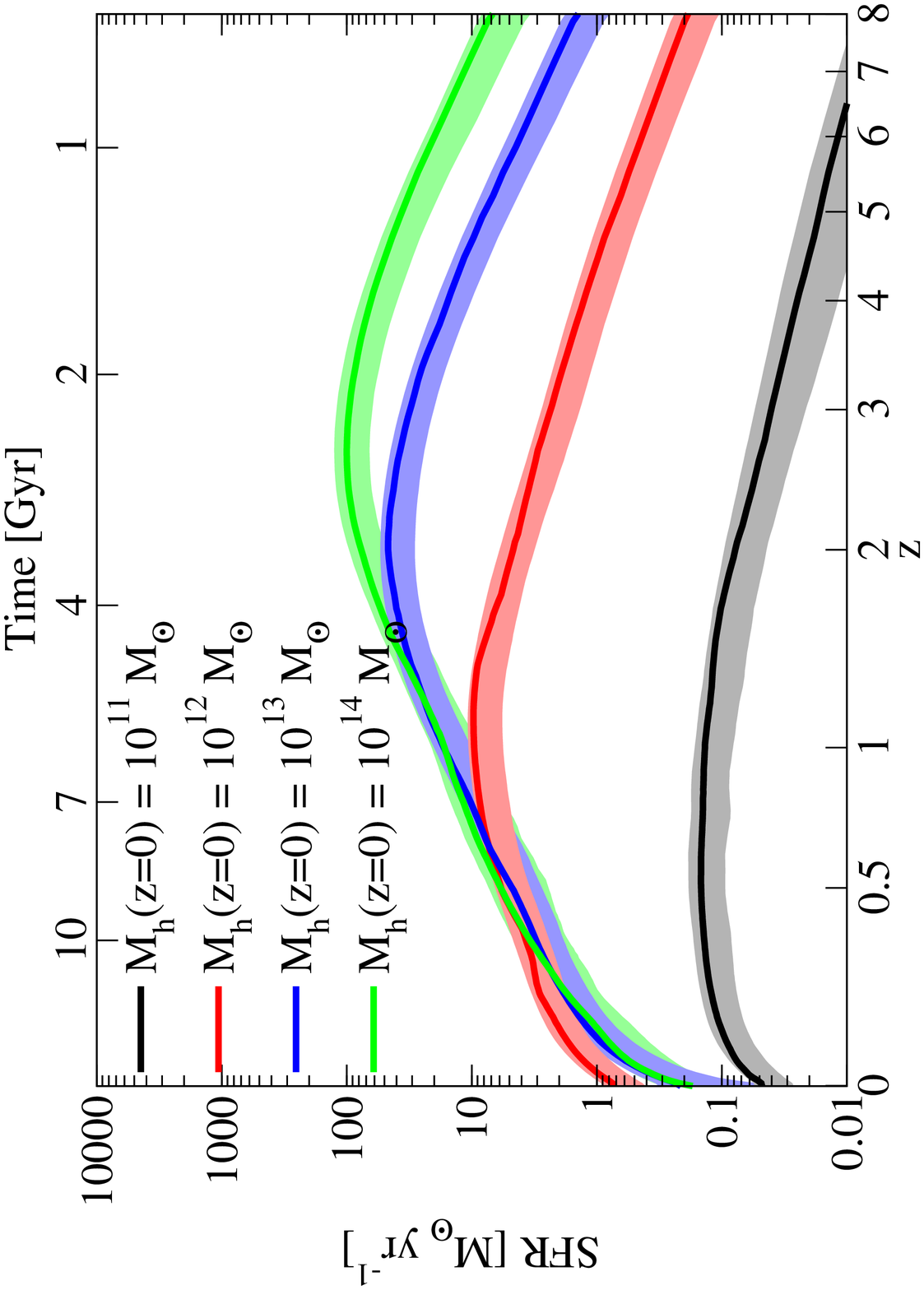}\\
\caption{\textbf{Left} panel: Average star formation rates for the galaxies in halos at a given halo mass and redshift (lines).
Shaded regions indicate the one-sigma posterior distribution.
\textbf{Right} panel: Average star formation histories as a function of halo mass and redshift (lines).
Shaded regions indicate the one-sigma posterior distribution.
Histories for $10^{15}\Msun$ halos are not shown as they are very similar to those for $10^{14}\Msun$ halos.}
\label{f:sfr_sfh2}
\plotgrace{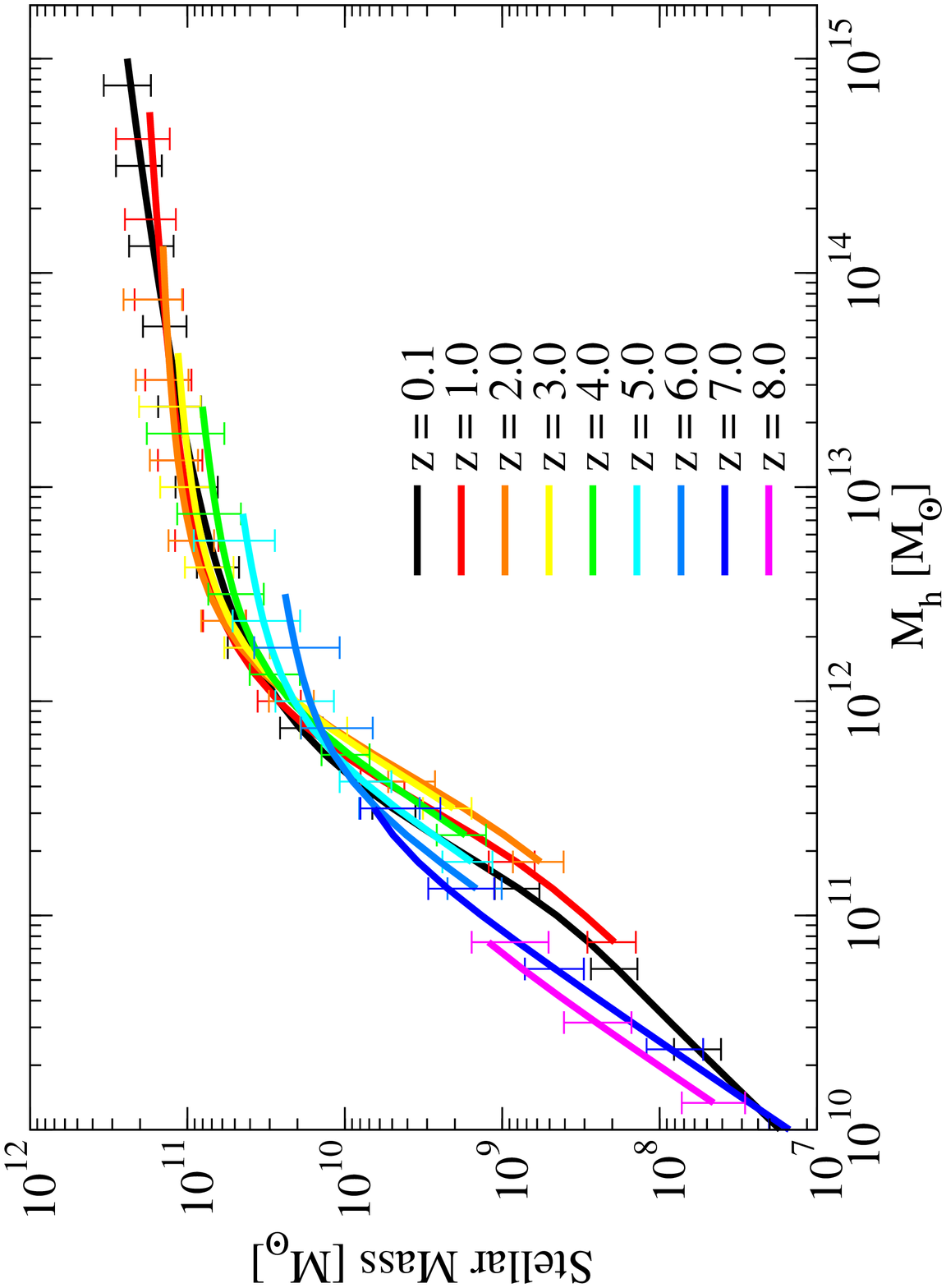}
\plotgrace{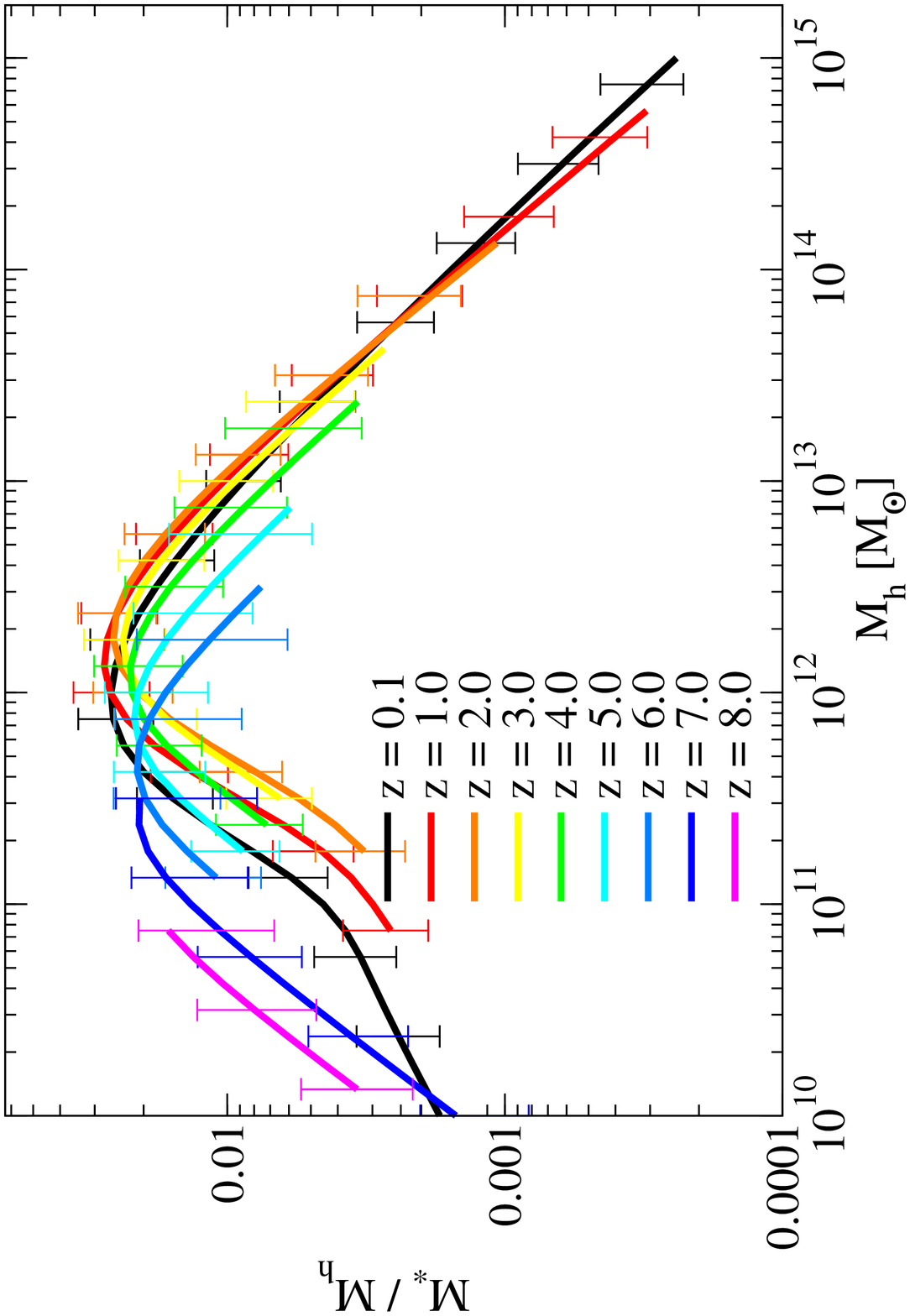}\\
\caption{\textbf{Left} panel: Evolution of the derived stellar mass as a function of halo mass.  In each case, the lines show the mean values for central galaxies.  These relations also characterize the satellite galaxy population if the horizontal axis is interpreted as the halo mass at the time of accretion.  Error bars include both systematic and statistical uncertainties, calculated for a fixed cosmological model (see \S \ref{s:sim} for details). \textbf{Right} panel: Evolution of the derived stellar mass fractions ($M_\ast / M_h$) as a function of halo mass.}
\label{f:smhm}
\end{figure*}

\begin{figure}
\vspace{-6ex}
\plotgrace{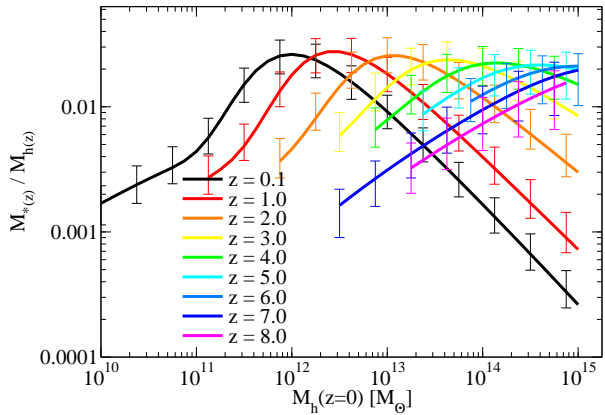}
  \caption{Evolution of the derived stellar mass fractions ($M_\ast(z) / M_h(z)$) as a function of halo mass at the present day.  More massive halos used to have a significantly larger fraction of mass in stars, but the peak star formation efficiency has remained relatively constant to the present day.}
\label{f:smhm2}
\end{figure}

\begin{figure*}
\vspace{-3ex}
\plotgrace{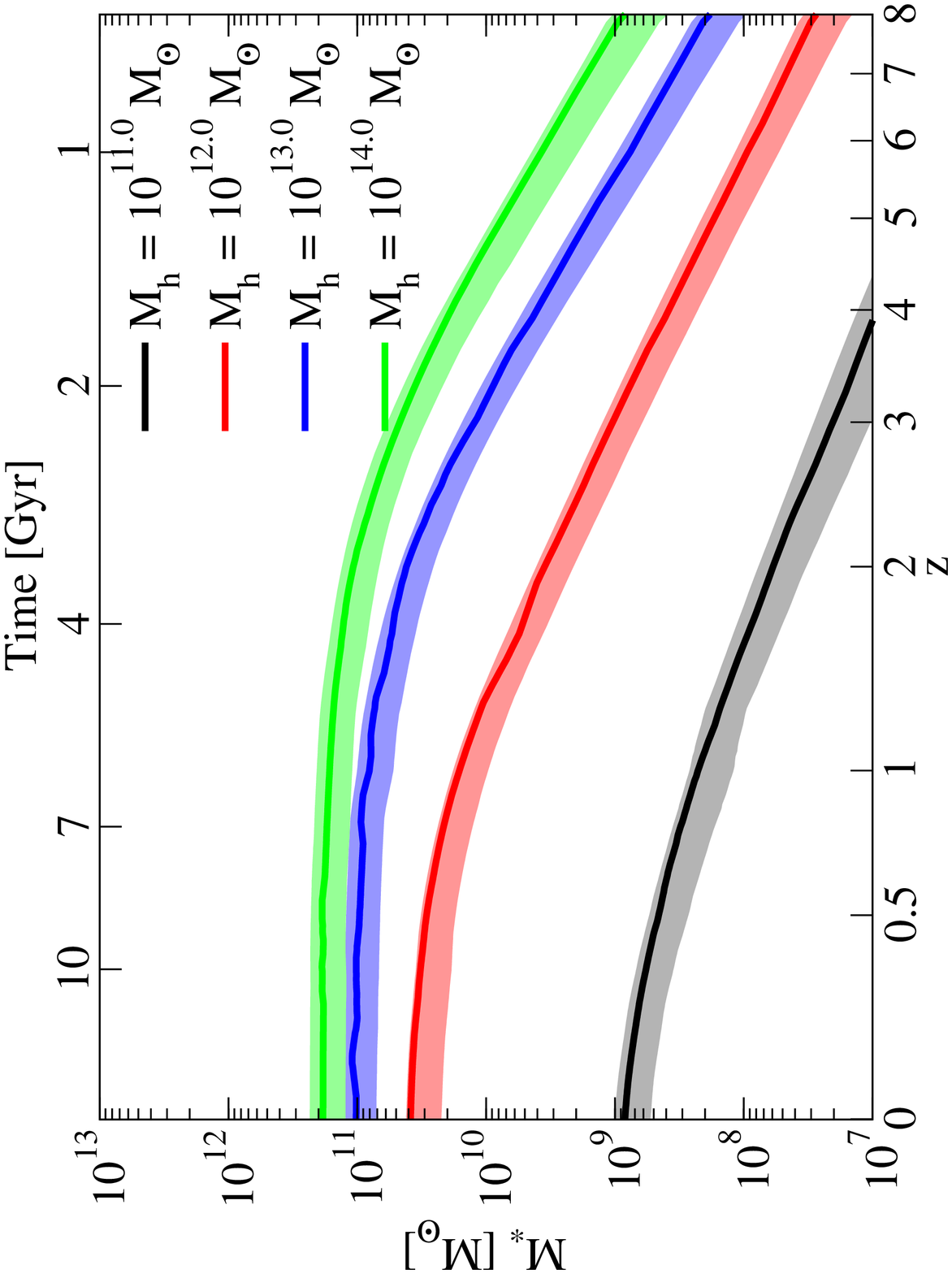}\plotgrace{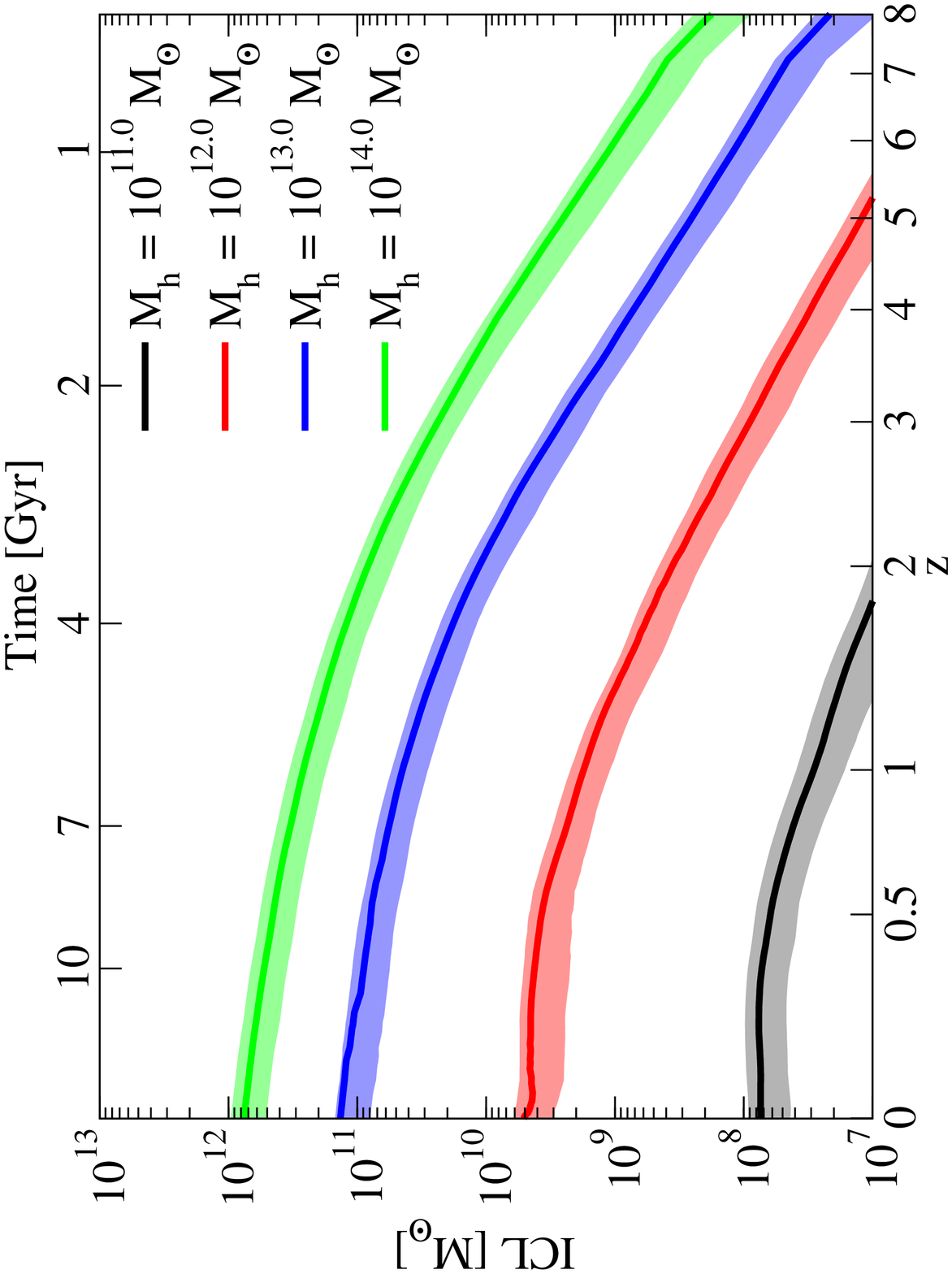}\\[-5ex]
\caption{\textbf{Left} panel: Stellar mass growth histories of galaxies of different masses.  Lines shown the 
amount of stellar mass remaining in central galaxies at the present day that was in place (in any progenitor galaxy) at a given redshift for our best fit model. \textbf{Right} panel: Amount of stellar mass in the intracluster light (ICL) remaining at the present day in place at a given redshift.  Note that plots for $10^{15}\Msun$ halos are not shown as they are nearly identical to those for $10^{14}\Msun$ halos.   Shaded regions in both panels shown the one-sigma posterior distribution. }
\label{f:sm_icl}
\end{figure*}

\begin{figure}
\plotgrace{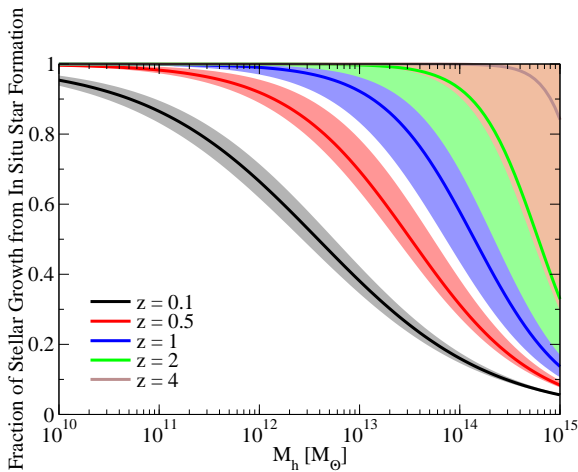}
\caption{The fraction of stellar mass growth in galaxies due to \textit{in situ} star formation (as opposed to growth by galaxy-galaxy mergers) as a function of halo mass and redshift.}
\label{f:sfr_frac}
\end{figure}

\begin{figure*}[h]
\plotgrace{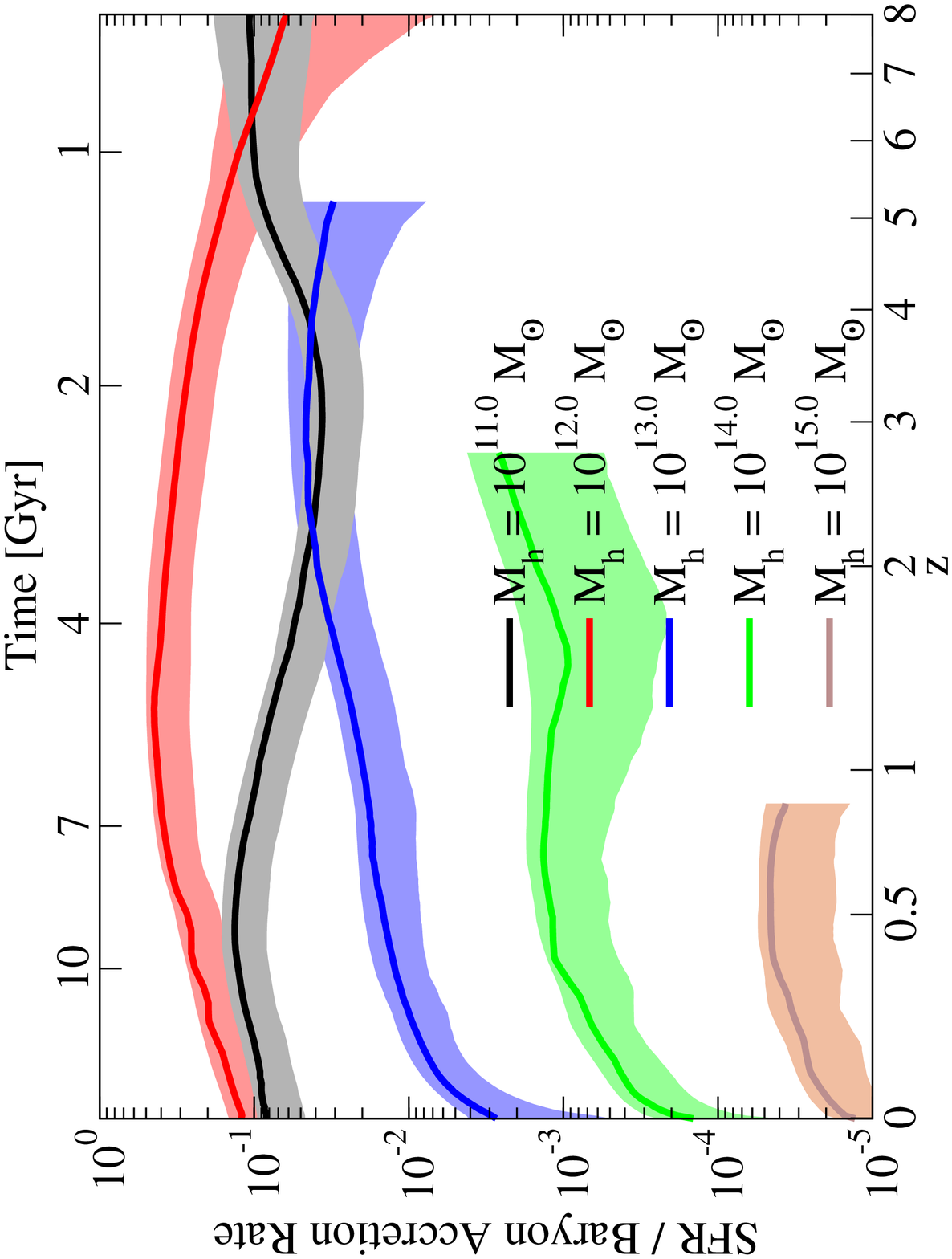}\plotgrace{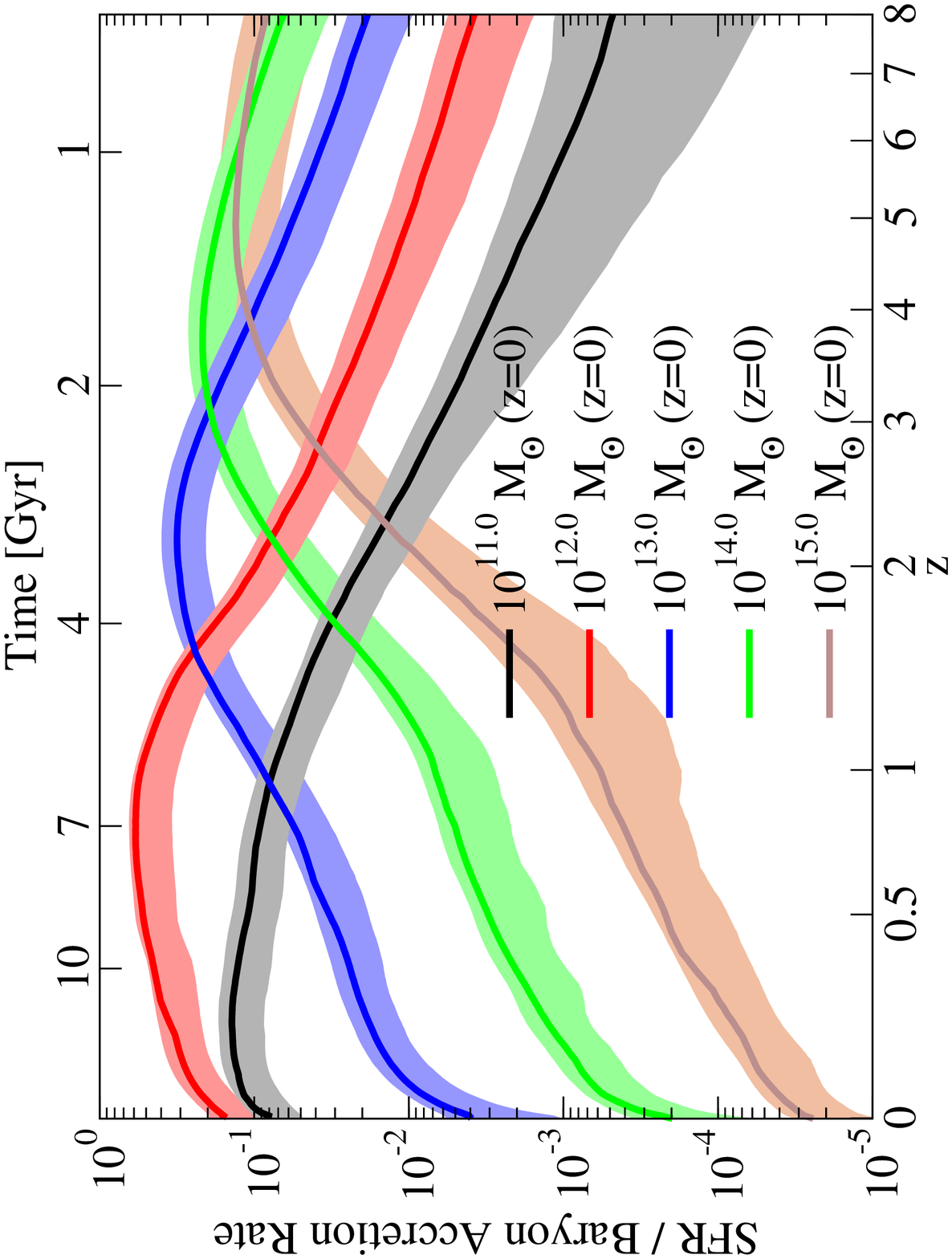}
\caption{\textbf{Left} panel: Ratio of the average SFR to the average baryon accretion rate ($f_b dM/dt$) for halos as a function of halo mass.  \textbf{Right} panel: same, except as a function of halo mass at $z=0$ (i.e., ratio of SFR to baryon accretion rate for the progenitors of present-day halos).
Lines shown the best fit model 
and shaded regions show the one-sigma posterior distribution.
} 
\label{f:smhist}
\plotgrace{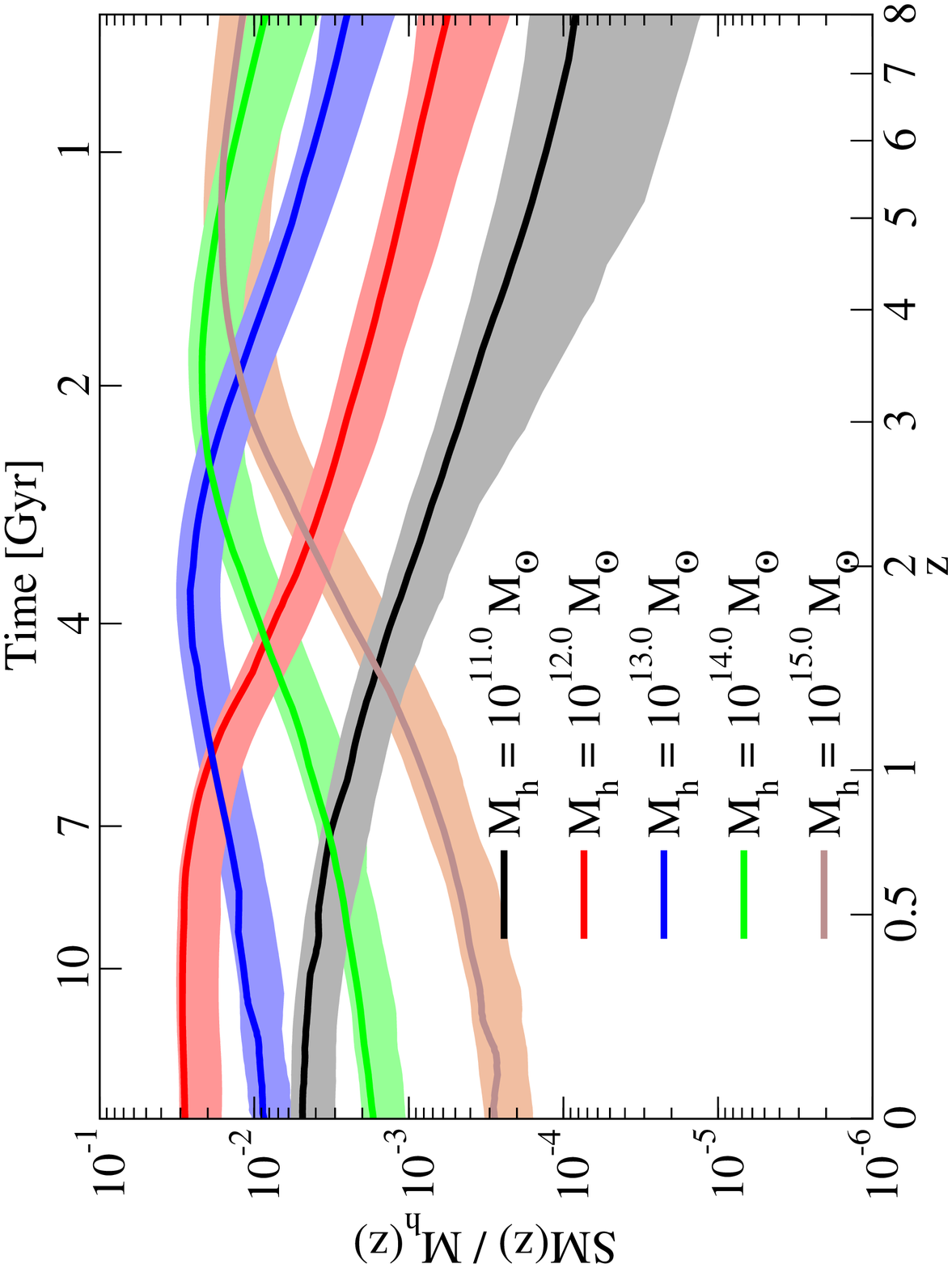} \plotgrace{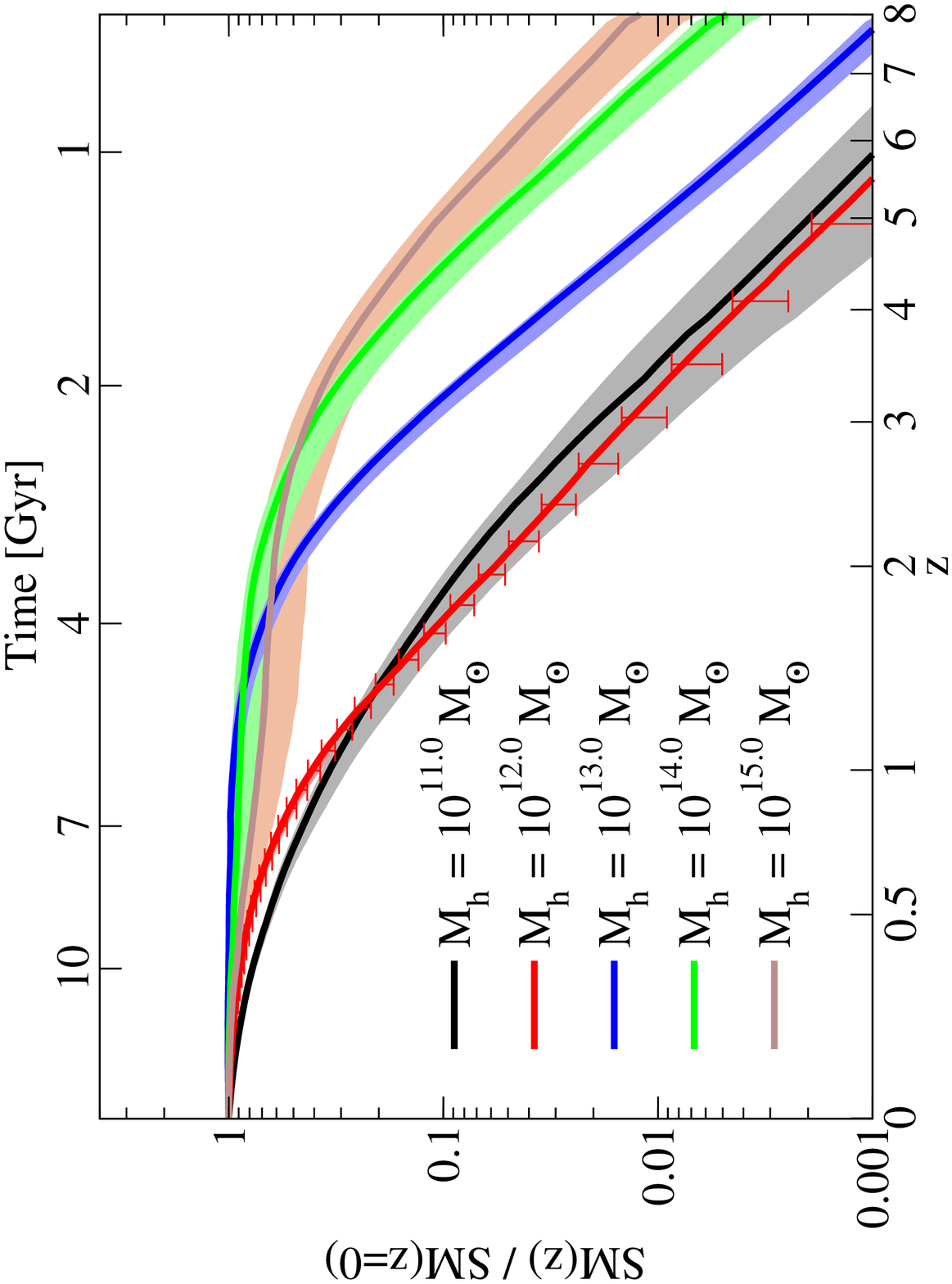} 
\caption{  \textbf{Left} panel: Ratio of the stellar mass to halo mass at a given redshift as a function of the halo mass at $z=0$.  \textbf{Right} panel: Ratio of the stellar mass at a given redshift to the stellar mass in the descendent galaxy today, as a function of the halo mass at $z=0$.  Lines shown the best fit model  and shaded regions show the one-sigma posterior distribution.
}
\label{f:smhist2}
\plotgrace{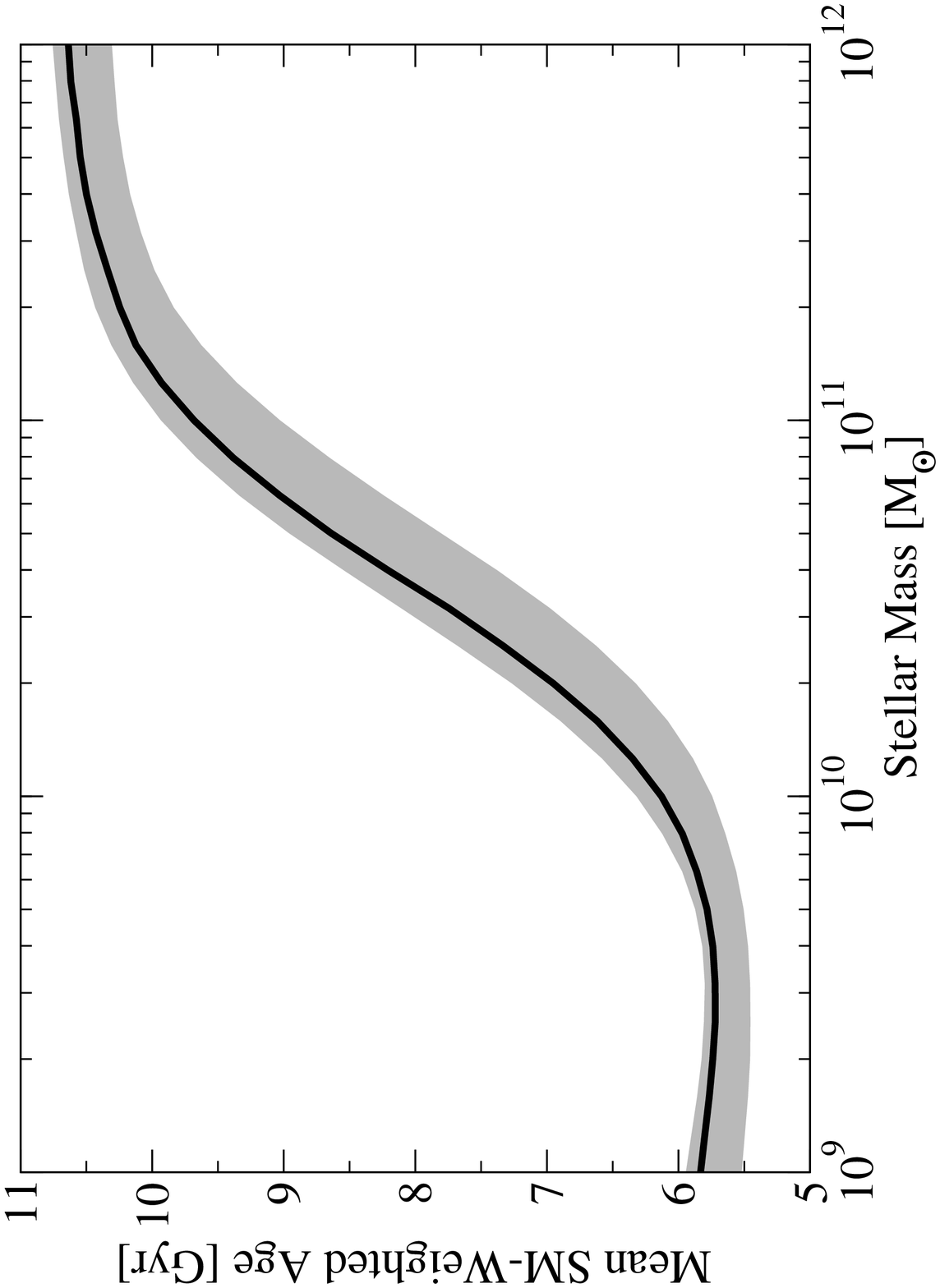}
\plotgrace{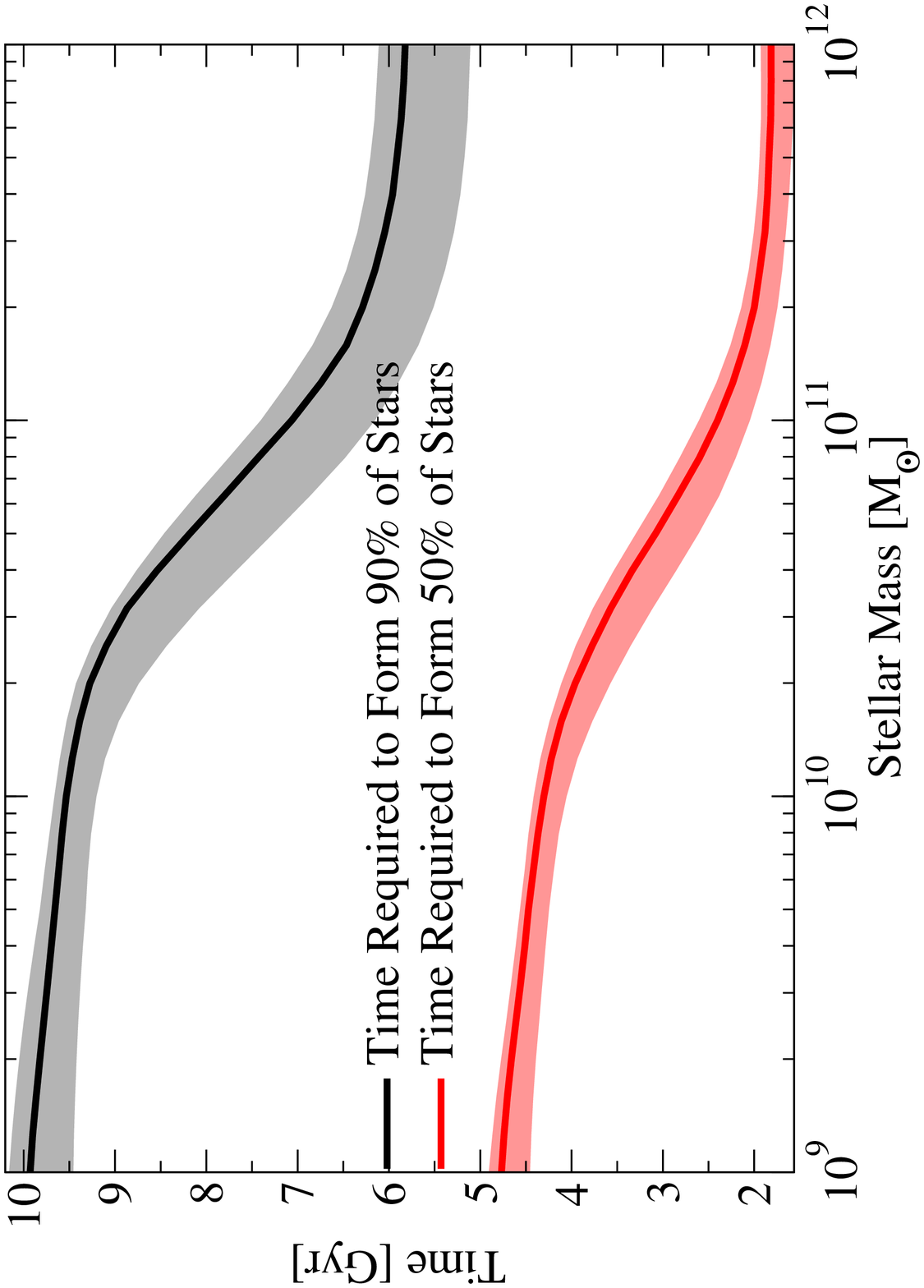}
\caption{\textbf{Left} panel: Average age of presently-existing stars in halos as a function of stellar mass.  \textbf{Right} panel: Time required to form 50\% and 90\% of presently-existing stars as a function of stellar mass.  Lines shown the best fit model 
and shaded regions show the one-sigma posterior distribution.}
\label{f:smhist3}
\end{figure*}

\section{Results}

\label{s:results}

The method presented above results in a posterior distribution for the set of parameters describing models that match observed stellar mass functions, specific star formation rates, and cosmic star formation rates from $z=0$ to $z=8$.  All data results in this paper are available for download online.\footnote{{\tt http://www.peterbehroozi.com/data.html}}  Our best-fitting parameters with one-sigma limits are as follows:\\

Intrinsic Parameters:
\begin{eqnarray}
\nu &=& \exp(-4a^2)\nonumber\\
\log_{10}(\epsilon) &=& -1.777^{+0.133}_{-0.146} + (-0.006^{+0.113}_{-0.361}(a-1) + (-0.000^{+0.003}_{-0.104})z)\nu +\nonumber\\&& -0.119^{+0.061}_{-0.012}(a-1)\nonumber\\
\log_{10}(M_1) &=& 11.514^{+0.053}_{-0.009} + (-1.793^{+0.315}_{-0.330}(a-1) + (-0.251^{+0.012}_{-0.125})z)\nu\nonumber\\
\alpha &=& -1.412^{+0.020}_{-0.105} + (0.731^{+0.344}_{-0.296}(a-1))\nu\nonumber\\
\delta &=& 3.508^{+0.087}_{-0.369} + (2.608^{+2.446}_{-1.261}(a-1) + -0.043^{+0.958}_{0.071}z)\nu\nonumber\\
\gamma &=& 0.316^{+0.076}_{-0.012} + (1.319^{+0.584}_{-0.505}(a-1) + 0.279^{+0.256}_{-0.081}z)\nu\nonumber\\
\log_{10}(M_{h,ICL}) &=& 12.515^{+0.050}_{-0.429} + (-2.503^{-0.202}_{-2.078})(a-1)\nonumber\\
\rho_{0.5} &=& 0.799^{+0.028}_{-0.355}\nonumber
\end{eqnarray}
Systematic Parameters:
\begin{eqnarray}
\mu &=& -0.020^{+0.168}_{-0.096} + 0.081^{+0.078}_{-0.036}(a-1)\nonumber\\
\kappa &=& 0.045^{+0.110}_{-0.051} + (-0.155^{+0.133}_{-0.133})(a-1)\nonumber\\
\xi &=& 0.218^{+0.011}_{-0.033} + -0.023^{+0.052}_{-0.068}(a-1)\nonumber\\
\sigma &=& 0.070 + 0.061^{+0.017}_{-0.008}(z-0.1)\nonumber\\
c_i(z) &=& 0.273^{+0.103}_{-0.222}(1+\exp(1.077^{+3.502}_{-0.099}-z))^{-1}\nonumber\\
b &=& 0.823^{+0.043}_{-0.629}\nonumber
\end{eqnarray}

Our total $\chi^2$ error for the best-fit model from all sources (observational and theoretical) is 245.  For the number of observational data points we use (628), the nominal reduced $\chi^2$ is 0.4.  While the true number of degrees of freedom is not easily computed (note for example that covariance matrices are not available for most stellar mass functions in the literature), this suggests that our best fit is a reasonable match to the data.  This is shown visually in Fig.\ \ref{f:constraints}, which shows the evolution of the stellar mass function and the posterior distribution for the observed cosmic star formation rate and in Fig.\ \ref{f:constraints2}, which shows the posterior distribution of specific star formation rates.

As discussed in \S \ref{s:methodology}, each model in the distribution contains complete information on the average stellar content and formation history of halos as a function of mass and redshift.  Herein, we focus on a few interesting implications, leaving more extensive coverage to a follow-up paper.  In \S\ref{s:sfh_results}, we present derived star formation rates and star formation histories as a function of halo mass and redshift.  Next, in \S\ref{s:smhm_results}, we discuss constraints on the stellar mass to halo mass relation, both as a function of historical halo mass and as a function of halo mass at $z=0$.     In \S\ref{s:sm_icl}, we compare the trajectories of the stellar mass and intracluster light (ICL) buildup in galaxies.  Then, in \S\ref{s:bce}, we discuss instantaneous baryon conversion efficiencies and how they relate to integrated baryon conversion efficiencies as well as stellar ages and formation times in \S\ref{s:sm_ages}.   In \S\ref{s:comparison}, we compare our main results to those obtained by previous studies.  We discuss the main effects of the uncertainties we have modeled in \S\ref{s:maj_uncertainties}; finally, in \S\ref{s:new_fits}, we present new fitting formulae for individual galaxy star formation histories relevant to observers.

\subsection{Star Formation Rates and Histories}

\label{s:sfh_results}

We show derived star formation rates as a function of halo mass in the left panel of Fig.\ \ref{f:sfr_sfh}, and the corresponding star formation histories for galaxies at $z=0$ in the right panel.  As discussed in \S\ref{s:sm_icl}, the contribution from merging galaxies to the central galaxy is small, so star formation histories for galaxies at $z=0$ trace the star formation rate as a function of their progenitor's halo mass.  Also, for this reason, star formation histories for galaxies at a given redshift $z_g$ are nearly the same as for the the $z=0$ stellar populations, except truncated at $z=z_g$.\footnote{Note that this also requires that the halo accretion histories are similar between the $z=z_g$ progenitors of $z=0$ halos and all similar halos of the same mass as the progenitors at $z=z_g$; however, this has been shown to be the case in \cite{McBride09}.}

We show similar plots with one-sigma uncertainties in Fig.\ \ref{f:sfr_sfh2}.  The left-hand panel demonstrates that the star formation rate at fixed halo mass has been monotonically decreasing since very early redshifts.  This rate of decrease is different for different halo masses.  At moderate to high redshifts ($z>2$), larger halo masses generically have larger average star formation rates.  However, at lower redshifts, the highest mass halos ($M_h\gtrsim 10^{14} \Msun$) become so inefficient that they have lower star formation rates than group-scale ($10^{13}\Msun$) halos or Milky-Way sized ($10^{12}\Msun$) halos.

From the perspective of individual galaxies, it is more illuminating to look at the star formation history in the right panel of Fig.\ \ref{f:sfr_sfh2}.  Because halos continually gain mass over time, and do so more rapidly at early redshifts, the star formation history for galaxies is not monotonically decreasing.  Instead, it \textit{increases} rapidly with time, approximately as a power law in time.  Depending on present-day halo mass, the galaxy's star formation rate reaches a peak at a redshift between $z=0.5$ to $z=2.5$ (higher redshifts for higher halo masses) and then decreases until the present day.  The rate of decrease depends again on the halo mass, with high halo masses shutting off more rapidly than lower halo masses.  Cluster-scale ($M_h\gtrsim 10^{14} \Msun$) halos form most of their stars rapidly, at early times, whereas galaxies in Magellanic Cloud-scale halos ($10^{11}\Msun$) form stars over an extended period of time (see also \S \ref{s:bce}).

\subsection{The Stellar Mass -- Halo Mass Relation}

\label{s:smhm_results}

We show constraints on stellar mass --- halo mass (SMHM) relation from $z=0$ to $z=8$ in the left panel of Fig.\ \ref{f:smhm} and on the stellar mass --- halo mass ratio in the right panel.  As seen in our previous work \citep{Behroozi10}, there is a strong peak in the stellar mass to halo mass ratio at around $10^{12}\Msun$ to at least $z\sim 4$ and a weaker peak still visible to $z\sim 8$.  While the location of the peak appears to move to higher masses with increasing redshift \citep[consistent with][]{Leauthaud11}, the abundance of massive halos is also falling off with increasing redshift.

In terms of dwarf galaxies ($M_h\sim10^{10}\Msun$), we only have constraints from observations at $z=0$.  These have been the subject of recent interest due to the finding of higher-than-expected stellar mass to halo mass ratios in dwarf galaxies around the Milky Way \citep{BK12}.  However, these expectations have been set largely by the assumption that the stellar mass to halo mass ratio remains a scale-free power law below $10^{11}\Msun$.  As seen in Fig.\ \ref{f:smhm}, the low mass power-law behavior is broken below $10^{11}\Msun$, corresponding with an upturn in the stellar mass function below $10^{8.5}\Msun$ \citep{Baldry08} (this result has also been seen by Kravtsov, in prep).  This underscores the danger of assuming that faint dwarfs obey the same physical scaling relations as Magellanic Cloud-scale galaxies; moreover, it also is a strong argument against fitting the stellar mass -- halo mass relation with a double power law (see discussion in Appendix \ref{a:dp_law}).

Concerning the range of allowed SMHM relations, the observational systematics are large enough that our results are marginally consistent with an unchanging SMHM relation from $z=6$ to $z=0$.  Nonetheless, the feature with strongest significance is a gradual decrease in stellar mass in the median $10^{11}\Msun$ halo from $z=0$ to $z=2$, followed by an increase again for redshifts $z>6$; also potentially indicated is an increase in stellar mass in the median $M_h > 10^{13}\Msun$ halo from $z=0$ to $z=2$.  The best fit  SMHM relations at $z=7$ and $z=8$ are significantly different than at lower reshifts, with more stellar mass per unit halo mass.  However, concerns about the reliability of the stellar mass functions at those redshifts (see \S \ref{s:sm_func}) urge caution in interpreting the physical meaning of this result.

A useful perspective on these results can be obtained by considering the historical stellar mass to halo mass ratio of halos, as shown in  Fig.\ \ref{f:smhm2}.  Despite the large systematic uncertainties, it is clear that halos go through markedly different phases of star formation.  This evolution is most apparent for massive halos, as observations have been able to probe the properties of the progenitor galaxies all the way to $z=8$.  Specifically, high-redshift progenitors of today's brightest cluster galaxies ($M_h \sim 10^{14}\Msun$ were relatively efficient in converting baryons to stars---comparable to the most efficient galaxies today.  However, between redshifts $2-3$, their efficiencies peaked, and thereafter they began to form stars less rapidly than their host halos were accreting dark matter.  At the present day, such galaxies have an integrated star formation efficiency that is two orders of magnitude less than at their peak.  The picture is less clear for progenitors of lower-mass galaxies because current observations cannot probe their progenitors as far back.  Nonetheless, their  apparent behavior in Fig.\ \ref{f:smhm} of rising to a peak efficiency and later falling is consistent with all available data.

\subsection{Stellar Mass and Intracluster Light Growth Histories}

\label{s:sm_icl}
Our model constrains the buildup of stars in the intracluster light (see definition in \S \ref{s:icl}) purely from observational galaxy data and measurements of the halo-halo merger rate in simulations (see also \citealt{Watson12} for an alternate method).  In our best-fitting model, only 5\% of stellar mass in mergers for $10^{14}\Msun$ halos is allowed to be deposited onto the central galaxy since $z=1$, and only 10\% for $10^{13}\Msun$ halos.  For Milky Way-sized and smaller halos, this number rises rapidly to 70-80\%.  Yet, due to the sharply decreasing stellar mass to halo mass ratio for lower-mass halos, most of the incoming stellar mass will be in (rare) major mergers.  Central galaxies are therefore relatively uncontaminated by stars from smaller recently-merged satellites.  At higher redshifts, however, larger fractions of the stellar mass in merging galaxies are allowed to be deposited onto the central galaxy.

In Fig.\ \ref{f:sm_icl}, we show the amount of galaxies' present-day stellar mass and intracluster light (ICL)/halo stars that was in place at a given redshift. The left-hand panel (stellar mass) shows that almost all stars in the central galaxy in present-day cluster-scale halos were in place at $z=2$.  However, since that time, a large number of their satellites have been disrupted into the ICL.  Thus, for cluster scale halos, the stellar mass in the ICL exceeds the stellar mass in the central galaxy by a factor of 4-5, consistent with observations \citep[e.g.,][]{Gonzalez05}.  We note that our model predicts what may seem to be a large ICL fraction for Milky Way-sized galaxies ($10^{12}\Msun$).  However, the Milky Way is a special case.  It has not had a major merger for $\sim$ 10-11 Gyr \citep{Hammer07}; however, as noted above, only major mergers can contribute substantially to the ICL.  A major merger 11 Gyr ago would have, however, contributed less than 3\% of the present-day stellar mass of the Milky Way into the ICL; allowing for passive stellar evolution, this would result in less than 2\% of the luminosity of the Milky Way coming from the intrahalo light, in excellent agreement with observations \citep{purcell-etal-07}.

Finally, in Fig.\ \ref{f:sfr_frac}, we show the inferred fraction of stellar mass growth in galaxies coming from \textit{in situ} star formation (as opposed to galaxy-galaxy mergers) as a function of halo mass and redshift.  At all redshifts greater than 1, the vast majority of stellar mass growth is from star formation, and hence stellar mass functions alone may be used to infer galaxy star formation rates.  However, the stellar mass growth rate of high-mass ($M_h>10^{13}\Msun$), low-redshift halos is such that it cannot be explained entirely by \textit{in situ} star formation.  For these halos, stellar mass functions alone cannot be used to infer their star formation rates (see also the discussion in \S \ref{s:discussion} and in Appendix \ref{a:models}).  Interestingly, in the local universe the transition between merger-dominated growth and star formation growth in the local Universe is roughly the mass of the Milky Way --- low mass galaxies are dominated by star formation, and galaxies in halos more massive than $10^{12}\Msun$ are dominated by merging.

\subsection{Baryon Conversion Efficiencies}

\label{s:bce}

Assuming that the baryon accretion rate for a halo is equal to the universal baryon fraction ($f_b$, which is 0.17 in our cosmology) times the overall halo mass accretion rate, we can calculate the instantaneous baryon conversion efficiency implied by our models.  In the left-hand panel of Fig.\ \ref{f:smhist}, we show the instantaneous conversion efficiency at fixed halo mass.  Remarkably, this efficiency is always highest for Milky-Way sized halos ($10^{12}\Msun$), corresponding well with the peak in the stellar mass to halo mass ratio seen in Fig.\ \ref{f:smhm}.  Moreover, this efficiency (20-40\%) is constant to within a factor of two over a remarkably large redshift range, suggesting that the efficiency is only a weak function of accretion rate in halos of this mass.  For more massive halos, there is a decrease in the baryon conversion efficiency following redshift 2-3.  This suggests that the ability of accreted material to cool onto star-forming regions becomes impaired at lower redshifts for such halos.  However, this decrease is gradual, taking place over many Gyr (see also \citealt{BehrooziEvolution}).  Rather than an abrupt change in the character of infalling gas, this may suggest that the decreasing density and accretion rate of gas make it gradually more difficult for cold clumps to form, especially in the presence of an active galactic nucleus.

On the right-hand side of Fig.\ \ref{f:smhist}, we show the historical baryon conversion efficiency for progenitors of $z=0$ halos.  For massive halos at $z=0$, their conversion efficiency climbs steeply towards higher redshifts as their halo mass falls to roughly $10^{12}\Msun$; once their halo mass drops below that value, they then become less and less efficient at very high redshifts.  For less-massive halos, such as those that are $10^{12}\Msun$ at $z=0$, their conversion efficiency has been increasing from early times to the present day.

On the left-hand side of Fig.\ \ref{f:smhist2}, we show extrapolated historical stellar mass to halo mass ratios---proportional to integrated baryon conversion efficiencies---similar to the bottom panel of Fig.\ \ref{f:smhm}.  Notably, stellar mass to halo mass ratios also peak when halos reach $10^{12}\Msun$, indicative of steeply falling star formation efficiencies at higher and lower masses.  It would appear that the maximum integrated stellar mass efficiency is around 20-40\% at all redshifts.

\subsection{Stellar Ages}

\label{s:sm_ages}

In the right-hand panel of Fig.\ \ref{f:smhist2}, we show the historical stellar mass in progenitors of halos at $z=0$ relative to the present-day stellar mass.  This should not be confused with Fig.\ \ref{f:sm_icl}, which shows what amount of the currently-remaining stellar mass was in place at a given redshift; these differ mainly because of passive stellar evolution (massive stars that formed sufficiently long ago will burn out by the present day).  

These data allow us to derive stellar mass-weighted ages as a function of stellar mass, shown in the left-hand panel of Fig.\ \ref{f:smhist3}.  These ages are consistent with the stellar populations in massive galaxies (and halos) forming at very early times and having little star formation continuing to the present day.  Less-massive galaxies have younger average stellar ages, consistent with the ongoing star formation seen in the star formation histories for such galaxies in Fig.\ \ref{f:sfr_sfh2}.  While we have little information on the progenitors of galaxies less massive than $10^{9}\Msun$, there is evidence that the average  ages of the stellar populations may increase for such galaxies.  Specifically, as shown in the right-hand panel of Fig.\ \ref{f:smhist3}, the time required to form 50\% and 90\% of the galaxy's stars increases towards lower stellar masses, indicating that the star formation histories become more and more flat (constant), on average.  A perfectly flat star formation history would result in an average age of about half the age of the universe, or roughly 6.7 Gyr; this is somewhat higher than that seen for $10^{9}\Msun$ galaxies.  On the other hand, for individual dwarf galaxies, the star formation history is likely to be stochastic, which may result in a large scatter around this average value.

For the most massive galaxies today ($\sim10^{12}\Msun$), the time required to form 50\% of their stars was extremely short---only 1--2 Gyr.  This in turn translates to a high star formation rate, on the order of 200-1000 $\Msun$ yr$^{-1}$, which is consistent with observations of ultra-luminous infrared galaxies (ULIRGs; \citealt{Magnelli12,Michalowski10,Michalowski10b,Daddi05,Chapman04}).

\begin{figure*} \plotlargegrace{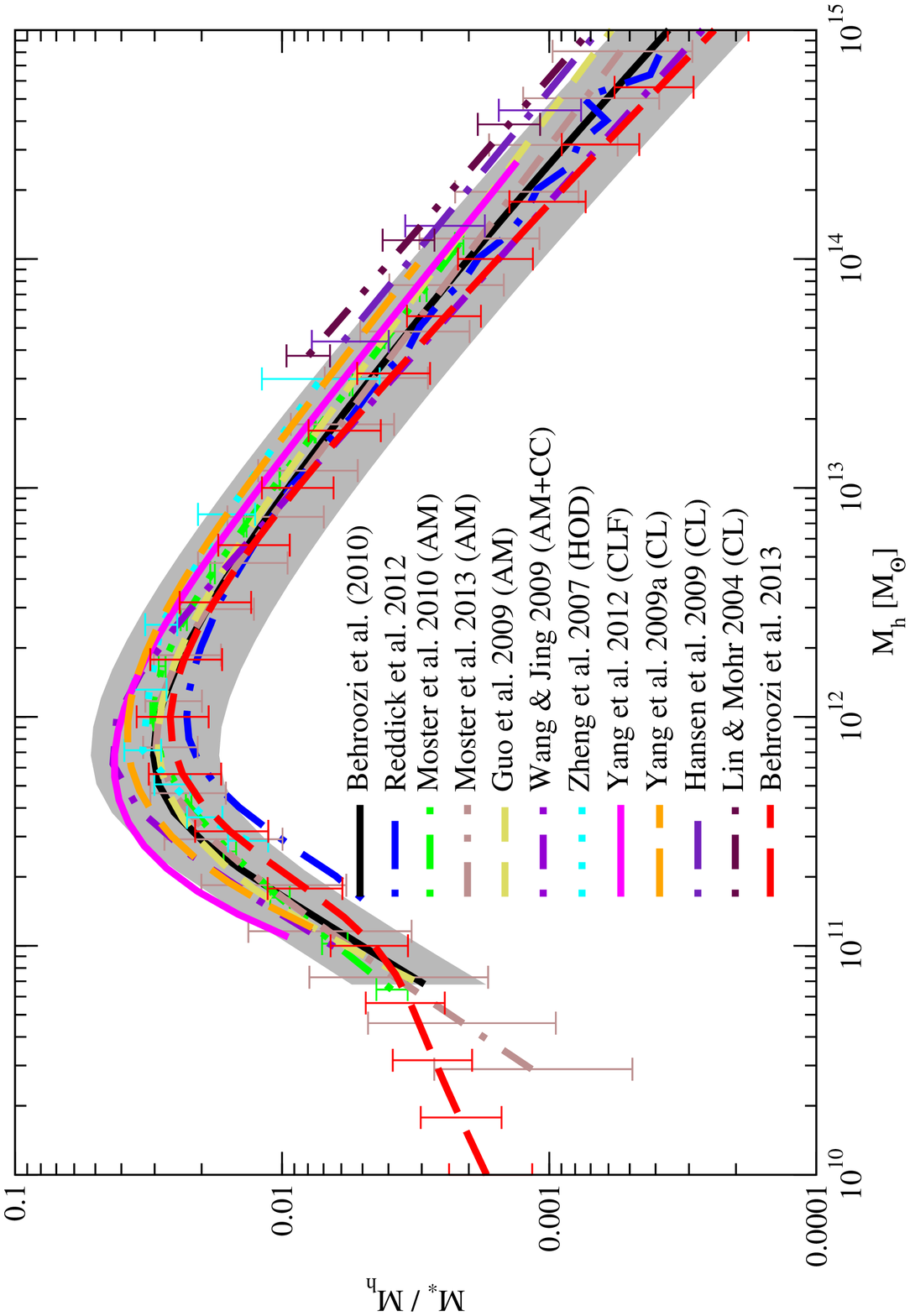} \caption{Comparison of our best-fit model at $z=0.1$ to previously published results.  Results compared include those from our previous work \citep{Behroozi10}, from abundance matching \citep{Moster12,Reddick12,moster-09,Guo-09,Wang09}, from HOD/CLF modeling \citep{Zheng07,Yang11}, and from cluster catalogs \citep{yang-08,Hansen09,LinMohr04}.  Grey shaded regions correspond to the 68\% confidence contours of \cite{Behroozi10}. The one-sigma posterior distribution for our model is shown by the red error bars.}
\label{f:comp_z0}
\plotgrace{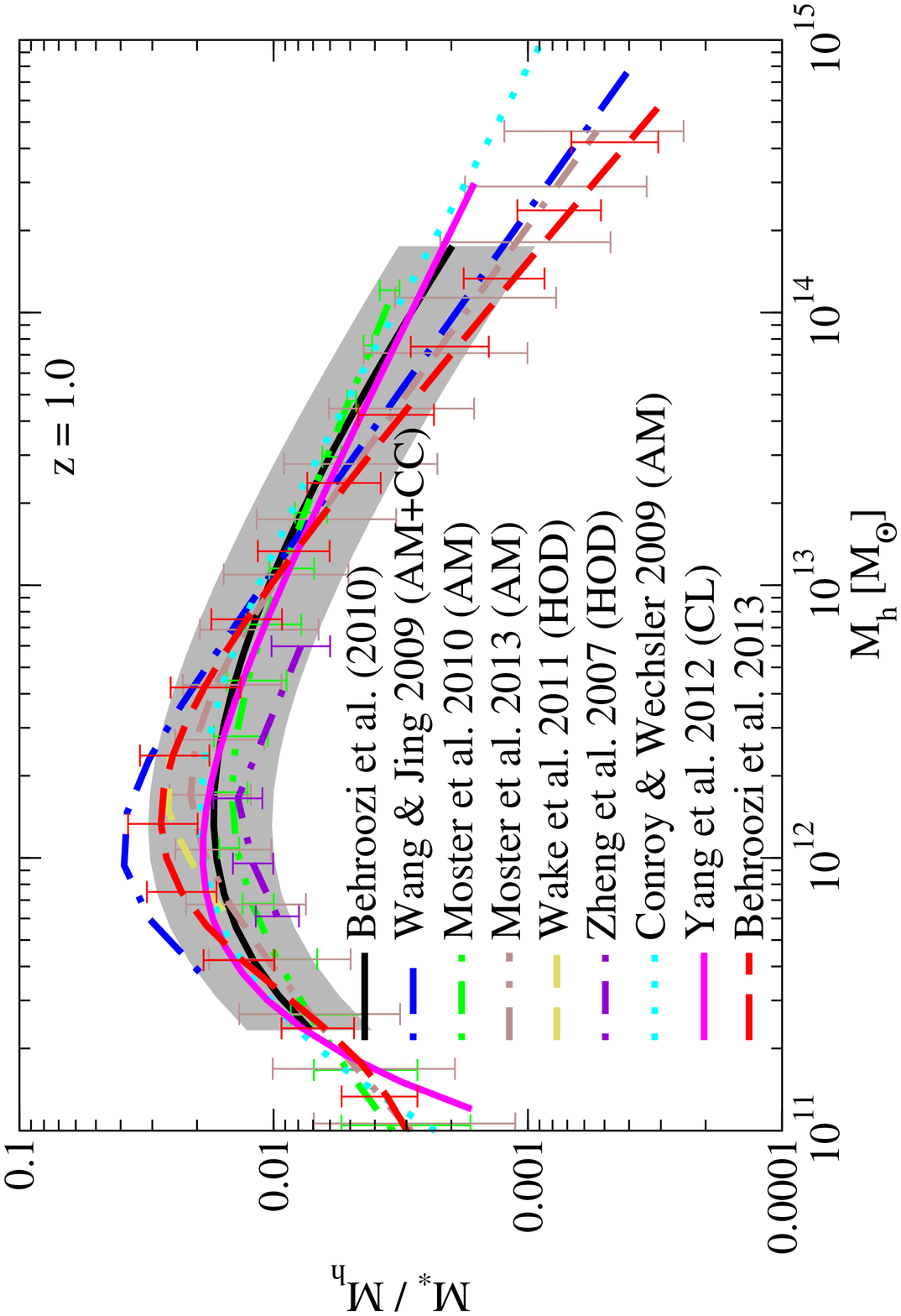}\plotgrace{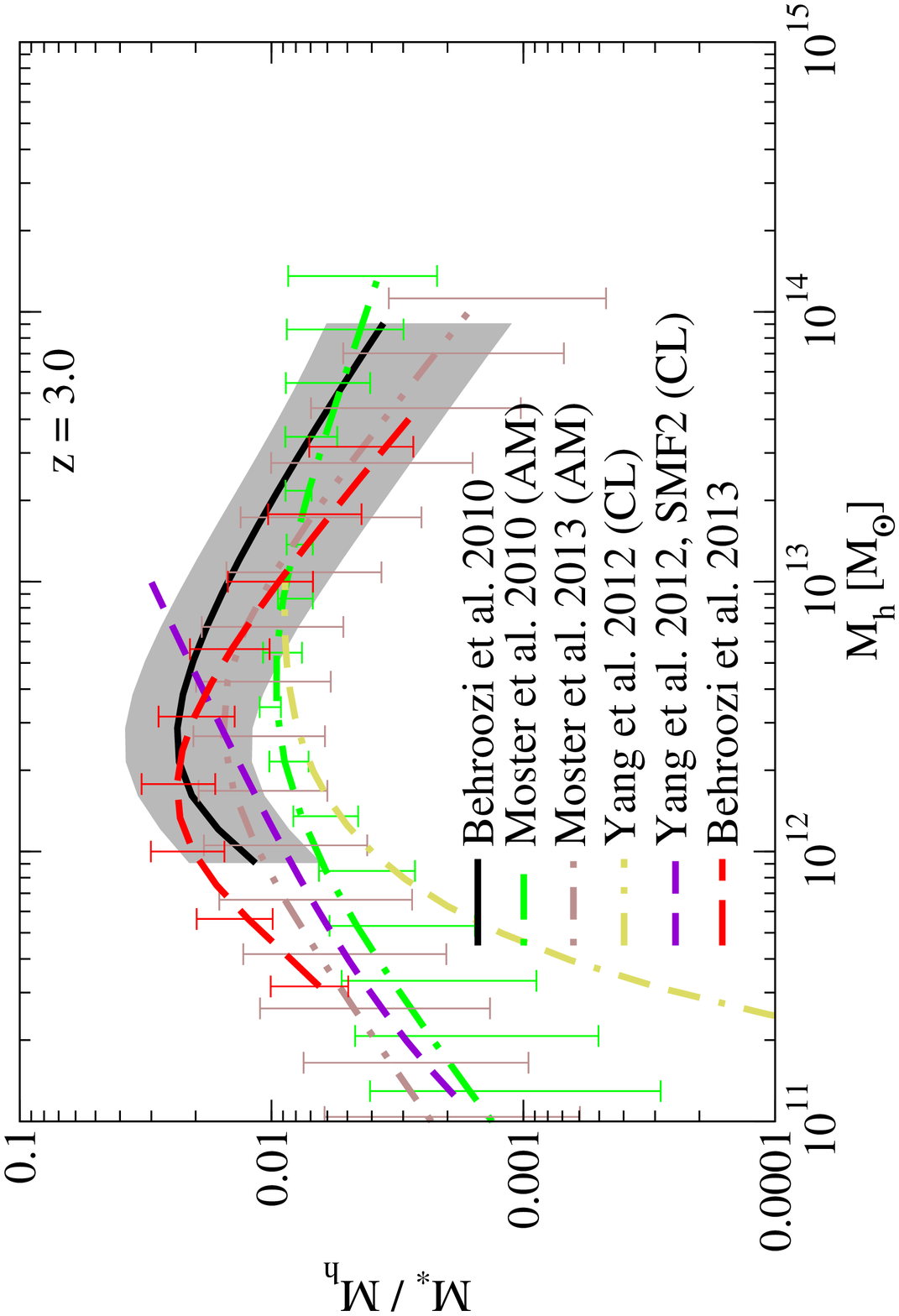}
\caption{Comparison of our best-fit model at $z=1.0$ and $z=3.0$ to previously published results.  Results compared include those from our previous work \citep{Behroozi10}, from abundance matching \citep{Moster12,moster-09,cw-08,Wang09}, and from HOD/CLF modeling \citep{Zheng07,Yang11,Wake11}.  \cite{Yang11} reports best fits for two separate stellar mass functions, and we include both at $z=3.0$.  Grey shaded regions correspond to the 68\% confidence contours of \cite{Behroozi10}.}
\label{f:comp_z1}
\end{figure*}

\subsection{Comparison to Other Results}

\label{s:comparison}

\begin{figure}[h]
\vspace{-6ex}
\plotgrace{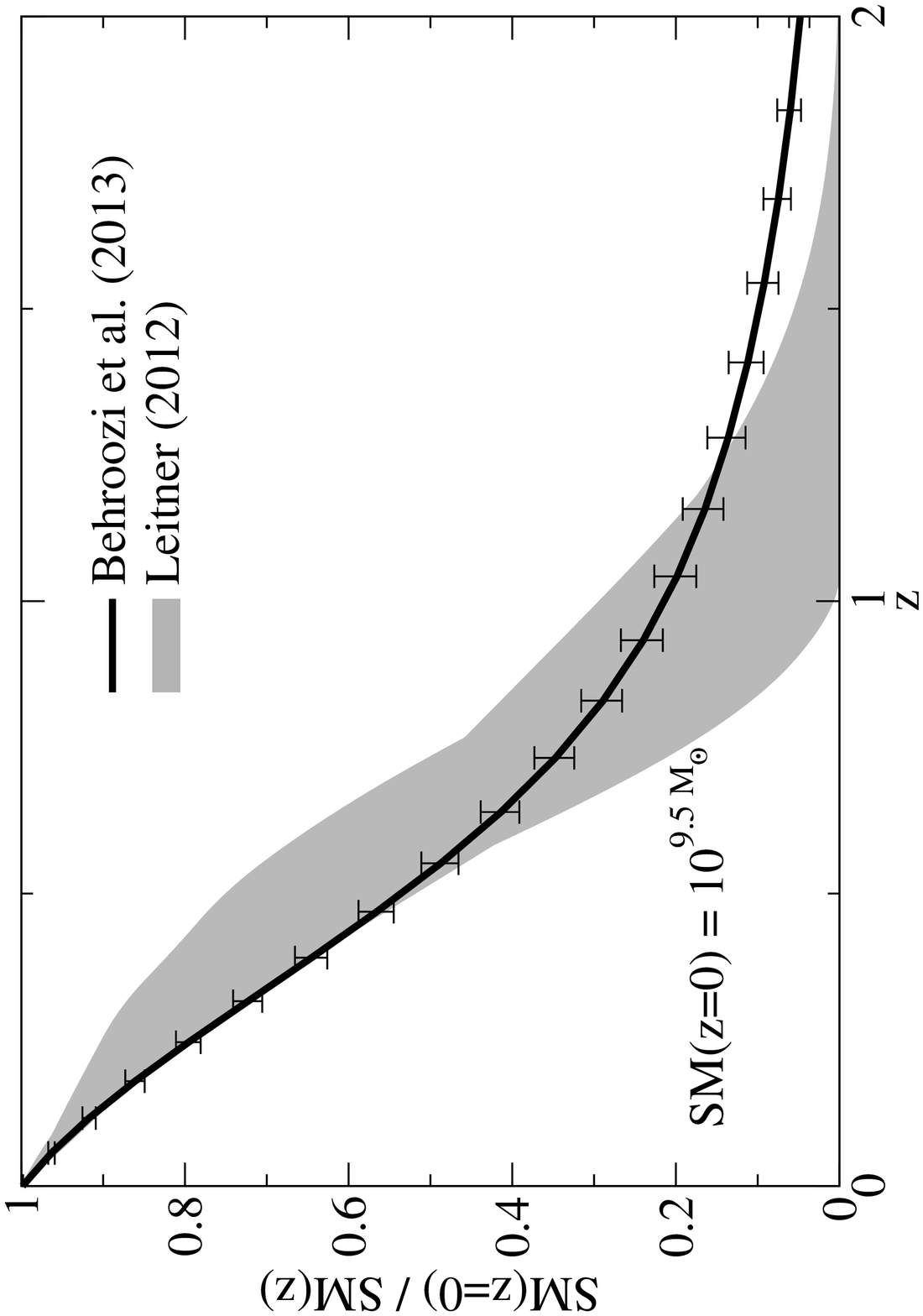}\\[-6ex]
\plotgrace{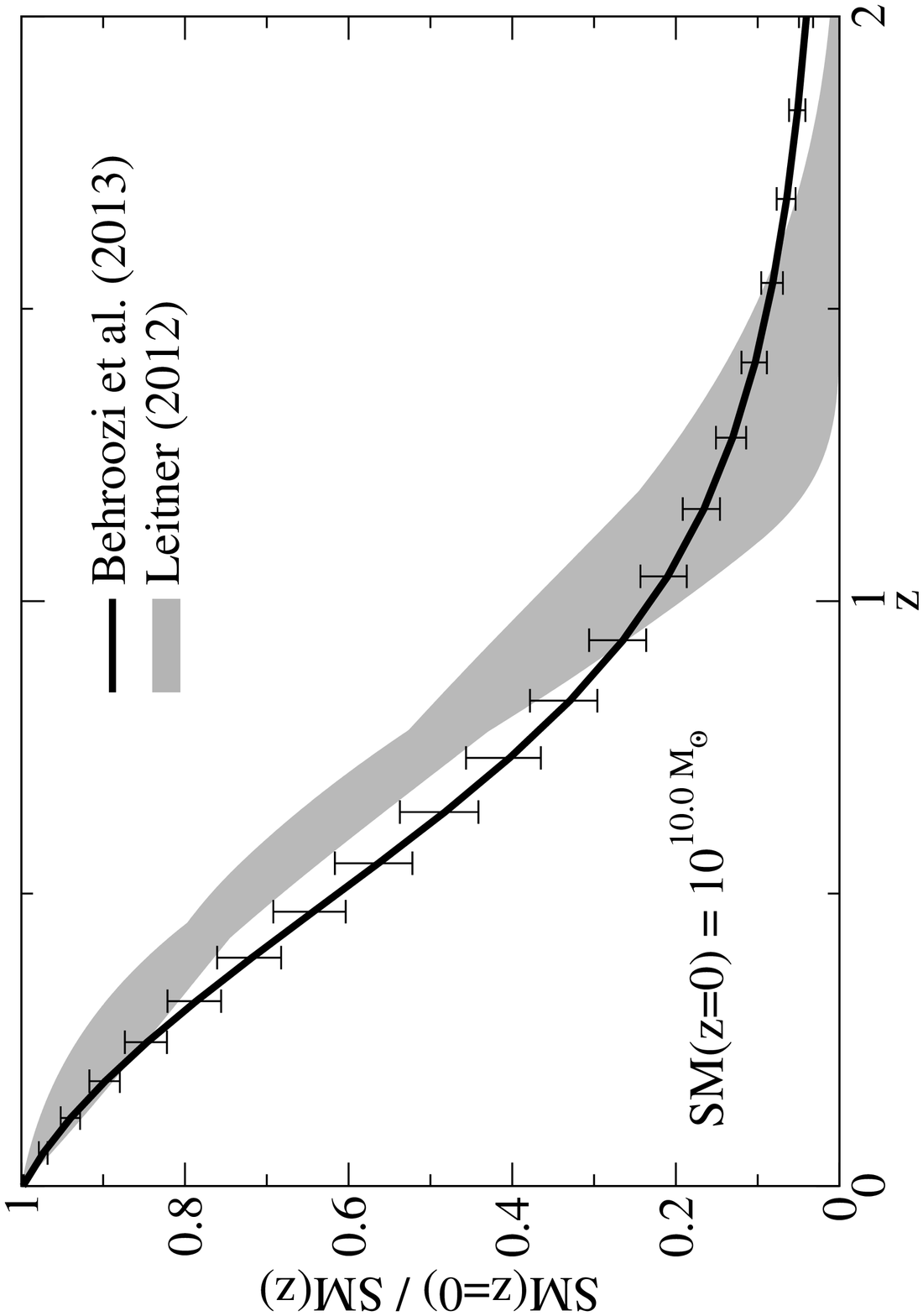}\\[-4ex]
\caption{Comparison of our best-fit model's stellar mass histories to those of \cite{Leitner11} (lines; shaded regions
show the one-sigma posterior distribution of our model).  Note that the 
\cite{Leitner11} results only consider star forming galaxies; ours show the average histories for all galaxies.}
\label{f:sm_tracks}
\end{figure}

We show a comparison of our best-fit results for the stellar mass to halo mass ratio at $z=0.1$ to previously-published results in Fig.\ \ref{f:comp_z0}, and we show comparisons at $z=1.0$ and $z=3.0$ in Fig.\ \ref{f:comp_z1}.  Where possible, conversions to our assumed cosmology and halo mass definition have been applied.  The results in this work are almost identical to our previous results in \cite{Behroozi10}, with the exception of a deviation at low halo masses due to the updated stellar mass functions used.  We also compare to the new constraint of \cite{Reddick12}, which uses additional input from the correlation function and conditional stellar mass function as measured by SDSS.  There is a slight discrepancy between the two results for small masses due to the different satellite fractions obtained using $\mpeak$ and $\vpeak$ (still within our systematic errors), but the \cite{Reddick12} result is only well constrained by additional data in the range $10^{12}$--$10^{14}\Msun$.

Comparisons to other data sets at $z \le 1.0$ \citep{moster-09,Guo-09,Wang09,Zheng07,yang-08,Hansen09,LinMohr04} are discussed extensively in \cite{Behroozi10}, so we do not repeat that discussion here.  The $z=3$ comparison is notable because it illustrates the large discrepancies that can occur when different stellar mass functions are used at $z\gtrsim 3$.  For example, \cite{Yang11} perform modeling for two separate high-$z$ stellar mass functions (\citealt{perezgonzalez-2008} and \citealt{Drory05}).   Part of the discrepancy between the two sets of results may be due to a somewhat restrictive redshift fit, and part of it may be due to systematic biases in stellar masses at high redshifts.

In Fig.\ \ref{f:sm_tracks}, we show a comparison of galaxy stellar mass histories between our results and those of \cite{Leitner11}.  A direct comparison is difficult because \cite{Leitner11} only considers star-forming galaxies.  However, according to Eq.\ \ref{e:passive}, the active fraction for $10^{9.5}\Msun$ galaxies is 90\%, and the active fraction for $10^{10}\Msun$ galaxies is 65\%---although most of the passive $10^{10}\Msun$ galaxies have only recently become passive.  Comparison of the stellar mass histories for those galaxies is thus somewhat more feasible.  Our stellar mass histories agree remarkably well with those in \cite{Leitner11} except at recent times for $10^{10}\Msun$ galaxies (where our model includes more passive galaxies) and at early times for $10^{9.5}\Msun$ galaxies, where observations do not constrain properties of their progenitors.

\begin{figure}
\vspace{-6ex}
\plotgrace{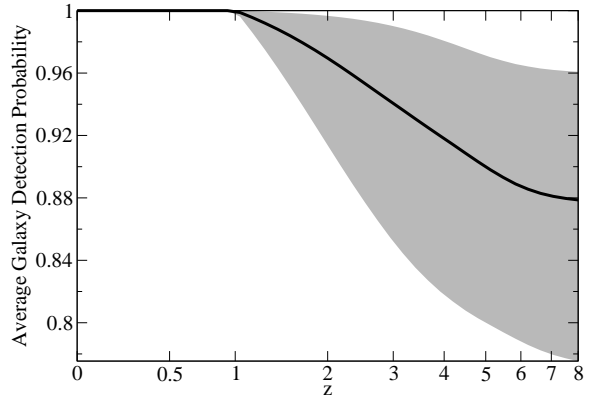}\\[-4ex]
\caption{Allowed range for the completeness (i.e., probability of galaxy detection) of the surveys we use as a function of redshift; the black line shows the median value, and the grey bands show 68\% confidence contours. This average applies only to galaxies above authors' stated completeness limits.  No strong evidence is seen for missing large numbers of high-redshift galaxies due to burstiness or dustiness.}
\label{f:systematics}
\end{figure}

\subsection{Systematic Uncertainties}

\label{s:maj_uncertainties}

Several aspects of the allowed parameter range are notable.  Most importantly, there is no necessity for a large number of galaxies to be missed (due to dustiness or burstiness) at high redshift; indeed, the allowed incompleteness is in general less than 30\%, as shown in the top panel of Fig.\ \ref{f:systematics}.  This would imply that most high-redshift surveys are not missing a large fraction (>50\%) of galaxies above their nominal detection thresholds.

The allowed parameter range is consistent with having both $\mu=0$ and $\kappa=0$ at $z=0$; i.e., no systematic offsets in stellar mass are necessary at $z=0$.  This is partially by design: the parametrization of the SMHM was chosen so that it could fit the $z=0$ constraints without need for $\mu$ or $\kappa$; moreover, the cosmic and specific star formation rates are derivative constraints, meaning that they only constrain how the SMHM evolves with time.   While the main fits favor some redshift evolution in the systematic parameters, it is still possible to reasonably fit all observations while disallowing all systematic corrections (i.e., assuming that the stellar masses and star formation rates used in this paper are on average correct and that all the surveys were 100\% complete).  Details of this fit are presented in Appendix \ref{a:models}.  We also present discussions therein of how the allowed parameter space varies when certain data sets are not used to constrain the fit; specifically, excluding $z>6$ data; excluding cosmic star formation constraints; and excluding all star formation constraints.

\subsection{Fits to Star Formation Histories}

\label{s:new_fits}

\begin{figure*}
\vspace{-6ex}
\plotgrace{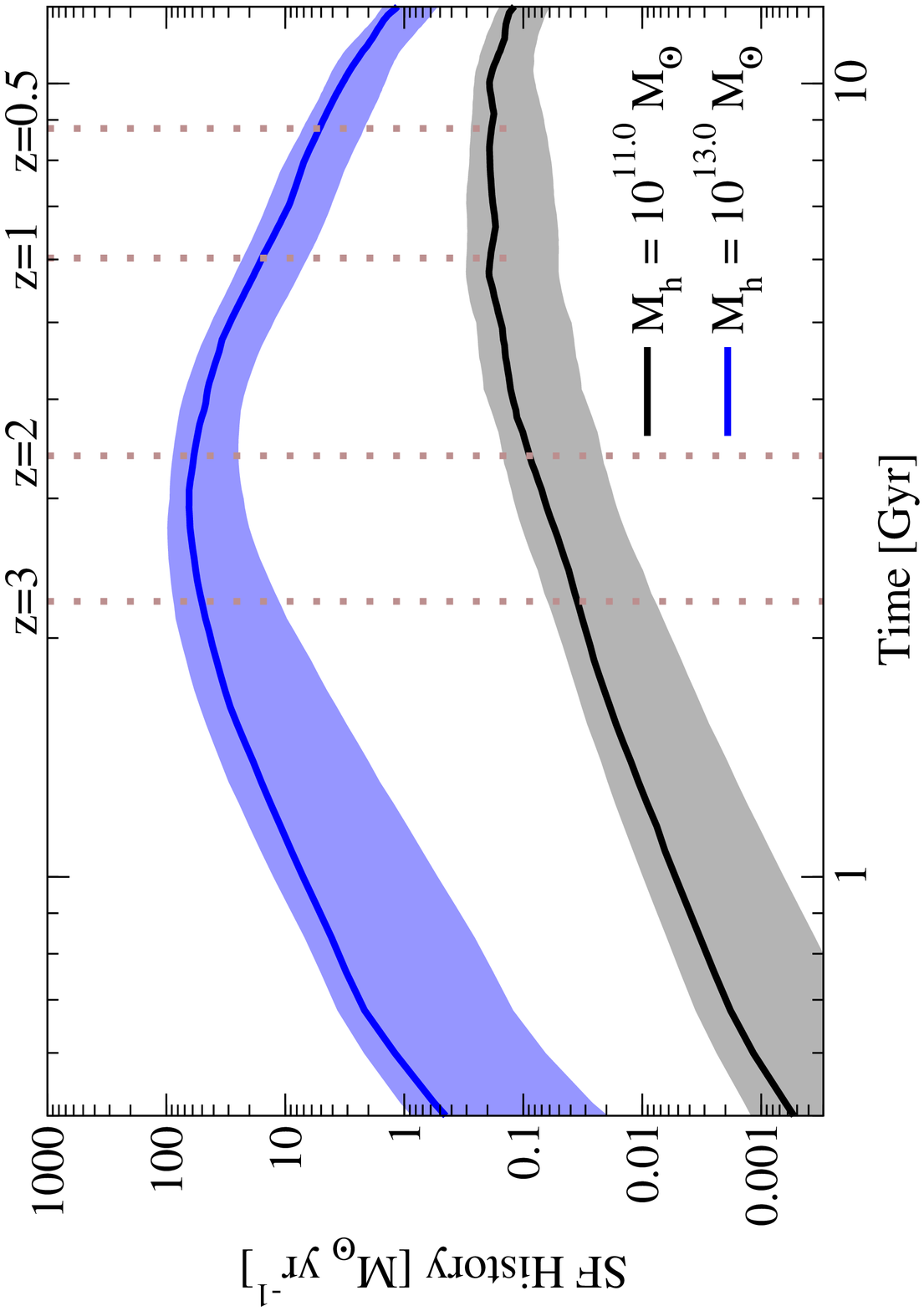} \plotgrace{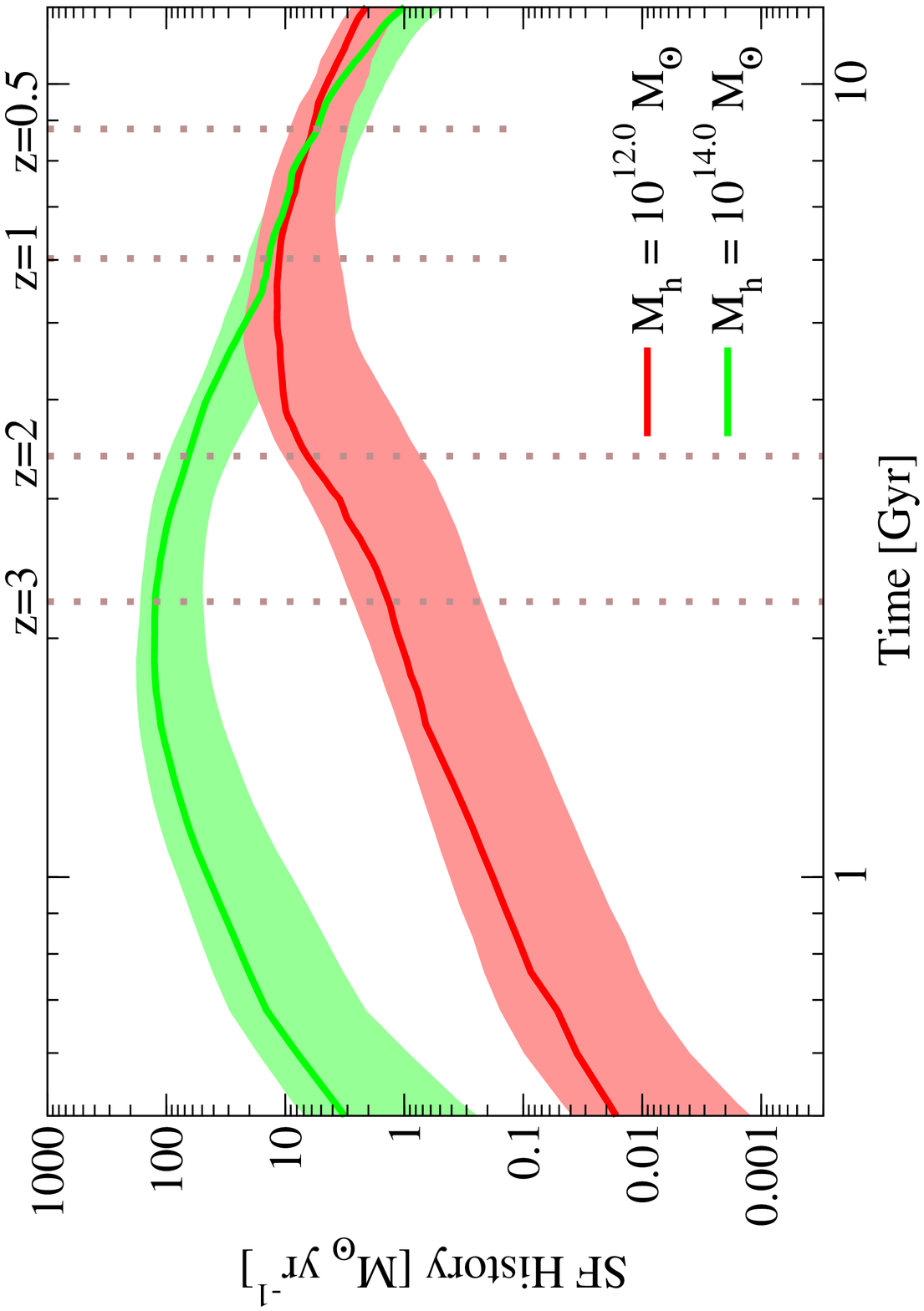}\\[-4ex]
\caption{Featureless, rising power law star formation histories are appropriate for $z>3$; however, for $z<3$, galaxies have peaks in their star formation rates that depend on halo mass.  These figures show constraints on individual star formation histories of galaxies in halos of mass $10^{11} - 10^{14}\Msun$ as a function of time from the beginning of the universe to $z=0.1$.  These halo masses correspond to stellar masses of $\sim10^{9}$, $3\times10^{10}\Msun$, $10^{11}\Msun$, and $2\times 10^{11}\Msun$ at $z=0$, respectively.}
\label{f:sfh_fits}
\end{figure*}

\begin{figure*}
\vspace{-6ex}
\plotgrace{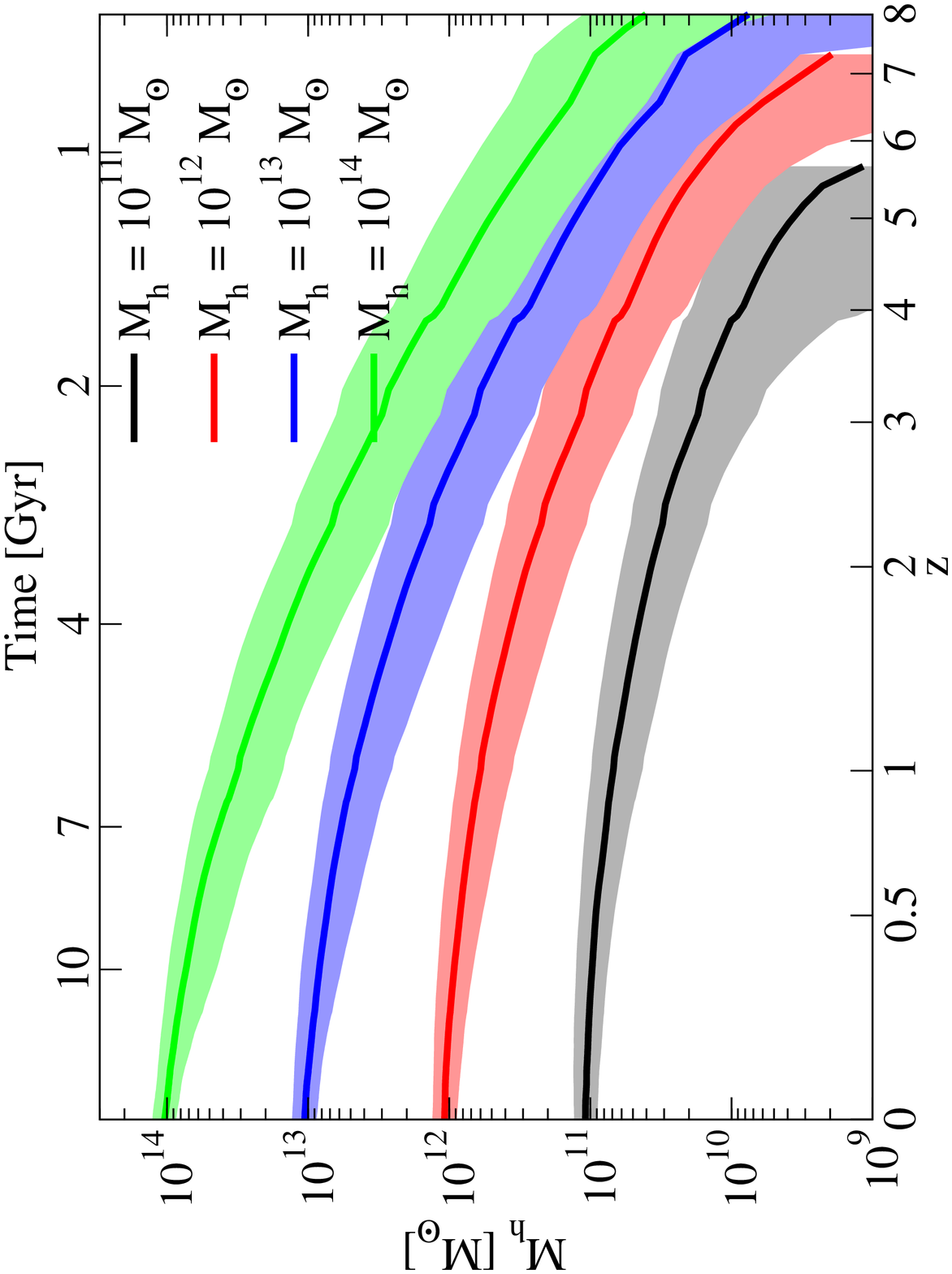}\plotgrace{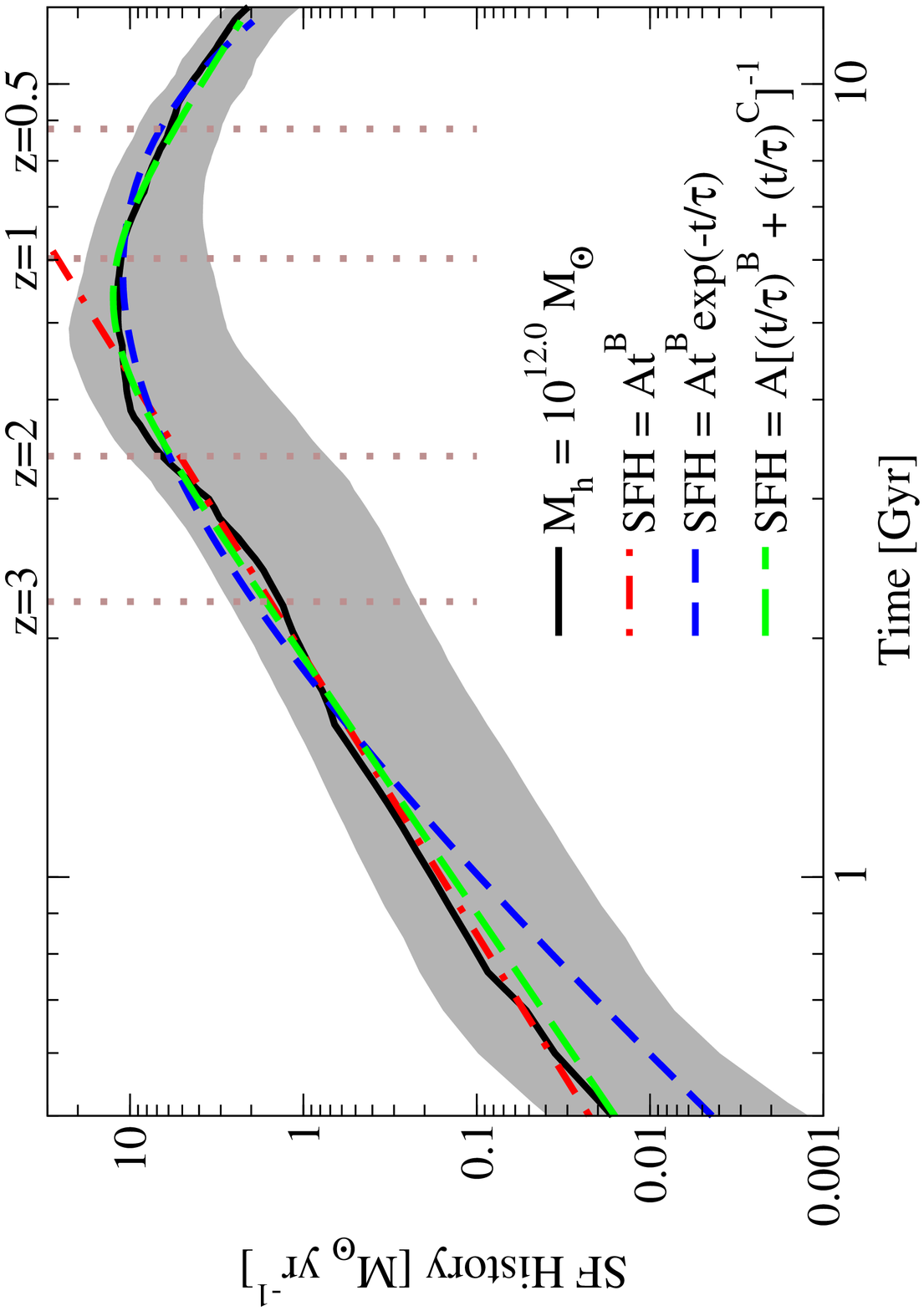}\\[-4ex]
\caption{\textbf{Left} panel: Median mass accretion histories for halos in narrow mass bins (0.25 dex).  The shaded regions contain 68\% of the spread in mass accretion histories for different halos. \textbf{Right} panel: Constraints on individual star formation histories for galaxies in $10^{12}\Msun$ halos, compared with fits from Eq.\ \ref{e:sfh_highz}-\ref{e:sfh_ok}.}
\label{f:mah_sfht}
\end{figure*}

The most commonly used star formation history fitting formula is a declining exponential with time ($SFH(t)\propto e^{-t/\tau}$), used by almost all the observational papers cited in this work (see references in Tables \ref{t:smf_data} and \ref{t:ssfr_data}). 
Various modifications to this form have been suggested (e.g., $t e^{-t/\tau}$ or $e^{+t/\tau}$ in \citealt{Maraston10}), and there has been increasing recent evidence that increasing star formation histories are more appropriate for massive galaxies at high redshifts.

Our best constraints in this study are on \textit{average} star formation histories for halos at a given mass.  Thus, we can place the strongest priors on the expected form of stacked galaxy star formation histories.  However, we can also place some weaker constraints on what individual galaxy star formation histories should look like.

Individual galaxies' histories can differ from the average star formation histories for three reasons.  The first is stochasticity in the star formation rate on short timescales relative to the dynamical time of the halo, due to feedback effects as well as shot noise in the star formation rate.  This variation cannot be modeled with our current approach, so we caution that the constraints on individual galaxy histories that we derive should be considered as smoothed over a period of at least 500 Myr.  If enough color bands are available to avoid degeneracies with other fitting parameters, observers may consider adding the possibility of recent starbursts to their models to account for the fact that they can significantly alter observed galaxy colors.

The next reason for deviations is sustained deficits in the star formation rate for satellite galaxies compared to field galaxies \citep[e.g.][]{Wetzel11}.  As shown in Fig.\ \ref{f:mf_satfit} (see also e.g. \citealt{Kravtsov04}), the satellite fraction of halos falls both with increasing mass and increasing redshift.  This is not a practical concern for $z>3$, where the satellite fraction is less than 20\% for all currently-observable galaxies ($M_\ast > 10^{9.5}\Msun$).  At lower redshifts, the observation that satellites cease to form stars relatively quickly after accretion \citep{Wetzel11} means that declining star formation histories may continue to be reasonable fits for satellite galaxies, even though most other galaxies of the same mass may be better fit by rising star formation histories.

Finally, we note that differences can arise in individual galaxy star formation histories because of the scatter in stellar masses at a given halo mass, as well as the scatter in mass accretion histories for halos (see Fig.\ \ref{f:mah_sfht}).  We can attempt to model these effects by sampling random mass accretion histories from the Bolshoi simulation and then sampling random stellar mass growth histories as allowed by the scatter in stellar mass at fixed halo mass (i.e., choosing random stellar mass offsets at $z=8$, $z=1.0$, and  $z=0$ and using spline interpolation at intermediate times).  Our results, expressed as a function of time since the beginning of the universe, are shown in Fig.\ \ref{f:sfh_fits}.

These results have dramatically larger error bars than those in Fig.\ \ref{f:sfr_sfh2} on account of the spread in mass accretion histories for halos.  Nonetheless, some basic conclusions can be drawn.  Star formation histories for $z>3$ galaxies increase with time.
Although there are different ways to parametrize these histories, a straightforward one is a direct power law form:
\begin{equation}
\label{e:sfh_highz}
SFH(t) = A t^B\;\textrm{(for $z>3$)}.
\end{equation}
This is also equally capable of fitting the average star formation rates in Fig.\ \ref{f:sfr_sfh2} at $z>3$.  Our result is similar to that of \cite{Papovich11}; however, we find steeper slopes ($B \sim 3-4$) than theirs ($B \sim 1.7$) because they ignore all effects of mergers that occur from $z=8$ to $z=3$.

At lower redshifts, there is a mass-dependent turnover after which the star formation history begins to decline.  This happens at $z\sim 2-3$ for $10^{11}\Msun$ galaxies, making a declining exponential at $z=0$ a reasonable fit.  However, it happens as late as $z=0.5$ for $10^{9.5}\Msun$ galaxies---meaning that a declining exponential is \textit{never} a good fit for these galaxies unless they are satellites.  The best fit in general for all the constraints on individual histories presented here is a double power law:
\begin{equation}
\label{e:sfh_general}
SFH(t) = A \left[\left(\frac{t}{\tau}\right)^B + \left(\frac{t}{\tau}\right)^{-C}\right]^{-1};
\end{equation}
typical values of $B$ and $C$ range from $1$ to $5$.  It may often be the case that the available data is insufficient to constrain all three shape parameters.  In this case, a hybrid of the exponential decline and power law rise still provides a reasonable fit:
\begin{equation}
\label{e:sfh_ok}
SFH(t) = A t^B \exp(-t/\tau).
\end{equation}

\section{Discussion}

\label{s:discussion}

The existence of a ``cold mode'' of gas accretion that allows efficient star formation at high redshifts and shuts off for massive galaxies past $z\sim 2$ has been predicted in hydrodynamical simulations for the past decade \citep{Birnboim03}.  In this work, we observe a transition after $z\sim3$, where galaxies in massive halos become progressively less efficient at forming stars, even after correcting for slowing accretion rates since that time (Fig. \ref{f:smhist}, right panel).  At fixed halo mass (Fig. \ref{f:smhist}, left panel), the transition is less severe, but there has been a clear reduction in the conversion efficiency since $z\sim 3$ to the present day.  This suggests a picture in which dense infalling gas can cool to the galaxy disk efficiently at high redshifts; however, at lower redshifts, the infalling gas becomes sparse enough that it is susceptible either to shock heating or feedback from an active galactic nucleus.

Another important feature of our results is that merging galaxies' stars almost never reach the central galaxy in massive halos ($<10\%$ since $z=1$ for halos larger than $10^{13}\Msun$).  This claim is not new \citep{purcell-etal-07,Conroy07}, but it is reassuring that we duplicate this conclusion with high confidence.  This is in stark contrast to what is asserted in \citep{Moster12} (i.e., 80\% of merging stars reach the central galaxy); however, their assertion is based on simulations of Milky-Way sized halos, for which we find that indeed $\sim 70\%$ of merging galaxies' stars reach the central galaxy.  Thus, the fraction of merging stars that reach the central galaxy is likely a strong function of halo mass.

It is worth mentioning that the SFRs derived in \cite{Moster12} appear reasonable despite this issue.  We note that \cite{Moster12} assumes that satellite stellar mass is fixed at the epoch of accretion, which results in less stellar mass in satellites than in this work.  Additionally, \cite{Moster12} assumes a higher value of the scatter in observed stellar mass vs. true stellar mass ($\sigma$) than we do, which comes from comparing technique-to-technique scatter in recovering observed stellar masses.  This results in a much larger allowed evolution for massive galaxies between $z=1$ and $z=0$, which is very sensitive to the value of $\sigma$ \citep{Behroozi10}.  The combination of these two assumptions (less stellar mass in satellites and greater allowed evolution in the stellar mass of massive galaxies) roughly cancels the effect on the SFR of assuming that more stellar mass in satellites is allowed to merge into the central galaxy.  We note, however, that our satellite stellar masses are already on the low side of what is allowed by clustering constraints \citep{Reddick12}, leaving little room to reduce them further as is done in the \cite{Moster12} analysis.  In addition, the scatter in the difference between two stellar mass estimation techniques tends to overestimate the scatter between any one technique and the true value by a factor of $\sim\sqrt{2}$.  This factor is consistent with other independent estimates of $\sigma$, for example, by evaluating the posterior distribution of allowed recovered stellar masses \citep{Kajisawa09,Conroy09}.  Thus, both the actual evolution in the stellar mass of massive galaxies and the fraction of stars in mergers which reach the central galaxy may not be as large the values implied by the \cite{Moster12} analysis.

We note further that the amount of stellar mass that was contained in disrupted galaxies can easily reach five times the amount of stellar mass in the central galaxy for high-mass halos (Fig.\ \ref{f:sm_icl}), consistent with observations in \cite{Gonzalez05}.  Regardless of uncertainties in determining stellar masses, there is no possible way to funnel any significant fraction of this mass into the central galaxy without badly violating observed limits on the number density of massive galaxies.  Rather, stars from merging galaxies must end up in a more extended distribution to avoid confusion with the central galaxy, a topic that will be explored in depth in a future paper (Behroozi et al., in prep).  From the opposite perspective, galaxy stellar populations contain mostly stars from the most-massive progenitor, with some contamination from major mergers.    Surveys of stars within a galaxy's optical radius (such as for the Milky Way) thus capture a potentially limited range of different star-forming environments, at least for stars formed since $z=1$.  A wider range of environments can be sampled by considering the ICL and halo stars in general.

The most efficient halo mass for converting baryons into stars appears to be close to $10^{12}\Msun$, regardless of redshift.  The progenitors of today's massive halos ($\gtrsim 10^{14}\Msun$) passed through this mass quickly, and as a result formed most of their stars in a short time: a mere 1--2 Gyr (Fig.\ \ref{f:smhist3}).  Given that baryon to star conversion becomes so inefficient above and below this halo mass, much of the shape of the star formation histories seen in this paper (Fig.\ \ref{f:sfr_sfh2}) could be explained by considering the duration of time that a given halo's mass accretion history spends close to this most efficient halo mass.  That is to say, the peakiness of the star formation history depends on how quickly the halo mass accretion history passes through halos of mass $\sim 10^{12}\Msun$.  As lower-mass halos have slower relative accretion rates, they have flatter star formation histories; e.g., Milky Way-sized halos have lingered in this mass range from $z\sim 1$, and so have the current highest stellar mass to halo mass ratio, as well as star formation histories that track their halo mass accretion rates well at late times.

Low-mass halos ($10^{11}\Msun$) show an unusual behavior in terms of their baryon conversion efficiency.  In the left panel of Fig.\ \ref{f:smhist}, it would appear that the instantaneous baryon conversion efficiency of $10^{11} \Msun$ halos is high at high redshifts ($z>5$), low at intermediate redshifts ($2<z<5$), and then high again at low redshifts.  While this behavior matches what is seen in the integrated baryon conversion efficiency for such halos (Fig.\ \ref{f:smhist2}, left panel), the corresponding galaxies are often at the faint edge of the completeness limit for stellar mass surveys.  For that reason, observational biases may be the most likely source of this behavior.

Our results have importance for future observational studies.  The current levels of systematic errors in the inference of stellar masses and star formation rates are the dominant systematic uncertainty in this work.  As discussed in \S \ref{s:data}, stellar mass functions at $z>3$ can disagree by up to 0.5 dex, well beyond any errors expected from sample variance.  Star formation rates are equally concerning, both at high redshifts and at low redshifts, especially in terms of specific star formation rates.  As a result, systematic errors are the single most important aspect preventing better understanding of the formation histories of massive galaxies.  As discussed in \S \ref{s:new_fits}, we have provided new fitting formulas for galaxy star formation histories, which should help reduce one of the largest sources of systematic error especially for low-mass and high-redshift galaxies.
New studies of stellar mass functions and star formation rates from deep multiwavelength surveys like CANDELS 
\citep{Koekemoer11, Grogin11}, analyzed with self-consistent star formation histories and assumptions about systematic
errors, will provide a large step forward in constraining the model presented here.

\section{Conclusions}

\label{s:conclusions}

We have presented a comprehensive analysis of average star formation rates and histories in galaxies from $z=0$ to $z=8$,
and their connection to the underlying growth and merging of dark matter halos, along with a treatment of the inherent uncertainties.  Our approach provides a self-consisent picture of the growth of galaxies over this epoch in the cosmological
context of $\Lambda$CDM structure formation. The model is able to match the evolution of cosmic star formation rates, specific star formation rates, and the stellar mass function of galaxies over the last 13.2 billion years of cosmic time.
Our main findings are as follows:
\begin{enumerate}
\item Halos of mass $\sim 10^{12}\Msun$ appear to be the most efficient at forming stars at every epoch, with baryon conversion efficiencies of 20-40\% over nearly the entire redshift range of this study.  Halos at higher and lower masses are less efficient by orders of magnitude, especially at low redshifts.
\item The baryon conversion efficiency of more massive halos is still reasonably high (10-20\%) until $z=2-3$, after which time it takes a steep downturn.  This is consistent with expectations of reduced cold mode accretion onto the corresponding galaxies.
\item We have characterized the fraction of stars in galaxies due to merging vs star formation.  At $z>1$, galaxy buildup
in all but the most massive galaxies was strongly dominated by star formation.  At the present day, the transition between
merger-dominated buildup (at high mass) and star formation-dominated buildup (at low mass) occurs at roughly the mass
of the Milky Way.
\item We confirm previous results in support of most merging galaxies' stars being disrupted in the ICL for massive halos.  We find, however, that the fraction of merging stars that reach the central galaxy may increase strongly with decreasing halo mass, to about 70\% for Milky Way-sized halos.
\item For massive galaxies ($M_\ast > 10^{11}\Msun$), we predict that at least as much stellar mass is present in the ICL as in the galaxy itself out to intermediate redshifts, possibly as high as $z\sim 1$ for the largest clusters (although this depends on the exact cut between ICL and galaxy stellar mass).
\item Star formation histories peak at $z \sim 3$ for massive galaxies and are increasing nearly to the present day for small galaxies.  We have presented new fitting formulas for galaxy star formation histories: a featureless power law for high redshifts (Eq.\ \ref{e:sfh_highz}), and more general formulas valid for all redshifts (Eq.\ \ref{e:sfh_general} and \ref{e:sfh_ok}).
\item Upturns in the stellar mass function at low stellar masses mean that dwarf galaxies ($M_\ast < 10^{8.5}\Msun$) have a higher stellar mass to halo mass ratio than that predicted by extrapolating results from more massive galaxies.  The upturn is most robustly observed at low redshifts, but there are also indications at higher redshifts for a steepening slope to the SMF at low stellar masses.  The stellar mass to halo mass relation thus cannot be fit with a double power law alone.
\item Systematic errors remain at high levels ($\sim 0.2-0.3$ dex in stellar masses, $0.15-0.3$ dex in SFRs), and disagreement between galaxy abundances for high-redshift surveys ($z>3$) is substantial.
\item Galaxies with stellar masses below $10^{10}\Msun$ have the least well-constrained properties in this study.   This could be improved by future surveys and by incorporating information on the star formation histories of local galaxies.
\end{enumerate}

\acknowledgments
Support for this work was provided by an HST Theory grant.  Program number HST-AR-12159.01-A was provided by NASA through a grant from the Space Telescope Science Institute, which is operated by the Association of Universities for Research in Astronomy, Incorporated, under NASA contract NAS5-26555.  This research was also supported in part by the National Science Foundation under Grant No. NSF PHY11-25915, through a grant to KITP during the program ``First Galaxies and Faint Dwarfs'',
and through a ARCS Foundation Graduate Fellowship to PSB.

We thank Anatoly Klypin and Joel Primack for access to the Bolshoi and MultiDark simulations and Michael Busha and the LasDamas collaboration for access to the Consuelo simulation.  The latter was run using computational resources at SLAC.  We thank John Moustakas for access to the PRIMUS stellar mass functions prior to publication.  This work has benefitted from many interesting conversations with Tom Abel, Steve Allen, Andrew Benson, Andreas Berlind, Kevin Bundy, Joanne Cohn, Sandy Faber, Dusan Keres, David Koo, Andrey Kravtsov, Alexie Leauthaud, Avi Loeb, Yu Lu, Chung-Pei Ma, John Moustakas, Joel Primack, Rachel Reddick, Beth Reid, Daniel Stark, Chuck Steidel, Tommaso Treu, David Wake, Andrew Wetzel, Frank van den Bosch, Alexey Vikhlinin, and Naoki Yoshida, We thank Yu Lu, Ari Maller, Simone Wienmann, and Anatoly Klypin for helpful comments on an earlier draft, and Surhud More and Michal Micha\l{}owski for catching typographical mistakes.  We additionally the anonymous referee for many suggestions which improved this paper.

\bibliography{master_bib}

\begin{thebibliography}{153}
\expandafter\ifx\csname natexlab\endcsname\relax\def\natexlab#1{#1}\fi

\bibitem[{{Baldry} {et~al.}(2008){Baldry}, {Glazebrook}, \&
  {Driver}}]{Baldry08}
{Baldry}, I.~K., {Glazebrook}, K., \& {Driver}, S.~P. 2008, \mnras, 388, 945

\bibitem[{{Behroozi} {et~al.}(2010){Behroozi}, {Conroy}, \&
  {Wechsler}}]{Behroozi10}
{Behroozi}, P.~S., {Conroy}, C., \& {Wechsler}, R.~H. 2010, \apj, 717, 379

\bibitem[{{Behroozi} {et~al.}(2012){Behroozi}, {Loeb}, \&
  {Wechsler}}]{BehrooziUnbound}
{Behroozi}, P.~S., {Loeb}, A., \& {Wechsler}, R.~H. 2012, arXiv:1208.0334

\bibitem[{{Behroozi} {et~al.}(2013{\natexlab{a}}){Behroozi}, {Wechsler}, \&
  {Conroy}}]{BehrooziEvolution}
{Behroozi}, P.~S., {Wechsler}, R.~H., \& {Conroy}, C. 2013{\natexlab{a}},
  \apjl, 762, L31

\bibitem[{{Behroozi} {et~al.}(2013{\natexlab{b}}){Behroozi}, {Wechsler}, \&
  {Wu}}]{Rockstar}
{Behroozi}, P.~S., {Wechsler}, R.~H., \& {Wu}, H.-Y. 2013{\natexlab{b}}, \apj,
  762, 109

\bibitem[{{Behroozi} {et~al.}(2013{\natexlab{c}}){Behroozi}, {Wechsler}, {Wu},
  {Busha}, {Klypin}, \& {Primack}}]{BehrooziTree}
{Behroozi}, P.~S., {Wechsler}, R.~H., {Wu}, H.-Y., {Busha}, M.~T., {Klypin},
  A.~A., \& {Primack}, J.~R. 2013{\natexlab{c}}, \apj, 763, 18

\bibitem[{{Berlind} \& {Weinberg}(2002)}]{Berlind02}
{Berlind}, A.~A., \& {Weinberg}, D.~H. 2002, \apj, 575, 587

\bibitem[{{Bernardi} {et~al.}(2010){Bernardi}, {Shankar}, {Hyde}, {Mei},
  {Marulli}, \& {Sheth}}]{Bernardi10}
{Bernardi}, M., {Shankar}, F., {Hyde}, J.~B., {Mei}, S., {Marulli}, F., \&
  {Sheth}, R.~K. 2010, \mnras, 404, 2087

\bibitem[{{Berrier} {et~al.}(2006){Berrier}, {Bullock}, {Barton}, {Guenther},
  {Zentner}, \& {Wechsler}}]{Berrier06}
{Berrier}, J.~C., {Bullock}, J.~S., {Barton}, E.~J., {Guenther}, H.~D.,
  {Zentner}, A.~R., \& {Wechsler}, R.~H. 2006, \apj, 652, 56

\bibitem[{{B{\'e}thermin} {et~al.}(2012){B{\'e}thermin}, {Dor{\'e}}, \&
  {Lagache}}]{Bethermin12}
{B{\'e}thermin}, M., {Dor{\'e}}, O., \& {Lagache}, G. 2012, \aap, 537, L5

\bibitem[{{Birnboim} \& {Dekel}(2003)}]{Birnboim03}
{Birnboim}, Y., \& {Dekel}, A. 2003, \mnras, 345, 349

\bibitem[{{Blanton} \& {Roweis}(2007)}]{blanton-roweis-07}
{Blanton}, M.~R., \& {Roweis}, S. 2007, \aj, 133, 734

\bibitem[{Borgani \& Kravtsov(2011)}]{Borgani11}
Borgani, S., \& Kravtsov, A. 2011, Advanced Science Letters, 4, 204

\bibitem[{{Bouwens} {et~al.}(2012){Bouwens}, {Illingworth}, {Oesch}, {Franx},
  {Labb{\'e}}, {Trenti}, {van Dokkum}, {Carollo}, {Gonz{\'a}lez}, {Smit}, \&
  {Magee}}]{Bouwens11b}
{Bouwens}, R.~J., {et~al.} 2012, \apj, 754, 83

\bibitem[{{Bouwens} {et~al.}(2011){Bouwens}, {Illingworth}, {Oesch},
  {Labb{\'e}}, {Trenti}, {van Dokkum}, {Franx}, {Stiavelli}, {Carollo},
  {Magee}, \& {Gonzalez}}]{Bouwens11}
---. 2011, \apj, 737, 90

\bibitem[{{Boylan-Kolchin} {et~al.}(2012){Boylan-Kolchin}, {Bullock}, \&
  {Kaplinghat}}]{BK12}
{Boylan-Kolchin}, M., {Bullock}, J.~S., \& {Kaplinghat}, M. 2012, \mnras, 422,
  1203

\bibitem[{{Bradley} {et~al.}(2012){Bradley}, {Trenti}, {Oesch}, {Stiavelli},
  {Treu}, {Bouwens}, {Shull}, {Holwerda}, \& {Pirzkal}}]{BORG12}
{Bradley}, L.~D., {et~al.} 2012, \apj, 760, 108

\bibitem[{{Brammer} {et~al.}(2011){Brammer}, {Whitaker}, {van Dokkum},
  {Marchesini}, {Franx}, {Kriek}, {Labb{\'e}}, {Lee}, {Muzzin}, {Quadri},
  {Rudnick}, \& {Williams}}]{Brammer11}
{Brammer}, G.~B., {et~al.} 2011, \apj, 739, 24

\bibitem[{{Bruzual} \& {Charlot}(2003)}]{bc-03}
{Bruzual}, G., \& {Charlot}, S. 2003, \mnras, 344, 1000

\bibitem[{{Bryan} \& {Norman}(1998)}]{mvir_conv}
{Bryan}, G.~L., \& {Norman}, M.~L. 1998, \apj, 495, 80

\bibitem[{{Bullock} {et~al.}(2002){Bullock}, {Wechsler}, \&
  {Somerville}}]{Bullock02}
{Bullock}, J.~S., {Wechsler}, R.~H., \& {Somerville}, R.~S. 2002, \mnras, 329,
  246

\bibitem[{{Caputi} {et~al.}(2011){Caputi}, {Cirasuolo}, {Dunlop}, {McLure},
  {Farrah}, \& {Almaini}}]{Caputi11}
{Caputi}, K.~I., {Cirasuolo}, M., {Dunlop}, J.~S., {McLure}, R.~J., {Farrah},
  D., \& {Almaini}, O. 2011, \mnras, 413, 162

\bibitem[{Chabrier(2003)}]{chabrier-2003-115}
Chabrier, G. 2003, Publications of the Astronomical Society of the Pacific,
  115, 763

\bibitem[{{Chapman} {et~al.}(2004){Chapman}, {Smail}, {Windhorst}, {Muxlow}, \&
  {Ivison}}]{Chapman04}
{Chapman}, S.~C., {Smail}, I., {Windhorst}, R., {Muxlow}, T., \& {Ivison},
  R.~J. 2004, \apj, 611, 732

\bibitem[{{Col{\'{\i}}n} {et~al.}(1999){Col{\'{\i}}n}, {Klypin}, {Kravtsov}, \&
  {Khokhlov}}]{Colin99}
{Col{\'{\i}}n}, P., {Klypin}, A.~A., {Kravtsov}, A.~V., \& {Khokhlov}, A.~M.
  1999, \apj, 523, 32

\bibitem[{{Conroy} \& {Gunn}(2010)}]{Conroy10}
{Conroy}, C., \& {Gunn}, J.~E. 2010, \apj, 712, 833

\bibitem[{{Conroy} {et~al.}(2009){Conroy}, {Gunn}, \& {White}}]{Conroy09}
{Conroy}, C., {Gunn}, J.~E., \& {White}, M. 2009, \apj, 699, 486

\bibitem[{{Conroy} \& {Wechsler}(2009)}]{cw-08}
{Conroy}, C., \& {Wechsler}, R.~H. 2009, \apj, 696, 620

\bibitem[{Conroy {et~al.}(2006)Conroy, Wechsler, \& Kravtsov}]{conroy:06}
Conroy, C., Wechsler, R.~H., \& Kravtsov, A.~V. 2006, \apj, 647, 201

\bibitem[{{Conroy} {et~al.}(2007){Conroy}, {Wechsler}, \&
  {Kravtsov}}]{Conroy07}
{Conroy}, C., {Wechsler}, R.~H., \& {Kravtsov}, A.~V. 2007, \apj, 668, 826

\bibitem[{{Conroy} {et~al.}(2010){Conroy}, {White}, \& {Gunn}}]{conroy-09}
{Conroy}, C., {White}, M., \& {Gunn}, J.~E. 2010, \apj, 708, 58

\bibitem[{{Crocce} {et~al.}(2006){Crocce}, {Pueblas}, \&
  {Scoccimarro}}]{Crocce06}
{Crocce}, M., {Pueblas}, S., \& {Scoccimarro}, R. 2006, \mnras, 373, 369

\bibitem[{{Cucciati} {et~al.}(2012){Cucciati}, {Tresse}, {Ilbert}, {Le
  F{\`e}vre}, {Garilli}, {Le Brun}, {Cassata}, {Franzetti}, {Maccagni},
  {Scodeggio}, {Zucca}, {Zamorani}, {Bardelli}, {Bolzonella}, {Bielby},
  {McCracken}, {Zanichelli}, \& {Vergani}}]{Cucciati11}
{Cucciati}, O., {et~al.} 2012, \aap, 539, A31

\bibitem[{{Daddi} {et~al.}(2005){Daddi}, {Dickinson}, {Chary}, {Pope},
  {Morrison}, {Alexander}, {Bauer}, {Brandt}, {Giavalisco}, {Ferguson}, {Lee},
  {Lehmer}, {Papovich}, \& {Renzini}}]{Daddi05}
{Daddi}, E., {et~al.} 2005, \apjl, 631, L13

\bibitem[{{Daddi} {et~al.}(2007){Daddi}, {Dickinson}, {Morrison}, {Chary},
  {Cimatti}, {Elbaz}, {Frayer}, {Renzini}, {Pope}, {Alexander}, {Bauer},
  {Giavalisco}, {Huynh}, {Kurk}, \& {Mignoli}}]{Daddi07}
---. 2007, \apj, 670, 156

\bibitem[{{Donahue} {et~al.}(2010){Donahue}, {Bruch}, {Wang}, {Voit}, {Hicks},
  {Haarsma}, {Croston}, {Pratt}, {Pierini}, {O'Connell}, \&
  {B{\"o}hringer}}]{Donahue10}
{Donahue}, M., {et~al.} 2010, \apj, 715, 881

\bibitem[{{Drory} {et~al.}(2009){Drory}, {Bundy}, {Leauthaud}, {Scoville},
  {Capak}, {Ilbert}, {Kartaltepe}, {Kneib}, {McCracken}, {Salvato}, {Sanders},
  {Thompson}, \& {Willott}}]{Drory09}
{Drory}, N., {et~al.} 2009, \apj, 707, 1595

\bibitem[{{Drory} {et~al.}(2005){Drory}, {Salvato}, {Gabasch}, {Bender},
  {Hopp}, {Feulner}, \& {Pannella}}]{Drory05}
{Drory}, N., {Salvato}, M., {Gabasch}, A., {Bender}, R., {Hopp}, U., {Feulner},
  G., \& {Pannella}, M. 2005, \apjl, 619, L131

\bibitem[{Dunkley {et~al.}(2005)Dunkley, Bucher, Ferreira, Moodley, \&
  Skordis}]{dunkley-2005}
Dunkley, J., Bucher, M., Ferreira, P.~G., Moodley, K., \& Skordis, C. 2005,
  \mnras, 356, 925

\bibitem[{{Dunne} {et~al.}(2009){Dunne}, {Ivison}, {Maddox}, {Cirasuolo},
  {Mortier}, {Foucaud}, {Ibar}, {Almaini}, {Simpson}, \& {McLure}}]{Dunne09}
{Dunne}, L., {et~al.} 2009, \mnras, 394, 3

\bibitem[{{Eddington}(1913)}]{Eddington13}
{Eddington}, A.~S. 1913, \mnras, 73, 359

\bibitem[{{Eddington}(1940)}]{Eddington40}
{Eddington}, Sir, A.~S. 1940, \mnras, 100, 354

\bibitem[{{Fakhouri} {et~al.}(2010){Fakhouri}, {Ma}, \&
  {Boylan-Kolchin}}]{Fakhouri10}
{Fakhouri}, O., {Ma}, C.-P., \& {Boylan-Kolchin}, M. 2010, \mnras, 406, 2267

\bibitem[{{Feulner} {et~al.}(2008){Feulner}, {Gabasch}, {Goranova}, {Hopp}, \&
  {Bender}}]{Feulner08}
{Feulner}, G., {Gabasch}, A., {Goranova}, Y., {Hopp}, U., \& {Bender}, R. 2008,
  in Relativistic Astrophysics Legacy and Cosmology - Einstein's, ed.
  {B.~Aschenbach, V.~Burwitz, G.~Hasinger, \& B.~Leibundgut}, 310--+

\bibitem[{{Firmani} \& {Avila-Reese}(2010)}]{Firmani10}
{Firmani}, C., \& {Avila-Reese}, V. 2010, \apj, 723, 755

\bibitem[{{Gavazzi} {et~al.}(2007){Gavazzi}, {Treu}, {Rhodes}, {Koopmans},
  {Bolton}, {Burles}, {Massey}, \& {Moustakas}}]{Gavazzi07}
{Gavazzi}, R., {Treu}, T., {Rhodes}, J.~D., {Koopmans}, L.~V.~E., {Bolton},
  A.~S., {Burles}, S., {Massey}, R.~J., \& {Moustakas}, L.~A. 2007, \apj, 667,
  176

\bibitem[{{Gonzalez} {et~al.}(2005){Gonzalez}, {Zabludoff}, \&
  {Zaritsky}}]{Gonzalez05}
{Gonzalez}, A.~H., {Zabludoff}, A.~I., \& {Zaritsky}, D. 2005, \apj, 618, 195

\bibitem[{{Gonzalez} {et~al.}(2012){Gonzalez}, {Bouwens}, {llingworth},
  {Labbe}, {Oesch}, {Franx}, \& {Magee}}]{Gonzalez12}
{Gonzalez}, V., {Bouwens}, R., {llingworth}, G., {Labbe}, I., {Oesch}, P.,
  {Franx}, M., \& {Magee}, D. 2012, arXiv:1208.4362

\bibitem[{{Gonz{\'a}lez} {et~al.}(2011){Gonz{\'a}lez}, {Labb{\'e}}, {Bouwens},
  {Illingworth}, {Franx}, \& {Kriek}}]{Gonzalez10}
{Gonz{\'a}lez}, V., {Labb{\'e}}, I., {Bouwens}, R.~J., {Illingworth}, G.,
  {Franx}, M., \& {Kriek}, M. 2011, \apjl, 735, L34

\bibitem[{{Grazian} {et~al.}(2011){Grazian}, {Castellano}, {Koekemoer},
  {Fontana}, {Pentericci}, {Testa}, {Boutsia}, {Giallongo}, {Giavalisco}, \&
  {Santini}}]{Grazian10}
{Grazian}, A., {et~al.} 2011, \aap, 532, A33

\bibitem[{{Grogin} {et~al.}(2011)}]{Grogin11}
{Grogin}, N.~A., {et~al.} 2011, \apjs, 197, 35

\bibitem[{{Guo} {et~al.}(2010){Guo}, {White}, {Li}, \&
  {Boylan-Kolchin}}]{Guo-09}
{Guo}, Q., {White}, S., {Li}, C., \& {Boylan-Kolchin}, M. 2010, \mnras, 404,
  1111

\bibitem[{Haario {et~al.}(2001)Haario, Saksman, \& Tamminen}]{Haario01}
Haario, H., Saksman, E., \& Tamminen, J. 2001, Bernoulli, 7, pp. 223

\bibitem[{{Hammer} {et~al.}(2007){Hammer}, {Puech}, {Chemin}, {Flores}, \&
  {Lehnert}}]{Hammer07}
{Hammer}, F., {Puech}, M., {Chemin}, L., {Flores}, H., \& {Lehnert}, M.~D.
  2007, \apj, 662, 322

\bibitem[{{Hansen} {et~al.}(2009){Hansen}, {Sheldon}, {Wechsler}, \&
  {Koester}}]{Hansen09}
{Hansen}, S.~M., {Sheldon}, E.~S., {Wechsler}, R.~H., \& {Koester}, B.~P. 2009,
  \apj, 699, 1333

\bibitem[{{Hopkins} \& {Beacom}(2006)}]{Hopkins06b}
{Hopkins}, A.~M., \& {Beacom}, J.~F. 2006, \apj, 651, 142

\bibitem[{{Jenkins}(2010)}]{Jenkins10}
{Jenkins}, A. 2010, \mnras, 403, 1859

\bibitem[{{Kajisawa} {et~al.}(2009){Kajisawa}, {Ichikawa}, {Tanaka}, {Konishi},
  {Yamada}, {Akiyama}, {Suzuki}, {Tokoku}, {Uchimoto}, {Yoshikawa}, {Ouchi},
  {Iwata}, {Hamana}, \& {Onodera}}]{Kajisawa09}
{Kajisawa}, M., {et~al.} 2009, \apj, 702, 1393

\bibitem[{{Kajisawa} {et~al.}(2010){Kajisawa}, {Ichikawa}, {Yamada},
  {Uchimoto}, {Yoshikawa}, {Akiyama}, \& {Onodera}}]{Kajisawa10}
{Kajisawa}, M., {Ichikawa}, T., {Yamada}, T., {Uchimoto}, Y.~K., {Yoshikawa},
  T., {Akiyama}, M., \& {Onodera}, M. 2010, \apj, 723, 129

\bibitem[{{Karim} {et~al.}(2011){Karim}, {Schinnerer},
  {Mart{\'{\i}}nez-Sansigre}, {Sargent}, {van der Wel}, {Rix}, {Ilbert},
  {Smol{\v c}i{\'c}}, {Carilli}, {Pannella}, {Koekemoer}, {Bell}, \&
  {Salvato}}]{Karim11}
{Karim}, A., {et~al.} 2011, \apj, 730, 61

\bibitem[{{Kistler} {et~al.}(2009){Kistler}, {Y{\"u}ksel}, {Beacom}, {Hopkins},
  \& {Wyithe}}]{Kistler09}
{Kistler}, M.~D., {Y{\"u}ksel}, H., {Beacom}, J.~F., {Hopkins}, A.~M., \&
  {Wyithe}, J.~S.~B. 2009, \apjl, 705, L104

\bibitem[{{Klypin} {et~al.}(2011){Klypin}, {Trujillo-Gomez}, \&
  {Primack}}]{Bolshoi}
{Klypin}, A.~A., {Trujillo-Gomez}, S., \& {Primack}, J. 2011, \apj, 740, 102

\bibitem[{{Koekemoer} {et~al.}(2011)}]{Koekemoer11}
{Koekemoer}, A.~M., {et~al.} 2011, \apjs, 197, 36

\bibitem[{{Komatsu} {et~al.}(2009){Komatsu}, {Dunkley}, {Nolta}, {Bennett},
  {Gold}, {Hinshaw}, {Jarosik}, {Larson}, {Limon}, {Page}, {Spergel},
  {Halpern}, {Hill}, {Kogut}, {Meyer}, {Tucker}, {Weiland}, {Wollack}, \&
  {Wright}}]{wmap5}
{Komatsu}, E., {et~al.} 2009, \apjs, 180, 330

\bibitem[{{Komatsu} {et~al.}(2011){Komatsu}, {Smith}, {Dunkley}, {Bennett},
  {Gold}, {Hinshaw}, {Jarosik}, {Larson}, {Nolta}, {Page}, {Spergel},
  {Halpern}, {Hill}, {Kogut}, {Limon}, {Meyer}, {Odegard}, {Tucker}, {Weiland},
  {Wollack}, \& {Wright}}]{wmap7}
---. 2011, \apjs, 192, 18

\bibitem[{Kravtsov \& Klypin(1999)}]{kravtsov_klypin:99}
Kravtsov, A., \& Klypin, A. 1999, \apj, 520, 437

\bibitem[{Kravtsov {et~al.}(2004)Kravtsov, Berlind, Wechsler, Klypin,
  Gottloeber, Allgood, \& Primack}]{Kravtsov04}
Kravtsov, A.~V., Berlind, A.~A., Wechsler, R.~H., Klypin, A.~A., Gottloeber,
  S., Allgood, B., \& Primack, J.~R. 2004, \apj, 609, 35

\bibitem[{Kravtsov {et~al.}(1997)Kravtsov, Klypin, \&
  Khokhlov}]{kravtsov_etal:97}
Kravtsov, A.~V., Klypin, A.~A., \& Khokhlov, A.~M. 1997, \apj, 111, 73

\bibitem[{{Labb{\'e}} {et~al.}(2010){Labb{\'e}}, {Gonz{\'a}lez}, {Bouwens},
  {Illingworth}, {Franx}, {Trenti}, {Oesch}, {van Dokkum}, {Stiavelli},
  {Carollo}, {Kriek}, \& {Magee}}]{Labbe10}
{Labb{\'e}}, I., {et~al.} 2010, \apjl, 716, L103

\bibitem[{{Labbe} {et~al.}(2012){Labbe}, {Oesch}, {Bouwens}, {Illingworth},
  {Magee}, {Gonzalez}, {Carollo}, {Franx}, {Trenti}, {van Dokkum}, \&
  {Stiavelli}}]{Labbe12}
{Labbe}, I., {et~al.} 2012, arXiv:1209.3037

\bibitem[{{Le Borgne} {et~al.}(2009){Le Borgne}, {Elbaz}, {Ocvirk}, \&
  {Pichon}}]{LeBorgne09}
{Le Borgne}, D., {Elbaz}, D., {Ocvirk}, P., \& {Pichon}, C. 2009, \aap, 504,
  727

\bibitem[{{Leauthaud} {et~al.}(2012){Leauthaud}, {Tinker}, {Bundy}, {Behroozi},
  {Massey}, {Rhodes}, {George}, {Kneib}, {Benson}, {Wechsler}, {Busha},
  {Capak}, {Cort{\^e}s}, {Ilbert}, {Koekemoer}, {Le F{\`e}vre}, {Lilly},
  {McCracken}, {Salvato}, {Schrabback}, {Scoville}, {Smith}, \&
  {Taylor}}]{Leauthaud11}
{Leauthaud}, A., {et~al.} 2012, \apj, 744, 159

\bibitem[{{Lee} {et~al.}(2011){Lee}, {Dey}, {Reddy}, {Brown}, {Gonzalez},
  {Jannuzi}, {Cooper}, {Fan}, {Bian}, {Glikman}, {Stern}, {Brodwin}, \&
  {Cooray}}]{Lee11}
{Lee}, K.-S., {et~al.} 2011, \apj, 733, 99

\bibitem[{{Lee} {et~al.}(2012){Lee}, {Ferguson}, {Wiklind}, {Dahlen},
  {Dickinson}, {Giavalisco}, {Grogin}, {Papovich}, {Messias}, {Guo}, \&
  {Lin}}]{Lee11b}
---. 2012, \apj, 752, 66

\bibitem[{{Lee} {et~al.}(2009){Lee}, {Giavalisco}, {Conroy}, {Wechsler},
  {Ferguson}, {Somerville}, {Dickinson}, \& {Urry}}]{Lee09}
{Lee}, K.-S., {Giavalisco}, M., {Conroy}, C., {Wechsler}, R.~H., {Ferguson},
  H.~C., {Somerville}, R.~S., {Dickinson}, M.~E., \& {Urry}, C.~M. 2009, \apj,
  695, 368

\bibitem[{{Leitner}(2012)}]{Leitner11}
{Leitner}, S.~N. 2012, \apj, 745, 149

\bibitem[{{Lin} \& {Mohr}(2004)}]{LinMohr04}
{Lin}, Y.-T., \& {Mohr}, J.~J. 2004, \apj, 617, 879

\bibitem[{{Lu} {et~al.}(2012){Lu}, {Mo}, {Katz}, \& {Weinberg}}]{Lu12}
{Lu}, Y., {Mo}, H.~J., {Katz}, N., \& {Weinberg}, M.~D. 2012, \mnras, 421, 1779

\bibitem[{{Lucatello} {et~al.}(2005){Lucatello}, {Gratton}, {Beers}, \&
  {Carretta}}]{Lucatello05}
{Lucatello}, S., {Gratton}, R.~G., {Beers}, T.~C., \& {Carretta}, E. 2005,
  \apj, 625, 833

\bibitem[{{Ly} {et~al.}(2011{\natexlab{a}}){Ly}, {Lee}, {Dale}, {Momcheva},
  {Salim}, {Staudaher}, {Moore}, \& {Finn}}]{Ly10}
{Ly}, C., {Lee}, J.~C., {Dale}, D.~A., {Momcheva}, I., {Salim}, S.,
  {Staudaher}, S., {Moore}, C.~A., \& {Finn}, R. 2011{\natexlab{a}}, \apj, 726,
  109

\bibitem[{{Ly} {et~al.}(2011{\natexlab{b}}){Ly}, {Malkan}, {Hayashi},
  {Motohara}, {Kashikawa}, {Shimasaku}, {Nagao}, \& {Grady}}]{Ly11}
{Ly}, C., {Malkan}, M.~A., {Hayashi}, M., {Motohara}, K., {Kashikawa}, N.,
  {Shimasaku}, K., {Nagao}, T., \& {Grady}, C. 2011{\natexlab{b}}, \apj, 735,
  91

\bibitem[{{Magnelli} {et~al.}(2011){Magnelli}, {Elbaz}, {Chary}, {Dickinson},
  {Le Borgne}, {Frayer}, \& {Willmer}}]{Magnelli11}
{Magnelli}, B., {Elbaz}, D., {Chary}, R.~R., {Dickinson}, M., {Le Borgne}, D.,
  {Frayer}, D.~T., \& {Willmer}, C.~N.~A. 2011, \aap, 528, A35+

\bibitem[{{Magnelli} {et~al.}(2012){Magnelli}, {Lutz}, {Santini}, {Saintonge},
  {Berta}, {Albrecht}, {Altieri}, {Andreani}, {Aussel}, {Bertoldi},
  {B{\'e}thermin}, {Bongiovanni}, {Capak}, {Chapman}, {Cepa}, {Cimatti},
  {Cooray}, {Daddi}, {Danielson}, {Dannerbauer}, {Dunlop}, {Elbaz}, {Farrah},
  {F{\"o}rster Schreiber}, {Genzel}, {Hwang}, {Ibar}, {Ivison}, {Le Floc'h},
  {Magdis}, {Maiolino}, {Nordon}, {Oliver}, {P{\'e}rez Garc{\'{\i}}a},
  {Poglitsch}, {Popesso}, {Pozzi}, {Riguccini}, {Rodighiero}, {Rosario},
  {Roseboom}, {Salvato}, {Sanchez-Portal}, {Scott}, {Smail}, {Sturm},
  {Swinbank}, {Tacconi}, {Valtchanov}, {Wang}, \& {Wuyts}}]{Magnelli12}
{Magnelli}, B., {et~al.} 2012, \aap, 539, A155

\bibitem[{{Mandelbaum} {et~al.}(2006){Mandelbaum}, {Seljak}, {Kauffmann},
  {Hirata}, \& {Brinkmann}}]{mandelbaum-06}
{Mandelbaum}, R., {Seljak}, U., {Kauffmann}, G., {Hirata}, C.~M., \&
  {Brinkmann}, J. 2006, \mnras, 368, 715

\bibitem[{{Maraston} {et~al.}(2010){Maraston}, {Pforr}, {Renzini}, {Daddi},
  {Dickinson}, {Cimatti}, \& {Tonini}}]{Maraston10}
{Maraston}, C., {Pforr}, J., {Renzini}, A., {Daddi}, E., {Dickinson}, M.,
  {Cimatti}, A., \& {Tonini}, C. 2010, \mnras, 407, 830

\bibitem[{{Marchesini} {et~al.}(2009){Marchesini}, {van Dokkum}, {F{\"o}rster
  Schreiber}, {Franx}, {Labb{\'e}}, \& {Wuyts}}]{marchesini-2008}
{Marchesini}, D., {van Dokkum}, P.~G., {F{\"o}rster Schreiber}, N.~M., {Franx},
  M., {Labb{\'e}}, I., \& {Wuyts}, S. 2009, \apj, 701, 1765

\bibitem[{{Marchesini} {et~al.}(2010){Marchesini}, {Whitaker}, {Brammer}, {van
  Dokkum}, {Labb{\'e}}, {Muzzin}, {Quadri}, {Kriek}, {Lee}, {Rudnick}, {Franx},
  {Illingworth}, \& {Wake}}]{Marchesini10}
{Marchesini}, D., {et~al.} 2010, \apj, 725, 1277

\bibitem[{{Mar{\'{\i}}n} {et~al.}(2008){Mar{\'{\i}}n}, {Wechsler}, {Frieman},
  \& {Nichol}}]{Marin08}
{Mar{\'{\i}}n}, F.~A., {Wechsler}, R.~H., {Frieman}, J.~A., \& {Nichol}, R.~C.
  2008, \apj, 672, 849

\bibitem[{{McBride} {et~al.}(2009){McBride}, {Fakhouri}, \& {Ma}}]{McBride09}
{McBride}, J., {Fakhouri}, O., \& {Ma}, C.-P. 2009, \mnras, 398, 1858

\bibitem[{{McLure} {et~al.}(2011){McLure}, {Dunlop}, {de Ravel}, {Cirasuolo},
  {Ellis}, {Schenker}, {Robertson}, {Koekemoer}, {Stark}, \&
  {Bowler}}]{McLure11}
{McLure}, R.~J., {et~al.} 2011, \mnras, 418, 2074

\bibitem[{{Micha{\l}owski} {et~al.}(2010{\natexlab{a}}){Micha{\l}owski},
  {Hjorth}, \& {Watson}}]{Michalowski10}
{Micha{\l}owski}, M., {Hjorth}, J., \& {Watson}, D. 2010{\natexlab{a}}, \aap,
  514, A67

\bibitem[{{Micha{\l}owski} {et~al.}(2010{\natexlab{b}}){Micha{\l}owski},
  {Watson}, \& {Hjorth}}]{Michalowski10b}
{Micha{\l}owski}, M.~J., {Watson}, D., \& {Hjorth}, J. 2010{\natexlab{b}},
  \apj, 712, 942

\bibitem[{{More} {et~al.}(2009){More}, {van den Bosch}, {Cacciato}, {Mo},
  {Yang}, \& {Li}}]{more-09}
{More}, S., {van den Bosch}, F.~C., {Cacciato}, M., {Mo}, H.~J., {Yang}, X., \&
  {Li}, R. 2009, \mnras, 392, 801

\bibitem[{{Mortlock} {et~al.}(2011){Mortlock}, {Conselice}, {Bluck}, {Bauer},
  {Gr{\"u}tzbauch}, {Buitrago}, \& {Ownsworth}}]{Mortlock11}
{Mortlock}, A., {Conselice}, C.~J., {Bluck}, A.~F.~L., {Bauer}, A.~E.,
  {Gr{\"u}tzbauch}, R., {Buitrago}, F., \& {Ownsworth}, J. 2011, \mnras, 413,
  2845

\bibitem[{{Moster} {et~al.}(2013){Moster}, {Naab}, \& {White}}]{Moster12}
{Moster}, B.~P., {Naab}, T., \& {White}, S.~D.~M. 2013, \mnras, 428, 3121

\bibitem[{{Moster} {et~al.}(2010){Moster}, {Somerville}, {Maulbetsch}, {van den
  Bosch}, {Macci{\`o}}, {Naab}, \& {Oser}}]{moster-09}
{Moster}, B.~P., {Somerville}, R.~S., {Maulbetsch}, C., {van den Bosch}, F.~C.,
  {Macci{\`o}}, A.~V., {Naab}, T., \& {Oser}, L. 2010, \apj, 710, 903

\bibitem[{{Moustakas} {et~al.}(2013){Moustakas}, {Coil}, {Aird}, {Blanton},
  {Cool}, {Eisenstein}, {Mendez}, {Wong}, {Zhu}, \& {Arnouts}}]{Moustakas12}
{Moustakas}, J., {et~al.} 2013, arXiv:1301.1688

\bibitem[{{Muzzin} {et~al.}(2009){Muzzin}, {Marchesini}, {van Dokkum},
  {Labb{\'e}}, {Kriek}, \& {Franx}}]{Muzzin09}
{Muzzin}, A., {Marchesini}, D., {van Dokkum}, P.~G., {Labb{\'e}}, I., {Kriek},
  M., \& {Franx}, M. 2009, \apj, 701, 1839

\bibitem[{{Nagamine} {et~al.}(2006){Nagamine}, {Ostriker}, {Fukugita}, \&
  {Cen}}]{Nagamine06}
{Nagamine}, K., {Ostriker}, J.~P., {Fukugita}, M., \& {Cen}, R. 2006, \apj,
  653, 881

\bibitem[{{Neyrinck} {et~al.}(2004){Neyrinck}, {Hamilton}, \&
  {Gnedin}}]{Neyrinck04}
{Neyrinck}, M.~C., {Hamilton}, A.~J.~S., \& {Gnedin}, N.~Y. 2004, \mnras, 348,
  1

\bibitem[{{Noeske} {et~al.}(2007){Noeske}, {Weiner}, {Faber}, {Papovich},
  {Koo}, {Somerville}, {Bundy}, {Conselice}, {Newman}, {Schiminovich}, {Le
  Floc'h}, {Coil}, {Rieke}, {Lotz}, {Primack}, {Barmby}, {Cooper}, {Davis},
  {Ellis}, {Fazio}, {Guhathakurta}, {Huang}, {Kassin}, {Martin}, {Phillips},
  {Rich}, {Small}, {Willmer}, \& {Wilson}}]{Noeske07}
{Noeske}, K.~G., {et~al.} 2007, \apjl, 660, L43

\bibitem[{{Papovich} {et~al.}(2011){Papovich}, {Finkelstein}, {Ferguson},
  {Lotz}, \& {Giavalisco}}]{Papovich11}
{Papovich}, C., {Finkelstein}, S.~L., {Ferguson}, H.~C., {Lotz}, J.~M., \&
  {Giavalisco}, M. 2011, \mnras, 412, 1123

\bibitem[{{P{\'e}rez-Gonz{\'a}lez} {et~al.}(2008){P{\'e}rez-Gonz{\'a}lez},
  {Rieke}, {Villar}, {Barro}, {Blaylock}, {Egami}, {Gallego}, {Gil de Paz},
  {Pascual}, {Zamorano}, \& {Donley}}]{perezgonzalez-2008}
{P{\'e}rez-Gonz{\'a}lez}, P.~G., {et~al.} 2008, \apj, 675, 234

\bibitem[{{Purcell} {et~al.}(2007){Purcell}, {Bullock}, \&
  {Zentner}}]{purcell-etal-07}
{Purcell}, C.~W., {Bullock}, J.~S., \& {Zentner}, A.~R. 2007, \apj, 666, 20

\bibitem[{{Rawle} {et~al.}(2012){Rawle}, {Edge}, {Egami}, {Rex}, {Smith},
  {Altieri}, {Fiedler}, {Haines}, {Pereira}, {P{\'e}rez-Gonz{\'a}lez},
  {Portouw}, {Valtchanov}, {Walth}, {van der Werf}, \& {Zemcov}}]{Rawle12}
{Rawle}, T.~D., {et~al.} 2012, \apj, 747, 29

\bibitem[{{Reddick} {et~al.}(2012){Reddick}, {Wechsler}, {Tinker}, \&
  {Behroozi}}]{Reddick12}
{Reddick}, R.~M., {Wechsler}, R.~H., {Tinker}, J.~L., \& {Behroozi}, P.~S.
  2012, arXiv:1207.2160

\bibitem[{{Reddy} {et~al.}(2012){Reddy}, {Pettini}, {Steidel}, {Shapley},
  {Erb}, \& {Law}}]{Reddy12}
{Reddy}, N.~A., {Pettini}, M., {Steidel}, C.~C., {Shapley}, A.~E., {Erb},
  D.~K., \& {Law}, D.~R. 2012, \apj, 754, 25

\bibitem[{{Reddy} \& {Steidel}(2009)}]{Reddy09}
{Reddy}, N.~A., \& {Steidel}, C.~C. 2009, \apj, 692, 778

\bibitem[{{Riebe} {et~al.}(2011){Riebe}, {Partl}, {Enke}, {Forero-Romero},
  {Gottloeber}, {Klypin}, {Lemson}, {Prada}, {Primack}, {Steinmetz}, \&
  {Turchaninov}}]{Riebe11}
{Riebe}, K., {et~al.} 2011, arXiv:1109.0003

\bibitem[{{Robotham} \& {Driver}(2011)}]{Robotham11}
{Robotham}, A.~S.~G., \& {Driver}, S.~P. 2011, \mnras, 413, 2570

\bibitem[{{Rujopakarn} {et~al.}(2010){Rujopakarn}, {Eisenstein}, {Rieke},
  {Papovich}, {Cool}, {Moustakas}, {Jannuzi}, {Kochanek}, {Rieke}, {Dey},
  {Eisenhardt}, {Murray}, {Brown}, \& {Le Floc'h}}]{Rujopakarn10}
{Rujopakarn}, W., {et~al.} 2010, \apj, 718, 1171

\bibitem[{{Salim} {et~al.}(2007){Salim}, {Rich}, {Charlot}, {Brinchmann},
  {Johnson}, {Schiminovich}, {Seibert}, {Mallery}, {Heckman}, {Forster},
  {Friedman}, {Martin}, {Morrissey}, {Neff}, {Small}, {Wyder}, {Bianchi},
  {Donas}, {Lee}, {Madore}, {Milliard}, {Szalay}, {Welsh}, \& {Yi}}]{Salim07}
{Salim}, S., {et~al.} 2007, \apjs, 173, 267

\bibitem[{{Salmi} {et~al.}(2012){Salmi}, {Daddi}, {Elbaz}, {Sargent},
  {Dickinson}, {Renzini}, {Bethermin}, \& {Le Borgne}}]{Salmi12}
{Salmi}, F., {Daddi}, E., {Elbaz}, D., {Sargent}, M.~T., {Dickinson}, M.,
  {Renzini}, A., {Bethermin}, M., \& {Le Borgne}, D. 2012, \apjl, 754, L14

\bibitem[{{Schaerer} \& {de Barros}(2010)}]{Schaerer10}
{Schaerer}, D., \& {de Barros}, S. 2010, \aap, 515, A73+

\bibitem[{{Shankar} {et~al.}(2006){Shankar}, {Lapi}, {Salucci}, {De Zotti}, \&
  {Danese}}]{Shankar06}
{Shankar}, F., {Lapi}, A., {Salucci}, P., {De Zotti}, G., \& {Danese}, L. 2006,
  \apj, 643, 14

\bibitem[{{Shim} {et~al.}(2009){Shim}, {Colbert}, {Teplitz}, {Henry}, {Malkan},
  {McCarthy}, \& {Yan}}]{Shim09}
{Shim}, H., {Colbert}, J., {Teplitz}, H., {Henry}, A., {Malkan}, M.,
  {McCarthy}, P., \& {Yan}, L. 2009, \apj, 696, 785

\bibitem[{{Smol{\v c}i{\'c}} {et~al.}(2009){Smol{\v c}i{\'c}}, {Schinnerer},
  {Zamorani}, {Bell}, {Bondi}, {Carilli}, {Ciliegi}, {Mobasher}, {Paglione},
  {Scodeggio}, \& {Scoville}}]{Smolcic09}
{Smol{\v c}i{\'c}}, V., {et~al.} 2009, \apj, 690, 610

\bibitem[{{Sobral} {et~al.}(2013){Sobral}, {Smail}, {Best}, {Geach}, {Matsuda},
  {Stott}, {Cirasuolo}, \& {Kurk}}]{Sobral12}
{Sobral}, D., {Smail}, I., {Best}, P.~N., {Geach}, J.~E., {Matsuda}, Y.,
  {Stott}, J.~P., {Cirasuolo}, M., \& {Kurk}, J. 2013, \mnras, 428, 1128

\bibitem[{{Springel}(2005)}]{Springel05}
{Springel}, V. 2005, \mnras, 364, 1105

\bibitem[{{Stark} {et~al.}(2009){Stark}, {Ellis}, {Bunker}, {Bundy}, {Targett},
  {Benson}, \& {Lacy}}]{Stark09}
{Stark}, D.~P., {Ellis}, R.~S., {Bunker}, A., {Bundy}, K., {Targett}, T.,
  {Benson}, A., \& {Lacy}, M. 2009, \apj, 697, 1493

\bibitem[{{Stark} {et~al.}(2013){Stark}, {Schenker}, {Ellis}, {Robertson},
  {McLure}, \& {Dunlop}}]{Stark12}
{Stark}, D.~P., {Schenker}, M.~A., {Ellis}, R., {Robertson}, B., {McLure}, R.,
  \& {Dunlop}, J. 2013, \apj, 763, 129

\bibitem[{{Tadaki} {et~al.}(2011){Tadaki}, {Kodama}, {Koyama}, {Hayashi},
  {Tanaka}, \& {Tokoku}}]{Tadaki11}
{Tadaki}, K.-I., {Kodama}, T., {Koyama}, Y., {Hayashi}, M., {Tanaka}, I., \&
  {Tokoku}, C. 2011, \pasj, 63, 437

\bibitem[{{Tasitsiomi} {et~al.}(2004){Tasitsiomi}, {Kravtsov}, {Wechsler}, \&
  {Primack}}]{Tasitsiomi04}
{Tasitsiomi}, A., {Kravtsov}, A.~V., {Wechsler}, R.~H., \& {Primack}, J.~R.
  2004, \apj, 614, 533

\bibitem[{{Tinker} {et~al.}(2008){Tinker}, {Kravtsov}, {Klypin}, {Abazajian},
  {Warren}, {Yepes}, {Gottl{\"o}ber}, \& {Holz}}]{tinker-umf}
{Tinker}, J., {Kravtsov}, A.~V., {Klypin}, A., {Abazajian}, K., {Warren}, M.,
  {Yepes}, G., {Gottl{\"o}ber}, S., \& {Holz}, D.~E. 2008, \apj, 688, 709

\bibitem[{{Trenti} \& {Stiavelli}(2008)}]{Trenti08}
{Trenti}, M., \& {Stiavelli}, M. 2008, \apj, 676, 767

\bibitem[{{Tumlinson}(2007{\natexlab{a}})}]{Tumlinson07a}
{Tumlinson}, J. 2007{\natexlab{a}}, \apj, 665, 1361

\bibitem[{{Tumlinson}(2007{\natexlab{b}})}]{Tumlinson07b}
---. 2007{\natexlab{b}}, \apjl, 664, L63

\bibitem[{{Twite} {et~al.}(2012){Twite}, {Conselice}, {Buitrago}, {Noeske},
  {Weiner}, {Acosta-Pulido}, \& {Bauer}}]{Twite12}
{Twite}, J.~W., {Conselice}, C.~J., {Buitrago}, F., {Noeske}, K., {Weiner},
  B.~J., {Acosta-Pulido}, J.~A., \& {Bauer}, A.~E. 2012, \mnras, 420, 1061

\bibitem[{{Vale} \& {Ostriker}(2004)}]{Vale04}
{Vale}, A., \& {Ostriker}, J.~P. 2004, \mnras, 353, 189

\bibitem[{{Vale} \& {Ostriker}(2006)}]{Vale06}
---. 2006, \mnras, 371, 1173

\bibitem[{{van der Burg} {et~al.}(2010){van der Burg}, {Hildebrandt}, \&
  {Erben}}]{vdBurg10}
{van der Burg}, R.~F.~J., {Hildebrandt}, H., \& {Erben}, T. 2010, \aap, 523,
  A74+

\bibitem[{{van Dokkum}(2008)}]{vanDokkum08}
{van Dokkum}, P.~G. 2008, \apj, 674, 29

\bibitem[{{van Dokkum} \& {Conroy}(2010)}]{Dokkum10}
{van Dokkum}, P.~G., \& {Conroy}, C. 2010, \nat, 468, 940

\bibitem[{{Wake} {et~al.}(2011){Wake}, {Whitaker}, {Labb{\'e}}, {van Dokkum},
  {Franx}, {Quadri}, {Brammer}, {Kriek}, {Lundgren}, {Marchesini}, \&
  {Muzzin}}]{Wake11}
{Wake}, D.~A., {et~al.} 2011, \apj, 728, 46

\bibitem[{{Wang} {et~al.}(2013){Wang}, {Farrah}, {Oliver}, {Amblard},
  {B{\'e}thermin}, {Bock}, {Conley}, {Cooray}, {Halpern}, {Heinis}, {Ibar},
  {Ilbert}, {Ivison}, {Marsden}, {Roseboom}, {Rowan-Robinson}, {Schulz},
  {Smith}, {Viero}, \& {Zemcov}}]{Wang12}
{Wang}, L., {et~al.} 2013, \mnras

\bibitem[{{Wang} \& {Jing}(2010)}]{Wang09}
{Wang}, L., \& {Jing}, Y.~P. 2010, \mnras, 402, 1796

\bibitem[{{Watson} {et~al.}(2012){Watson}, {Berlind}, \& {Zentner}}]{Watson12}
{Watson}, D.~F., {Berlind}, A.~A., \& {Zentner}, A.~R. 2012, \apj, 754, 90

\bibitem[{{Wechsler} {et~al.}(2002){Wechsler}, {Bullock}, {Primack},
  {Kravtsov}, \& {Dekel}}]{Wechsler02}
{Wechsler}, R.~H., {Bullock}, J.~S., {Primack}, J.~R., {Kravtsov}, A.~V., \&
  {Dekel}, A. 2002, \apj, 568, 52

\bibitem[{{Weinmann} {et~al.}(2011){Weinmann}, {Neistein}, \&
  {Dekel}}]{Weinmann11}
{Weinmann}, S.~M., {Neistein}, E., \& {Dekel}, A. 2011, \mnras, 417, 2737

\bibitem[{{Weinmann} {et~al.}(2012){Weinmann}, {Pasquali}, {Oppenheimer},
  {Finlator}, {Mendel}, {Crain}, \& {Macci{\`o}}}]{Weinmann12}
{Weinmann}, S.~M., {Pasquali}, A., {Oppenheimer}, B.~D., {Finlator}, K.,
  {Mendel}, J.~T., {Crain}, R.~A., \& {Macci{\`o}}, A.~V. 2012, \mnras, 426,
  2797

\bibitem[{{Wetzel} {et~al.}(2012){Wetzel}, {Tinker}, \& {Conroy}}]{Wetzel11}
{Wetzel}, A.~R., {Tinker}, J.~L., \& {Conroy}, C. 2012, \mnras, 424, 232

\bibitem[{{Wetzel} \& {White}(2010)}]{wetzel-09}
{Wetzel}, A.~R., \& {White}, M. 2010, \mnras, 403, 1072

\bibitem[{{Whitaker} {et~al.}(2012){Whitaker}, {van Dokkum}, {Brammer}, \&
  {Franx}}]{Whitaker12}
{Whitaker}, K.~E., {van Dokkum}, P.~G., {Brammer}, G., \& {Franx}, M. 2012,
  \apjl, 754, L29

\bibitem[{{White} {et~al.}(2007){White}, {Zheng}, {Brown}, {Dey}, \&
  {Jannuzi}}]{White07}
{White}, M., {Zheng}, Z., {Brown}, M.~J.~I., {Dey}, A., \& {Jannuzi}, B.~T.
  2007, \apjl, 655, L69

\bibitem[{{Wilkins} {et~al.}(2008){Wilkins}, {Trentham}, \&
  {Hopkins}}]{Wilkins08}
{Wilkins}, S.~M., {Trentham}, N., \& {Hopkins}, A.~M. 2008, \mnras, 385, 687

\bibitem[{{Yang} {et~al.}(2009{\natexlab{a}}){Yang}, {Mo}, \& {van den
  Bosch}}]{yang-08}
{Yang}, X., {Mo}, H.~J., \& {van den Bosch}, F.~C. 2009{\natexlab{a}}, \apj,
  695, 900

\bibitem[{{Yang} {et~al.}(2009{\natexlab{b}}){Yang}, {Mo}, \& {van den
  Bosch}}]{yang-09}
---. 2009{\natexlab{b}}, \apj, 693, 830

\bibitem[{{Yang} {et~al.}(2012){Yang}, {Mo}, {van den Bosch}, {Zhang}, \&
  {Han}}]{Yang11}
{Yang}, X., {Mo}, H.~J., {van den Bosch}, F.~C., {Zhang}, Y., \& {Han}, J.
  2012, \apj, 752, 41

\bibitem[{{Yang} {et~al.}(2003)}]{Yang03}
{Yang}, X., {et~al.} 2003, \mnras, 339, 1057

\bibitem[{{Yoshida} {et~al.}(2006){Yoshida}, {Shimasaku}, {Kashikawa}, {Ouchi},
  {Okamura}, {Ajiki}, {Akiyama}, {Ando}, {Aoki}, {Doi}, {Furusawa},
  {Hayashino}, {Iwamuro}, {Iye}, {Karoji}, {Kobayashi}, {Kodaira}, {Kodama},
  {Komiyama}, {Malkan}, {Matsuda}, {Miyazaki}, {Mizumoto}, {Morokuma},
  {Motohara}, {Murayama}, {Nagao}, {Nariai}, {Ohta}, {Sasaki}, {Sato},
  {Sekiguchi}, {Shioya}, {Tamura}, {Taniguchi}, {Umemura}, {Yamada}, \&
  {Yasuda}}]{Yoshida06}
{Yoshida}, M., {et~al.} 2006, \apj, 653, 988

\bibitem[{{Zheng} {et~al.}(2012){Zheng}, {Postman}, {Zitrin}, {Moustakas},
  {Shu}, {Jouvel}, {H{\o}st}, {Molino}, {Bradley}, {Coe}, {Moustakas},
  {Carrasco}, {Ford}, {Ben{\'{\i}}tez}, {Lauer}, {Seitz}, {Bouwens},
  {Koekemoer}, {Medezinski}, {Bartelmann}, {Broadhurst}, {Donahue}, {Grillo},
  {Infante}, {Jha}, {Kelson}, {Lahav}, {Lemze}, {Melchior}, {Meneghetti},
  {Merten}, {Nonino}, {Ogaz}, {Rosati}, {Umetsu}, \& {van der Wel}}]{Zheng12}
{Zheng}, W., {et~al.} 2012, \nat, 489, 406

\bibitem[{{Zheng} {et~al.}(2007{\natexlab{a}}){Zheng}, {Bell}, {Papovich},
  {Wolf}, {Meisenheimer}, {Rix}, {Rieke}, \& {Somerville}}]{Zheng07}
{Zheng}, X.~Z., {Bell}, E.~F., {Papovich}, C., {Wolf}, C., {Meisenheimer}, K.,
  {Rix}, H.-W., {Rieke}, G.~H., \& {Somerville}, R. 2007{\natexlab{a}}, \apjl,
  661, L41

\bibitem[{{Zheng} {et~al.}(2007{\natexlab{b}}){Zheng}, {Coil}, \&
  {Zehavi}}]{zheng-07}
{Zheng}, Z., {Coil}, A.~L., \& {Zehavi}, I. 2007{\natexlab{b}}, \apj, 667, 760

\end{thebibliography}

\appendix

\section{Impact of Assuming Average Star Formation Histories on Derived Star Formation Rates}

\label{a:av_afh}

As mentioned in the introduction to \S \ref{s:methodology}, there is a potential error in our method because we propagate average star formation histories for halos at a given mass and redshift, as opposed to individual star formation histories.  We discuss effects from both a distribution in star formation rates for galaxies at a given halo mass and from a distribution of halo mass accretion histories.

Bimodality has been observed in the star formation rates of galaxies at fixed stellar mass, partially but not exclusively driven by quenched satellite galaxies \citep[see, e.g.,][]{Wetzel11}.  This may appear to be a problem for our method because the rate of specific mass loss (see \S\ref{s:smhist}) will be different depending on whether a given galaxy is active or passive.  Remarkably, however, our method is robust to any distribution of star formation rates in galaxies of the same stellar mass.  Because our method only depends on the \textit{total} stellar mass buildup in halos of a given mass $M_h$ at any given time, the fact that some galaxies get a larger share of that buildup than others is not a concern.  If certain galaxies sustain a higher growth rate for a long time, then the \textit{scatter} in stellar mass at fixed halo mass will certainly increase; however, the possibility of scatter changing with time is an effect that we include explicitly in our model.  Bimodality in star formation rates is only a problem if it results in a distribution of {\em stellar masses} at fixed halo mass that cannot be reasonably modeled by a log-normal distribution (i.e., the distribution we assume in this work).  So far, no evidence has emerged for any other distribution to be preferred (see comments in \S \ref{s:smhm} where $\xi$ is defined).

There is more possibility for error because we do not account for correlations between the star formation history (SFH) and the mass accretion history of a halo.  For example, a rapidly accreting halo might have a more recent star formation history than other halos at the same mass.  In this case, when it transitions from one mass bin to another, we may use the incorrect SFH in Eq.\ \ref{e:sfh_tensor}.  This effect is balanced by halos that accrete more slowly than other halos of the same mass, which cancels the effect to first order.

It is worth working through a more quantitative example to verify this assertion.  Consider a population of galaxies all with the same stellar mass $M_\ast = 10^{10}\Msun$ today.  Assume that each galaxy has had a constant SSFR for its entire lifetime (i.e., an exponential growth history), and moreover that the distribution of the SSFRs is log-normal with width 0.3 dex (similar to that seen in \citealt{Salim07}) and median $10^{-9}$ yr$^{-1}$.  The average (as opposed to median) star formation rate for this population is 12.7 $\Msun$ yr$^{-1}$; given the star formation histories just specified, this translates into an average rate of change of stellar mass of 9.3 $\Msun$ yr$^{-1}$ over the next 100 Myr.  If, on the other hand, we had simply assumed that all of the galaxies had star formation histories corresponding to a constant SSFR of $10^{-9}$ yr$^{-1}$, we would interpret a buildup in stellar mass of 9.3 $\Msun$ yr$^{-1}$ as implying an instantaneous star formation rate of 13.4 $\Msun$ yr$^{-1}$, an error of 6\%.  Thus, we conclude that the errors on the average quantities considered here introduced  by lumping together star formation histories for halos with different mass accretion histories are strongly subdominant to existing observational uncertainties.

\section{Merging Galaxies and Mass Accretion}

\label{a:mergers}

Because of passive stellar evolution, calculating the amount of stellar mass remaining in a galaxy (required for determining the star formation rate in Eq.\ \ref{e:sfr_schem}), one must know the entire star formation history.  For a large simulation, keeping track of the individual stellar mass histories (covering hundreds of timesteps) for millions of galaxies repeated over several million realizations of the stellar mass -- halo mass relation is inefficient.  Instead, since we are interested in only the average star formation rate as a function of mass, we take an elegant (although approximate) shortcut.

Instead of keeping track of every halo/galaxy individually, we instead keep track of average star formation histories at each timestep in bins of halo mass (which is a rank-3 tensor, $SFH_{t,m}^{t_{form}}$; $t$ is the time now, $t_{form}$ is the timestep at which the stellar mass was formed, and $m$ is the halo mass).  For central halos, we use the virial mass (as discussed in \S \ref{s:sim}); for satellite halos, we use the mass of the largest progenitor (the peak mass) in the satellite's mass accretion history.  At every timestep, we calculate the number of halos that transition between one mass bin and another ($T_{t,m_n}^{m_p}$), i.e., the number of halos that had a mass $m_p$ at timestep $t$ and a mass $m_n$ at timestep $t+1$ (see Appendices \ref{a:tinker} and \ref{a:mah}).  Similarly, we calculate the merger rates for subhalos at each timestep as a function of the subhalo peak mass and the target host halo mass ($M_{t,m_n}^{m_p}$), as discussed in Appendix \ref{a:disruption}.  Keeping track of star formation histories then becomes an exercise in tensor multiplication, which is fast regardless of the number of halos in the simulation.

To convert Eq.\ \ref{e:sfr_schem} into a full tensor equation, we need a small amount of additional notation.  The stellar mass at a particular timestep $t$ and halo mass $m$ becomes $SM_{t,m}$, the star formation rate at a particular timestep and halo mass becomes $SFR_{t,m}$, and the number counts at a given timestep for halos in a given mass bin becomes $N_{t,m}$.  In addition, we denote the difference in stellar mass from the previous timestep by $\Delta SM_{t,m}$.  Thus, Eq.\ \ref{e:sfr_schem} formally becomes:
\begin{eqnarray}
\label{e:full_sfr}
\Delta SM_{t,m} &=& \left(SM_{t,m} - \sum_{t_{form}} SFH_{t,m}^{t_{form}}(1-C_\mathrm{loss}(t, t_{form} - \Delta t, t_{form}))\right)\nonumber\\
SFR_{t,m} & = & \frac{f_{SFR}(m,t)\Delta SM_{t,m}}{\Delta t (1-C_\mathrm{loss}(t, t-\Delta t,t))}\nonumber\\
C_\mathrm{loss}(t_{now}, t_a,t_b) & = & |t_b-t_a|^{-1} \int_{t_a}^{t_b} f_\mathrm{loss}(t_{now}-x) dx,
\end{eqnarray}
where $C_\mathrm{loss}(t_{now}, t_a,t_b)$ is the average stellar mass loss fraction (a.k.a., ``recycling'' fraction) at the present time for stars that formed at a constant rate from $t_a$ to $t_b$.  (Note that the indices $t$ and $m$ are not meant to be summed over).

To calculate star formation histories, we need to know both internal star formation and the amount of stars accreted in mergers.  As discussed in \S \ref{s:icl}, purely definitional changes in what constitutes the boundary between a galaxy and the ICL can effectively shift stellar mass from the ICL into the galaxy.  Not knowing the relative fraction of each, we assume that merging stellar mass from satellites takes priority over stellar mass from the ICL.  For convenience, we define $SM'_{t,m}$ and $ICL'_{t,m}$ be the total stellar mass in merging satellites and the total stellar mass in the main progenitors' ICL, respectively; we also define $ICLH_{t,m}^{t_{form}}$ to be the star formation history of the ICL (analogously to $SFH_{t,m}^{t_{form}}$).  The new star formation histories will then be the sum of the old star formation histories from all progenitors, with the addition of new stars formed at the latest timestep and material from mergers and the ICL:
\begin{eqnarray}
\label{e:sfh_tensor}
SFH_{t+\Delta t,m}^{t_{form}} & =& \sum_{m_p} \left[SFH_{t,m_p}^{t_{form}} \left(T_{t,m}^{m_p} + C^{merge}_{t,m} M_{t,m}^{m_p}\right) + ICLH_{t,m_p}^{t_{form}} T_{t,m}^{m_p} C^{def}_{t,m} \right] (N_{t+\Delta t,m})^{-1} + SFR_{t,m} \delta_{t}^{t_{form}} \Delta t\\
C^{merge}_{t,m} &=& \min\left(1,\frac{(1-f_{SFR}(m,t))\Delta SM_{t+\Delta t,m}}{SM'_{t,m}}\right) \\
C^{def}_{t,m} &=& \max\left(0, \frac{(1-f_{SFR}(m,t))\Delta SM_{t+\Delta t,m}-SM'_{t,m}}{ICL'_{t,m}}\right) \\
SM'_{t,m} &=& \sum_{m_p} SFH_{t,m_p}^{t_{form}} M_{t,m}^{m_p} (1-C_\mathrm{loss}(t+\Delta t, t_{form} - \Delta t, t_{form}))(N_{t+\Delta t,m})^{-1}\\
ICL'_{t,m} &=& \sum_{m_p} ICLH_{t,m_p}^{t_{form}} T_{t,m}^{m_p} (1-C_\mathrm{loss}(t+\Delta t, t_{form} - \Delta t, t_{form}))(N_{t+\Delta t,m})^{-1}.
\end{eqnarray}
(Again, the indices $t$ and $m$ are not meant to be summed over; $\delta_{t}^{t_{form}}$ is the standard identity tensor: $\delta_{t}^{t_{form}} = 1$ if $t=t_{form}$ and $0$ otherwise).  Note that, in this present analysis, we do not separate the star formation histories of satellites and centrals.  As discussed in \S \ref{s:new_fits}, satellites at high redshifts ($z>1$) tend to have short infall timescales, so they have roughly similar star formation histories as centrals.  At later times, this approximation is less valid; however, we can nonetheless correctly recover the average star formation history across all halos (centrals and satellites) at a given halo mass, as discussed in Appendix \ref{a:av_afh}.

Calculating the ICL formation history as well as the total amount of stars in the ICL is similar:
\begin{eqnarray}
ICLH_{t+\Delta t,m}^{t_{form}} & = & \sum_{m_p} \left( SFH_{t,m_p}^{t_{form}} (1-C^{merge}_{t,m}) M_{t,m}^{m_p} + ICLH_{t,m_p}^{t_{form}} T_{t,m}^{m_p} (1-C^{def}_{t,m}) \right) (N_{t+\Delta t,m})^{-1} \\
ICL_{t,m} & = & \sum_{t_{form}} ICLH_{t,m_p}^{t_{form}} (1-C_\mathrm{loss}(t, t_{form} - \Delta t, t_{form})).
\end{eqnarray}
Note that the merging satellite ICL is not deposited into the central galaxy's ICL, which is the same approach taken by \cite{Conroy07} and \cite{purcell-etal-07}.  Because the ICL of a merging satellite halo will get stripped long before the stars in its associated galaxy, the stripped satellite's ICL will end up in a much more extended, faint distribution than the host's ICL.

This formal specification is equivalent to a discrete differential equation for the stellar mass history.  It is thus important to properly specify the boundary conditions; that is, the stellar mass history at $t=0$.  In the absence of constraints on stellar mass functions above $z \approx 9$, we extrapolate the fit to the stellar mass -- halo mass relation to $z = 15$ and assume a uniform star formation rate prior to that redshift.  While incorrect in detail at $z=15$, this choice has little effect on the star formation rates or histories of galaxies at later (and more visible) times.

\section{Calculating Observables}

\label{a:observables}

Given the large number of equations in our framework, it is appropriate to describe exactly how a point in the parameter space of the cosmic stellar mass history (Eqs. \ref{e:cosmic_sfh}-\ref{e:scatter}) and uncertainties (Eqs. \ref{e:syst_mu}-\ref{e:f_sfr}) gets converted into quantities comparable to published observations.  The three observables in our case are stellar mass functions, specific star formation rates, and the cosmic star formation rate density.  We deal with each in turn.

\subsection{The Observed Stellar Mass Function}

Equation \ref{e:cosmic_sfh} gives the median stellar mass as a function of halo mass for the entire redshift range we consider.  If no scatter between stellar mass and halo mass were present, the conversion from the halo mass function to the fiducial stellar mass function ($\phi_{fid}$) would be straightforward application of the chain rule:
\begin{equation}
\phi_\mathrm{fid}(M_*) = \frac{dN}{d \log_{10} M_\ast} = \frac{dN}{d\log_{10} M_h(M_\ast)} \left(\frac{d\log_{10} M_\ast}{d\log_{10} M_h}\right)^{-1}.
\end{equation}
In other words, the stellar mass function is equal to the halo mass function divided by the logarithmic derivative of Eq.\ \ref{e:cosmic_sfh}.  As discussed in \cite{Behroozi10}, this may be convolved with the scatter in observed stellar mass at fixed halo mass ($P(\Delta_{M_\ast})$) to give the true stellar mass function:
\begin{equation}
\phi_\mathrm{true} (M_\ast) = \int_{-M_\ast}^{\infty} \phi_\mathrm{fid}(M_\ast + \Delta_{M_\ast}) P(\Delta_{M_\ast})d\Delta_{M_\ast}.
\end{equation} 
In our model, both the scatter in true stellar mass at fixed halo mass (Eq.\ \ref{e:scatter}) and the scatter in observed stellar mass at fixed true stellar mass (Eq.\ \ref{e:psf}) are log-normal distributions; hence $P(\Delta_{M_\ast})$ is a log-normal distribution with width equal to the sum in quadrature of the widths of the two sources of scatter ($\sqrt{\sigma^2 + \xi^2}$).

To convert from the true stellar mass function to the observed stellar mass function, we need to correct not only for observational completeness (Eq.\ \ref{e:completeness}), but also for systematic biases in the measured stellar masses (Eqs. \ref{e:syst_mu} and \ref{e:syst_kappa}).  This latter correction is made somewhat trickier by the fact that we only know the passive fraction as a function of the measured stellar mass.  The true passive fraction, $f_\mathrm{passive,true}(M_\ast)$ is related to the observed passive fraction by:
\begin{equation}
f_\mathrm{passive,obs}(M_{\ast,\mathrm{meas}}) = \frac{\phi_\mathrm{true}(M_{\ast,\mathrm{meas}}10^{-\mu}) f_\mathrm{passive,true}(M_{\ast,\mathrm{meas}}10^{-\mu})}{\phi_\mathrm{true}(M_{\ast,\mathrm{meas}}10^{-\mu}) f_\mathrm{passive,true}(M_{\ast,\mathrm{meas}}10^{-\mu}) + \phi_\mathrm{true}(M_{\ast,\mathrm{meas}}10^{-\mu-\kappa}) (1-f_\mathrm{passive,true}(M_{\ast,\mathrm{meas}}10^{-\mu-\kappa}))}.
\end{equation}
In the case that $\kappa=0$, then this equation reduces to $f_\mathrm{passive,true}(M_{\ast,\mathrm{meas}}10^{-\mu}) = f_\mathrm{passive,obs}(M_{\ast,\mathrm{meas}})$.  Otherwise, this equation becomes a recurrence relation:
\begin{equation}
\frac{f_\mathrm{passive,true}(M_{\ast,\mathrm{meas}}10^{-\mu})}{1-f_\mathrm{passive,true}(M_{\ast,\mathrm{meas}}10^{-\mu-\kappa})} = \frac{\phi_\mathrm{true}(M_{\ast,\mathrm{meas}}10^{-\mu-\kappa})}{\phi_\mathrm{true}(M_{\ast,\mathrm{meas}}10^{-\mu})}\left[(1-f_\mathrm{passive,obs}(M_{\ast,\mathrm{meas}}))^{-1}-1\right].
\end{equation}
Solving this equation is possible, but using the result has disadvantages.  Specifically, calculating $f_\mathrm{passive,true}$ couples the values of ill-constrained regions of the stellar mass function to well-constrained regions through this recurrence relation.  The values for $\alpha$---the faint-end slope of the SMHM relation--and $\kappa$ will then be constrained by the MCMC algorithm attempting to improve fits to the well-constrained regions of the stellar mass function, and will lose their physical significance.

We therefore compromise by decoupling the true passive fraction from the value of $\kappa$.  We take $f_\mathrm{passive,true}(M_{\ast,\mathrm{meas}}10^{-\mu}) = f_\mathrm{passive,obs}(M_{\ast,\mathrm{meas}})$, the exact solution for $\kappa =0$.  This results in a change to the definition of $\kappa$ so that it no longer corresponds to an exact offset for active galaxy stellar masses; however, the correlation between $\kappa$ and the average offset implied for active galaxies should still be high.  With this approach, it is straightforward to calculate the measured stellar mass function:
\begin{equation}
\phi_\mathrm{meas}(M_{\ast,\mathrm{meas}}) =  f_\mathrm{passive,true}(M_{\ast,\mathrm{meas}}10^{-\mu})\phi_\mathrm{true}(M_{\ast,\mathrm{meas}}10^{-\mu}) +  \phi_\mathrm{true}(M_{\ast,\mathrm{meas}}10^{-\mu-\kappa})(1-f_\mathrm{passive,true}(M_{\ast,\mathrm{meas}}10^{-\mu-\kappa})).
\end{equation}
Or, in terms of the observed passive fraction,
\begin{equation}
\phi_\mathrm{meas}(M_{\ast,\mathrm{meas}}) =  f_\mathrm{passive,obs}(M_{\ast,\mathrm{meas}})\phi_\mathrm{true}(M_{\ast,\mathrm{meas}}10^{-\mu}) +  \phi_\mathrm{true}(M_{\ast,\mathrm{meas}}10^{-\mu-\kappa})(1-f_\mathrm{passive,obs}(M_{\ast,\mathrm{meas}}10^{-\kappa})).
\end{equation}

Then, to correct for observational completeness, we multiply by the appropriate factor:
\begin{equation}
\phi_\mathrm{obs}(M_\mathrm{\ast,\mathrm{meas}}) = c(z) \phi_\mathrm{meas}(M_\mathrm{\ast,\mathrm{meas}}).
\end{equation}

Finally, as surveys usually return the stellar mass function over a range of redshifts (e.g., $z_1 < z < z_2$), we must perform one last integration  to determine the survey mass function:
\begin{equation}
\phi_\mathrm{survey}(M_{\ast,\mathrm{meas}}) = \frac{\int_{z_1}^{z_2} \phi_\mathrm{obs}(M_{\ast,\mathrm{meas}})dV_c(z)}{V_c(z_2) - V_c(z_1)},
\end{equation}
where $V_c(z)$ is the comoving volume out to redshift $z$.

\subsection{Calculating the Observed Specific Star Formation Rates and Cosmic Star Formation Rate Density}
\label{s:obs_ssfr_csfr}

Eq.\ \ref{e:full_sfr} gives the average star formation rate for halos as a function of halo mass and redshift.  Eq.\ \ref{e:cosmic_sfh} gives the median stellar mass for halos as a function of halo mass and redshift.  Ignoring scatter, the fiducial SSFR is simply:
\begin{equation}
SSFR_\mathrm{fid}(M_\ast) = \frac{SFR(M_h(M_\ast))}{M_\ast}.
\end{equation}
Scatter in the observed stellar mass at fixed halo mass results in contributions from lower- and higher-mass halos, weighted by number density.  If there is no correlation between stellar mass and star formation rate beyond the correlation between halo mass and star formation rate, then the specific star formation rate would be as follows:
\begin{equation}
SSFR(M_\ast)_\mathrm{nc} = M_\ast^{-1} \phi_\mathrm{true}^{-1}(M_\ast)\int_{-M_\ast}^{\infty} \phi_\mathrm{fid}(M_\ast + \Delta_{M_\ast}) SFR(M_h(M_{\ast}+\Delta_{M_\ast})) P(\Delta_{M_\ast})d\Delta_{M_\ast}.\\
\end{equation}
However, with a nonzero covariance ($\rho$) between SFR and stellar mass at fixed halo mass this equation is amended to become:
\begin{eqnarray}
\label{e:ssfr}
SSFR(M_\ast)_\mathrm{true} & =&  M_\ast^{-1} \phi_\mathrm{true}^{-1}(M_\ast)\int_{-M_\ast}^{\infty} \phi_\mathrm{fid}(M_\ast + \Delta_{M_\ast}) (SFR(M_h(M_{\ast}+\Delta_{M_\ast}))10^{\rho \Delta_{M_\ast}'}) P(\Delta_{M_\ast})d\Delta_{M_\ast}\nonumber\\
\Delta_{M_\ast}' &=& \log_{10}\left(\frac{M_\ast + \Delta_{M_\ast}}{M_\ast \exp(0.5(\xi \ln 10)^2)}\right).
\end{eqnarray}
The formula for $\Delta_{M_\ast}'$ is necessary because $M_h(M_\ast + \Delta_{M_\ast})$ gives the halo mass at which the \textit{median} stellar mass is $M_\ast + \Delta_{M_\ast}$; however, the covariance between SFR and SM is calculated relative to the \textit{average} stellar mass at fixed halo mass.  The exponential in the denominator ($\exp(0.5(\xi \ln 10)^2)$) is exactly the ratio of the average value to the median value of a log-normal distribution with standard deviation $\xi$.

There are restrictions on the plausible range of covariance coefficient $\rho$.  At low redshifts, a multiple regression analysis on group catalogue data from the SDSS \citep{Wetzel11} suggests that, at fixed halo mass, the average star formation rate rises by 0.23 dex per 1 dex increase in stellar mass.  At high redshifts (e.g., $z > 6$), the covariance will be almost perfect (1 dex rise in SFR per 1 dex rise in stellar mass) because of the short times that galaxies have had to form stars.  Between these two times, it is reasonable to assume that the covariance increases with increasing redshift, so we adopt the following formula to interpolate between $z \sim 0$ and $z \gtrsim 6$:
\begin{equation}
\label{e:rho_def}
\rho(a) = 1 + (4\rho_{0.5} -3.23)a + (2.46-4\rho_{0.5})a^2.
\end{equation}
This formula for $\rho(a)$ gives $\rho(0) = 1$, $\rho(1) = 0.23$, and $\rho(0.5) = \rho_{0.5}$, where $\rho_{0.5}$ is a free parameter in our analysis.

By comparison, the total (cosmic) star formation rate density is straightforward to calculate; it is simply the sum of all star formation in all halos:
\begin{equation}
\label{e:csfr}
CSFR_\mathrm{true} = \int_{-\infty}^{\infty} SFR(M_h) \frac{dN}{d\log M_h} d\log M_h.
\end{equation}

To calculate the measured quantities (as opposed to true quantities), we adopt the same conversion between measured stellar mass and true stellar mass as for the stellar mass function.  However, we do not impose additional corrections for active galaxies; by definition, because recent star formation is more evident in such galaxies, it is somewhat easier to recover the star formation rate, as opposed to the overall stellar mass.  Therefore, we only apply a uniform systematic correction and do so in the same way as for stellar masses:
\begin{equation}
SFR_\mathrm{meas}(M_h) = SFR_\mathrm{true}10^{\mu}.
\end{equation}
To calculate $SSFR_\mathrm{meas}$ and $CSFR_\mathrm{meas}$, we would therefore replace $SFR_\mathrm{true}$ with $SFR_\mathrm{meas}$ in Eqs.\ \ref{e:ssfr} and \ref{e:csfr}.  There is one more complicating factor; namely the burstiness / dustiness of high-redshift galaxies.  To account for this, we add additional factors to calculate the observed SSFR and CSFR:
\begin{eqnarray}
SSFR_\mathrm{obs}(M_\ast) & = & \frac{1}{1-b(1-c(z))} SSFR_\mathrm{meas}(M_\ast)\nonumber\\
CSFR_\mathrm{obs} & = & [1 - (1-b)(1-c(z))] CSFR_\mathrm{meas}.
\label{e:burstiness_correction}
\end{eqnarray}

\section{The Use of Double Power Laws to Fit the Stellar Mass -- Halo Mass Relation}

\label{a:dp_law}
\begin{figure}
\vspace{-6ex}
\begin{tabular}{rr}
\plotminigrace{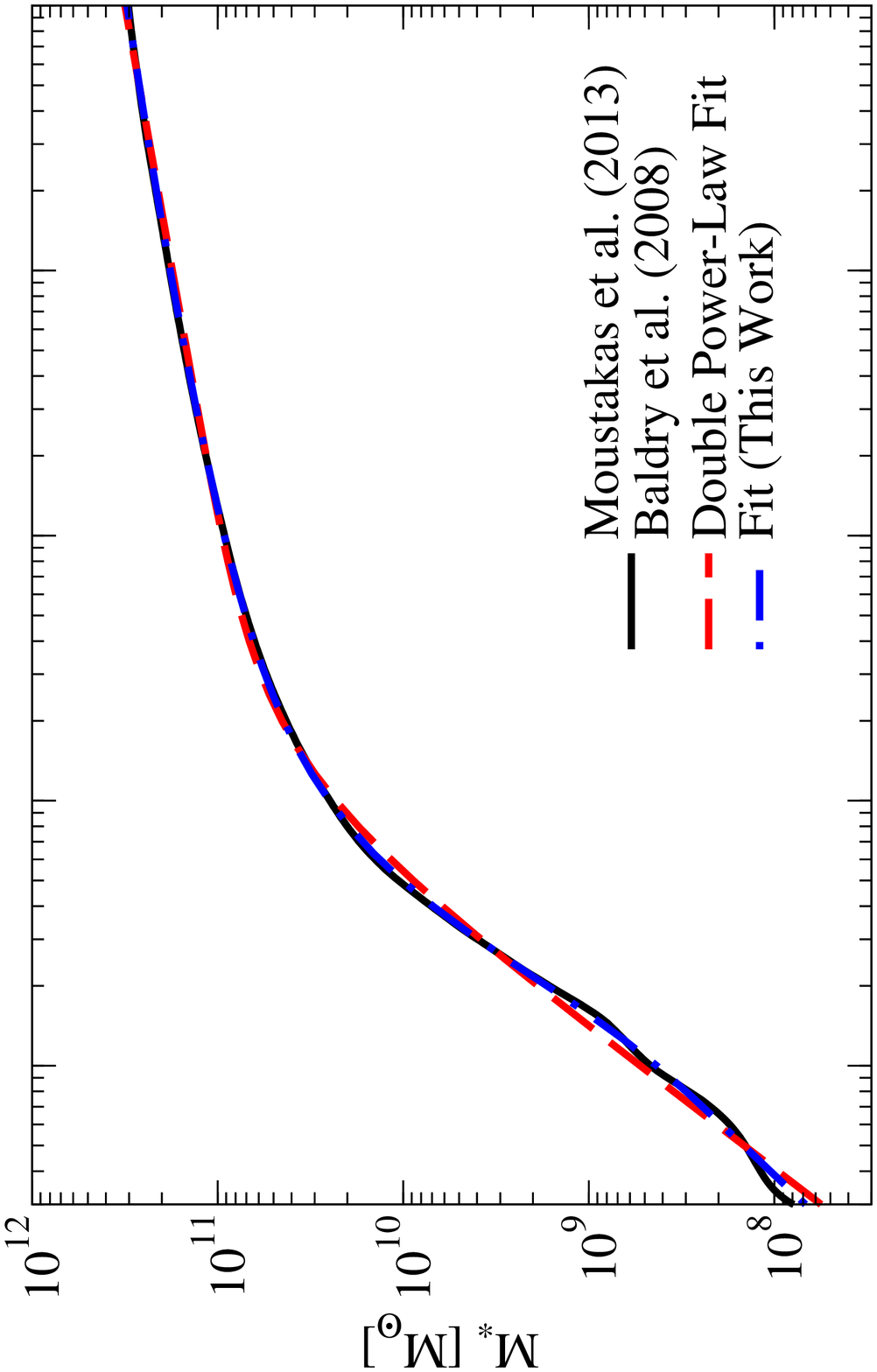} & \plotminigrace{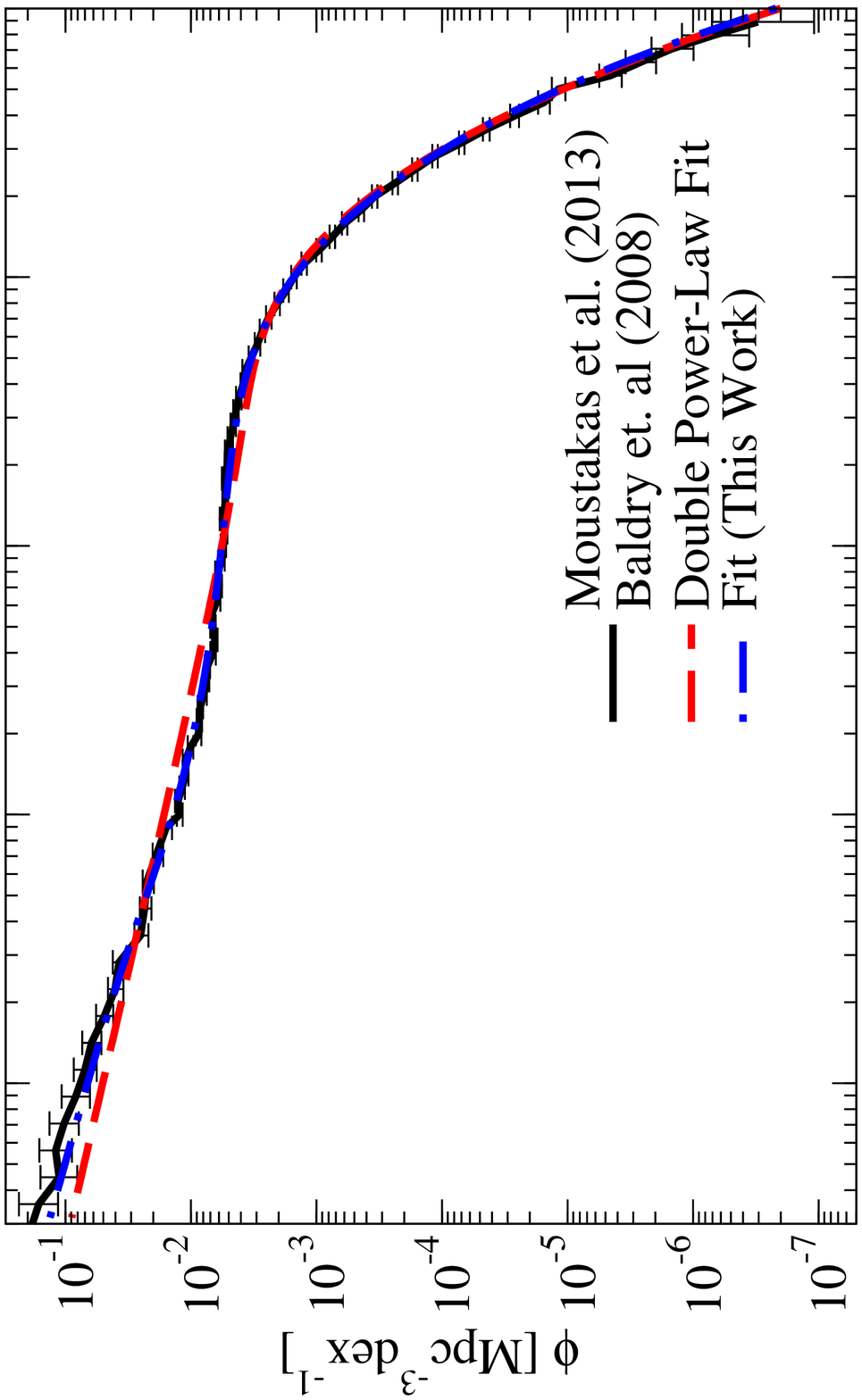}\\[-18ex]
\plotminigrace{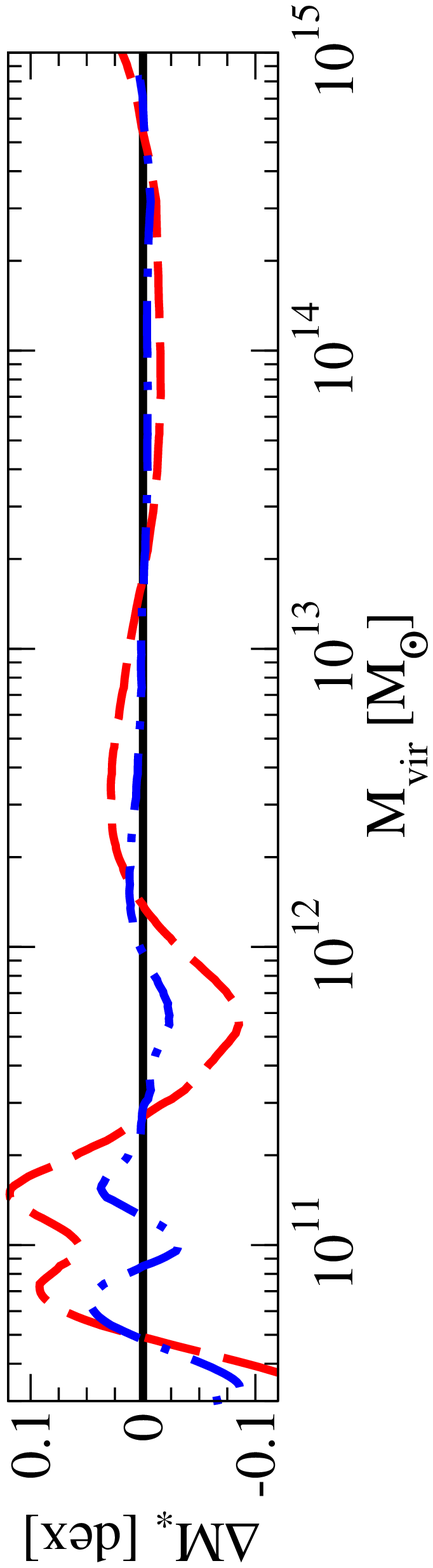} & \plotminigrace{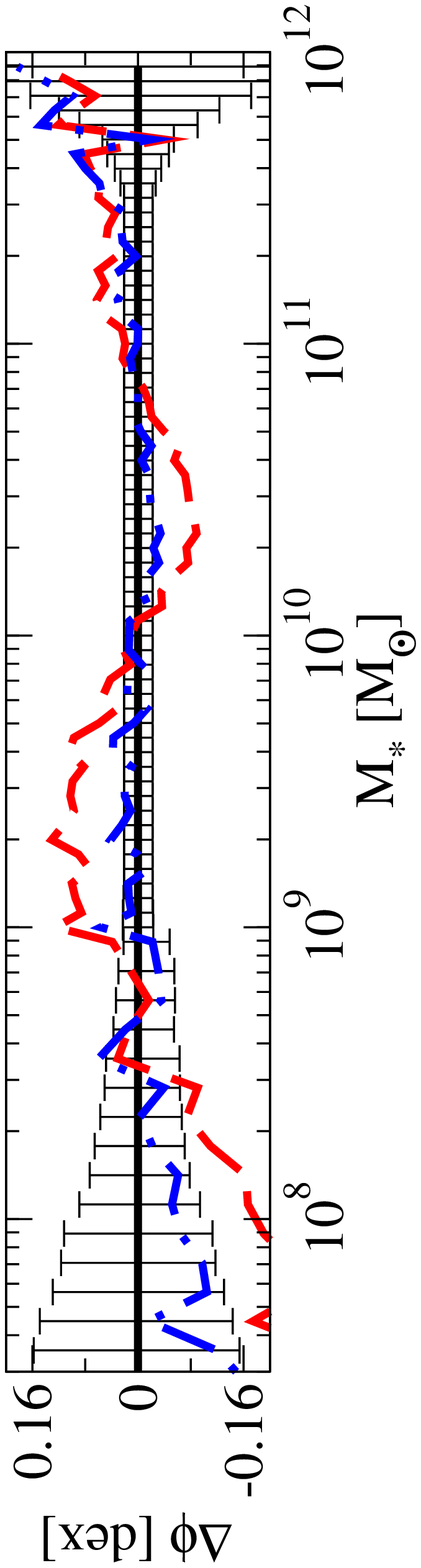}\\[-30ex]
\end{tabular}
\caption{\textbf{Left} panel: The stellar mass--halo mass relation as obtained by deconvolution of observed data \citep{Moustakas12,Baldry08} with an assumed scatter in observed stellar mass at fixed halo mass of 0.2 dex; the dashed and dashed-dotted lines show a double power law fit and the five-parameter fit used in this work.  The double power law fit differs by up to 0.1 dex from the observationally-derived relation.  The effects on the stellar mass function are shown in the \textbf{right} panel; namely, the fits shown in the left panel have been convolved with 0.2 dex scatter as well as with the halo mass function so as to compare to the original stellar mass functions.  Again, differences of 0.1 dex are seen in the double power law case, which is well outside of the reported error bars.}
\label{f:fit_compare}
\end{figure}

A double power law has been used to fit the stellar mass (or luminosity) to halo mass relation at least since \cite{Yang03}.  While using it is simple and convenient, the quantity and quality of galaxy observations have improved dramatically since its first use.  To test its continued usefulness, we have used abundance matching to extract the implied stellar mass -- halo mass (SMHM) relation from our low-redshift stellar mass functions \citep{Moustakas12,Baldry08}.  To account for the presence of scatter in observed stellar mass at fixed halo mass, we use a method based on Richardson-Lucy deconvolution to extract the median stellar mass as a function of halo mass.  We assume a combined scatter of 0.2 dex.

The deconvolved SMHM relation is shown in Fig. \ref{f:fit_compare}.  The figure demonstrates that a double power-law can approximate the relation, but can be discrepant by up to 0.1 dex.  Adding one more parameter improves the fit substantially, to a mean offset of less than 0.025 dex in our case.  The discrepancies in the SMHM relation translate directly to discrepancies in the assumed stellar mass function.  This may be seen in the right-hand panel of Fig.\ \ref{f:fit_compare}; the fits for the SMHM relations have been used to populate the halo mass function and reconvolved with 0.2 dex scatter to determine the resulting stellar mass functions.  While the five-parameter fit that we use is largely within the error bars of the observational data, the double power law fit gives results that are again discrepant by up to 0.1 dex or 4$\sigma$ from the observed stellar mass function.

It is tempting to ignore this issue: after all, the size of the systematic uncertainties in stellar mass is significantly larger ($\sim$0.25 dex).  However, if it is indeed the case that the deviations from the power law are driven by systematic uncertainties, then part of the value of the physical interpretation of the double power law is lost: the high-mass and low-mass slopes are now contaminated by using the incorrect parametrization.  In our five-parameter fit, the high-mass dependence and low-mass slope are somewhat more insulated from unparametrized systematic errors in the transition region, leading to a more fair estimation of the high- and low-mass behavior of the SMHM relation.

Naturally, no fit will perfectly reproduce the original function.  Special caution is necessary when using a MCMC method where a parametrization to tightly-constrained data is accompanied by additional systematic or nuisance parameters.  This is because the MCMC method may try to recruit the additional parameters to improve the fit, rather than for their intended purpose (i.e., to cover the range of allowed uncertainties).  As shown in Fig.\ \ref{f:fit_compare}, our chosen parametrization can reproduce the local SMF to within 0.02 dex in the vast majority of cases.  Hence, to avoid overfitting bias, we reduce the magnitude of errors by 0.02 dex at each data point before calculating the $\chi^2$ value used with our MCMC method.  At $z<0.2$, this has the impact of decoupling the systematic uncertainty parameters from the detailed shape of the SMF; at higher redshifts, there is no effect, as the typical size scale of the errors for observational data is already 0.1--0.2 dex.

\section{The Redshift Evolution of the Stellar Mass -- Halo Mass Relation from Abundance Matching}

\label{a:ab_matching}

\begin{figure}
\vspace{-6ex}
\plotminigrace{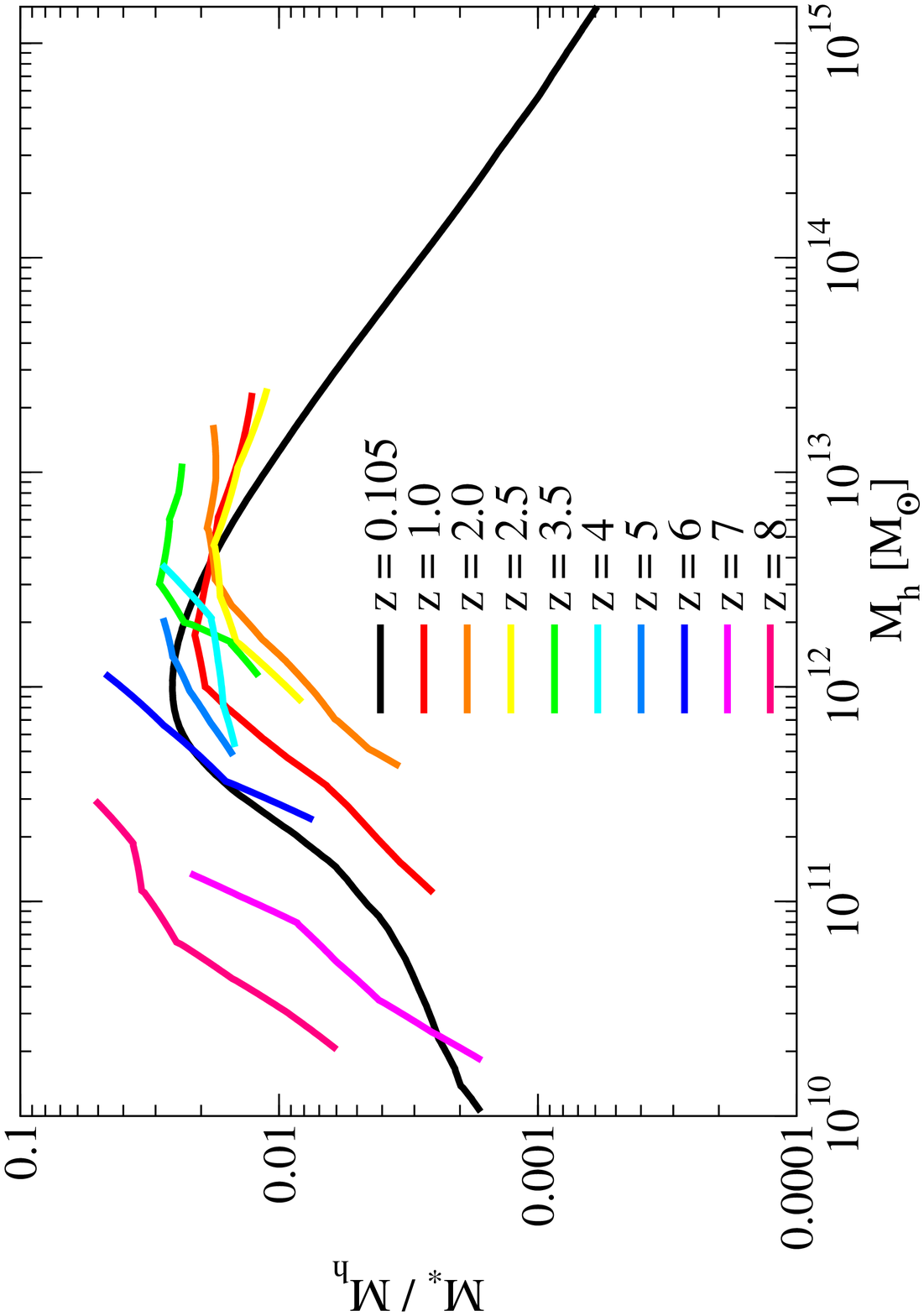}
\plotminigrace{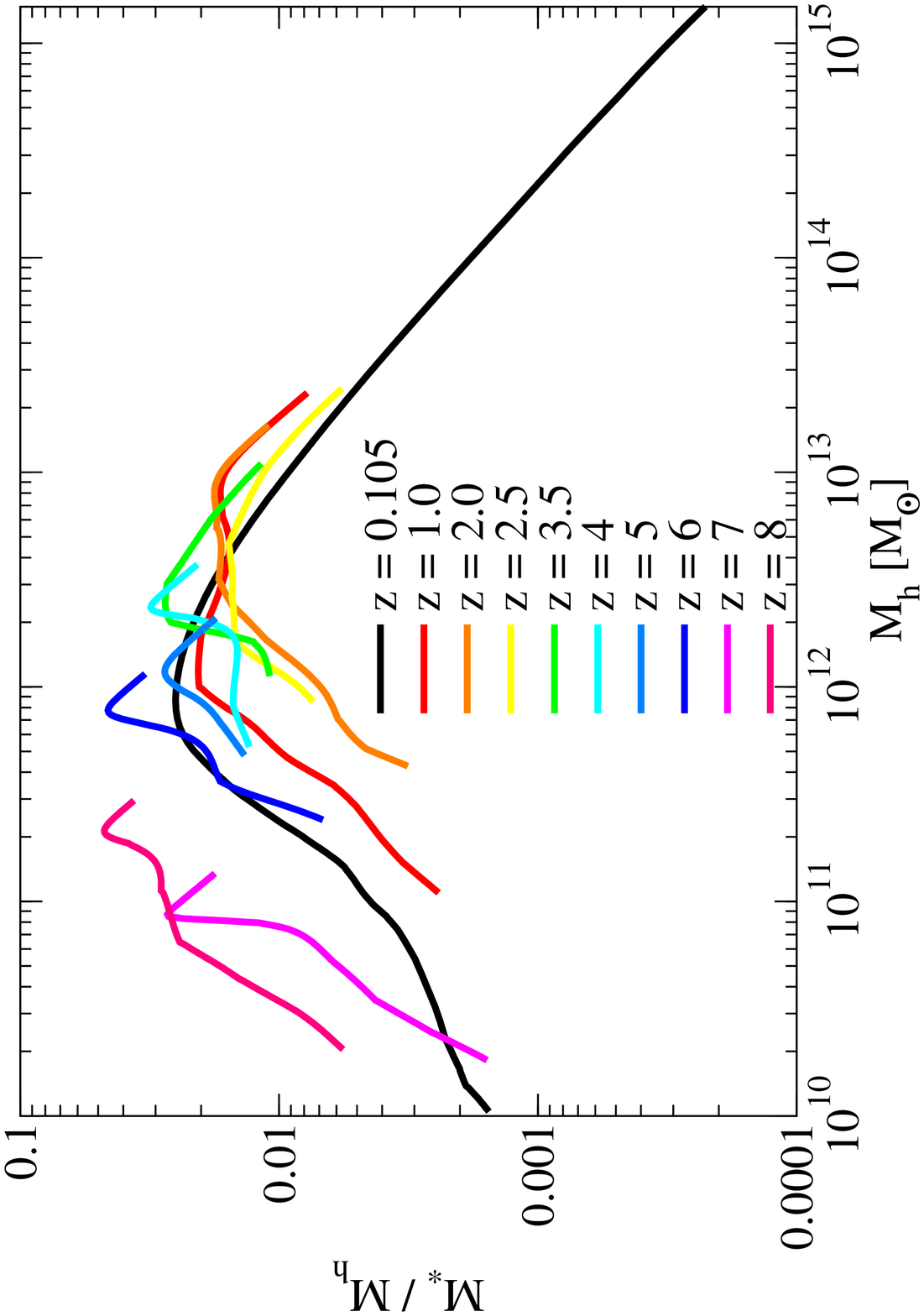}
\caption{\textbf{Left} panel: the stellar mass to halo mass ratio inferred via direct abundance matching of stellar mass functions (Table \ref{t:smf_data}) to halo mass functions from $z=0$ to $z=8$, with zero scatter.  \textbf{Right} panel: same, with deconvolution of 0.16 dex intrinsic scatter (in stellar mass at fixed halo mass) and $(0.07+0.04z)$ dex of measurement scatter (in measured stellar mass at fixed true stellar mass) applied.  Note that deconvolution applied to noisy data (e.g., SMFs) amplifies the noise, which is why features in the left panel appear smoother than those in the right panel.}
\label{f:ab_smmr} 
\end{figure}

As a check of the accuracy of our analysis pipeline, it is instructive to compare with a direct abundance matching approach.  In Fig.\ \ref{f:ab_smmr}, we show results for both direct abundance matching (left panel) and abundance matching including an assumed intrinsic ($\xi$) and measurement ($\sigma$) scatter that match our assumed fiducial defaults.  Notably, several features present in our main analysis are evident in this simpler approach:

\begin{enumerate}
\item The SMHM relation for low-mass halos ($M_h < 10^{11}\Msun$) at $z=0$ deviates from a power law.
\item The halo mass at which the integrated star formation is most efficient rises by 0.5 dex from $10^{12}\Msun$ at $z=0$ to $10^{12.5}\Msun$ at $z=2$.
\item Above $z=2$ it is difficult to locate the halo mass at which the peak integrated efficiency occurs because surveys at high redshifts do not have sufficient volume to probe the exponential tail of the stellar mass function (if one exists).  If the UV luminosity function is any indication, however, the $z=8$ results suggest that the peak may occur at lower masses above $z=2$.
\item There appears to be a turnaround in the SMHM relation near $z\sim 2$.   The integrated efficiency of $M_h\sim 10^{11.5}\Msun$ halos falls from $z=0$ to $z=2$, but then increases at higher redshifts.
\end{enumerate}
At high redshifts, it is difficult to tell what, if any, evolution there is in the shape of the SMHM relation.  However, it is clear that either the efficiency or the characteristic halo mass scale is evolving.  If we constrain the shape parameters in our fitting formula for the SMHM relation ($\alpha,\delta,$ and $\gamma$) to the $z=0$ values, we find that high redshifts strongly prefer a halo mass scale that scales as a power law with redshift, but have comparatively little implied evolution in the efficiency at high redshifts.  For that reason, when we extrapolate to high redshifts unconstrained by data in our main analysis ($z>8.5$), we assume that $\epsilon,\alpha,\delta,$ and $\gamma$ do not evolve beyond $z=8.5$, but that the characteristic halo mass $M_1$ continues to evolve according to the redshift scaling we have adopted.

\comment{
\section{Uncertainties in Observationally-Derived Star Formation Rates}

\label{a:sfr_uncertainties}

\subsection{Uncertainties in UV/H$\alpha$ SFRs}

\label{s:uv_errs}

In principle, rest-frame ultraviolet luminosities probe the abundance of short-lived, massive OB stars and thus (for a given IMF) represent an excellent tracer of recent star formation.  In practice, especially for massive galaxies, heavy dust obscuration can reprocess a substantial fraction of the emitted UV radiation into the infrared.  This substantially reddens the galaxy, also posing problems for galaxy identification using the Lyman-Break technique.  At very high redshifts ($z>6$), however, there is some evidence \citep{Labbe10} that the effects of dust reprocessing may not be as significant, perhaps because galaxies have not had time to form significant dust lanes.

The presence of dust must be analyzed for and removed from each individual galaxy being studied.  Red galaxy colors may be caused by any combination of dust obscuration, old stellar populations, and high redshifts, and so detailed modeling of the galaxy in many color bands is necessary to recover the galaxy redshift and dust obscuration (which both influence the derived UV luminosity) in an unbiased way.  This recovery is thus subject to many of the same systematics relevant to recovering galaxy stellar masses; however, the systematics can in some cases affect the recovered SFR and stellar mass in opposite ways.  For example, if the estimated stellar population is too old, the galaxy may be assigned a lower dust correction to account for the galaxy colors.  The net effect may result in an unchanged galaxy stellar mass (as the inferred stellar mass-to-light ratio will be too high, but the inferred dust-corrected luminosity will be too low), but the star formation rate will definitely be underestimated due to the error in dust correction.  At high redshifts ($z>2$) or when limited individual galaxy data is available at longer wavelengths, authors will often use a single UV correction factor based on stacked galaxy samples or on extrapolations from lower redshifts, which can carry substantial systematic uncertainties \citep{Labbe10,Magnelli11,Robotham11}.

H$\alpha$ luminosities also trace massive star formation; this happens because massive stars cause photoionization of surrounding HII regions, which then produce Lyman $\alpha$ emission due to recombination.  Like UV luminosities, H$\alpha$ luminosities must be dust-corrected, but the magnitude of this correction (up to $\sim$ a factor of 3) is somewhat less than that for UV luminosities \citep{Twite12}.  H$\alpha$ luminosities probe star formation over a significantly shorter timescale (5-20 Myr) than UV luminosities \citep{Leroy12}; intrinsic variability in star formation rates on those timescales can therefore cause large amounts of scatter in H$\alpha$ luminosities with respect to UV luminosities.  This problem especially complicates H$\alpha$ measurements at high redshifts.  Because spectroscopy of large samples of galaxies beyond $z\sim 1.0$ is difficult, most high-redshift H$\alpha$ studies have relied on narrow-band filters \citep[e.g.,][]{Tadaki11} to select H$\alpha$ luminosities at a single redshift.  Since the volume probed is low, this results in only a handful of galaxy star formation rates; this in combination with the high variance in individual rates means that sample variance can cause significant issues with accuracy for average H$\alpha$ SFRs.

One commonly-overlooked systematic error in optical SFRs comes from the bias introduced by separately fitting the dust model for each galaxy.  As the dust model for each galaxy can never be derived exactly, individual galaxy SFRs will have some scatter relative to the true galaxy SFRs.  One way to estimate the shape of this scatter is to compare two different SFR estimates with different assumptions; for example, \cite{Salim07} finds log-normal scatter consistent with intrinsic scatter of 0.3 dex.  In calibrating one estimator to another (or to observations), authors typically make sure that the \textit{median} offsets are zero; however, they often fail to notice that the \textit{expected mean} offset is nonzero because the log-normal distribution is skewed in linear space.  Specifically, a log-normal distribution with median value 1 and scatter $\sigma$ (in dex) has an expected mean value of
\begin{equation}
\label{e:median_to_average}
\langle x \rangle = \exp(0.5(\sigma\ln 10)^2).
\end{equation}
For a log-normal distribution with 0.3 dex scatter, this corresponds to an overall bias of +0.1 dex.  This correction does not apply to UV studies that apply a median dust correction to all galaxies or to those that already correct for this bias.

\subsection{Uncertainties in IR/FIR SFRs}

\label{s:ir_errs}

Some studies seek to avoid this problem entirely by focussing on IR emission only, especially at rest-frame $24\mu$m.  This wavelength is heavily radiated by dusty star formation regions \citep{Calzetti07}; it is low-frequency enough to avoid expected contamination from polycyclic aromatic hydrocarbons \citep{Calzetti07,Calzetti10} and high enough to avoid expected IR contamination from sources in the Milky Way \citep{Shupe98}.  An obvious disadvantage of this approach is that low-metallicity and low-luminosity galaxies do not have enough dust to reprocess all the UV luminosity, leading to the potential to underestimate SFRs unless another indicator (such as UV or H$\alpha$) is used \citep{Calzetti10}.  The calibration of IR estimates of SFRs is also somewhat variable, as estimates of SFRs are calibrated to SEDs of star-forming regions, and thus somewhat dependent on the exact choice of stellar libraries used at the 20\% level \citep{Rujopakarn10}.  

Just as important, an often-overlooked uncertainty in IR-only studies is how the redshift determinations can bias the recovered luminosities.  High-redshift surveys often have limited spectroscopic data, and so require a way to convert galaxy photometry to a redshift to determine the absolute IR magnitude.  Furthermore, spectroscopic coverage is generally biased towards more luminous sources.  While the photometric redshift calibration is fixed to agree on average with the spectroscopic redshift, there is nonetheless a degeneracy between non-cosmological sources of reddening (e.g., dust obscuration and the stellar population) and the redshift of a galaxy, especially for lower-luminosity objects where even the photometry is poor. The photometric redshift estimate thus carries an implicit set of assumptions about star formation no matter what method is used (e.g., neural networks, SED templates).  These assumptions may then cause systematic biases in the conversion of apparent IR luminosities into absolute ones as a function of e.g., galaxy dust or age; thus, an implicit dependence on star formation rate modeling assumptions is \textit{unavoidable}, with significant consequences for the recovered normalization of the SFR as a function of redshift.

\subsection{Uncertainties in 1.4 GHz SFRs}

\label{s:radio_errs}

The largest uncertainty in using radio-derived star formation rates is in the calibration.  Competing calibrations have been proposed by e.g., \cite{Condon02}, \cite{Yun01} and \cite{Bell03b}, which differ by a factor of 0.3 dex (see also \citealt{Dunne09}).  The main reason for this discrepancy is that the radio-SFR calibration is determined by dividing the radio flux by another measure of the star formation rate, such as the FIR luminosity, which subjects the method to the same overall systematics as affect any other measurement.  Additionally, while 1.4 GHz radiation is believed to be largely unaffected by dust or other complicating factors that interfere with lower-wavelength emissions, \cite{Bell03b} found a linear correlation between 1.4 GHz luminosity and IR luminosity, even down to the faintest galaxies, where the dust contribution (and thus the correlation between IR luminosity and star formation) should be less important.  \cite{Bell03b} derived a 1.4GHz to SFR conversion with an increased contribution from low-luminosity galaxies, although the overall normalization is less than that in \cite{Condon02}.  As for the calibration at higher redshifts, \cite{Sargent10} find no need for evolution in the radio-SFR conversion, based on a steady FIR-radio correlation out to $z \sim 2$.

A secondary systematic concerns the fraction of radio emissions coming from Active Galactic Nuclei (AGN).  \cite{deVries07} argue that, at low redshifts, AGN contribute only 10\% of the total radio luminosity in galaxies, with star formation responsible for the remaining 90\%.  \cite{Reddy05} find a luminosity-dependent AGN fraction of up to $\sim$30 \% in $z=2$ $K$-selected galaxies; however, in their sample as a whole, they find that the correction for AGN contamination is only 13\%.
}

\section{Calculating Uncertainties in Cosmic and Specific Star Formation Rates}
\label{a:csfr_ssfr_uncert}

\begin{table}
\begin{center}
\caption{Empirical Fit to Cosmic Star Formation Rates}
\label{t:csfr_fit}
\begin{tabular}{rcccc}
\hline
\hline
Publication & $z_0$ & $A$ & $B$ & $C$\\
\hline
This Work & 1.243 & -0.997 & 0.241 & 0.180\\
\cite{Hopkins06b} & 0.840 & -1.311 & 0.085 & 0.143\\
\hline
\end{tabular}
\end{center}
\tablecomments{Fits according to Eq.\ \ref{e:csfr_fit} for the collected cosmic star formation rate data in Table \ref{t:csfr_data} (``This Work'') and the data in \cite{Hopkins06b}, all corrected to a \cite{chabrier-2003-115} IMF.  \textbf{Note that these are fits
to observational data, not fits to our best-fit model.}  The quality of the fits may be seen in Fig.\ \ref{f:csfr_comp}.  The constant $C$ has units of $\Msun$ yr$^{-1}$ comoving Mpc$^{-3}$; all other constants are unitless.  Systematic errors are estimated in Table \ref{t:csfr_errs}.}
\end{table}

\begin{table}
\begin{center}
\caption{Error Estimates for the Cosmic Star Formation Rate}
\label{t:csfr_errs}
\begin{tabular}{ccc}
\hline
\hline
Redshift Range & Inter-Publication & This Work - HB06\\
\hline
$0.025 - 0.5$ & 0.13 & -0.08 \\ 
$0.5 - 0.9$ & 0.13 & -0.12 \\ 
$0.9 - 1.5$ & 0.17 & -0.07 \\ 
$1.7 - 3$ & 0.19 & -0.05 \\ 
$3 - 8.0$ & 0.27 & -0.33 \\ 
\hline
\end{tabular}
\end{center}
\tablecomments{``Inter-publication'' refers to the standard deviation in dex between published data points (from references in Table \ref{t:csfr_data}) in redshift bins after correction for the evolution of the cosmic star formation rate (Eq.\ \ref{e:csfr_fit}); the error in the error estimate for each redshift bin is of order 25\%.  ``This Work - HB06'' refers to the difference in dex between the best fit (Eq.\ \ref{e:csfr_fit}) to data points in Table \ref{t:csfr_data} and the best fit to data points in \cite{Hopkins06b} over the same redshift bins; this is a qualitative estimate of how much assumptions about calculating the cosmic star formation rate have changed over the past decade.}
\end{table}

For cosmic star formation rates (SFRs), we selected papers from the astro-ph arXiv that
\begin{enumerate}
\item Were posted within the past 6 years (2006-2012).
\item Contained the word ``cosmic'' and either the word ``SFR'' or the words ``star'' and ``formation'' in the title or abstract.
\item Contained an estimate of the cosmic star formation rate from observations.
\item Were not superseded by later publications.
\item Contained a reference to the initial mass function (IMF) used.
\end{enumerate}
We have in addition excluded constraints from long gamma-ray burst SFR estimates for this work (e.g., \citealt{Kistler09}); we feel that not enough data points are available at this time to independently calibrate the evolution of known systematic biases (e.g., higher GRB rates in low-luminosity galaxies) with redshift.  The selected papers are summarized in Table \ref{t:csfr_data}.  From these, we calculate the inter-publication variance in five redshift bins with 16 data points each.  To take into account variation of the cosmic SFR in bins, we subtract a double-power law fit to the data points in Fig.\ \ref{f:csfr_comp}:
\begin{equation}
\label{e:csfr_fit}
CSFR(z) = \frac{C}{10^{A(z-z_0)} + 10^{B(z-z_0)}},
\end{equation}
where $A$, $B$, $C$, and $z_0$ are constants given in Table \ref{t:csfr_fit}; adding more parameters did not make the quality of the fit substantially better.  Our results are given in Table \ref{t:csfr_errs}; we find the average systematic error to range from 0.13 dex at $z=0$ to 0.27 dex for $z>3$.

\begin{figure}
\vspace{-6ex}
\plotminigrace{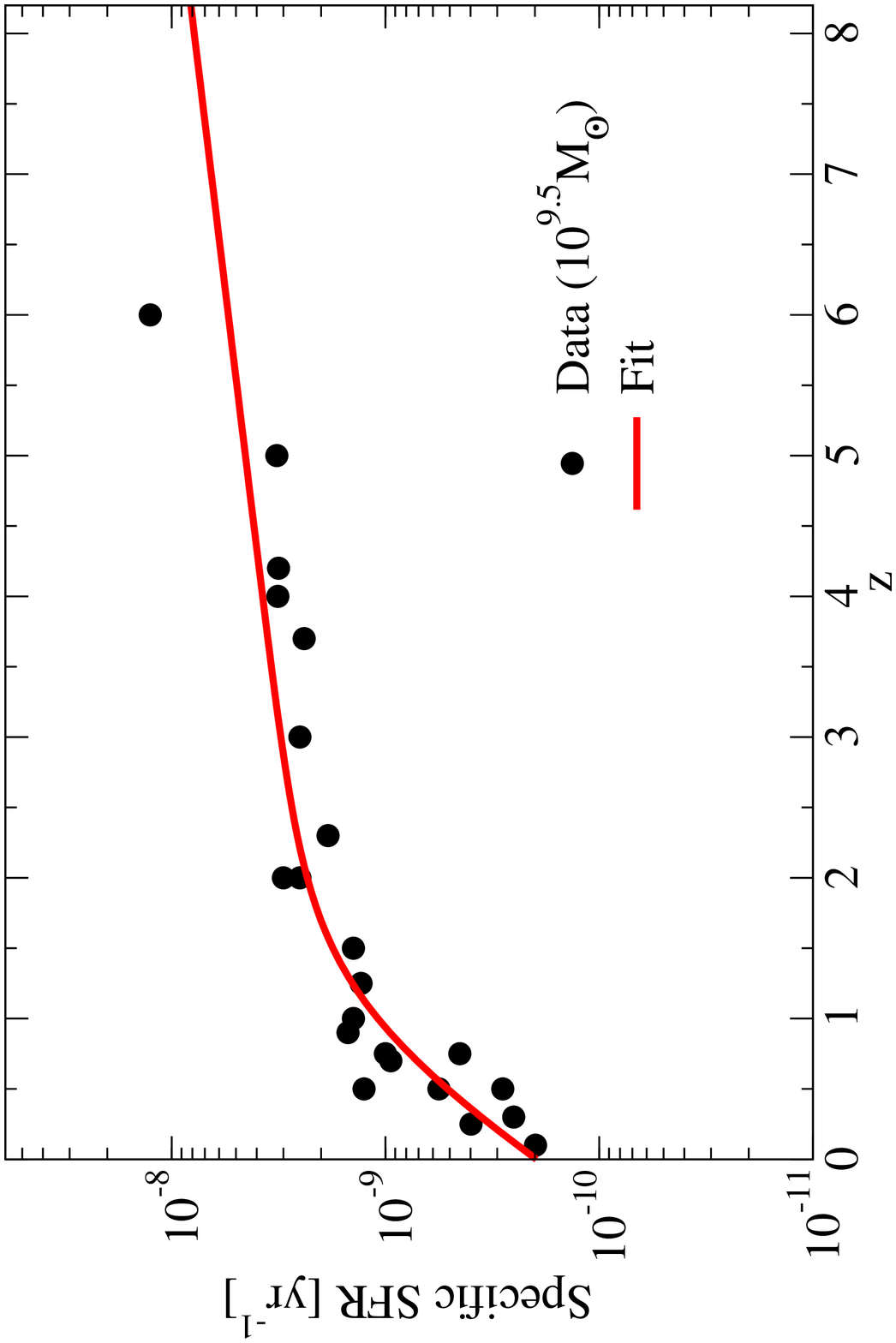}\plotminigrace{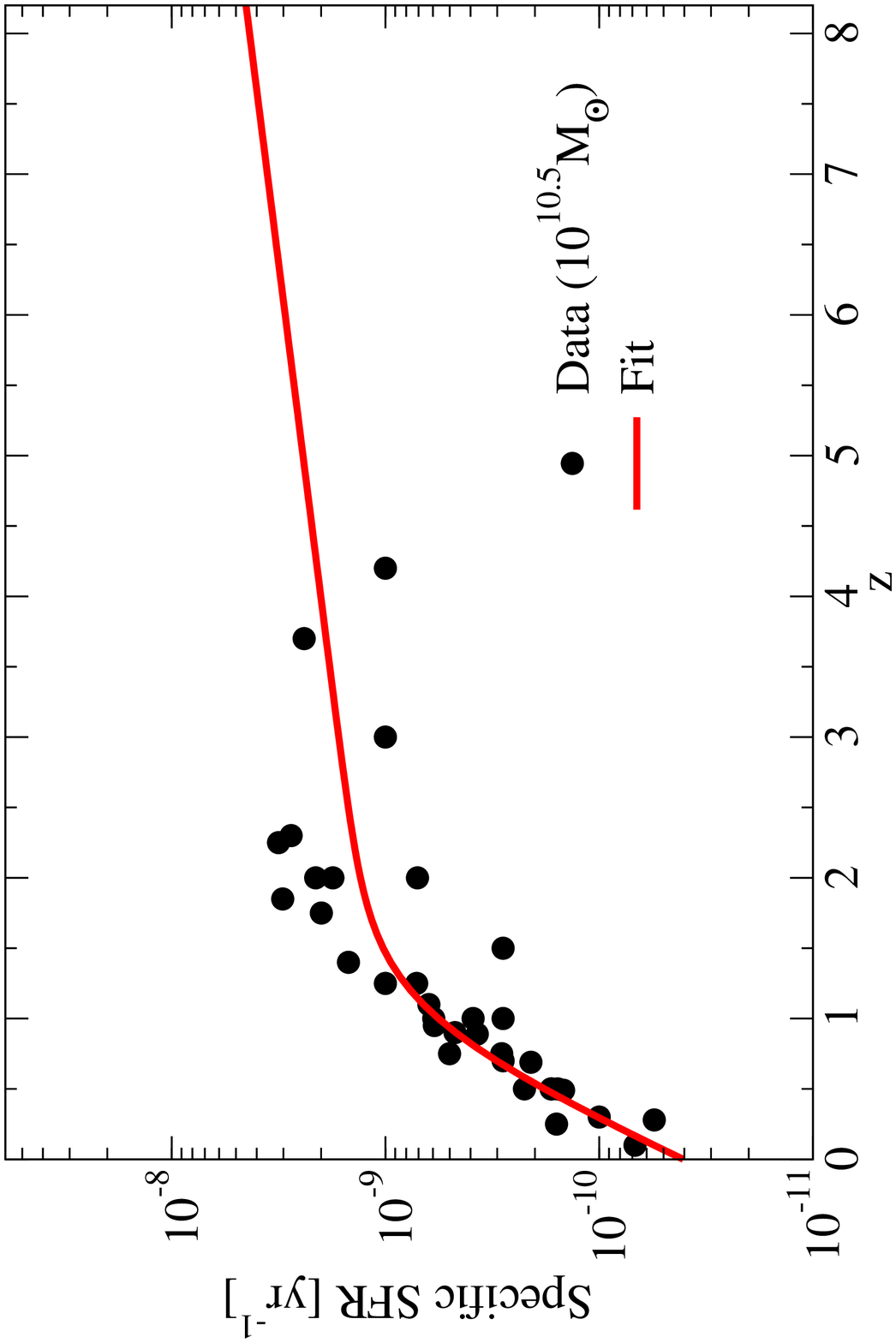}\\[-4ex]
\caption{Plots of the specific star formation rate data assembled in this study (see Table \ref{t:ssfr_data}) for galaxies of stellar mass $10^{9.5}\Msun$ (top) and $10^{10.5}\Msun$ (bottom).  Double power-law fits (Eq.\ \ref{e:csfr_fit}) to the data sets are taken from Table \ref{t:ssfr_fits}.  Error bars are not shown since authors often dramatically underestimate the true magnitude of their systematic errors (see \S \ref{s:ssfr}).}
\label{f:ssfr_comp}
\end{figure}

\begin{table}
\begin{center}
\caption{Empirical Fits for Specific Star Formation Rates}
\label{t:ssfr_fits}
\begin{tabular}{ccccc}
\hline
\hline
Stellar Mass & $z_0$ & $A$ & $B$ & $C$ \\
\hline
$10^{9} \Msun$ & 1.000* & -1.028 & -0.060 & $3.135 \times 10^{-9}$ \\
$10^{9.5} \Msun$ & 1.000* & -0.993 & -0.080 & $2.169 \times 10^{-9}$ \\
$10^{10} \Msun$ & 1.000* & -1.219 & -0.023 & $1.873 \times 10^{-9}$ \\
$10^{10.5} \Msun$ & 1.000* & -1.426 & -0.083 & $1.129 \times 10^{-9}$ \\
\hline
\end{tabular}
\end{center}
\tablecomments{Fits according to Eq.\ \ref{e:csfr_fit} for the collected specific star formation rate data in Table \ref{t:ssfr_data}, all corrected to a \cite{chabrier-2003-115} IMF.  The quality of the fits may be seen in Fig.\ \ref{f:ssfr_comp}.  \textbf{Note that these are fits to observational data, not fits to our best-fit model.}  Asterisks denote that the parameter in question ($z_0$) was consistent with 1.0 when allowed to vary; hence, all fits were constrained to have $z_0=1$ so as to allow better comparison between the evolution of the remaining fit parameters with stellar mass. The constant $C$ has units of yr$^{-1}$; all other constants are unitless.  Systematic errors are estimated in Table \ref{t:ssfr_errs}.}
\end{table}

\begin{table}
\begin{center}
\caption{Error Estimates for Specific Star Formation Rates}
\label{t:ssfr_errs}
\begin{tabular}{ccccc}
\hline
\hline
Redshift Range & $10^{9}\Msun $ & $10^{9.5}\Msun$ & $10^{10}\Msun$ & $10^{10.5} \Msun$\\
\hline
$0.025 - 0.5$ & - & 0.39 & 0.31 & 0.29\\
$0.5-0.9$ & - & 0.32 & 0.31 & 0.27\\
$0.9-1.7$ & - & 0.30 & 0.19 & 0.29\\
$1.7-3$ & - & - & 0.24 & 0.35\\
$3-8$ & 0.22 & 0.22 & 0.21 & 0.31\\
\hline
Average & \multicolumn{4}{c}{0.28}\\
\hline
\end{tabular}
\end{center}
\tablecomments{The standard deviation in dex between published data points (from references in Table \ref{t:ssfr_data}) in redshift bins after correction for the evolution of the specific star formation rate (Eq.\ \ref{e:csfr_fit}).  For reasons mentioned in \S \ref{s:ssfr}, the average does not include the estimates in the $z=3-8$ redshift range.  The error in the error estimate for each bin is of order 45\%; the error on the average error is of order 12\%.  Errors are not estimated when there are fewer than four references in a given stellar mass and redshift range.}
\end{table}

For SSFRs we conducted a very similar literature search.  In this case, the selection criteria on astro-ph were papers that were posted within the past 6 years (2006-2012); contained the word ``specific'' and either the word ``SFR'' or the words ``star'' and ``formation'' in the title or abstract; contained an estimate of the specific star formation rate as a function of stellar mass, both as calculated from observations; were not superseded by later publications; contained a reference to the initial mass function (IMF) used; and did not limit themselves to a specific class of galaxies (sub-mm, ULIRGs, satellites, star-forming only, etc.).  The selected papers are summarized in Table \ref{t:ssfr_data}.

To estimate uncertainties, we interpolate data for each publication's tabulated redshift ranges to obtain (if possible) estimates of the SSFR at four stellar masses ($10^{9}$, $10^{9.5}$, $10^{10}$ and $10^{10.5}\Msun$).  We divide the SSFR data into the same five redshift bins as for the cosmic SFR data and estimate the systematic errors from the inter-publication variance.  To exclude the effects of evolution in the SSFR over a redshift bin, we again subtract a double-power-law fit (Eq.\ \ref{e:csfr_fit}) with parameters determined separately for each stellar mass in Table \ref{t:ssfr_fits}; the quality of the fit is shown in Fig.\ \ref{f:ssfr_comp}.  It is worth remarking that \cite{Weinmann11} conduct a similar analysis of SSFRs at $M_\ast = 10^{9.5}\Msun$ and find a ``plateau'' at high redshifts; while the data are consistent with a plateau, they are also consistent with \textit{not} having a plateau: indeed, any range of shallow slopes at high redshifts would be consistent with the data collected here.

Our results for the inter-publication variance as a function of stellar mass and redshift are shown in Table \ref{t:ssfr_errs}.  Because only UV surveys can probe low stellar masses at high redshifts ($z>3$), the published SSFRs for low stellar masses at high redshift are more in agreement (0.22 dex scatter) than for low stellar masses at low redshift (0.32 dex scatter)!  This strongly suggests that the inter-publication variance in these bins does not represent the true systematic uncertainties.  Excluding high-redshift bins ($z>3$), there are no obvious systematic trends in the inter-publication variance, either with redshift or stellar mass.  Indeed, a one-way analysis of variance (ANOVA) test delivers no strong evidence that the inter-publication variances differ across bins ($p=0.6$).  This conclusion is not affected by merging bins by stellar mass or by redshift, and the size of the effect is too small to be measurable without adding more data.  We adopt the fiducial model of constant errors across stellar masses and redshifts, where the error is determined by merging all redshift and stellar mass bins except for the high-redshift bins ($z>3$) which are known underestimates, we would estimate the average systematic error in SSFRs to be 0.28 dex.  

Finally, we mention a commonly-overlooked systematic error in optical SFRs, which comes from the bias introduced by separately fitting the dust model for each galaxy.  As the dust model for each galaxy can never be derived exactly, individual galaxy SFRs will have some scatter relative to the true galaxy SFRs.  One way to estimate the shape of this scatter is to compare two different SFR estimates with different assumptions; for example, \cite{Salim07} finds log-normal scatter consistent with intrinsic scatter of 0.3 dex.  In calibrating one estimator to another (or to observations), authors typically make sure that the \textit{median} offsets are zero; however, they often fail to notice that the \textit{expected mean} offset is nonzero because the log-normal distribution is skewed in linear space.  Specifically, a log-normal distribution with median value 1 and scatter $\sigma$ (in dex) has an expected mean value of
\begin{equation}
\label{e:median_to_average}
\langle x \rangle = \exp(0.5(\sigma\ln 10)^2).
\end{equation}
For a log-normal distribution with 0.3 dex scatter, this corresponds to an overall bias of +0.1 dex.  This correction does not apply to UV studies that apply a median dust correction to all galaxies or to those that already correct for this bias.

\section{Corrections to and Uncertainties in the Tinker Mass Function}

\label{a:tinker}

The halo mass function for distinct halos presented by \cite{tinker-umf} was calibrated for a wide array of simulations from $z=0$ to $z=2.5$.  The present work requires a mass function from $z=0$ to $z>8$ including satellite halos, so some additional corrections must be made.  In addition, as new studies \citep{Reddick12} suggest that abundance matching to halo properties
at their peak value before stripping offers a better match to $z=0$ clustering data, calibration for the quantities of interest 
using $M_\mathrm{peak}$ must also be applied. 

\begin{figure}
\vspace{-6ex}
\plotminigrace{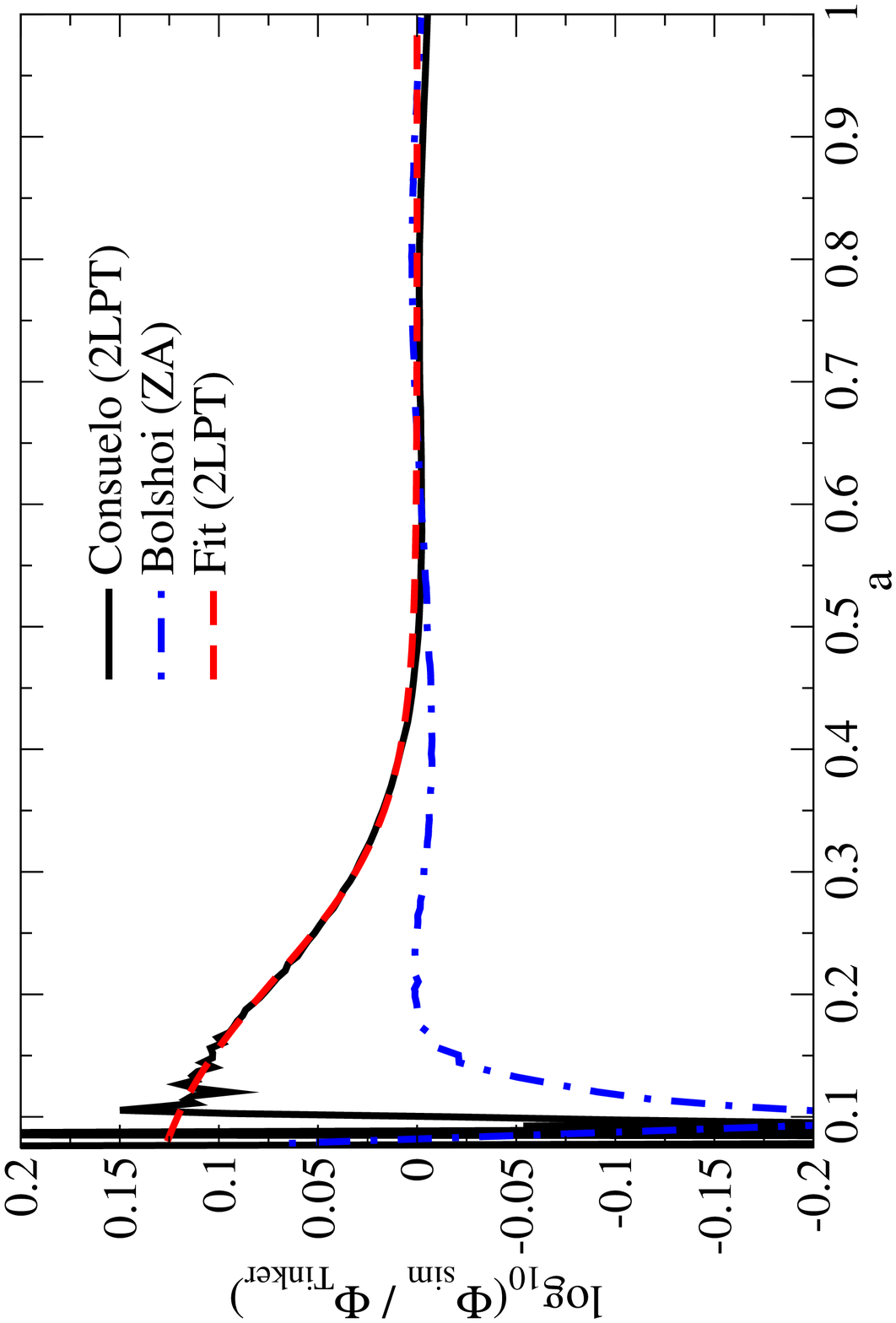}
\plotminigrace{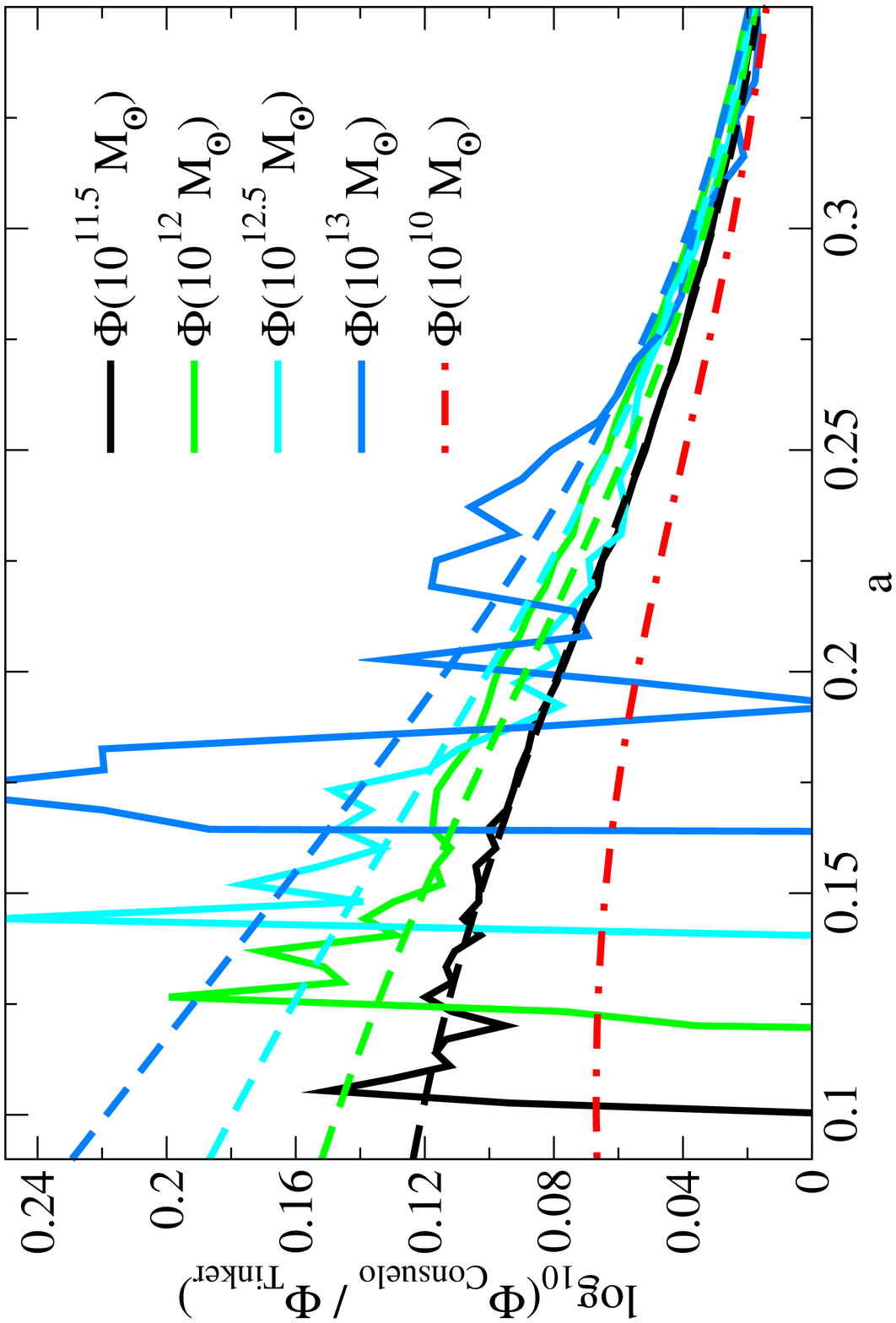}\\[-4ex]
\caption{\textbf{Left:} A comparison between the \cite{tinker-umf} mass function and two simulations with different initial conditions (\texttt{ZA}: Zel'dovich Approximation; \texttt{2LPT}: second-order Lagrangian Perturbation Theory).  Specifically, the plot shows the ratio of the cumulative central halo mass function to $10^{11.5}\Msun$ ($\Phi(10^{11.5}\Msun)$) in the simulations to that for the Tinker MF.  Small mass-dependent corrections (Eq.\ \ref{e:consuelo_corr}) were made to renormalize the Consuelo mass function to match the Tinker MF at $z=0$, but no such corrections were necessary for Bolshoi.  The Tinker MF, which was calibrated using ZA simulations, agrees remarkably well with Bolshoi over a wide range of redshifts; however, it is noticeably different from the Consuelo simulation for $z>3$.  As several authors \citep{Crocce06,Jenkins10} have shown improved convergence for 2LPT over ZA initial conditions, we use the 2LPT normalization as the basis for our corrective fit (Eq.\ \ref{e:tinker_highz}). \textbf{Right:} the mass-dependence of the correction to the Tinker cumulative mass function from the Consuelo simulation.  Solid lines show direct results from the simulation; dashed lines show the fit in Eq.\ \ref{e:tinker_highz_mass}.  The dot-dashed line shows the fit in Eq.\ \ref{e:tinker_highz_mass} extrapolated down to $\Phi(10^{10}\Msun)$.}
\label{f:mf_highz}
\end{figure}

To test the high-redshift calibration of the mass function in \cite{tinker-umf}, we integrate the mass function in two simulations (Bolshoi and Consuelo) to $10^{11.5} \Msun$ and compare to the Tinker mass function for the appropriate cosmology.  A mass-dependent correction is necessary to normalize the $z=0$ mass function for Consuelo to the Tinker mass function to account for slight incompleteness in the Consuelo mass function at this level:
\begin{equation}
\label{e:consuelo_corr}
\log_{10}\left(\frac{\phi_\mathrm{Consuelo}(M)}{\phi_\mathrm{Tinker}(M)}\right) = -0.0008 -0.042\left(\frac{M}{10^{11.51}\Msun}\right)^{-0.69}
\end{equation}
The net effect of this correction on $\Phi(10^{11.5} \Msun)$ is an increase of 0.03 dex (7\%) compared to the raw simulation result for Consuelo.  No correction is necessary for the Bolshoi simulation.  As shown in Fig.\ \ref{f:mf_highz}, the redshift fit given in \cite{tinker-umf} matches Bolshoi to within 0.02 dex up to a redshift of 5-6.  However, Consuelo diverges from the Tinker mass function by 0.06 dex (15\%) already at $z=3$.  This discrepancy is due to the different initial conditions calculations used in the two simulations: Consuelo uses second-order Lagrangian Perturbation Theory (2LPT), whereas Bolshoi and the Tinker MF use the Zel'dovich approximation (1LPT).  The Zel'dovich approximation has been previously found to underestimate the early nonlinear collapse of overdense regions \citep{Crocce06,Jenkins10}, which is reconfirmed by the comparison in Fig.\ \ref{f:mf_highz}.  For that reason, we prefer to use the correction from Consuelo, parametrized as follows:
\begin{equation}
\label{e:tinker_highz}
\log_{10}\left(\frac{\Phi_\mathrm{true}(10^{11.5}\Msun)}{\Phi_\mathrm{Tinker}(10^{11.5}\Msun)}\right) = \frac{0.144}{1+\exp\left[14.79(a-0.213)\right]}.
\end{equation}
As shown in Fig.\ \ref{f:mf_highz}, this fit matches the correction from the Consuelo simulation to within 1\%.  However, as also shown in Fig.\ \ref{f:mf_highz}, there is a slight mass dependence to the correction; the magnitude of the correction is larger for more massive halos, as their nonlinear collapse begins earlier.  We parametrize the mass dependence as follows:
\begin{equation}
\label{e:tinker_highz_mass}
\log_{10}\left(\frac{\Phi_\mathrm{true}(M)}{\Phi_\mathrm{Tinker}(M)}\right) = \log_{10}\left(\frac{\Phi_\mathrm{true}(10^{11.5}\Msun)}{\Phi_\mathrm{Tinker}(10^{11.5}\Msun)}\right) \left(\frac{M}{10^{11.5}\Msun}\right)^\frac{0.5}{1+\exp(6.5a)}.
\end{equation}
This function is well-behaved, even when extrapolated down to low masses, as shown in Fig.\ \ref{f:mf_highz}.  

Parametrizing the corrections as we have done introduces several uncertainties.  The use of only a single simulation to calibrate the mass function introduces sample variance uncertainties, but these are in fact small for a box as large as Consuelo (420 Mpc $h^{-1}$); at low redshifts, the median offset from the Tinker mass function is 0.2\% (see Eq.\ \ref{e:consuelo_corr}), and even at redshift 6, the Poisson and sample variance errors combined are on the order of 1\% \citep{Trenti08} for $\Phi(10^{11.5}\Msun)$.  Much larger uncertainties come from the use of a mass-dependent incompleteness correction to the Consuelo mass function, because uncertainties in this correction are degenerate with the mass dependence of the normalization correction in Eq.\ \ref{e:tinker_highz_mass}.  If the incompleteness correction needed to be larger at high redshift, the evidence for the mass-dependence in the normalization correction would weaken.  We can take this as an estimate of the additional systematic uncertainties on the high-redshift mass function.  In terms of the impact on the halos in which high-redshift galaxies ($z=7-8$) are expected to reside, the number density of such objects would be underestimated by at most 0.05 dex (12\%), according to Fig.\ \ref{f:mf_highz}, if there were no mass-dependence in the normalization correction.  This is tiny in comparison to the uncertainties in the stellar masses and star formation rates at high redshift.  Indeed, even if the real universe corresponded exactly to the Tinker mass function, the total error introduced by using the corrections in this section (0.1 dex, or 26\%) would again be dwarfed by uncertainties in measuring galaxy stellar content.

\begin{figure}
\vspace{-6ex}
\plotminigrace{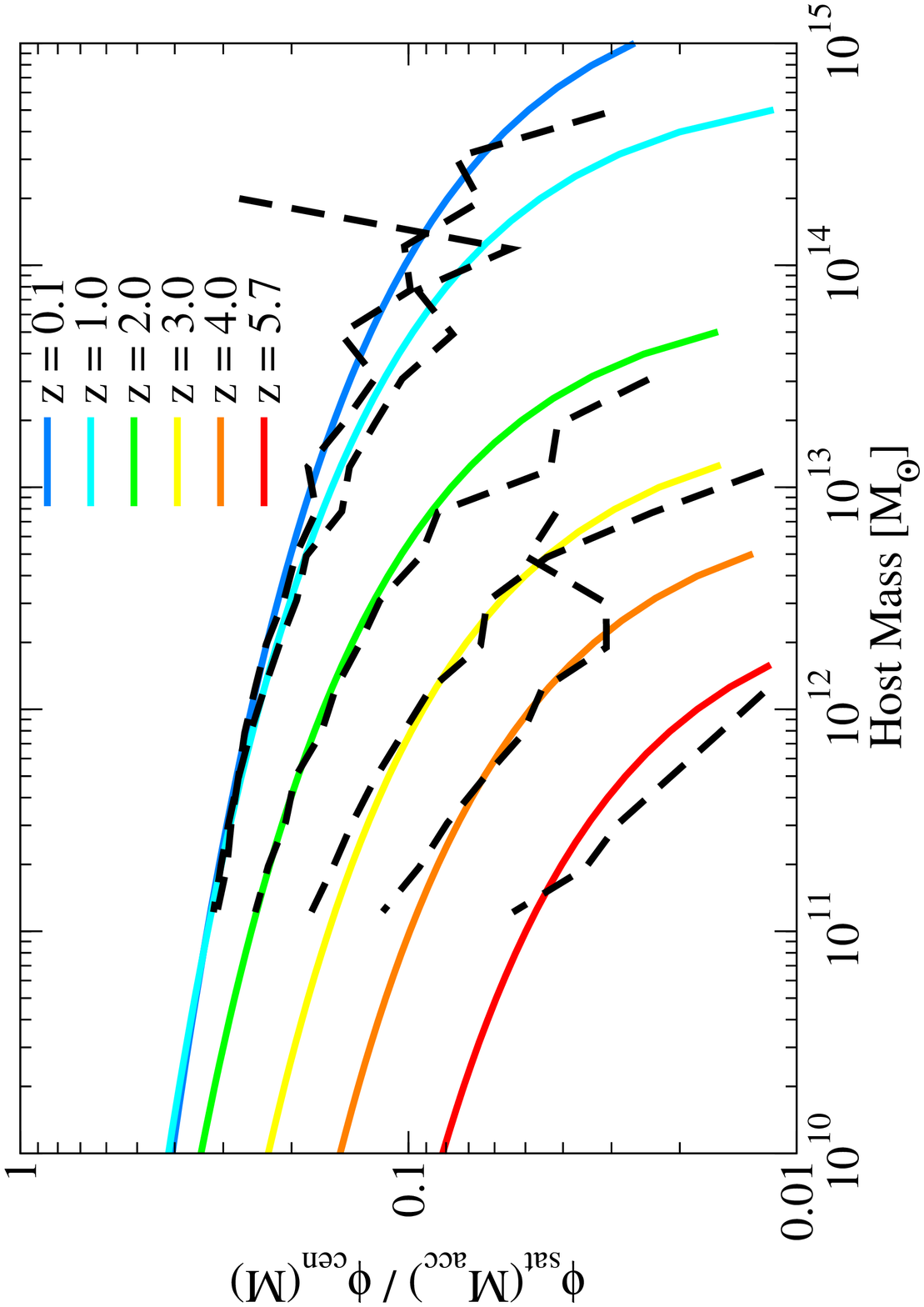} \plotminigrace{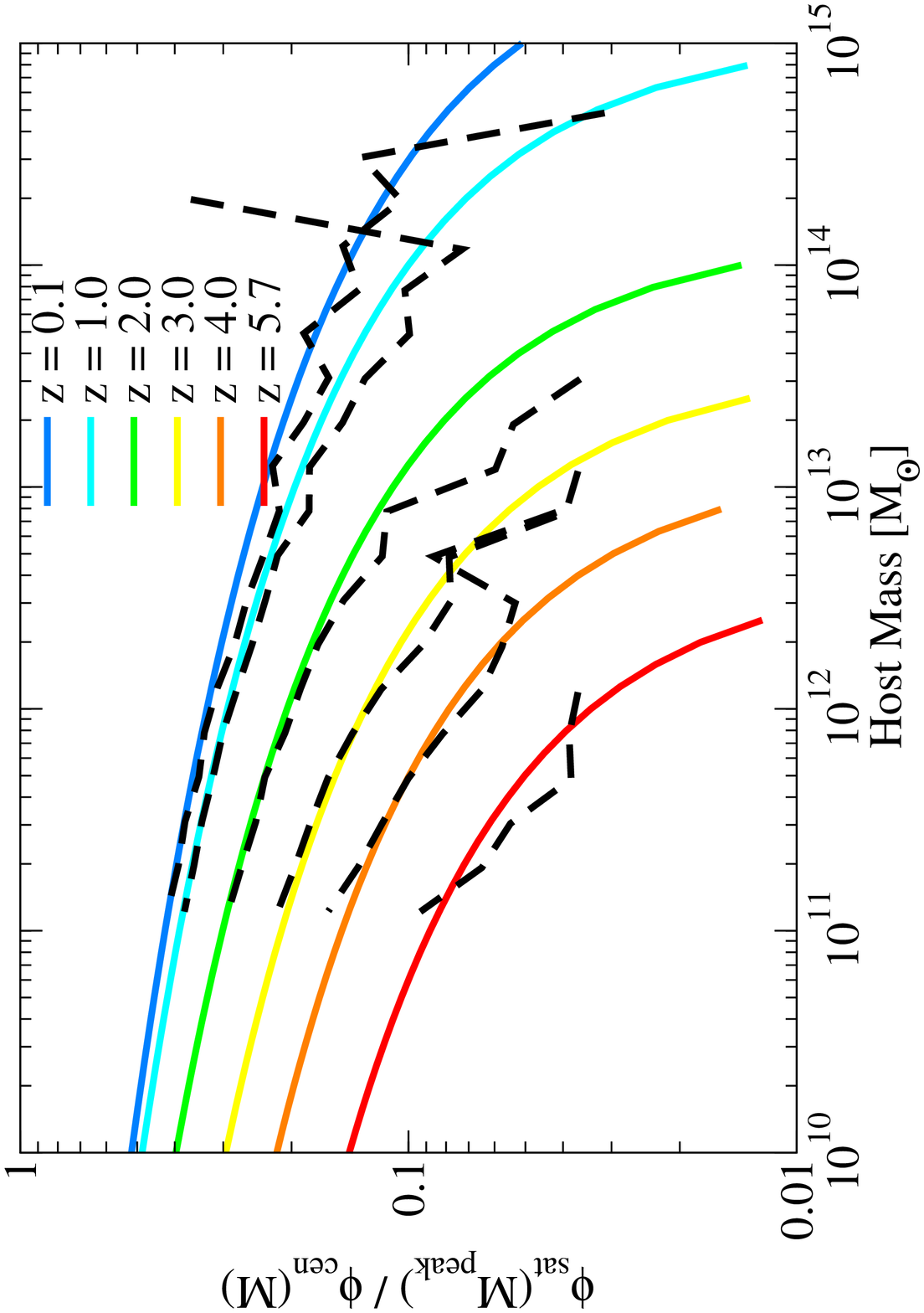}\\[-2ex]
\caption{Best fits to the evolution of the satellite fractions (Eqs.\ \ref{e:fit_sat_macc}-\ref{e:fit_sat_mpeak}) in Bolshoi for satellite accretion masses (left panel) and satellite peak masses (right panel), according to the simplified model of Eq.\ \ref{e:satfit}.  The colored lines correspond to the fitted satellite fractions; the black dashed lines correspond to the satellite fractions in Bolshoi.}
\label{f:mf_satfit}
\end{figure}

In terms of the satellite fraction, we determine fits using a simple, physically-motivated model.  It is an oft-observed feature of dark matter substructure that the number of satellite halos at a given mass scales approximately with the mass of the host --- the number density of satellites in hosts of mass $M_\mathrm{c}$ is proportional to $M_\mathrm{c}\phi_\mathrm{c}(M_\mathrm{c})$, where $\phi_c(M)$ is the mass function for central halos.  Another feature of dark matter substructure is that the number of satellites at a given satellite-to-host mass ratio is largely self-similar across host halo masses; this requires that the proportionality must scale inversely with the satellite mass. Therefore, we can write down a simple formula for the satellite mass function, $\phi_s$:
\begin{equation}
\label{e:phi_sat}
\phi_s(M) \sim \frac{C(a)}{M} \int_M^\infty M_\mathrm{c} \phi_\mathrm{c}(M_\mathrm{c})d\log_{10}M_\mathrm{c}
\end{equation}
where the proportionality factor, $C(a)$, is a function of scale factor only.  The low-mass limit of this function takes on an especially appealing form; as $\phi_c(M_\mathrm{c}) \sim \phi_0 M_c^{-1}$ for $M_c$ is less than the exponential cutoff scale, $M_\mathrm{cutoff}$, we find:
\begin{equation}
\frac{\phi_s(M)}{\phi_c(M)} \sim \frac{\frac{C(a)}{M} \int_M^{M_\mathrm{cutoff}} \phi_0 d\log_{10} M_c}{\phi_0 M^{-1}}
\end{equation}
which can be simplified to
\begin{equation}
\label{e:satfit}
\frac{\phi_s(M)}{\phi_c(M)} \sim C(a) \log_{10}\left(\frac{M_\mathrm{cutoff}(a)}{M}\right)
\end{equation}
Thus, the satellite fraction grows only logarithmically in the low-mass limit.

However, the picture in Eq.\ \ref{e:phi_sat} is not entirely complete; because they have later assembly times, massive clusters are more likely to be undergoing major mergers than smaller halos.  However, these major mergers also become absorbed more quickly, as the effects of dynamical friction are more pronounced in massive halos.  These two (opposite) effects combine to give a satellite fraction slightly larger at the high-mass end than Eq.\ \ref{e:phi_sat} would predict; moreover, this excess evolves nontrivially with redshift.  For that reason, we use the slightly more flexible form of Eq.\ \ref{e:satfit} and allow great freedom in the parametrization for both $C(a)$ and $M_\mathrm{cutoff}(a)$.

Fitting the satellite fraction to a simulation is made especially tricky by the fact that simulations are incomplete for satellites at higher masses than for central halos, even before selecting satellites on $M_\mathrm{peak}$ or $M_\mathrm{acc}$.  For that reason, Consuelo is unsuitable for determining the fraction of all but the most massive satellites ($M \sim 10^{13} \Msun$), which provides an especially poor handle on constraining the redshift evolution of the fit.  Bolshoi fares better, being complete for satellites down to $10^{11} \Msun$; yet, given its use of ZA initial conditions, it may not seem especially suitable for high-redshift calibration, either!

Yet, there are two arguments in its favor.  First, the satellite fraction is low at high redshift, so even large relative errors in the satellite fraction will translate into small errors on the overall mass fraction.  Second, the influence of the initial conditions on the satellite fraction are largely through the effects of assembly bias; however, recent work has shown that satellite richness (when tallied above a fixed cut in accretion or peak mass) is almost uncorrelated with formation time (Wu et al., in prep.).  The impact of the initial conditions on the satellite fraction is thus minor compared to the effect of the mass function normalization.

Therefore, using the Bolshoi simulation, we fit the satellite fraction from $z=0$ to $z=10$; when satellite masses are determined by $\macc$, we find:
\begin{eqnarray}
\label{e:fit_sat_macc}
log_{10}(C(a)) & = & -2.69+11.68 a-28.88 a^2+29.33 a^3-10.56 a^4\\
\log_{10}(M_\mathrm{cutoff}(a)) & = & 11.34+8.34 a-0.36 a^2-5.08 a^3+0.75 a^4
\end{eqnarray}
and find the following parametrization for the satellite fraction when satellite masses are determined by $M_\mathrm{peak}$:
\begin{eqnarray}
\log_{10}(C(a)) & = & -1.91+6.23 a-15.07 a^2+15.02 a^3-5.29 a^4\\
\log_{10}(M_\mathrm{cutoff}(a)) & = & 10.66+15.93 a-21.39 a^2+18.20 a^3-8.21 a^4
\label{e:fit_sat_mpeak}
\end{eqnarray}
Both of these fits in comparison to satellite fractions in Bolshoi are shown in Fig.\ \ref{f:mf_satfit}; they are accurate to within the estimated uncertainties in the mass function (5\%, \citealt{tinker-umf}).

\section{Halo Mass Accretion Rates}
\label{a:mah}

\begin{figure}
\vspace{-6ex}
\plotminigrace{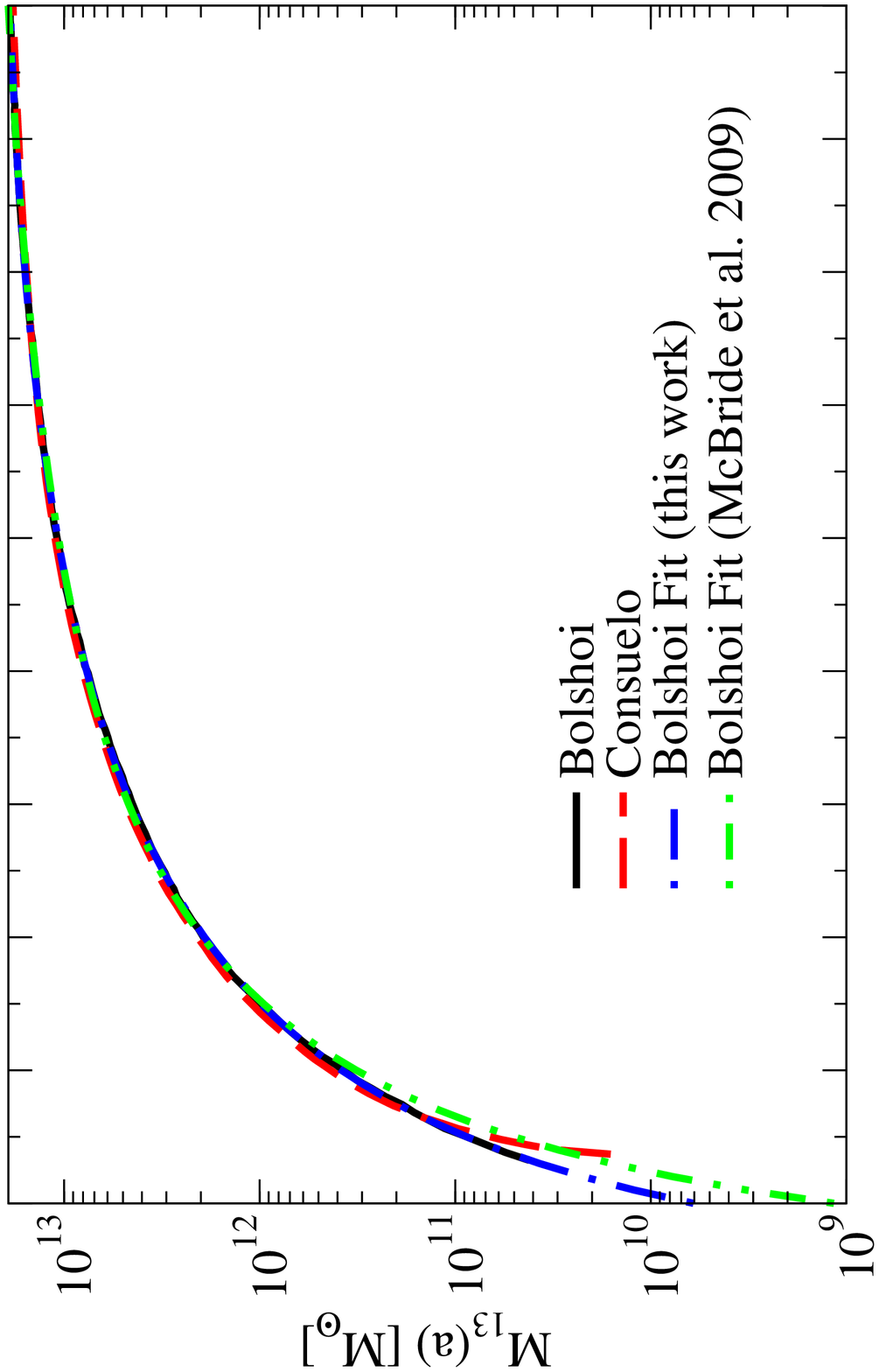}\plotminigrace{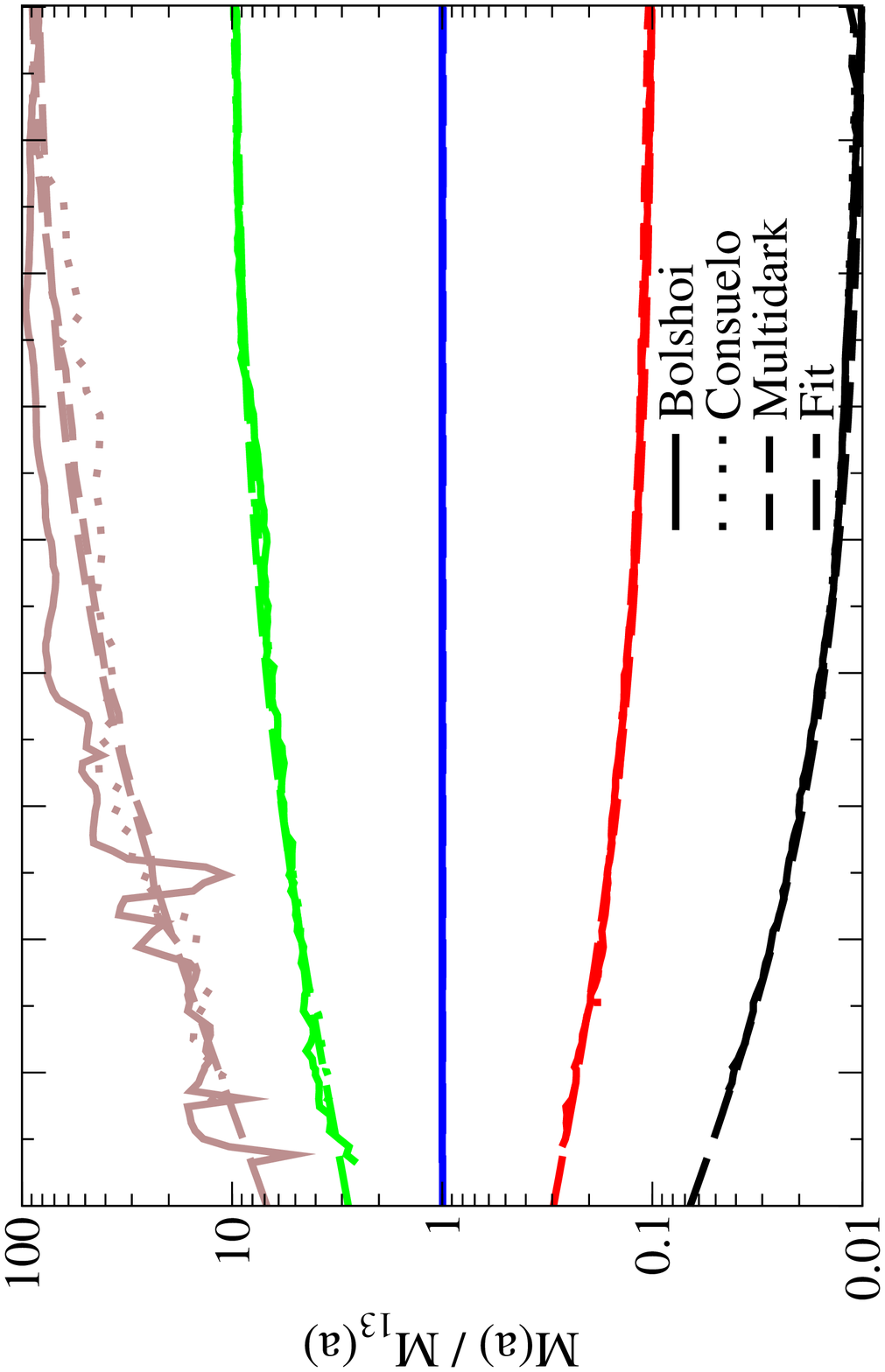}\\[-14ex]
\plotminigrace{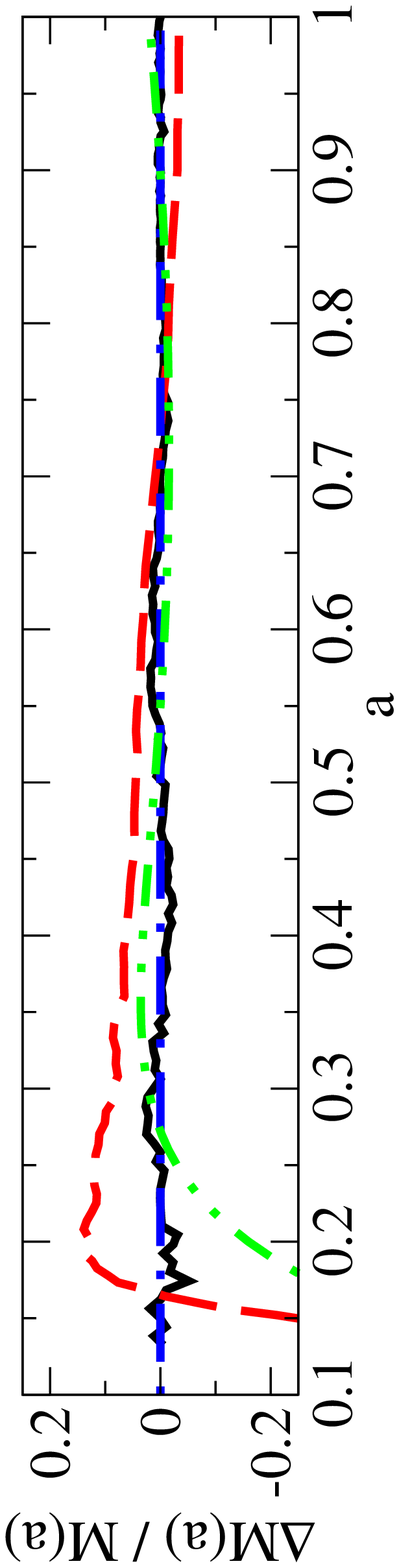}\plotminigrace{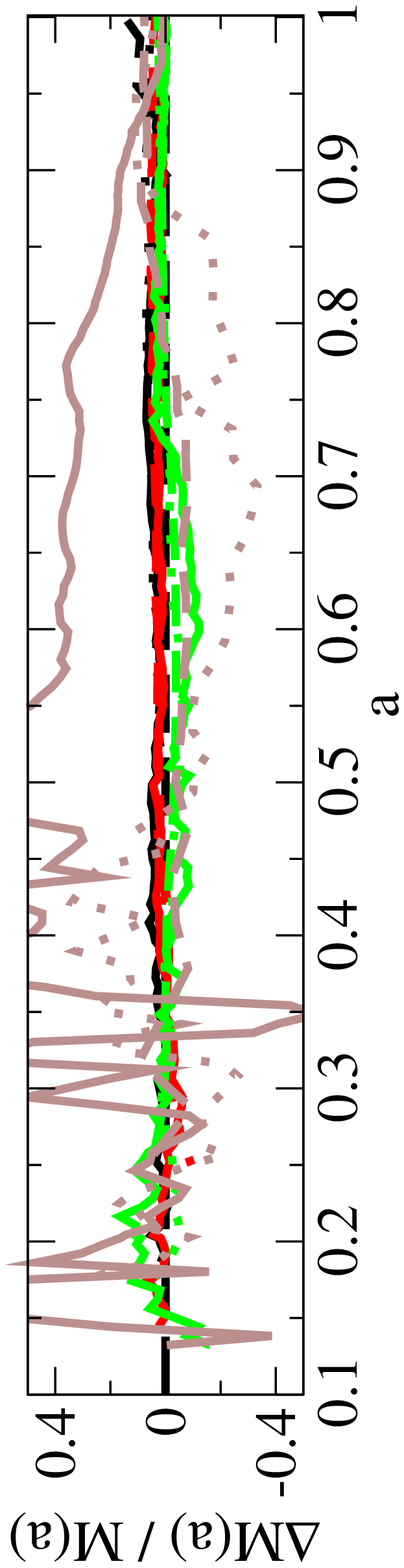}\\[-30ex]
\caption{Fits to median mass accretion histories (M$_\mathrm{vir}$) for central halos in Bolshoi, MultiDark, and Consuelo simulations.  Common functional forms for the mass accretion rates, such as the exponential fit of \cite{Wechsler02} or the exponential plus power-law fit of \cite{McBride09} fail to accurately capture mass accretion histories beyond $z=2$.  Moreover, at high redshifts, these common functional forms imply unphysical behavior---i.e., that progenitors of $10^{15}\Msun$ halos are on average less massive than progenitors of $10^{13}\Msun$ halos (see text), which is not at all the case.  The \textbf{left} panel shows mass accretion histories for $10^{13.15}$ to $10^{13.40}\Msun$ halos in Bolshoi and Consuelo, along with residuals (linear scale).  Bolshoi and Consuelo have different cosmologies, so it is expected that the mass accretion rates of halos differ slightly.  The \textbf{right} panel shows progenitor mass ratios (compared to progenitors of $10^{13}\Msun$ halos) for halos of mass $10^{11}$ to $10^{15}\Msun$ at $z=0$ for the Bolshoi, MultiDark, and Consuelo simulations, along with residuals.  In all cases, our fits to the mass accretion histories of halos for the Bolshoi cosmology are accurate to 5\% on average; the only exception is for the highest mass bins in the Bolshoi and Consuelo simulations, which contains only a few halos.}
\label{f:mah}
\end{figure}

The most commonly used fitting formula for halo mass accretion rates is that proposed by \cite{Wechsler02}, which suggests that halo mass growth is an exponential function in redshift ($M(z) \propto \exp(-cz)$).  As simulations improved, deviations from this simple formula were found, which led authors to propose more complicated formulas; e.g., $M(z) \propto (1+z)^b\exp(-cz)$ in \cite{McBride09}.  However, little attention has been given to the high-redshift behavior of these formulas, with the predictable result that they fail to accurately capture median mass accretion histories for halos as recently as $z\sim 3$ (see Fig.\ \ref{f:mah}, left panel).\footnote{Note that, for \textit{individual} halos, as opposed to ensemble medians or averages, the \cite{McBride09} formula provides a reasonable fit: high precision is not as necessary for cases when individual stochasticity in merger events results in higher deviations than those shown in the bottom-left panel of Fig.\ \ref{f:mah}.}

This limitation a problem for this study.  To motivate our choice of improved fitting formulas, we note an issue with \textit{all} proposed fitting formulas to date, which is their exponential dependence at high redshifts.  That is to say, the mass ratio of two halos $M_1$ and $M_2$ will asymptote to
\begin{equation}
\frac{M_1(z)}{M_2(z)} \to \frac{M_1(0)}{M_2(0)} \exp((c_2-c_1)z)
\end{equation}
This is problematic because more massive halos usually have steeper exponential dependencies (i.e., $M_1 < M_2$ implies $c_1 < c_2$).  If $M_1(z=0)<M_2(z=0)$, these formulas then predict that eventually the progenitor histories will cross and $M_1(z) > M_2(z)$.  Using the \cite{McBride09} formula, for example, with the MultiDark simulation, we find that it predicts the progenitors of $10^{15}\Msun$ halos to be less massive than the progenitors of $10^{14.75}\Msun$ halos as early as $z=5$.  However, this is not the case: in simulations, the hierarchy of average or median progenitor masses is maintained, even though the distance between them may decrease (see, for example, the right panel of Fig.\ \ref{f:mah}).

We therefore parametrize the mass accretion history of a single halo mass, and then parametrize the \textit{ratios} of progenitor masses for other halo masses, which entirely avoids the problem of crossing progenitor mass histories.  The mass bin we choose ($10^{13.15}-10^{13.40}\Msun$) is well-resolved in all three dark matter simulations at our disposal, and sufficient statistics are available even in the smallest simulation (Bolshoi) to be able to reduce scatter to an acceptable level.  Our resulting fit is:
\begin{eqnarray}
M(M_0,z) &=& M_{13}(z) 10^{f(M_0,z)}\\
M_{13}(z) &=& 10^{13.276} (1+z)^{3.00} (1+\frac{z}{2})^{-6.11}\exp(-0.503z)\Msun\\
f(M_0,z) & = & \log_{10}\left(\frac{M_0}{M_{13}(0)}\right)\frac{g(M_0,1)}{g(M_0,\frac{1}{1+z})}\\
g(M_0,a) & = & 1 + \exp(-4.651(a-a_0(M_0))\\
a_0(M_0) & = & 0.205 - \log_{10}\left[\left(\frac{10^{9.649}\Msun}{M_0}\right)^{0.18} + 1\right]
\end{eqnarray}
where $M(M_0,z)$ gives the median virial mass \citep{mvir_conv} for progenitors of halos with mass $M_0$ at $z=0$.  Despite its complexity, this seven-parameter fit (three parameters for $M_{13}(z)$, one parameter for $g(M_0,a)$, and three parameters for $a_0(M_0)$) provides an excellent fit to mass accretion histories for a wide range of halo masses and redshifts, as shown in Fig.\ \ref{f:mah}.

\section{Subhalo Merger/Disruption Rates}

\begin{figure}
\vspace{-6ex}
\plotminigrace{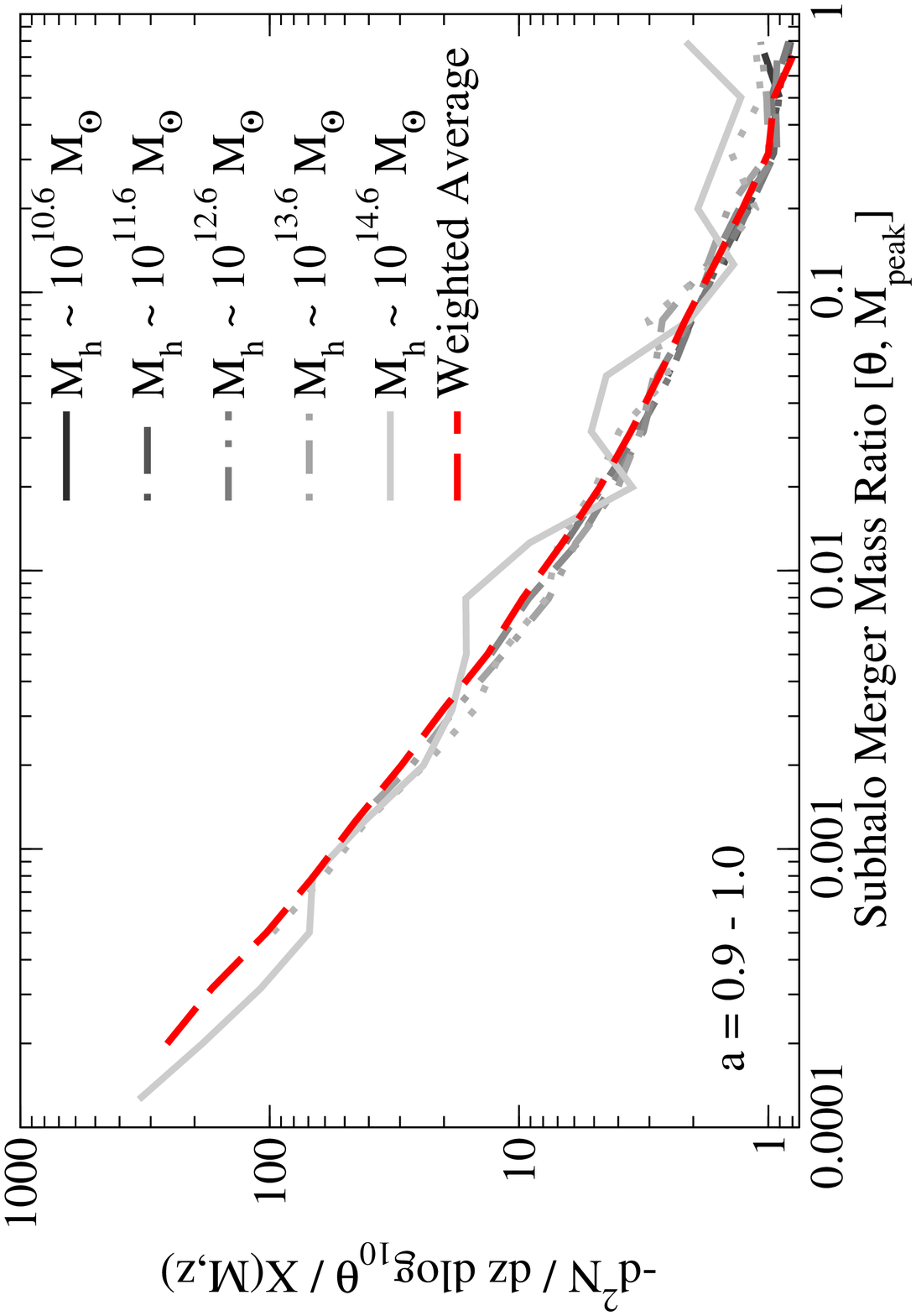}\plotminigrace{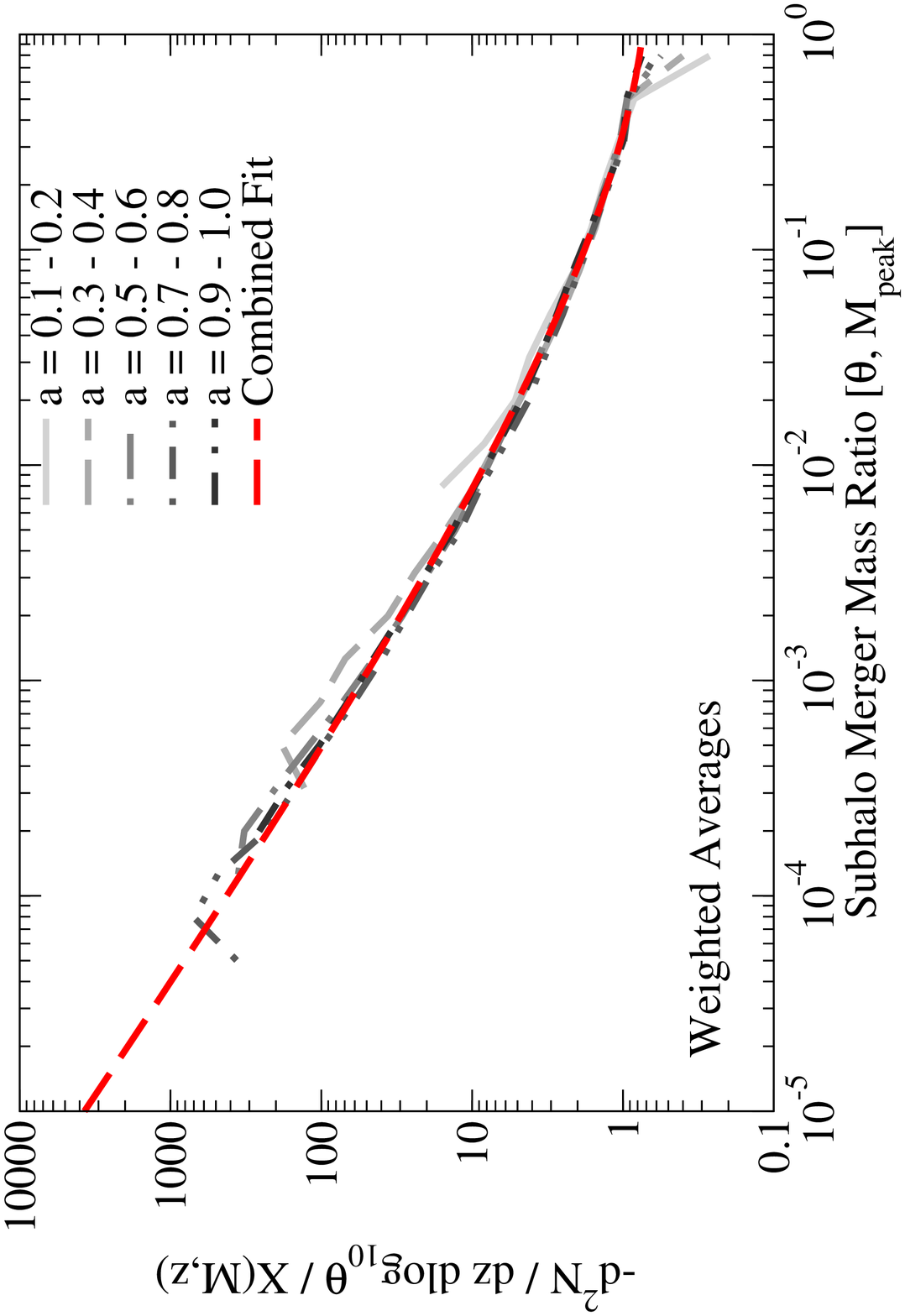}\\[-2ex]
\caption{The rate of subhalo disruption in host halos in the Bolshoi simulation with mass and redshift dependencies ($X(M,z)$) scaled out.  Results are presented in terms of the number of subhalos disrupted per host halo per unit redshift per log interval in satellite mass ratio.  The subhalo merger mass ratio ($\theta$) in these plots is the ratio of the subhalo peak mass (i.e., the highest mass any progenitor of the subhalo ever had) to the host halo mass.  The \textbf{left} panel shows the scaled subhalo disruption rate at $z\sim 0$ over a range of host halo masses, demonstrating excellent convergence.  The red dashed line in the left panel represents the average of scaled subhalo disruption rates as a function of subhalo mass ratio $\theta$, excluding halos for which subhalos of the given mass ratio would be below the mass resolution limit of the simulation ($10^{10}\Msun$).  The \textbf{right} panel shows the mass-averaged scaled subhalo disruption rate as a function of redshift, as well as the fit we adopt in this work.  The fit shows excellent conformance to the simulation data, except for very low subhalo mass ratios, where stochasticity in evaluating halo masses at low particle numbers artificially inflates peak halo masses (see text).}
\label{f:mergers}
\end{figure}

\label{a:disruption}

Halo mergers are important for our study because they impact the rate at which stellar mass is transfered from low-mass galaxies to high-mass galaxies as well as the rate at which stars are ejected into the intracluster light (ICL).  While there have been many studies of the rate at which halos become subhalos (see, e.g., \citealt{Fakhouri10}) in dark matter simulations, these studies are less relevant to our work because it is expected that satellite galaxies should last at least as long as the corresponding subhalo that hosts them \citep[e.g.][]{Yang11,Reddick12}.  Studying the subhalo destruction rate is more difficult because of issues with halo finding completeness when subhalos pass close to the centers of their hosts \citep{Rockstar,BehrooziTree}, and relatively few authors have attempted to correct for this issue in their work \citep[c.f.,][]{wetzel-09}.

However, with the advantage of a phase-space halo finder and a merger tree algorithm that corrects for such inconsistencies, we are in a unique position to provide a calibration of the merger rate (or  \textit{disruption} rate) of subhalos into host halos all the way back to the earliest redshifts in simulations.  We find that the functional form proposed in \cite{Fakhouri10} is an excellent fit to the subhalo disruption rate; however, the parameter values and redshift evolution are different due to the different merger definition, cosmology, and mass definition used.

Letting $M_h$ be the mass of the host halo and $\theta$ be the ratio of subhalo peak mass to host mass, we find that the disruption rate (expressed in terms of the decrement in number of subhalos per unit host halo per unit redshift per log interval in subhalo mass ratio) is:
\begin{eqnarray}
-\frac{d^2N}{dz d\log_{10}\theta}(M,\theta,z) & = & A(M,(1+z)^{-1})\theta^{-1}\exp(b(M,\theta)) \label{e:pm_merger_start}\\
A(M,a) & = & 0.0316 \left(1 + \frac{0.20\ln(a)}{1+a}\right) \left(\frac{M}{10^{12}\Msun}\right)^{0.03+0.05a}\\
b(M,\theta) & = & \left[10^{1.929} \left(\frac{M}{10^{12}\Msun}\right)^{0.11}\theta \right]^{0.2586} \label{e:pm_merger_end}
\end{eqnarray}

If $\theta$ is instead the ratio of subhalo accretion mass to host halo mass, we find instead a slightly simpler dependence:
\begin{eqnarray}
A_\mathrm{acc}(M,a) & = & 0.0139 \left(1 + \frac{0.25\ln(a)}{1+a}\right) \left(\frac{M}{10^{12}\Msun}\right)^{0.052+0.072a}\\
b_\mathrm{acc}(M,\theta) & = & \left[10^{3.091} \theta \right]^{0.1965}
\end{eqnarray}

We show the quality of the fits for the peak subhalo mass in Fig.\ \ref{f:mergers}.  For ease of comparing simulation data to the fit, we have scaled out host mass and redshift dependencies according to Eqs.\ \ref{e:pm_merger_start}-\ref{e:pm_merger_end}, so that all plotted lines are directly comparable to the subhalo disruption rate for $M=10^{12}\Msun$ at $z=0$.

Fig.\ \ref{f:mergers} shows that scaling in both mass and redshift of the subhalo disruption rate has been successfully captured in Eqs.\ \ref{e:pm_merger_start}-\ref{e:pm_merger_end}.  The remaining deviations are due to Poisson noise in the disruption rates for halo masses where Bolshoi has few halos (especially in the highest mass bin) and stochasticity in halo mass estimations for low-mass halos.  Specifically, as the peak mass captures the highest mass ever recorded in a halo's mass accretion history, greater variance in the halo mass estimator translates to artificially increased peak masses for low-mass halos.  This manifests as a slight upturn in the subhalo disruption rate close to the halo mass resolution limit of the simulation.

\section{Fits to Star Formation Rates and Histories Under Different Modeling Constraints}

\label{a:models}

\begin{figure}
\plotminigrace{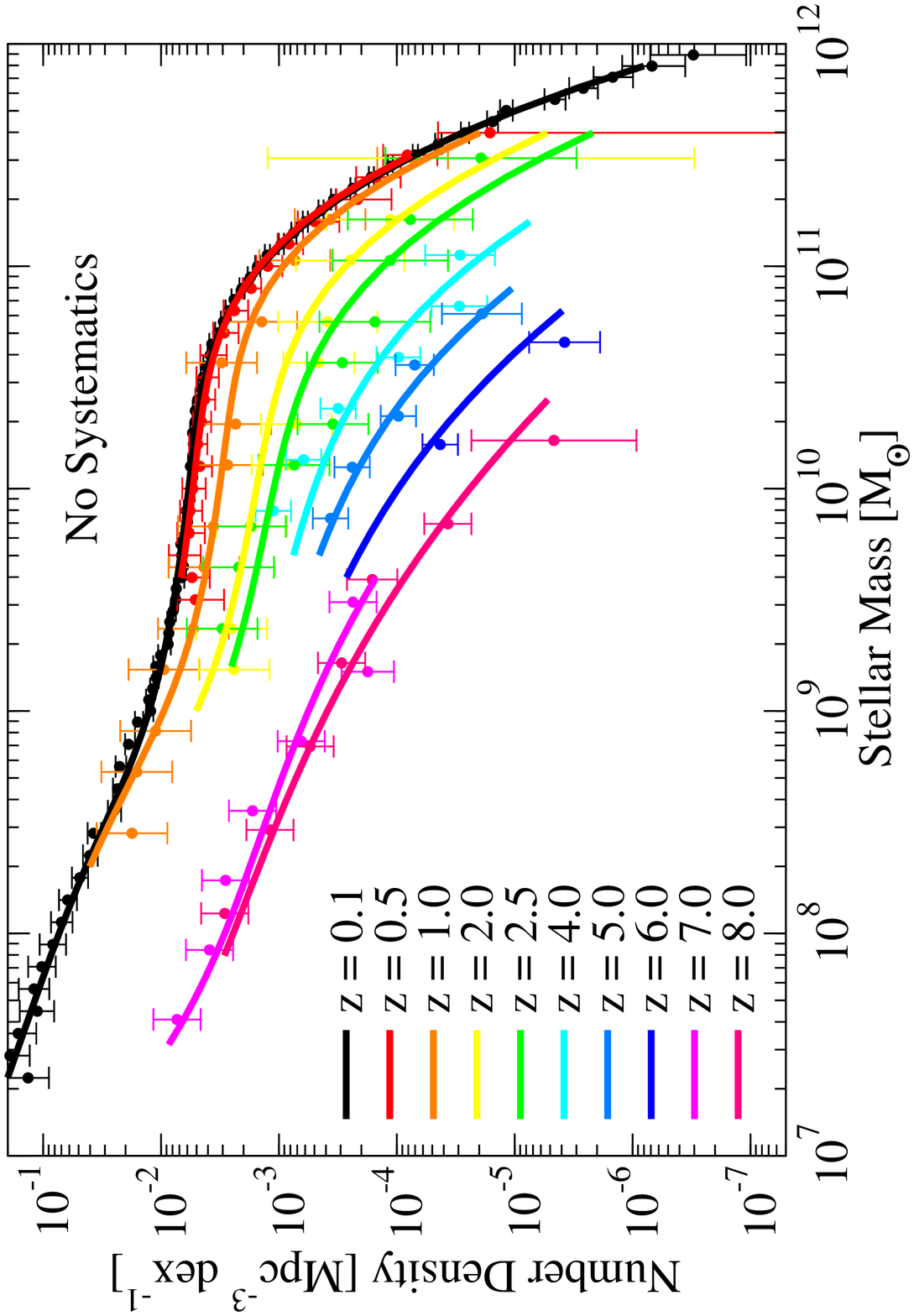}\plotminigrace{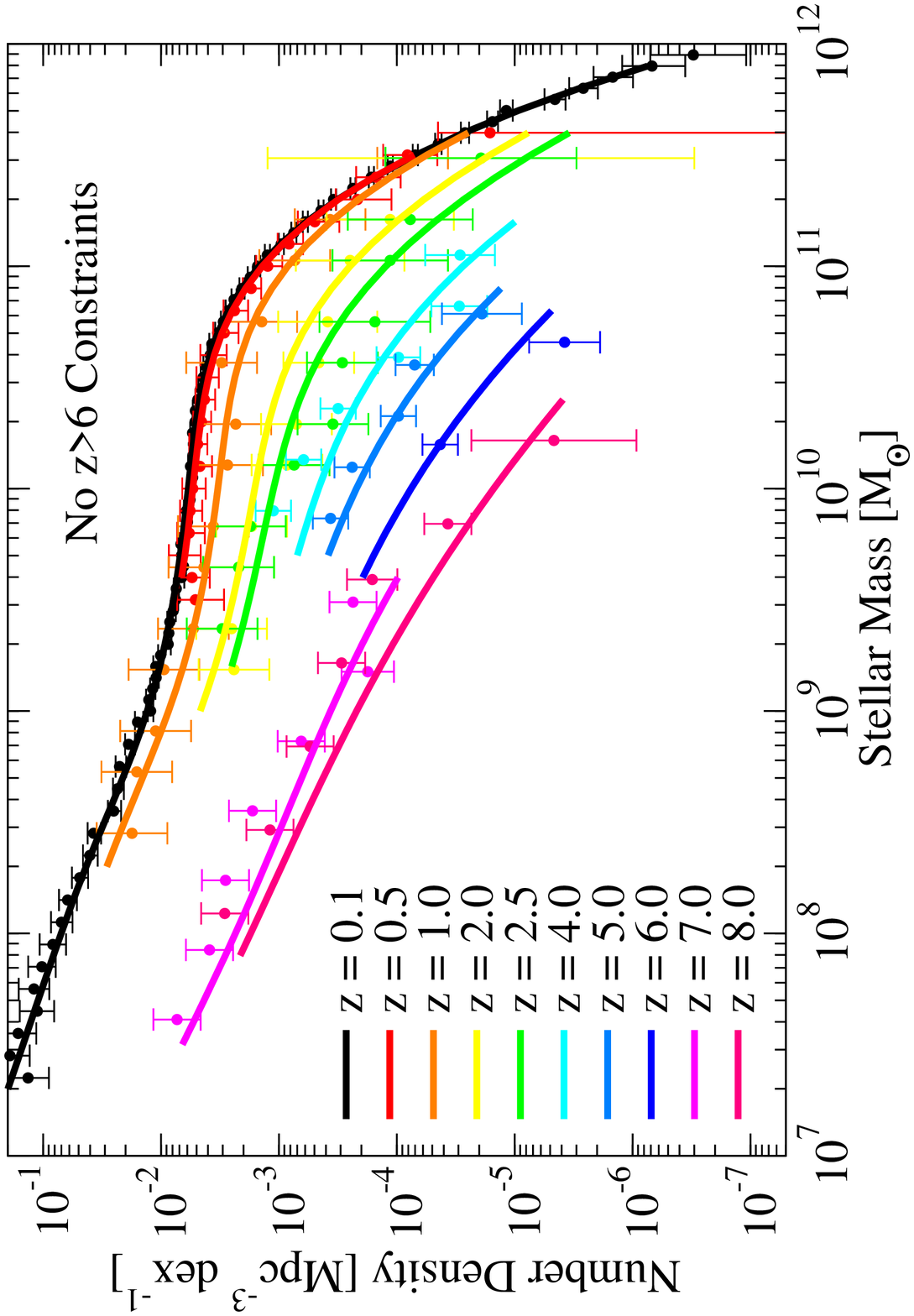}\\
\plotminigrace{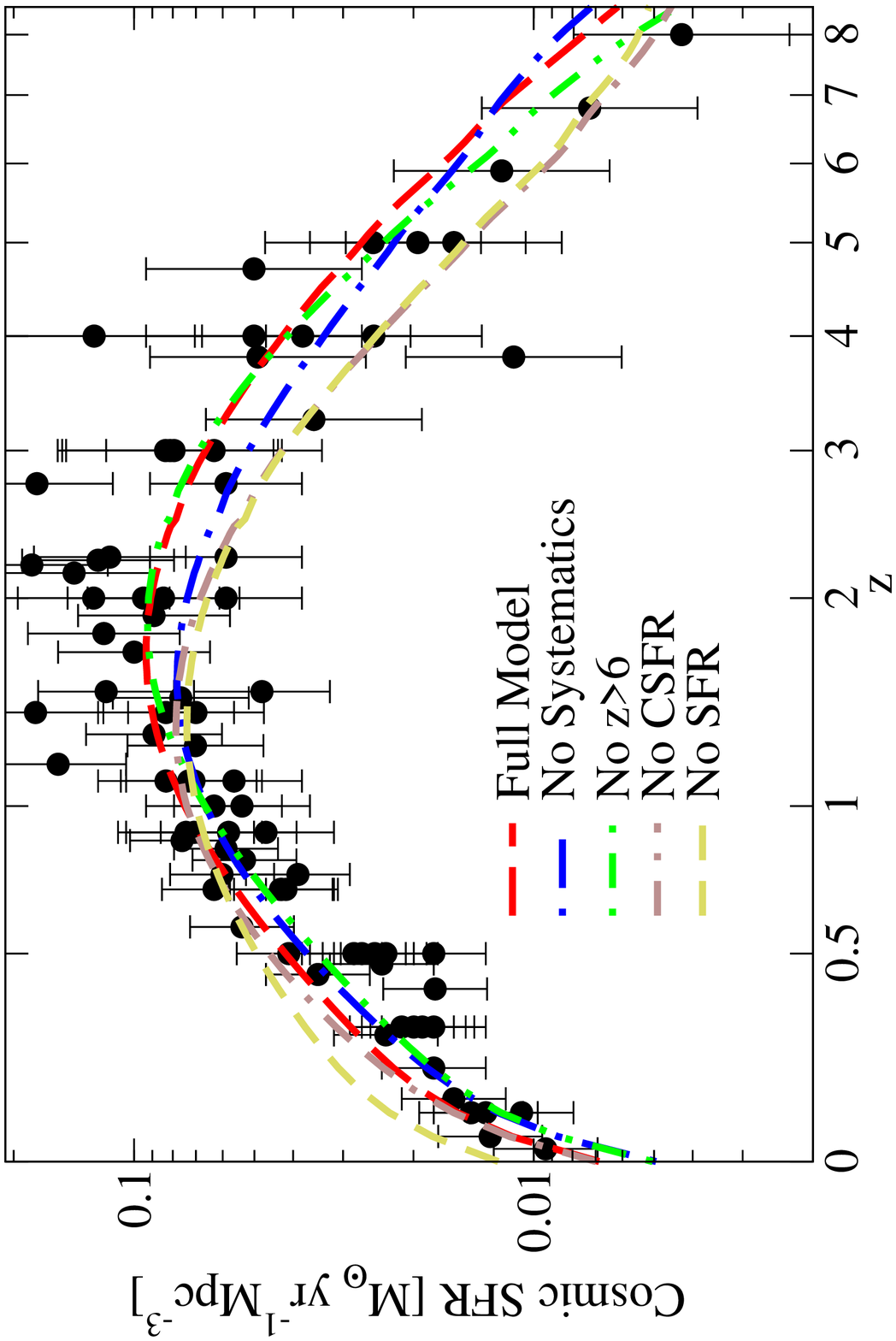}\plotminigrace{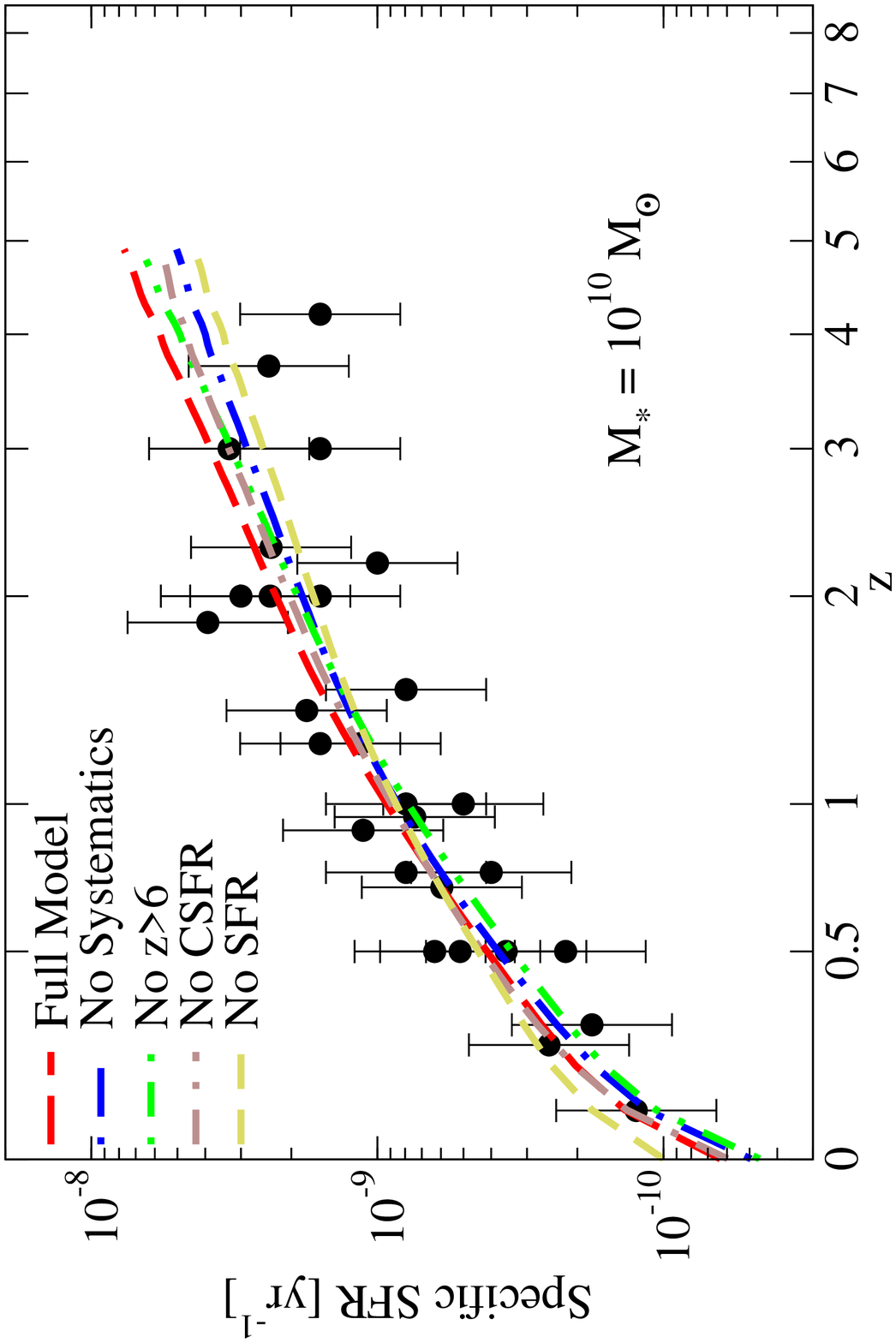}\\
\caption{Comparisons to observational data for the best fits of alternate models.  \textbf{Top left}: comparisons to observed stellar mass functions for the model without systematic uncertainties (i.e., $\mu=0$, $\kappa=0$, $\eta_\mathrm{cal} = \infty$, no incompleteness).  \textbf{Top right}: comparisons to stellar mass functions for a model excluding all constraints at $z>6$ (i.e., from\cite{Bouwens11,BORG12,Bouwens11b,Schaerer10,McLure11}). \textbf{Bottom left}: comparisons to observations of the cosmic star formation rate for the two models already mentioned, along with a model that excludes all cosmic star formation rate constraints (``No CSFR''; using SSFRs only), and a model that excludes all star formation rate constraints and systematic effects (``No SFR'').  \textbf{Bottom right}: comparisons with observed SSFRs for all five models.}
\label{f:alt_constraints}
\end{figure}

\begin{figure}
\plotminigrace{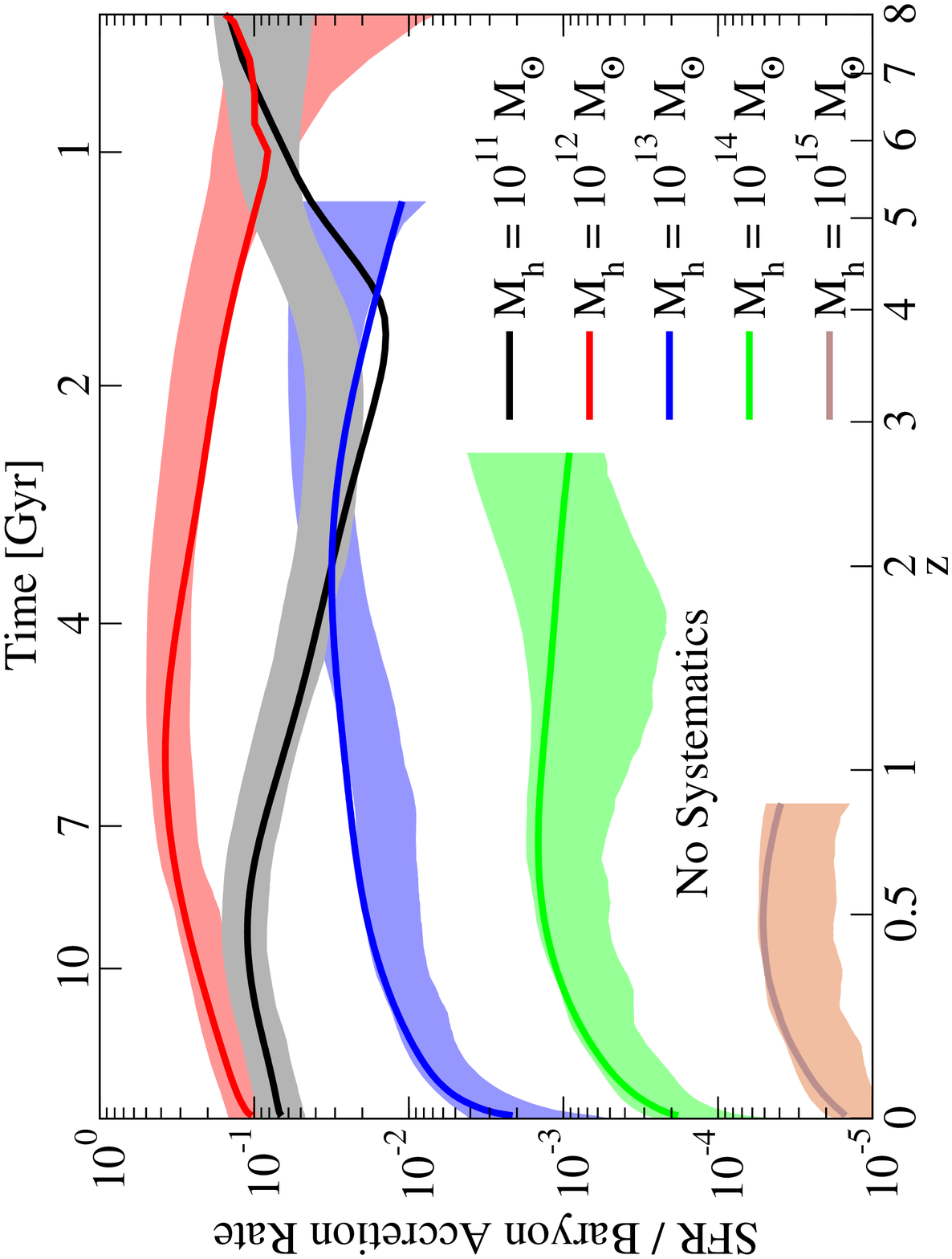}\plotminigrace{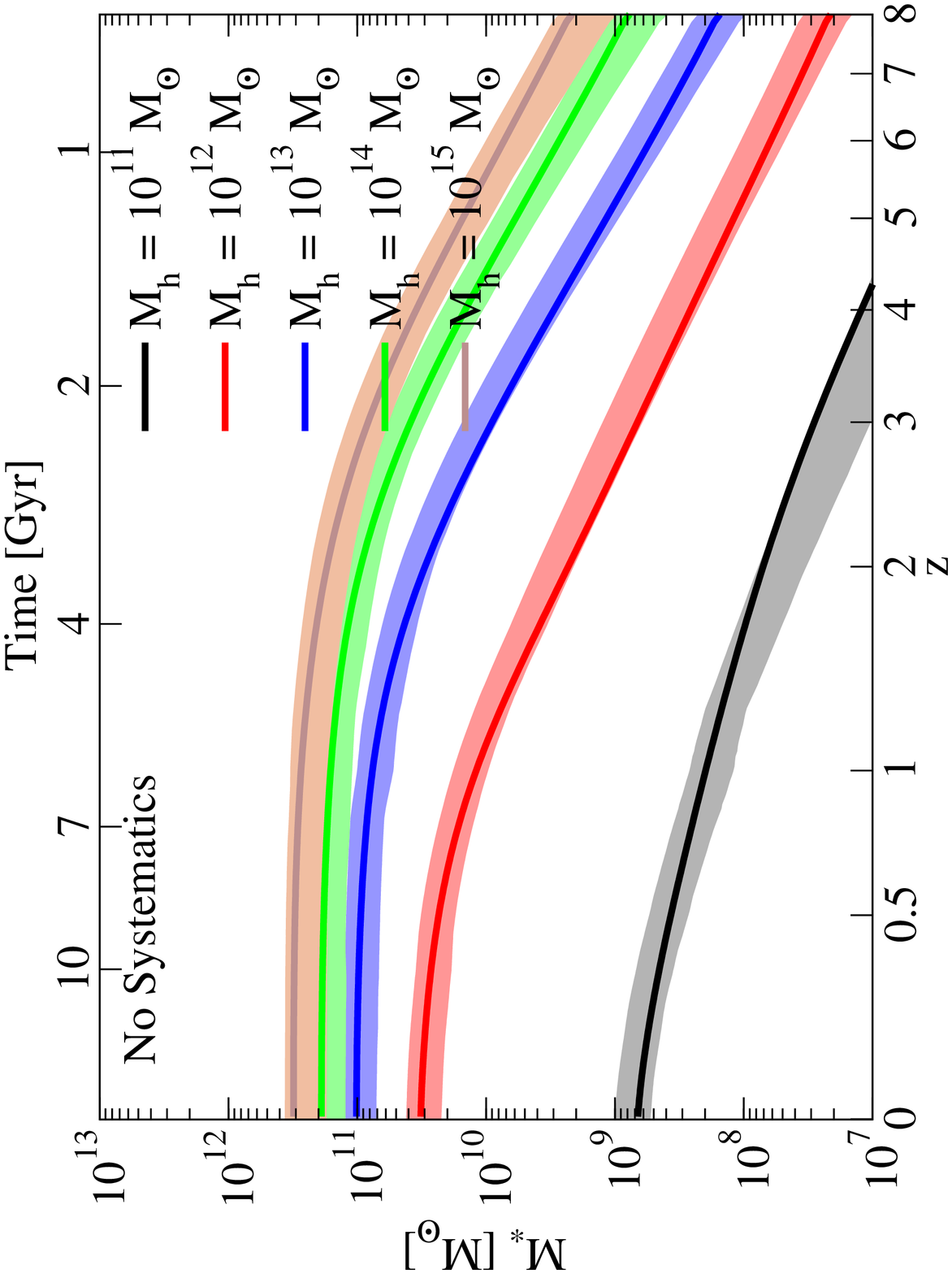}\\
\plotminigrace{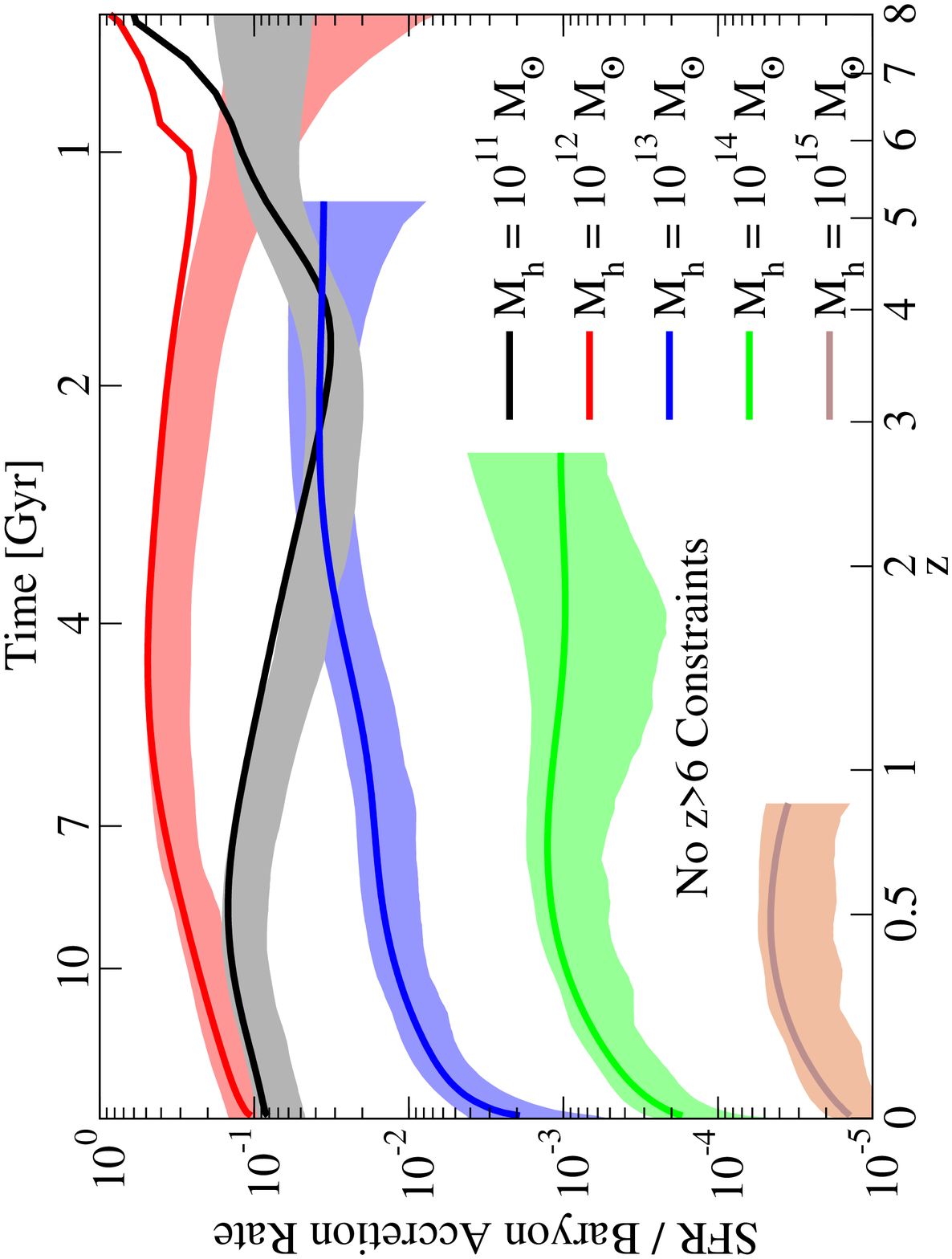}\plotminigrace{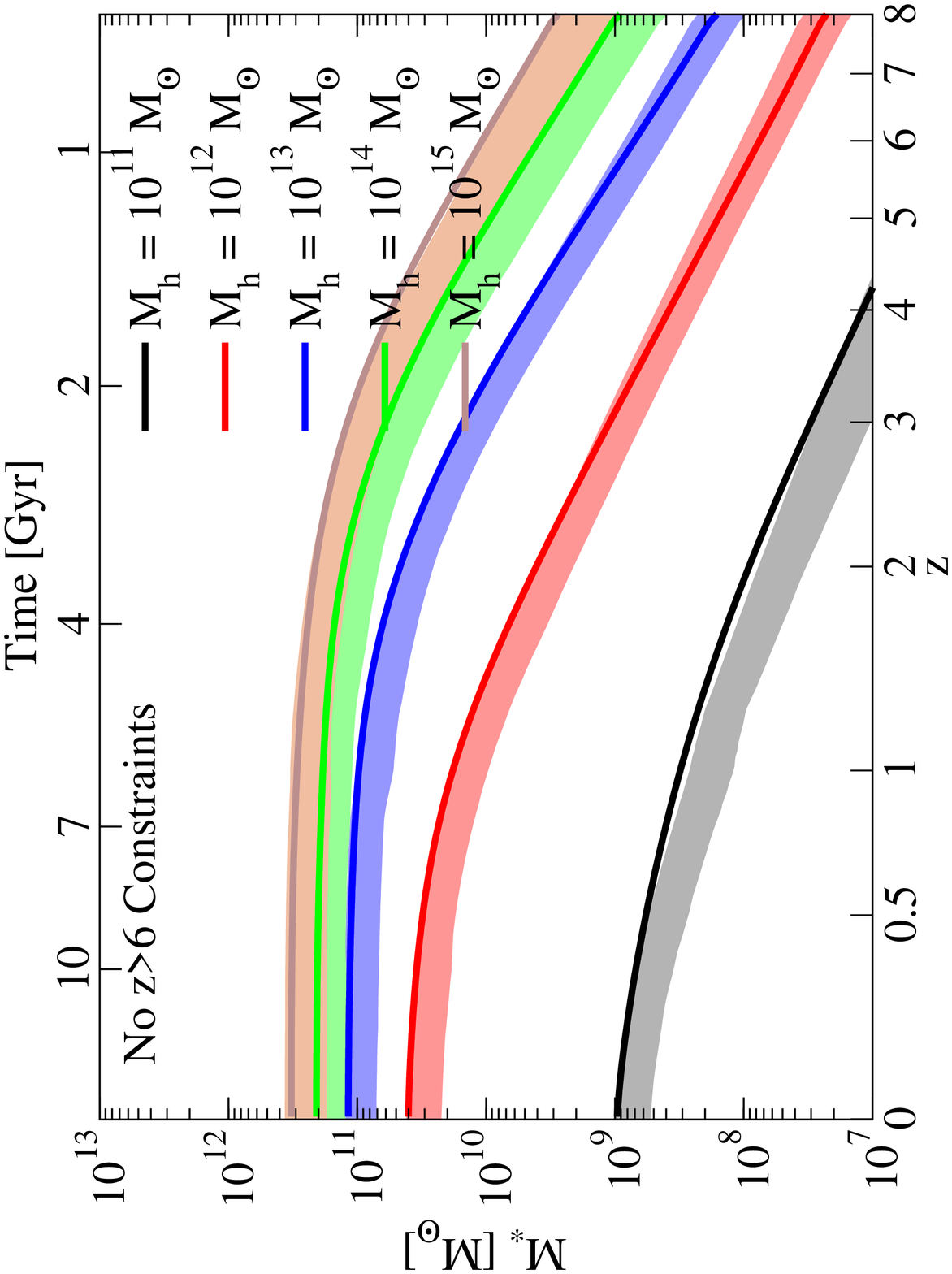}\\
\plotminigrace{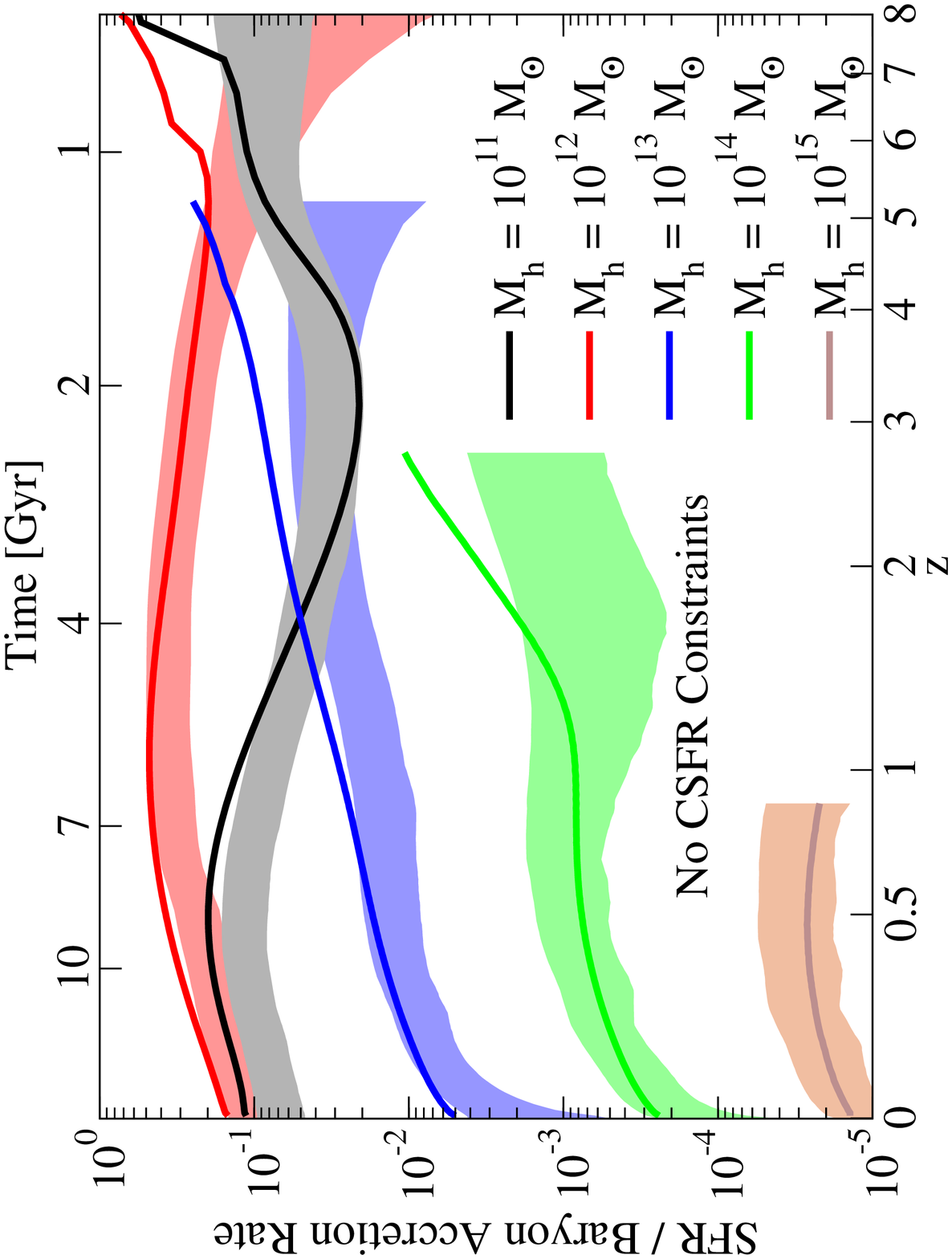}\plotminigrace{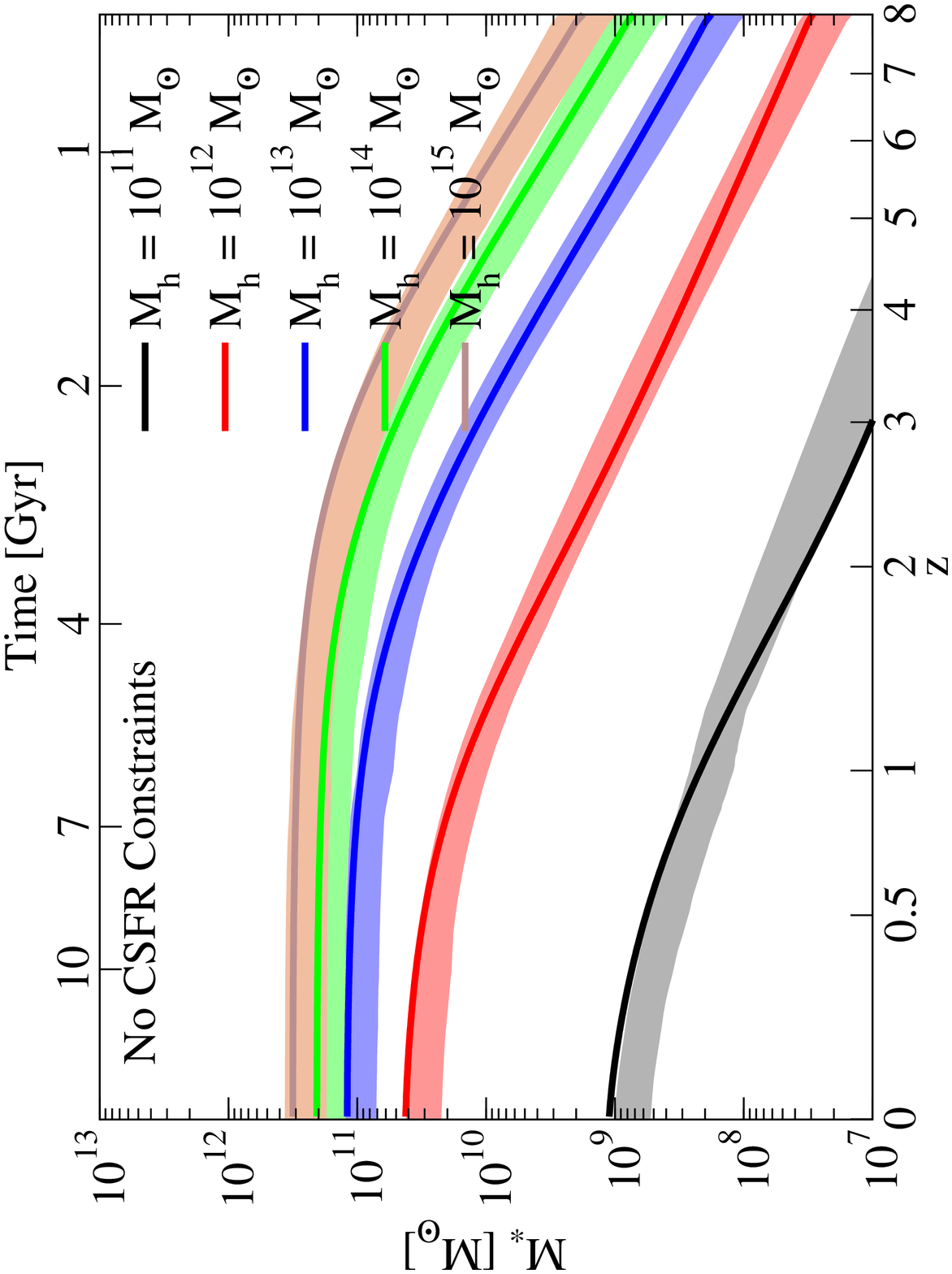}\\
\caption{Derived baryon conversion efficiencies and stellar mass histories for three alternate models (excluding all systematic uncertainties, excluding constraints at $z>6$, and excluding constraints from cosmic star formation rates); the shaded bands show the one-sigma distributions from the main model in this paper, and the solid lines show results from the alternate models.}
\label{f:alt_sfr}
\end{figure}

\begin{figure}
\plotminigrace{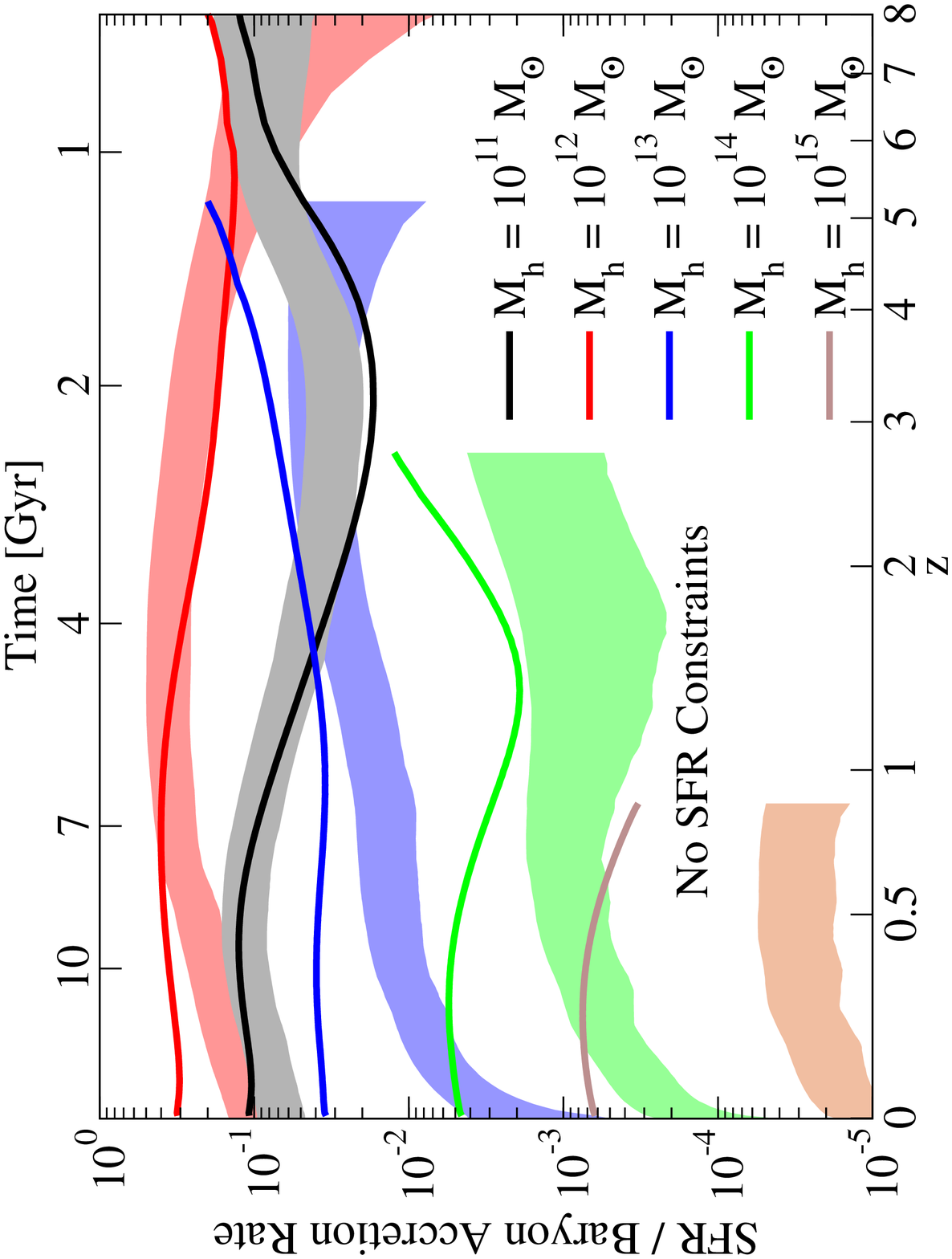}\plotminigrace{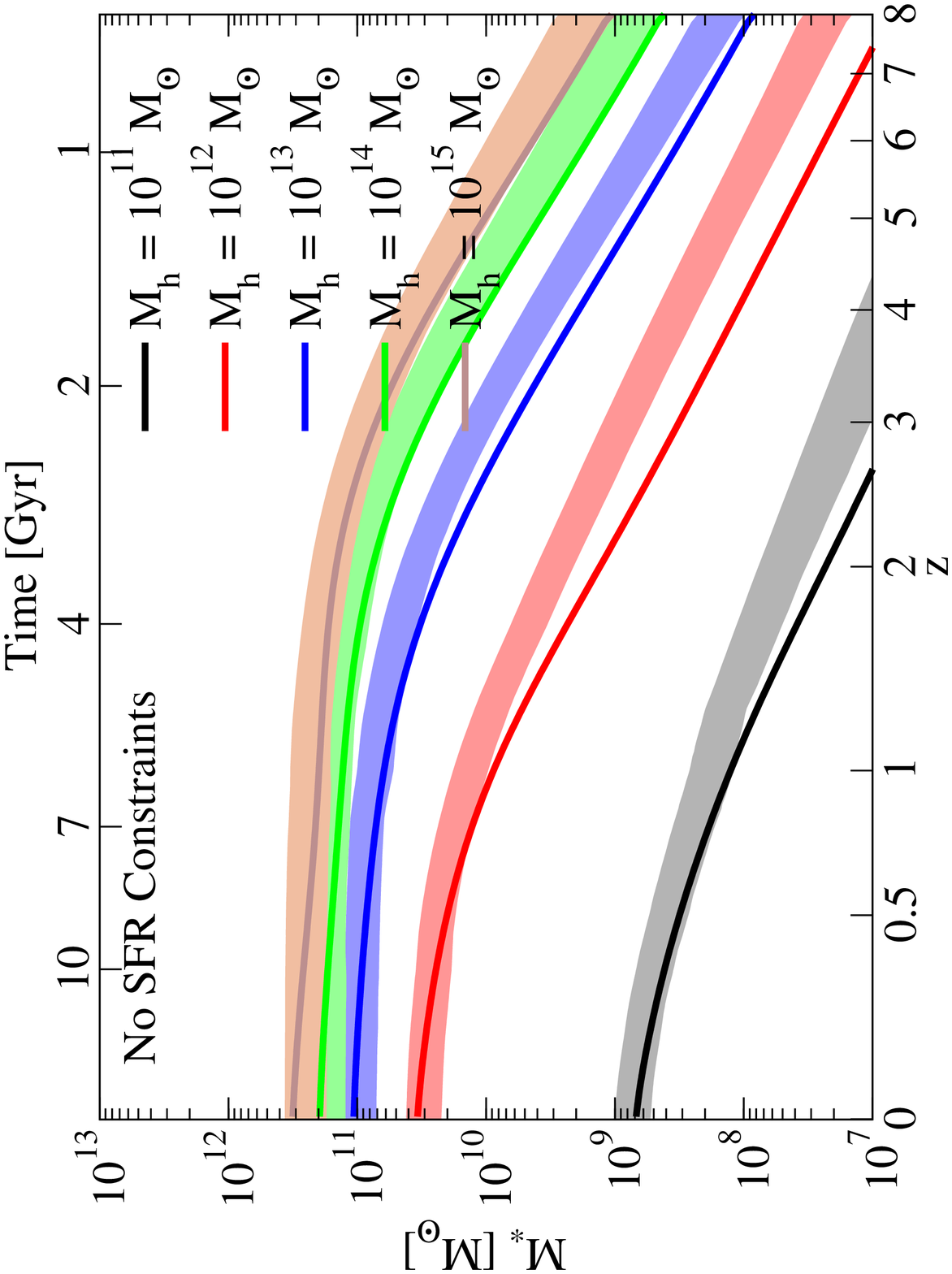}\\
\caption{Derived baryon conversion efficiencies and stellar mass histories for one alternate model (excluding all systematic uncertainties and constraints from star formation rates); the shaded bands show the one-sigma distributions from the main model in this paper, and the solid lines show results from the alternate model.}
\label{f:alt_sfr2}
\end{figure}

We consider four models with alternate sets of assumptions to the main analysis in this paper:
\begin{enumerate}
\item A model (``No Systematics'') that excludes all nuisance parameters related to errors in converting luminosities into stellar masses (i.e., $\mu=\kappa=0$, $\eta_\mathrm{cal} = \infty$) and to incompleteness ($A=0$).
\item A model (``No $z>6$ Constraints'') that excludes all data constraints at high redshift ($z>6$), including data in \cite{Bouwens11,BORG12,Schaerer10,McLure11}, and the data points above $z=6$ in \cite{Bouwens11b}.
\item A model (``No CSFR Constraints'') that excludes all constraints from observed cosmic star formation rates.
\item A model (``No SFR Constraints'') that excludes all constraints from observed star formation rates (both CSFRs and SSFRs) as well as excluding all systematic nuisance parameters as in the first alternate model (``No Systematics'').
\end{enumerate}
Comparisons with observed data are presented in Fig.\ \ref{f:alt_constraints}.  For ease of comparison across mass ranges, we show comparisons between derived baryon conversion efficiencies and stellar mass histories for the main model in this paper and the alternate models  in Figs.\ \ref{f:alt_sfr} and \ref{f:alt_sfr2}.

First, it may be verified that the model including no systematics succeeds remarkably well in matching all three kinds of observations; while it is potentially somewhat high at low redshifts for the CSFR, this is not unexpected on account of issues with the galaxy / ICL definition (\S \ref{s:icl}).  The No Systematics model has much tighter error bars on the stellar mass to halo mass ratios (not shown) than the full model at all redshifts, which suggests that systematics remain the single largest source of errors in determining stellar masses.  Fig.\ \ref{f:alt_sfr} is comparable to Fig.\ \ref{f:sfr_sfh} for the No Systematics model, except that at high redshifts and low halo masses, there is an upturn in the SFR that is not present in the full model.

The model excluding $z>6$ data gives almost identical results as the full model for $z\le6$, which is encouraging.  The ``predictions'' for high-redshift stellar mass functions are somewhat lower than in the best-fitting model, which results in a higher SFR (i.e., rate of stellar mass growth) over the period from $z\sim 7$ to $z\sim 6$. 

The model excluding all cosmic star formation constraints is remarkably similar to the full model in our analysis.  This is as expected: the specific star formation rate in combination with the stellar mass function gives a constraint on the total amount of star formation.  Yet, the CSFR data has tighter error bars than the SSFR in combination with the SMF, at least for $z<1$ (see \S \ref{s:csfr} and \ref{s:ssfr}), so it provides slightly better constraints at those redshifts.

Excluding all star formation rate data results in a good fit to the CSFR at high redshifts, where growth in stellar mass is rapid---i.e., where stellar mass functions at successive redshifts are significantly different even despite large errors.  At lower redshifts, the predicted cosmic star formation rate for the best-fitting model is discrepant from observations.  At these redshifts, the growth in stellar mass is poorly constrained because the intrinsic stellar mass functions are changing comparatively little, especially from $z=0.5$ to $z=0$.  A small difference in the choice of stellar mass function evolution at low redshifts therefore has a much larger impact on the relative error in star formation rates.  Thus, although the stellar mass to halo mass ratios are similar to the No Systematics model (and the abundance matching model in Appendix \ref{a:ab_matching}), the star formation rates derived using this model are unreliable.

\end{document}